\documentclass[prx,showpacs,footnoteinbib,twocolumn,
unsortedaddress]{revtex4-2}
\usepackage{amsmath,amssymb,epic,eepic}
\usepackage[binary-units]{siunitx}
\usepackage[utf8]{inputenc}
\usepackage{graphicx}
\usepackage{hyperref}
\usepackage[capitalise]{cleveref}
\usepackage{todonotes}
\usepackage{bbm}
\usepackage{tikz}
\usepackage{circuitikz}
\usepackage{braket}
\usepackage[caption=false]{subfig}
\usepackage{fontawesome}
\usetikzlibrary{%
        shapes.geometric,calc,intersections,%
        decorations.markings,decorations.text,decorations,%
        patterns,decorations.pathmorphing,fpu,%
                decorations.pathreplacing,%
        positioning,automata,shapes.multipart,circuits,%
        circuits.ee, circuits.ee.IEC, shapes.gates.ee,%
                arrows.meta,arrows,fadings,
}

\newcommand{\pascal}[1]%
{{\todo[inline,backgroundcolor=blue!10!white]{#1}
}}
\newcommand{\me}{\mathrm{e}}
\newcommand{\mi}{\mathrm{i}}
\newcommand{\md}{\mathrm{d}}

\newcommand{\Ge}{\mathcal{G}^\text{(e)}}
\newcommand{\Gh}{\mathcal{G}^{(h)}}
\newcommand{\DGe}{\Delta\mathcal{G}^\text{(e)}}
\newcommand{\Gee}{\mathcal{G}^\text{(2e)}}
\newcommand{\Ghh}{\mathcal{G}^\text{(2h)}}
\newcommand{\Geh}{\mathcal{G}^\text{(eh)}}
\newcommand{\Ghe}{\mathcal{G}^\text{(he)}}
\newcommand{\DGee}{\Delta\mathcal{G}^\text{(2e)}}
\newcommand{\DcGee}{\Delta_c\mathcal{G}^\text{(2e)}}
\newcommand{\DWGee}{\Delta_{\mathrm{W}}\mathcal{G}^\text{(2e)}}
\newcommand{\OGe}{\mathbf{G}^\text{(e)}}
\newcommand{\OGh}{\mathbf{G}^\text{(h)}}
\newcommand{\OGee}{\mathbf{G}^\text{(2e)}}
\newcommand{\OGhh}{\mathbf{G}^\text{(2h)}}
\newcommand{\OGeh}{\mathbf{G}^\text{(eh)}}
\newcommand{\OGhe}{\mathbf{G}^\text{(he)}}
\newcommand{\ODWGee}{\Delta_{\mathrm{W}}\mathbf{G}^\text{(2e)}}
\newcommand{\ODWGhh}{\Delta_{\mathrm{W}}\mathbf{G}^\text{(2h)}}
\newcommand{\OC}{\mathbf{C}}
\newcommand{\OId}{\mathbf{1}}
\newcommand{\We}{W^{(e)}}

\newcommand{\Wee}{W^{(2e)}}

\newcommand{\DWe}{\Delta W^{(e)}}

\newcommand{\DcWee}{\Delta_c W^{(2e)}}
\newcommand{\DWWee}{\Delta_{\text{W}} W^{(2e)}}

\newcommand{\SumOfSl}{\ket{(1_a,1_b)+(1_{a'},1_{b'})}}
\newcommand{\SlDeloc}{\ket{2\times(\text{deloc})}}%

\begin{document}

\title{Accessing electronic entanglement via photo-assisted two-electron
tomography}

\author{M. Paulet$^{1}$}
\thanks{These authors contributed equally to this work.}
\author{G. Rebora$^{1}$}
\thanks{These authors contributed equally to this work.}
\author{M. Bienioszek$^3$}
\author{G. Ménard$^2$}
\author{G. Fève$^2$}
\author{A. Feller$^3$}
\author{P. Degiovanni$^1$}

\today

\affiliation{(1) Univ Lyon, Ens de Lyon, Universit\'e Claude Bernard 
Lyon 1, CNRS, 
Laboratoire de Physique, F-69342 Lyon, France}
\affiliation{(2) Laboratoire de Physique de l’Ecole Normale Sup\'erieure,
ENS, Universit\'e PSL, CNRS, Sorbonne Universit\'e, Universit\'e de Paris,
F-75005 Paris, France}
\affiliation{(3) 
Laboratoire PhLam (UMR 8523), Université de Lille, 2 Av.~Jean Perrin, 59650 Villeneuve d'Ascq, France}

\begin{abstract}

Witnessing entanglement between electrons propagating in quantum
circuits requires accessing their two-electron coherences, which
encode the two-particle quantum correlations generated by Coulomb
interactions. Yet, experimental reconstruction of two-electron
coherence remains a major challenge.
We introduce photo-assisted two-electron tomography,
a quantum coherence reconstruction protocol from dc-currents and time-averaged
current correlations, routinely measured in electron quantum
optics experiments. Driving an energy filter with a small rf tone encodes
off-diagonal coherences in the frequency domain into the dc-current response, enabling
single-electron tomography without resorting to current-noise
measurements. Extending it to two channels, current
correlations after two photo-assisted filters reconstruct the
two-electron Wigner distribution 
and isolate its irreducible two-body part.
When applied to Coulomb-coupled quantum
Hall edge channels, the protocol reveals interaction-induced
energy entanglement. Signatures of this entanglement are visible through the 
violation of an entanglement witness based on Cauchy-Schwarz inequalities,
as well as Wigner function negativities, which remain resolved above the
thermal background for realistic device parameters.
A generalized Franson geometry further extends
the approach to intra-channel two-electron coherences. This opens a
direct pathway to detect
and characterize electronic entanglement in ballistic quantum circuits.%

\end{abstract}

\maketitle

\section*{Introduction}
\label{sec/introduction}

Entangling two propagating quantum excitations in a controlled way is a
cornerstone of on-the-fly quantum information processing. While photons are the
canonical platform for this purpose \cite{Slussarenko-2019-1}, electrons propagating in ballistic
solid-state conductors have emerged as a compelling alternative
\cite{Bertoni-2000-1,Ionicioiu-2001-1,Bertoni-2007-1}. This
has given rise to a field now known as electron quantum optics,
combining single-particle coherence with the possibility of on-chip
integration. Recent years have
seen remarkable progress in the generation, manipulation, and
characterization of single-electron excitations propagating in chiral
ballistic conductors. Key milestones
include single-electron tomography {\it via} Hong-Ou-Mandel (HOM)
interferometry \cite{Bisognin-2019-2} and {\it via} time-dependent filtering
\cite{Fletcher-2019-1} in AlGaAs/GaAs heterostructures. Complementary
advances in decoherence control, achieved through sample design
\cite{Duprez-2019-1} or through the use of graphene as a low-decoherence
material \cite{Wei-2017-1,Jo-2022-1,Deprez-2021-2}, strongly suggest
that electronic degrees of freedom can serve as reliable carriers of
quantum information, a paradigm known as electronic flying qubits
\cite{Edlbauer-2022-1,Bertoni-2000-1,Ionicioiu-2001-1}. A
landmark step in this direction was recently achieved in
Ref.~\cite{Assouline-2023-1} with the coherent
manipulation of a single electronic flying qubit carried by a Leviton
\cite{Dubois-2013-1} propagating along quantum Hall edge channels of a
high-mobility graphene monolayer, exploiting the valley degree of
freedom to encode quantum information.

Turning the electronic flying qubit paradigm into a practical reality
requires the demonstration of reliable, controlled entanglement between two
such qubits \cite{Edlbauer-2022-1}. Capacitive coupling between two
copropagating chiral electronic channels provides a natural mechanism
for implementing a controlled-phase gate on railroad qubits
\cite{Bertoni-2000-1}. This calls for new methods to detect and
quantify Coulomb-interaction-induced entanglement between two flying
electronic excitations in quantum conductors.

In quantum photonics, entanglement witnesses built from Glauber
coherences \cite{Glauber-1963-1} have been developed \cite{Wolk-2014-1}
and successfully applied to demonstrate entanglement generation by a
dc-biased Josephson junction \cite{Peugeot-2021-1}. This
entanglement witness has even been adapted to detect spin entanglement between top
quarks produced in $\SI{13}{\tera\electronvolt}$ $pp$ collisions at the
ATLAS detector \cite{Atlas-2024-1}. 

For electrons, current-correlation entanglement witnesses were pioneered
by Burkard and Loss \cite{Burkard-2003-1} who bounded the spin entanglement of two
electrons from the shot noise at a beam splitter. This idea was extended by
Giovannetti {\it et al.} \cite{Giovannetti-2006-2} to orbital and spin modes, where this shot
noise both witnesses the entanglement of a two-electron input state and
lower-bounds its entanglement of formation. These approaches provide a
single-measurement test of entanglement carried 
by discrete degrees of freedom such as spin or orbital labels.
Studying entanglement between fermionic modes \cite{Benatti-2020-1} 
generated within a
quantum circuit, in particular by Coulomb interactions, instead calls
for reconstructing the full inter-channel two-electron coherence
\cite{Thibierge-2016-1} in the
appropriate time-frequency mode basis \cite{Roussel-2020-1}. 
This is all the more necessary as an entanglement witness makes no
assumption about the origin of the entanglement it detects: its violation can stem from
single-particle delocalization through Pauli exchange as well as from
genuine two-body correlations, and only the anatomy of the reconstructed
two-electron coherence can tell them apart
(\cref{sec/g2-and-entanglement/G2-anatomy}).

The purpose of this paper is to introduce a new electronic tomography
protocol that grants access to single- and two-electron coherences
through routine transport measurements. This protocol consists in
a direct interferometric scheme that reconstructs off-diagonal
coherences in the frequency domain via photo-assisted transitions: a
small rf-drive induces virtual emission or absorption of single photons,
coherently mixing states at different energies so that their off-diagonal
coherence becomes encoded into a measurable dc-signal. 
Although photo-assisted tomography has recently been proposed and
demonstrated at the single-electron level in attosecond physics
\cite{Laurell-2022-1,Laurell-2025-1,Laurell-2025-2}, transposing this
principle to the chiral ballistic conductors of electron quantum optics
brings specific novelties.
First, unlike HOM tomography \cite{Bisognin-2019-2}---that
encodes single-electron coherences in low-frequency current noise
\cite{Roussel-2016-2}---our protocol does not require time-consuming
high-sensitivity measurements. It thus holds the promise of substantial
practical gains for single-electron tomography within electrical 
circuits. Second, our protocol extends naturally to two-electron
tomography, where time-averaged current correlations following
two photo-assisted filters reconstruct the full two-electron Wigner
distribution and isolate its irreducible two-body part, disentangling
genuine interaction-induced correlations 
from those carried by the single-particle coherence through 
Pauli exchange and interaction dressing.

This direction has not yet been explored in the attosecond physics
context. From a quantum signal processing perspective
\cite{Roussel-2016-1}, this constitutes a generalization of Franson
interferometry \cite{Franson-1989-1} to the energy domain in the spirit
of Ref.~\cite{Thibierge-2016-1}. Besides, electronic decoherence within the photo-assisted filters
only provides a minor correction (\cref{sec/protocols/decoherence}). Both protocols are built on average
currents and current correlations \cite{Parmentier-2010-1}, that are
routinely measured quantities in electronic interferometry
\cite{Ji-2003-1,Roulleau-2008-2} as well as in HBT and HOM experiments performed
with electrons \cite{Bocquillon-2012-1,Bocquillon-2013-1} or anyons
\cite{Bartolomei-2020-1}. This ensures that the required measurements are
within reach of existing setups.

Beyond the specific setups considered here, the ability to
reconstruct and decompose two-electron coherence may be of interest for
the broader effort to generate and certify entanglement in mesoscopic
conductors, for which it provides a source-agnostic characterization
tool. We return to these perspectives in \cref{sec/conclusion}.

This paper is structured as follows. \Cref{sec/g2-and-entanglement}
introduces two-electron coherences, the Cauchy-Schwarz entanglement
witness used throughout this work, and the anatomy of two-electron
coherence that motivates the rest of the paper. \Cref{sec/protocols}
presents the photo-assisted tomography protocol, from single-electron
tomography via dc-current measurements to the reconstruction of the
irreducible two-body part of two-electron coherence from current
correlations. \Cref{sec/2e-collision} applies it to Coulomb-coupled
chiral channels, and \cref{sec/Franson} extends it to intra-channel
two-electron coherences via a generalized Franson interferometer, which
also clarifies the connection between HOM experiments and two-electron
coherences.

\section{Two-electron coherence and electronic entanglement}
\label{sec/g2-and-entanglement}

\subsection{Entanglement witness from electronic coherences}
\label{sec/g2-and-entanglement/witness}

Entanglement in many-body systems can be probed through correlation
functions of suitably chosen operators. 
A useful example follows from a general Cauchy-Schwarz argument
\cite{Wolk-2014-1}. For two quantum systems $A$ and $B$ in a separable
state, with operators $(A_1, A_2)$ acting on $A$ and $(B_1, B_2)$
acting on $B$, one obtains
\begin{align}
	\begin{split}
		\big\lvert \langle A_1 A_2& B_1 B_2 \rangle_{\text{SEP}} \big\rvert^2\\
		&\leq \langle A_1 A^\dagger_1 B_2^\dagger B_2 \rangle_{\text{SEP}}
		\langle A^\dagger_2 A_2 B_1 B^\dagger_1 \rangle_{\text{SEP}}.
	\end{split}
	\label{eq/gen-crit}
\end{align}
A violation of \cref{eq/gen-crit} unambiguously signals entanglement 
between $A$ and $B$. As shown in Ref.~\cite{Wolk-2014-1}, it even signals 
a violation of positive partial transposition (PPT), which is a weaker property
than separability. By contrast, inequality \eqref{eq/gen-crit} being satisfied does
not mean that the state is not entangled. It is a non-linear entanglement
witness \cite{Guhne-2006-1,Terhal-2000-1} which has been used, for instance, to 
demonstrate photonic entanglement generated in the microwave domain by a dc-biased
Josephson junction \cite{Peugeot-2021-1}.

In the context of electron quantum optics, this witness can probe two-particle
entanglement by specifying $A_1=\psi_A^\dagger(1')$, $A_2=\psi_A(1)$, $B_1=
\psi_B^\dagger(2')$ and $B_2=\psi_B(2)$. \cref{eq/gen-crit} can then be expressed
in terms of two-electron electronic coherences as
\begin{align}
	\begin{split}
		\big\lvert \Gee_{AB}(1,2|&1',2') \big\rvert^2\\
		&\leq \Gee_{AB}(1,2'|1,2')\,\Gee_{AB}(1',2|1',2).
	\end{split}
	\label{eq/gen-crit/G2}
\end{align}
in which the inter-channel two electron coherence $\Gee_{AB}$ associated
to an electronic fluid described by the density matrix $\rho$ is defined by
\begin{align}
	\begin{split}
		\Gee_{AB}(1,2|&1',2')\\
		&= \mathrm{Tr}\left(\psi_B(2)\psi_A(1)\,\rho\,\psi_A^\dagger(1')
		\psi_B^\dagger(2')\right)\,.
	\end{split}
	\label{eq/G2/def}
\end{align}
Here $(1,2\vert1'2')$ can be times, frequencies or normalized single-particle
modes depending on the situation. \cref{eq/G2/def} is an instance of the 
second-order coherence function defined in \cite{Moskalets-2014-1,Thibierge-2016-1}, 
here specified for two channels $A$ and $B$. Appendix \ref{appendix/G2} recalls the basic
properties of two-electron coherences and their various representations while
Appendix \ref{appendix/G1} does the same for single-electron coherences.
Having an entanglement witness completely expressed in terms of two-electron coherences,
let us discuss archetypical examples in the light of typical experimental setups.

\subsection{Examples and setups}
\label{sec/g2-and-entanglement/examples-setups}

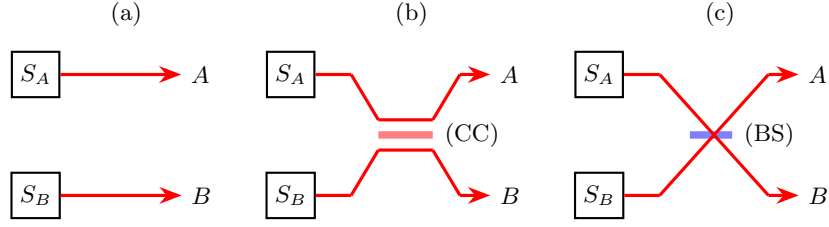
\begin{figure*}
	\begin{tikzpicture}[
        x=0.8cm,y=0.8cm,
        >=Stealth,
        thick,
        channel/.style={-Stealth, very thick},
        src/.style={draw, minimum width=0.6cm, minimum height=0.6cm},
        ]
        
        \node at (0.5, 2.) {(a)};

        \node[src] (S1)  at (-1., 1.) {$S_A$};
        \node[src] (S2) at (-1.,-1.) {$S_B$};

        \draw[channel,red] (S1.east)  -- ++(2,0) node[right,black] {$A$};
        \draw[channel,red] (S2.east)  -- ++(2,0) node[right,black] {$B$};
        
\end{tikzpicture}%
\hspace{5mm}
	\begin{tikzpicture}[
        x=0.8cm,y=0.8cm,
        >=Stealth,
        thick,
        channel/.style={-Stealth, very thick},
        src/.style={draw, minimum width=0.6cm, minimum height=0.6cm},
        coulomb/.style={draw, fill=black!80, minimum width=0.5cm, minimum height=0.07cm},
        ]

        \node at (0, 2.) {(b)};
        
        \node[src] (S1)  at (-2., 1.) {$S_A$};
        \node[src] (S2) at (-2.,-1.) {$S_B$};
        
        \draw[channel,-,red, line width=2.8pt, color=red!50]
			(-0.55,0) -- (0.35,0) node[right,black] {$(\text{CC})$};
        
        \draw[channel,-,red] (S1.east)  -- (-1., 1.);
        \draw[channel,-,red] (S2.east) -- (-1.,-1.);
        \draw[channel,-,red] (-1., 1.) -- (-0.55, 0.25);
        \draw[channel,-,red] (-1.,-1.) -- (-0.55,-0.25);

		\draw[channel,-,red] (-0.55, 0.25) -- (0.35, 0.25);
        \draw[channel,-,red] (-0.55,-0.25) -- (0.35,-0.25);
        
        \draw[channel,-,red] (0.35, 0.25) -- (0.8, 1.);
        \draw[channel,-,red] (0.35,-0.25) -- (0.8,-1.);
		\draw[channel,red] (0.8, 1.)  -- ++(0.5,0) node[right,black] {$A$};
        \draw[channel,red] (0.8,-1.)  -- ++(0.5,0)node[right,black] {$B$};
        
\end{tikzpicture}%
\hspace{5mm}
	\begin{tikzpicture}[
        x=0.8cm,y=0.8cm,
        >=Stealth,
        thick,
        channel/.style={-Stealth, very thick},
        src/.style={draw, minimum width=0.6cm, minimum height=0.6cm},
        qpc/.style={draw, fill=black!80, minimum width=0.07cm, minimum height=0.5cm},
        ]

        \node at (0, 2.) {(c)};
        
        \node[src] (S1)  at (-2., 1.) {$S_A$};
        \node[src] (S2) at (-2.,-1.) {$S_B$};
        
        \draw[channel,-,blue, line width=2.8pt, color=blue!50]
			(-0.5,0) -- (0.2,0) node[right,black] {$(\text{BS})$};
        
        \draw[channel,-,red] (S1.east)  -- (-1., 1.);
        \draw[channel,-,red] (S2.east) -- (-1.,-1.);
        \draw[channel,-,red] (-1., 1.) -- (0.8,-1);
        \draw[channel,-,red] (-1.,-1.) -- (0.8,1.);

		\draw[channel,red] (0.8, 1.)  -- ++(0.5,0) node[right,black] {$A$};
        \draw[channel,red] (0.8,-1.)  -- ++(0.5,0)node[right,black] {$B$};
        
\end{tikzpicture}%

	\caption{\label{fig/geometries} Three possible two-channel systems:
	(a) the two channels $A$ and $B$ are fed by independent sources
	$S_A$ and $S_B$; (b) the two channels $A$ and $B$, fed by the
	independent sources $S_A$ and $S_B$, are capacitively coupled via a
	finite length interaction region called the Coulomb coupler
	$(\text{CC})$; (c) HOM geometry: the two channels are fed by 
	independent sources but connected to an ideal electronic 
	beam-splitter $(\text{BS})$.}
\end{figure*}

We start with the example of two independent electronic sources
feeding two uncoupled electronic channels, as depicted in
\cref{fig/geometries}-(a). In this situation, one doesn't expect
\cref{eq/gen-crit/G2} to be violated, since a factorized many-body
electronic state $\rho_{AB} = \rho_A \wedge \rho_B$ is emitted.
To check it explicitly, let us start from the inter-channel
electronic coherences which, in the present case, satisfy:
\begin{subequations}
\label{eq/factorized-coh}
	\begin{align}
		\label{eq/factorized-coh/G1}
	\Ge_{AB}(1|1')&=0\\
		\label{eq/factorized-coh/G2}
		\Gee_{AB}(1,2|1',2')&=\Ge_A(1|1')\,\Ge_B(2|2').
	\end{align}
\end{subequations}
These reflect that the sources $S_A$ and $S_B$ are independent: 
\cref{eq/factorized-coh/G1} means that no single-electron excitation is delocalized between the
two channels and \cref{eq/factorized-coh/G2} is independent emission in
the single particle sector of each channel.
Using \cref{eq/factorized-coh/G2}, it is possible to rewrite both sides of
\cref{eq/gen-crit/G2} in terms of intra-channel single-electron coherences
$\Ge_A$ and $\Ge_B$. The two Cauchy-Schwarz inequalities for intra-channel single-electron coherence
\cite{Ferraro-2013-1}
\begin{equation}
	\left|\Ge_\alpha(a|a')\right|^2\leq \Ge_\alpha(a|a)\,\Ge_\alpha(a'|a')
\end{equation}
for $(\alpha,a,a')=(A,1,1')$ and $(B,2,2')$ (see also
\cref{appendix/positivity/G1}) 
then directly imply that \cref{eq/gen-crit/G2} is satisfied.

Violating \cref{eq/gen-crit/G2} requires a non-trivial state, such as
the superposition of two orthogonal Slater determinants with one excitation
localized on $A$ and one excitation localized on $B$:
\begin{align}
	\begin{split}
		&\SumOfSl\\
		&\qquad=\left(\alpha\,\psi_B^\dagger[\varphi_b]\psi_A^\dagger[\varphi_a]\right.
		+\left. \beta\, \psi_B^\dagger[\varphi_{b'}]\psi_A^\dagger[\varphi_{a'}]
		\right)\ket{\emptyset}.
	\end{split}
\end{align}
In this expression, the normalized single-particle wavepackets $\varphi_a$ and
$\varphi_{a'}$, as well as $\varphi_b$ and $\varphi_{b'}$ are mutually
orthogonal with $|\alpha|^2+|\beta|^2=1$. The state $\SumOfSl$
is a pure two-particle entangled state between $A$ and $B$. It
violates \cref{eq/gen-crit/G2} since, for $(1, 2\vert1', 2') = (\varphi_a,
\varphi_b\vert \varphi_{a'}, \varphi_{b'})$, the l.h.s. of \cref{eq/gen-crit/G2} equals $\vert \alpha
\beta^*\vert^2$ while the r.h.s. vanishes.
Note that, taking $\alpha=\beta=1/\sqrt{2}$, one obtains a Bell state for the railroad qubits
\cite{Ionicioiu-2001-1} built from modes $(a,b)$ and $(a',b')$.

Such a state $\SumOfSl$ cannot be generated solely by mixing two
independent single electron sources at a beam splitter
(\cref{fig/geometries}-(c)). It could, on the other hand, be
generated by interactions, as in the capacitively
coupled setup of \cref{fig/geometries}-(b), where 
a Coulomb coupler, a
non-linear electron quantum optics component, is used to capacitively
couple electrons.

However, the $\Gee_{AB}$-based entanglement witness presented in
\cref{eq/gen-crit/G2} can also detect entangled states arising from a
linear electron quantum optics setup. This is the case of
\cref{fig/geometries}-(c) when one assumes that each source
$S_A$ and $S_B$ emits one normalized electronic wavepacket, the two
being mutually orthogonal. The resulting state is a single
two-particle Slater determinant:
\begin{equation}
	\SlDeloc=d_1^\dagger\,d_2^\dagger\ket{\emptyset}.
\end{equation}
This state involves two excitations delocalized over the two channels:
\begin{subequations}
	\begin{align}
		d_1^\dagger &=\sqrt{T}\,\psi_A^\dagger[\varphi_1]-\mi\sqrt{R}\,\psi_B^\dagger[\varphi_1]\\
		d_2^\dagger &=\sqrt{T}\,\psi_B^\dagger[\varphi_2]-\mi\sqrt{R}\,\psi_A^\dagger[\varphi_2].
	\end{align}
\end{subequations}
Here, $R$ and $T$ denote the reflection and transmission probabilities
for the ideal electronic beam splitter, coming from amplitudes
$\sqrt{T}$ and $\mi\sqrt{R}$.

For $(1,2\vert1',2')=(\varphi_1,\varphi_2\vert\varphi_2,\varphi_1)$, \cref{eq/gen-crit/G2}'s
l.h.s. is equal to $(RT)^2$ while its r.h.s. vanishes because each
pair of $(A,B)$ modes is populated by a single delocalized electron. The
witness thus detects entanglement in the state $\SlDeloc$,
but this state is completely different from $\SumOfSl$:
it can be generated by a linear quantum optics
setup such as \cref{fig/geometries}-(c). 
By contrast, the state $\SumOfSl$ requires a non-linear
operation: either the electrons 
interact as in the Coulomb coupler setup of
\cref{fig/geometries}-(b) or one performs a measurement 
such as post-selecting
the $(1,1)$ sector of the state $\SlDeloc$. 
As expected, the entanglement witness
is agnostic to the genesis of the states on
which it is tested: both $\SlDeloc$ and $\SumOfSl$, which belong to
different realms of electron quantum optics, violate
\cref{eq/gen-crit/G2}.

These two examples show that the witness detects entanglement in states
of very different nature. Telling them apart calls for a closer look at
the anatomy of two-electron coherence, in order to separate its
contributions inherited from single-particle physics from the part
inherent to two-particle physics.

\subsection{Anatomy of the two-electron coherence}
\label{sec/g2-and-entanglement/G2-anatomy}

Wick's theorem applies to Gaussian states which naturally arise from the unitary
evolution of equilibrium reservoirs under linear (single particle) elements as,
for example, in the setup depicted on \cref{fig/geometries}-(c) assuming the two
sources $S_A$ and $S_B$ are time dependent single-particle scatterers.
In this case, the two-electron coherence is
expressed using the single-electron coherence
\cite{Moskalets-2014-1}:
\begin{align}
	\begin{split}
		\Gee_{\text{W}}&(1,2|1',2')\\
		&=\Ge(1|1')\,\Ge(2|2') - \Ge(1|2')\,\Ge(2|1').
	\end{split}
	\label{eq/basics/Wick/1}
\end{align}
The first term in the r.h.s. represents direct contributions whereas the
second term reflects quantum exchange.

In full generality, at the level of the second-order coherence, we can extract Wick's
contributions and define the Wick deviation of two-electron coherence by
\begin{equation}
	\Delta_{\text{W}}\Gee(1,2|1',2')=\left(\mathcal{G}^{(2e)}
	-\mathcal{G}^{(2e)}_{\text{W}}\right)(1,2|1',2').
	\label{eq/basics/Wick/2}
\end{equation}
It represents the deviation to the two-electron coherence that would
arise solely from the contribution of fermionic single-particle physics.
It is important to note that 
generically,
interactions contribute to $\Ge$ \cite{Ferraro-2014-1}. Consequently,
the decomposition of $\Gee$ as a sum of $\DWGee$ and the r.h.s. of
\cref{eq/basics/Wick/1} is not a decomposition into a non-interacting and
an interaction-induced part of $\Gee$. It should really be interpreted as
a decomposition into a term that can be attributed to single-electron coherence
(including its contributions from interactions) and an irreducible
two-particle contribution.
Notably, this single-particle contribution is not a mere background: even a
single injected electron yields a nonvanishing $\Gee$ as soon as its
single-electron coherence is mixed, as happens at finite
temperature~\cite{Moskalets-2018-3}. A nonvanishing $\Gee$ therefore does not by
itself signal genuine two-particle physics, which further motivates isolating
its irreducible part $\DWGee$.

Going back to the examples considered in \cref{sec/g2-and-entanglement/examples-setups},
the difference
between the state $\SlDeloc$ and the quantum
superposition $\SumOfSl$ of two Slater determinants now appears
clearly and can be expressed in terms that contain no reference to
their
specific genesis: for
the former
the violation of \cref{eq/gen-crit/G2} entirely
originates from the exchange contribution $\Ge_{AB}(1|2')\,\Ge_{BA}(2|1')$,
while $\DWGee_{AB}=0$. This is the only way for a Gaussian state to
violate the witness. By contrast, the state $\SumOfSl$ has no
inter-channel single-electron coherence ($\Ge_{AB}=0$) and the violation
of \cref{eq/gen-crit/G2} only comes from $\DWGee_{AB}$.
It is an irreducible two-particle effect which arises from the deviation
from Gaussianity associated with $\DWGee_{AB}\neq 0$. 

This distinction is directly reflected in the quantity reconstructed by
our tomography protocol. As will be shown in \cref{sec/protocols}, the
protocol reconstructs the Wick deviation $\DWGee_{AB}$, although current
correlations give access to the connected two-electron coherence
$\DcGee_{AB}$, obtained by subtracting from $\Gee_{AB}$ only the direct
term $\Ge_A\Ge_B$. The two differ by the exchange term:
\begin{subequations}
	\begin{align}
		\Delta_c&\Gee_{AB}(1,2|1',2')\\
		&= \Gee_{AB}(1,2|1',2') - \Ge_{A}(1|1')\Ge_{B}(2|2')
		\label{eq/basics/DWGee/1}\\
		&=\DWGee_{AB}(1,2|1',2')-\Ge_{AB}(1|2')\Ge_{BA}(2|1').
		\label{eq/basics/DWGee/2}
	\end{align}
	\label{eq/basics/DWGee}
\end{subequations}
A direct consequence is that, in any setup where $\Ge_{AB}=0$ (like in
\cref{fig/geometries}-(b) for example), the connected coherence coincides
with Wick's deviation. The tomography protocol will then give direct access
to the irreducible two-particle contribution $\DWGee_{AB}$. 
In the case where $\Ge_{AB}\neq 0$ (like in \cref{fig/geometries}-(c) for
example), reconstructing $\DWGee=\DcGee_{AB}+\Ge_{AB}\,\Ge_{BA}$ requires
not only accessing current correlations but also reconstructing $\Ge_{AB}$. 

Wrapping up our discussion of the relation between two-electron coherences and
electronic entanglement, the take-home message is that 
the witness certifies entanglement between the modes of channel
$A$ and the modes of channel $B$ (inter-channel entanglement)
\cite{Benatti-2020-1}, irrespective of its genesis: it responds to 
the full two-electron coherence
$\mathcal{G}^{(2e)}_{AB}$.
Both single-particle delocalization, through the exchange term 
$\mathcal{G}^{(e)}_{AB}\mathcal{G}^{(e)}_{BA}$,
and the irreducible two-body coherence $\DWGee_{AB}$ can be signaled by the
entanglement witness.
The two examples considered in \cref{sec/g2-and-entanglement/examples-setups} 
are thus of different nature yet both detected: it is the
anatomy of $\Gee_{AB}$ discussed in
\cref{sec/g2-and-entanglement/G2-anatomy}, not 
the witness, that tells which mechanism is at play. 

This sets the roadmap for the rest of this work. Since
\cref{eq/gen-crit/G2} makes no assumption about the distinction between single- and
intrinsic two-particle contributions, reconstructing the Wick deviation
$\DWGee_{AB}$ is crucial to fully understand what
precisely contributes to entanglement detection by the witness given in
\cref{eq/gen-crit/G2}. Section \ref{sec/protocols} thus introduces the
photo-assisted tomography protocol that renders single- and two-electron
coherences accessible from dc-current and current correlations, thereby allowing
to reconstruct $\DWGee_{AB}$. Section \ref{sec/2e-collision} applies it to
a minimal model of two capacitively coupled channels (\cref{fig/geometries}-(b)).
In this case, $\Ge_{AB}=0$ directly implies $\DWGee_{AB}=\DcGee_{AB}$
and provides
the cleanest signature of Coulomb-induced irreducible two-particle
entanglement. Section \ref{sec/Franson} then extends the tomography
protocol to
intra-channel two-electron coherences through a generalized Franson
interferometer.%

\section{Photo-assisted tomography protocols}
\label{sec/protocols}

In this section, we present a new protocol for reconstructing the full
structure of inter-channel two-electron coherence $\Gee_{AB}$ and,
following \cref{sec/g2-and-entanglement/G2-anatomy}, to isolate its two-particle irreducible part
$\DWGee_{AB}$ from single-particle contributions. 

\Cref{sec/protocols/strategy} presents the general strategy for two-electron tomography,
highlighting the nested two-stage scheme essential for a full
reconstruction procedure that enables isolating the two-particle
irreducible part. We also discuss the positivity constraints that the
electronic first and second order coherences have to verify and how we
reduce them to a practical set of necessary constraints.

The physical idea of photo-assisted filtering that effectively enables us
to perform tomographic reconstruction of two-electron coherence is
explained in
\cref{sec/protocols/principles}. 
Then, \cref{sec/protocols/G1} and \cref{sec/protocols/G2} present the
details of the protocol based on standard measurement techniques
(average dc current and low frequency current noise correlations) as
well as the details of the reconstruction procedure.  
Finally,
\cref{sec/protocols/decoherence} discusses how to 
mitigate electronic decoherence
effects within the photo-assisted filter.

\subsection{Reconstruction strategy and physical constraints}
\label{sec/protocols/strategy}

\subsubsection{A two-stage nested strategy}
\label{sec/protocols/strategy/two-stage}

The first stage is a single electron tomography: it reconstructs the full
single-electron coherence, both intra-channel $\Ge_A$ and $\Ge_B$ and,
when relevant, the inter-channel component $\Ge_{AB}$. It provides the
Wick contribution $\Gee_{AB,W}=\Ge_A\,\Ge_B-\Ge_{AB}\,\Ge_{BA}$ to 
$\Gee_{AB}$. 

The second stage is the two-electron tomography protocol {\it per se}:
it reconstructs the Wick deviation of two-electron coherence
$\DWGee_{AB}$ from current correlation measurements. For $T$-periodic
systems, $\DWGee_{AB}$ is expected to be the regular, irreducible
two-particle part of two-electron coherence, the singular part being the
Wick contribution determined from the first stage
(see \cref{appendix/G2/T-periodic}).
Combining the two stages yields the complete structure
for tomographic reconstruction up to two-electron coherence:
\begin{equation}
	\Gee_{AB}=\Ge_A\,\Ge_B-\Ge_{AB}\,\Ge_{BA}+\DWGee_{AB}
\end{equation}
in which every term in the r.h.s. is reconstructed from experimental
signals. 

Importantly, the reconstruction is nested. Current correlations give
access to the connected coherence $\DcGee_{AB}$ which, as noted in
\cref{sec/g2-and-entanglement/G2-anatomy}, differs from $\DWGee_{AB}$ by
the singular exchange term whenever $\Ge_{AB}\neq0$. This term, computed
from the $\Ge_{AB}$ reconstructed at the first stage, is subtracted at
the level of the experimental signal, so that the linear inverse problem
is posed directly for the regular part $\DWGee_{AB}$
(\cref{sec/protocols/G2/tomographic-reconstruction}). When $\Ge_{AB}=0$, as in the setup of
\cref{fig/geometries}-(b), the first stage reduces to intra-channel
single electron tomography while the second stage directly gives
$\DcGee_{AB}=\DWGee_{AB}$. Nevertheless, the two stages remain coupled 
through the positivity constraints discussed below.

\subsubsection{Reconstruction as a noisy linear inverse problem}
\label{sec/protocols/strategy/inverse-problem}

The tomography protocols considered here, which are based on
statistical measurement of observables such as the
average dc current or the time-averaged low frequency current
correlation, express these measured quantities as a
\emph{linear filtering} of the sought electronic coherence
(\cref{sec/protocols/principles,sec/protocols/G1,sec/protocols/G2}).
Recovering the coherence from the experimental data is therefore a \emph{linear inverse
problem}. Since the measured signal is unavoidably noisy, this inverse
problem is ill-posed: a direct deconvolution amplifies the noise and
generically returns an unphysical coherence. The inversion must therefore
be regularized using prior physical knowledge and by restricting the
reconstructed coherence to the domain of physically admissible
coherences.

This admissible domain is anchored in a physical requirement: the
positive semi-definiteness of all the reconstructed coherence
operators,
which expresses the non-negativity of the
probability of any detection outcome computed from them. 
How this requirement is enforced in practice is discussed in
\cref{sec/protocols/strategy/positivity}.

Given this admissible domain, the reconstruction proceeds differently in
two regimes. In the \emph{data-rich} regime considered here, the inversion
reduces to a linear deconvolution whose estimator is projected onto the
admissible domain~\cite{Book-Inverse-Problems,Bisognin-2019-2}, thereby
yielding a unique reconstructed coherence operator. By contrast, in the
\emph{data-sparse} regime, a complete Bayesian treatment is required,
producing a probability distribution for the reconstructed coherence
operators over the physically admissible domain~\cite{Blume-Kohout-2010-1,Ubbelohde-Kokainis-Kashcheyevs-PC}. 

\subsubsection{The positivity constraints}
\label{sec/protocols/strategy/positivity}

Let us now detail the positivity constraints and their (weaker) operational
versions relevant for the protocols presented in the present Section,
reviewed in \cref{appendix/positivity}. Physically, positivity of
order $n\geq 1$ electronic coherences follows from the positivity of the probability of
detection of any $n$-particle excitations: it is formulated as positive
semi-definiteness of $n$-particle coherence operators defined in
\cref{appendix/G1/operators,appendix/G2/operators} for the cases
considered here.

\paragraph{Single-particle level:}
Positivity must hold both for the electron
coherence, $\Ge\succeq0$, and for the hole coherence, $\Gh\succeq0$. Through
the relation between electron and hole coherences, the latter provides the
upper bound, yielding $1\succeq\Ge\succeq0$, {\it i.e.} mode occupation
probabilities in $[0,1]$. With two channels, this condition bears on the
\emph{complete} single-electron coherence over $A\oplus B$ including
the inter-channel block $\Ge_{AB}$. 

In practice, following Ref.~\cite{Bisognin-2019-2}, one works with a
smaller set of necessary but not sufficient constraints. At the single
particle level, these are the $[0,1]$ bounds on the occupation numbers
for normalized states and the Cauchy-Schwarz constraints summarized in
\cref{appendix/positivity/G1}.

\paragraph{Two-particle level:}
For the reconstruction considered here,
whose key output
is the Wick deviation of the inter-channel coherence $\DWGee_{AB}$, 
the simplest constraint follows from
the positivity of the two-electron coincidence probability for
normalized single particle states $\varphi_A$ in
channel $A$ and $\varphi_B$ in channel $B$. Writing
$n_\alpha(\varphi)\equiv\Ge_\alpha(\varphi|\varphi)$ for the
occupation of a mode $\varphi$ in channel $\alpha$, and using
$\Gee_{AB}=\Ge_A\,\Ge_B-\Ge_{AB}\,\Ge_{BA}+\DWGee_{AB}$, we can obtain
the following positivity condition:
\begin{align}
	\label{eq/tomo/G2-positivity}
	\Delta_W\mathcal{G}^{(2e)}_{AB}&(\varphi_A,\varphi_B|\varphi_A,\varphi_B)\ge 
	\nonumber \\
	&\left|\Ge_{AB}(\varphi_A|\varphi_B)\right|^2
	-\,n_A(\varphi_A)\,n_B(\varphi_B)\,.
\end{align}
\Cref{eq/tomo/G2-positivity} makes a key feature explicit: although the two stages rely on different
measurements, the positivity requirement at the two-particle level
re-couples them. This is the nesting of \cref{sec/protocols/strategy/two-stage} at the level of
positivity constraints. The admissible
Wick deviation of two-electron coherence reconstructed at the second stage is bounded in terms of the
single-electron occupations and inter-channel coherences $\Ge_{AB}$ reconstructed in the first. Enforcing
positivity of the full two-electron coherence thus requires both stages jointly.

More generally, the complete coherence spans the intra-channel
($AA$ and $BB$) and inter-channel ($AB$) sectors, and positivity may be
imposed on each of the four particle/hole combinations
($(ee)$, $(eh)$, $(he)$ and $(hh)$ depending on the type of excitation
detected) as discussed in \cref{appendix/G2/operators}. 
As at the single-particle
level, one then works with necessary constraints of Cauchy--Schwarz
type (\cref{appendix/positivity/G2}), among which the explicit
constraints on $\DWGee_{AB}$ derived in
\cref{appendix/positivity/G2/Wannier}.

\subsection{Photo-assisted filtering} 
\label{sec/protocols/principles}

The idea behind photo-assisted tomography protocols is to exploit photo-assisted
transitions to recombine two contributions from coherences at different energies
into detection at a given energy. Photo-assisted transitions coherently scatter
electrons at a given energy into a linear combination of electrons with energies
shifted by the energies of the emitted or absorbed photons. Considering photons
with individual energy $hf$, the photo-assisted scattering process can be
described at the single particle level by:
\begin{equation}     
	\label{eq/tomo-g2/photoassisted/1}     
	\ket{\omega}_{\text{in}}\mapsto \ket{\omega}_{\text{out}}= \sum_{n\in\mathbb{Z}}p_n\,\ket{\omega+2\pi nf},
\end{equation}
where $\ket{\omega}$ denotes a single particle state at energy
$\hbar\omega$ and the $p_n$ factors are called the photo-assisted
amplitudes. They depend on the coherent radiation applied to the system
and thus both on its amplitude and its phase $\varphi$. In quantum Hall
systems, these amplitudes do not depend on the energies of the incoming
electrons and are therefore indexed by $n$ only.

The moduli square of these amplitudes give the probabilities for
emission ($n<0$) and absorption ($n>0$) of $|n|$ photons. Since the
transformation $\ket{\omega}_{\text{in}}\mapsto
\ket{\omega}_{\text{out}}$ is unitary, these amplitudes satisfy the sum
rule
\begin{equation}         
	\label{eq/tomo-g2/photoassisted/2}     
	\sum_{k\in\mathbb{Z}}p_{n+k}\,p_k^*=\delta_{n,0}\,. 
\end{equation} 
The amplitude $p_0$ is the amplitude for an electron at a given energy
to be scattered without any energy change (elastic scattering
amplitude).

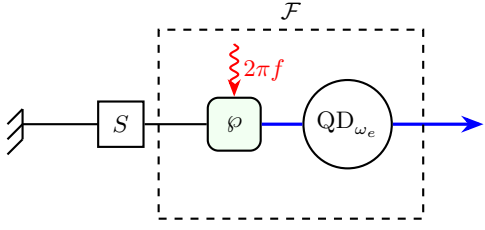
\begin{figure}     
	\begin{tikzpicture}[
	x=1cm,y=1cm,
	>=Stealth,
	thick,
	channel/.style={-Stealth, very thick},
	src/.style={draw, minimum width=0.6cm, minimum height=0.6cm},
	qpc/.style={draw, fill=black!80, minimum width=0.07cm, minimum height=0.5cm},
	filter/.style={draw, rounded corners, fill=green!5,
		minimum width=0.7cm, minimum height=0.7cm},
    photon/.style={
        draw=red,
        thick,
        ->,
        decorate,
        decoration={snake, amplitude=1.5pt, segment length=5pt, post length=2pt}
    }
	]
	
	\node[src] (S) at (0, 0) {$S$};
	\coordinate (G) at (-0.3,0.);   
	\draw[black, thick] (G) -- ++(-1,0) coordinate (Gright);
	\draw[black, thick] (Gright) -- ++(0,-0.2);
	\draw[black, thick] (Gright) -- ++(0,0.2);
  	\draw[black, thick] ($(Gright)+(0,0.2)$) -- ++(-0.2,-0.2);
	\draw[black, thick] ($(Gright)+(0,0)$) -- ++(-0.2,-0.2);
	\draw[black, thick] ($(Gright)+(0,-0.2)$) -- ++(-0.2,-0.2);
	
	\node[filter] (FB) at ( 1.5,0.) {$\wp$};

    \draw[photon] ($(FB.north)+(0,0.7)$) -- (FB.north) node[midway, right, color=red] {$2\pi f$};

	\draw[black,thick] (S.east) -- (FB.west);

	\node[circle, draw, minimum size=7mm, thick] (QDB) at ($(FB)+(1.5,0)$)
	{$\mathrm{QD}_{\omega_e}$};
	
	\draw[channel,-,blue] (FB.east) -- (QDB.west);
	node[left, blue] {\small 1out};
	\draw[channel,blue] (QDB.east) -- ++ (1.2,0.);

	\node[draw, minimum width=3.5cm, minimum height=2.5cm, dashed, label={above:$\mathcal{F}$}] (FF) at ($(FB)!0.5!(QDB)$) {};	
		
\end{tikzpicture}%
 
	\caption{\label{fig/photo-assisted-filter} Scheme of a
	photo-assisted filter $\mathcal{F}$ used to probe the output of a
	source $S$. A small sinusoidal classical drive at frequency $f$ is
	applied to the electrons within the radiation coupler $\wp$. The
	quantum dot $\text{QD}_{\omega_e}$ then filters the electrons at
	energy $\hbar\omega_e$. Ballistic propagation is assumed between
	these two stages (blue lines). The average dc current is measured at
	the output. This ideal photo-assisted filter $\mathcal{F}$,
	delimited by the dashed box, thus depends on $(\omega_e,f)$ as well
	as on the amplitude of the coherent tone at frequency $f$ applied to
	$\wp$.}
\end{figure}

We then combine such a photo-assisted process (denoted by $\wp$)  with
an electron detector placed after an energy filter selecting electrons
of energies $\hbar\omega_e$ with ballistic propagation between the two
(see Fig.~\ref{fig/photo-assisted-filter}). The detection operator on
the incoming electronic states is generated by back-propagating the
projection operator $\ket{\omega_e}\bra{\omega_e}$ that describes, at
the single-particle level, the detection of an electron at fixed energy
$\hbar\omega_e$: 
\begin{equation}         
	\label{eq/tomo-g2/photoassisted/3}     
	\Pi[\omega_e,\wp]=\sum_{(n_+,n_-)\in\mathbb{Z}^2} p_{n_+}\,p_{n_-}^*\ket{\omega_e-2\pi n_+f}\bra{\omega_e-2\pi n_-f}\,. 
\end{equation}
It is a projection operator that involves off-diagonal coherences in the
energy domain. Working in the single-particle sector with a reduced
density operator $\rho^{(1)}$ for a single-particle, the single-shot
detection probability $P_{click}$ is then given by
\begin{subequations}         
	\label{eq/tomo-g2/photoassisted/4}     
	\begin{align}         
		P_{\text{click}}&=\mathrm{Tr}\left(\rho^{(1)}\,\Pi[\omega_e,\wp] \right)\\         
		&=\sum_{(n_+,n_-)\in\mathbb{Z}} p_{n_+}p_{n_-}^*\rho^{(1)}_{\omega_e-2\pi n_-f,\omega_e-2\pi n_+f}\,. 
	\end{align} 
\end{subequations} 
This expression shows that tuning the selected detection energy
$\hbar\omega_e$ allows to fix the average energy
$\omega=\omega_e-\pi(n_++n_-)f$ over the two quantum paths, also called
the {\it diagonal parameter}. Also, depending on the photo-assisted
filter $\wp$, the detection probability is a linear combination of
off-diagonal coherences with {\it off-diagonal parameter} $\Omega=2\pi
(n_--n_+)f$, corresponding to the energy difference between the two
quantum paths.

\begin{figure}     
	\begin{tikzpicture}[
    x=1cm, y=1cm,
    >=Stealth,
    thick,
    channel/.style={-Stealth, very thick},
    pics/vaxis/.style={
        code={
            \draw[thick, ->] (0,0) -- (0,3) node[above] {$\omega$};
			\draw (-0.1, 2.5) -- (0.1, 2.5) node[left, xshift=-4pt]
			{$\omega_e+2\pi f$};
			\draw (-0.1, 1.5) -- (0.1, 1.5) node[left, xshift=-4pt] {$\omega_e$};
            \foreach \y in {0.5, 1.5, 2.5}
                \draw (-0.1, \y) -- (0.1, \y);
        }
    },%
    filter/.style={draw, rounded corners, fill=green!5,
        minimum width=1.0cm, minimum height=2.5cm},
    emitted/.style={
        ->,
        red,
        decorate,
        decoration={snake, amplitude=1.5pt, segment length=5pt, post length=3pt}
    }
    ]
    
    \draw (0, -1.5) pic (S) {vaxis}; 
    
	\node[filter] (FA) at (1.5,0) {};
	\node at ($(FA.center)+(0,-0.4)$) {$\wp$};

    \draw[blue,thick] (0,1) -- ($(FA.west)+(0,1)$);
    \draw[blue,thick] ($(FA.west)+(0,1)$) -- (FA.east);

    \draw[black,thick] (0, 0) -- (FA.west);

	\draw[emitted] ($(FA.east)+(-0.5,0.5)$) -- ++(1.2, 0.8)
		node[right] {$\omega_{\text{ph}}=2\pi f$};

    \draw[black,line width=1.5pt] (FA.east) -- ++(0.5,0);%
    \filldraw[black] ($(FA.east)+(0.5,0)$) circle (2pt);
    \draw[black,thick] (0.1,0) -- (FA.east);
        
\end{tikzpicture}%
     
	\caption{\label{fig/photoassisted-processes-example}     photo-assisted
	processes: recombination of a coherence between $\omega_+=\omega_e+2\pi f$
	and $\omega_-=\omega_e$ via the emission of a single photon of energy $hf$.
	The direct amplitude $p_-$ is associated with the blue path while the
	amplitude $p_0^*$ is associated with the black path. The thick black line
	ending with a black circle represents electronic detection at energy
	$\hbar\omega_e$.} 
\end{figure}
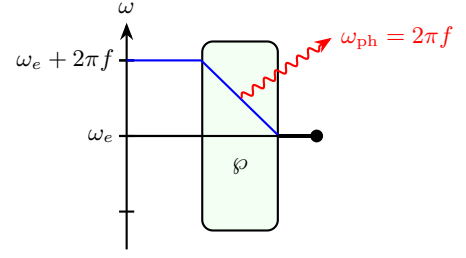

With $V_{\text{ac}}$ being the amplitude of the ac-drive, we expect - in
the case of a small amplitude monochromatic drive of the filter - that
the small dimensionless parameter $\varepsilon \propto
eV_{\text{ac}}/hf$ is present in the problem. 
The proportionality factor between $\varepsilon$
and $eV_{\text{ac}}/hf$ reflects the efficiency of the radiation coupler
as the frequency $f$ \cite{Souquet-Basiege-2024-1}. 
It thus depends on the specifics of the device. In
the small $\varepsilon$ limit, we expect single-photon emission and
absorption to be the dominant inelastic processes at first order in
$\varepsilon$: $p_{\pm 1}\simeq \mathcal{O}(\varepsilon)$ while
$p_0\simeq 1-\mathcal{O}(\varepsilon^2)$ with all $p_n$ for $|n|\geq 2$
being $\mathcal{O}(\varepsilon^2)$. There are four ways of combining
these processes in pairs involving exactly one amplitude $p_0$ (or its
complex conjugate) and one amplitude $p_\pm$ (or its complex conjugate).
Fig.~\ref{fig/photoassisted-processes-example} depicts one of these
processes associated with the pair of amplitudes $(p_{-1},p_0^*)$.

At first order in $\varepsilon$, the sum rule
\eqref{eq/tomo-g2/photoassisted/2} then leads to the following form for
the photo-assisted amplitudes:
\begin{subequations}     
	\label{eq/tomo-g2/photoassisted/6}     
	\begin{align}         
		p_{\pm 1}&=-\frac{\mi \varepsilon}{2}\,
		\me^{\mp\mi\varphi}+\mathcal{O}(\varepsilon^2)\\         
		p_0&=1%
		+\mathcal{O}(\varepsilon^2)
	\end{align} 
\end{subequations} 
in which a global phase has been absorbed in the redefinition of the
single-particle states. The remaining phase $\varphi$ can indeed be tuned by
adjusting the phase of the electromagnetic wave or classical voltage driving the
filtering device.%
Note that all other
photo-assisted amplitudes are $\mathcal{O}(\varepsilon^2)$.

Substituting this into Eq.~\eqref{eq/tomo-g2/photoassisted/4}, the linear
response of the detector in $\varepsilon$ then directly gives access to off
diagonal coherences:%
\begin{align}     
	\frac{\md P_{\text{click}}}{\md \varepsilon}&= \frac{\mi}{2}\, \left[\me^{\mi\varphi} 
	\left(\rho^{(1)}_{\omega_e,   \omega_e-2\pi f}-\rho^{(1)}_{\omega_e-2\pi f,\omega_e}\right)\right.\nonumber\\
	&\left.+\me^{-\mi\varphi}\left(\rho^{(1)}_{\omega_e,\omega_e+2\pi f} -\rho^{(1)}_{\omega_e+2\pi f,\omega_e}\right) \right] 
	\label{eq/tomo-g2/photoassisted/7} 
\end{align} 
Note that the r.h.s. contains four terms, corresponding to all possible ways to
recombine coherences at energies $(\omega_+,\omega_-)$ using only the amplitude
$p_0$ coupled to one amplitude $p_{\pm 1}$ for emitting or absorbing one photon
of frequency $f$. Equation \eqref{eq/tomo-g2/photoassisted/7} thus shows how to
obtain information on the single-particle reduced density matrix by counting
statistics after a photo-assisted filter. It is well suited for interpreting
experiments performed on electrons extracted from atoms and molecules using
attosecond pulses. This is to be compared with the single-electron tomography
method proposed \cite{Laurell-2022-1,Laurell-2025-1} and implemented in
Ref.~\cite{Laurell-2025-2}.

We will now develop this idea in the context of electron quantum optics, first
to  reconstruct single-electron coherence from average dc current measurement
and then to reconstruct inter-channel two-electron coherences from time-averaged
zero-frequency current correlations.

\subsection{Single-electron tomography} 
\label{sec/protocols/G1}

\subsubsection{Tomography equations}
\label{sec/protocols/G1/general}

Let us now consider how this photo-assisted filtering can be used to reconstruct
the single-electron coherence within a quantum Hall edge channel. Here, the
single-shot detection of electrons at energy $\hbar\omega_e$ is replaced by an
ideal electronic energy-filter for the electrons with transmission probability
$\mathcal{T}(\omega-\omega_e)$, where $\mathcal{T}(\omega)$ is peaked around
$\omega=0$. This could be realized using a quantum dot as in
Ref.~\cite{Altimiras-2010-1}. In this case, $\omega_e$ is controlled using the
gate voltage applied to the dot.

For a $T$-periodic electronic source, the electronic Wigner function
\cite{Ferraro-2013-1} (see \cref{appendix/G1} for details)
can be expanded as a Fourier series:
\begin{equation}     
	\label{eq/tomo-g2/photoassisted/8}     
	W^{(e)}(t,\omega)=\sum_{n\in\mathbb{Z}}W_n^{(e)}(\omega)\,\me^{-2\pi \mi nft},
\end{equation} 
where, since $W^{(e)}$ is a real function,
$W^{(e)}_n(\omega)^*=W^{(e)}_{-n}(\omega)$. The $n$-th harmonic corresponds to
an off-diagonal parameter $\Omega = 2 \pi n f$ in the frequency domain.

When the photo-assisted drive is switched off, the average net dc-current 
coming out of the electronic filter is given by 
\begin{equation}     
	\label{eq/tomo-g2/photoassisted/9}     
	\langle i_{\text{dc}}\rangle=-e\int \mathcal{T}(\omega-\omega_e)\,\DWe_{0}(\omega)\,\md\omega\,.
\end{equation} 
in which $\DWe_0(\omega)$ represents the excess electronic distribution compared
to the reference one at $\mu=0$. This spectroscopy technique has been used to
study relaxation and decoherence within quantum Hall edge channels in
\cite{Altimiras-2010-2,LeSueur-2010-1,Rodriguez-2020-1}.

\begin{figure} 
	\centering \begin{tikzpicture}
    \hspace{1em}

    \node at (0, 2) {\includegraphics[width=4cm]{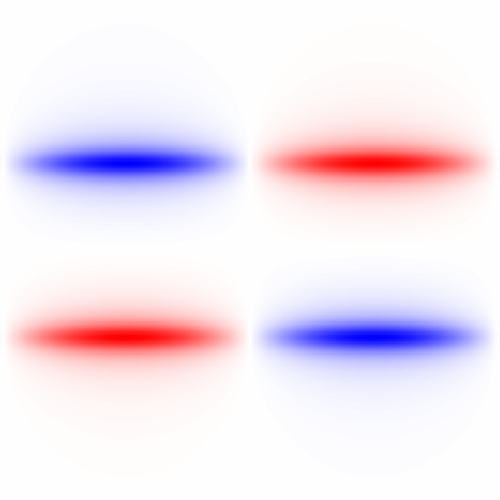}};

    \draw[-{Stealth}, line width=1.25pt] (-3, 0) -- (3, 0) node[scale=1.25, right] {$t$};
    \draw[-{Stealth}, line width=1.25pt] (0, -0.5) -- (0, 4) node[scale=1.25, above] {$\omega$};

    \draw[dashed, line width=0.75pt] (-2, 2.9) -- (-2, 0) node[below, scale=1.25] {$-T / 2$};
    \draw[dashed, line width=0.75pt] (2, 2.9) -- (2, 0) node[below, scale=1.25] {$T / 2$};
    \draw[{Stealth}-{Stealth}, line width=0.75pt] (0.7, 1.4) -- (0.7, 2.6) node[midway, scale=1.25, xshift=14pt] {$2 \pi n f$};
    \filldraw (0, 2) circle (2pt) node[left, scale=1.25] {$\omega_e$};
    \draw[{Stealth[reversed]}-{Stealth[reversed]}, line width=0.75pt] (-1, 1.6) -- (-1, 1) node[below, scale=1.25] {$\gamma$};

    \node at (4.5, 2) {\includegraphics[width=1.25cm]{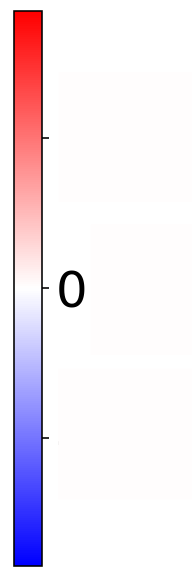}};

\end{tikzpicture}%
 
	\caption{\label{fig/g1_filter} Single-electron photo-assisted filtering
	protocol for the $n=1$ harmonic, representing the  action of
	\cref{eq/tomo-g2/photoassisted/G1-tomography} in the $(t,\omega)$ plane.} 
\end{figure}
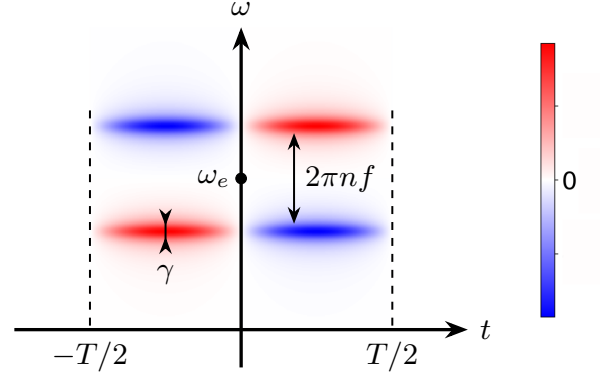

Probing the $n\neq 0$ harmonic requires applying a small drive at this
harmonic frequency $nf$. The response in the dc average current is then
proportional to the amplitude of the small drive. It involves four terms
corresponding to the four processes recombining off-diagonal coherences
via single-photon emission or absorption (see
Eq.~\eqref{eq/tomo-g2/photoassisted/7}). In terms of the Wigner
function's harmonics, the linear response can then be rewritten as:
\begin{align}         
	\frac{\md\langle i_{\text{dc}}\rangle}{\md\varepsilon_n} 
	&=\frac{e}{2}\int_{\mathbb{R}} \mathcal{T}^{(1)}(\omega-\omega_e;2\pi nf)\nonumber \\
	& \times\overline{\DWe(t,\omega)\,\sin(2\pi nft+\varphi_n)}^t \frac{\md\omega}{2\pi}         
	\label{eq/tomo-g2/photoassisted/result} 
\end{align}
in which $\varepsilon_n$ and $\varphi_n$ denote the small amplitude and the
phase parameters associated with the drive at frequency $nf$ and 
\begin{equation} 	
	\label{eq/tomo-g2/photoassisted/10} 	
	\mathcal{T}^{(1)}(\omega;\Omega)=\frac{1}{2} 
	\left[\mathcal{T}\left(\omega-\frac{\Omega}{2}\right)-\mathcal{T}\left(\omega+\frac{\Omega}{2}\right)\right]\,. 
\end{equation} 
Eqs.~\eqref{eq/tomo-g2/photoassisted/9} and \eqref{eq/tomo-g2/photoassisted/result} can then be recast 
as a linear problem by introducing the experimental signal $\mathbf{X}_n$ defined as 
\begin{subequations} 	
	\begin{align} 		
		\mathbf{X}_0&=\frac{\langle i_{\text{dc}}\rangle}{-e}\\
		\mathbf{X}_{n\neq 0} &=-\frac{1}{e}\,\frac{\md\langle i_{\text{dc}}\rangle}{\md\varepsilon_n} 	
	\end{align} 
\end{subequations} 
and the tomography kernel $\mathbf{H}_n$ defined by 
\begin{align} 	
	\label{eq/tomo-g2/photoassisted/11} 	
	\mathbf{H}_0(\omega) &= \mathcal{T}(-\omega)\\ 	
	\mathbf{H}_{n\neq 0}(\omega) &=\mathcal{T}^{(1)}(-\omega;2\pi nf) 
\end{align} 
so that the ideal tomography equation are convolution equations of the form  
\begin{equation} 	
	\label{eq/tomo-g2/photoassisted/G1-tomography} 	
	\mathbf{X}_n(\omega_e)=\left(\mathbf{H}_n\star \overline{\sin(2\pi 	nft+\varphi)\,\DWe}^t\right)(\omega_e) 
\end{equation} 
in which $\overline{\sin(2\pi nft+\varphi)\,\DWe}^t$ denotes the time average
over one period of $\sin(2\pi nft+\varphi)\,\DWe(t,\omega)$. Measurements
performed for two values of $\varphi$ are then necessary to extract the real
and imaginary parts of $\DWe_n(\omega)$ \cite{Bisognin-2019-2}.
\cref{fig/g1_filter} depicts the shape of the filter described by
\cref{eq/tomo-g2/photoassisted/G1-tomography} in the time-frequency plane.

\subsubsection{Comparison with HOM tomography}
\label{sec/protocols/G1/HOM-vs-PAF}

We now compare the ideal case tomography equations
\eqref{eq/tomo-g2/photoassisted/9} and \eqref{eq/tomo-g2/photoassisted/result}
to those arising from HOM tomography (see
Refs.~\cite{Ferraro-2013-1,Bisognin-2019-2}). To this purpose, we have derived
the HOM tomography equations in Appendix \ref{appendix/HOM-tomography}.  A
direct comparison of \cref{eq/tomo-g2/photoassisted/result} with
\cref{eq/HOM/A5} shows that the derivative of the noise signal in an HOM
experiment with respect to $\omega_{\text{dc}}=-eV_{\text{dc}}/\hbar$  has
exactly the same form as the linear response of the dc average current given by
\cref{eq/tomo-g2/photoassisted/result},  provided that one identifies 
\begin{equation} 	
	\label{eq/tomo-g2/photoassisted/12}
	\mathcal{T}(\omega)\longleftrightarrow \frac{\hbar}{4k_BT_{\text{el}}}\, 
	\frac{1}{\cosh^2\left(\frac{\hbar\omega}{2k_BT_{\text{el}}}\right)}\,.
\end{equation}
In other words, the width of the energy filter in the photo-assisted filtering
protocol plays the role of the thermal smearing in the HOM tomography protocol.

But the most important difference between the HOM protocol demonstrated in
Ref.~\cite{Bisognin-2019-2} and the photo-assisted protocol presented here lies
in the fact that the former requires a low frequency noise measurement whereas
the photo-assisted tomography protocol is based on average dc-current
measurements. The latter are much easier to perform and this is where a real
gain could be expected.

\subsubsection{The noisy linear inverse problem}
\label{sec/protocols/G1/inverse-problem}

In a real experiment, the measured quantity includes an experimental noise
$\mathbf{N}_n$ and the ideal tomography equation is replaced by the linear
signal processing equation 
\begin{equation} 	
	\label{eq/tomo-g2/photoassisted/G1-tomography-with-noise} 	
	\mathbf{X}_n(\omega_e)=\left(\mathbf{H}_n\star 
	\overline{\sin(2\pi nft+\varphi)\,\DWe}^t \right)(\omega_e)+\mathbf{N}_n 
\end{equation} 
in which the distribution of the random variable $\mathbf{N}_n$ reflects the
experimental noise on the measurement of $X_n$ (here this is the current noise).
Many techniques for this type of problems have been developed
\cite{Book-Inverse-Problems} and, in Ref.~\cite{Bisognin-2019-2}, a Bayesian
deconvolution method was used. Importantly, one has to ensure that the estimated
solution $(\DWe_n(\omega))_n$ satisfies the Cauchy-Schwarz constraints defined
in \cref{appendix/positivity/G1} so that the probability of 
emission of any electron or hole excitation computed from the reconstructed 
single-electron coherence is in the interval $[0,1]$.

As we will see now, the principle of photo-assisted single-electron tomography 
can be adapted to perform a two-electron tomography, which only
requires measurement of low frequency current correlations.

\subsubsection{Reconstructing $\Ge_{AB}$}
\label{sec/protocols/G1/inter-channel}

The intra-channel coherences $\mathcal{G}^{(e)}_A$ and $\mathcal{G}^{(e)}_B$ are
reconstructed by applying the protocol of
\cref{sec/protocols/G1/general}
to each channel. The inter-channel
coherence $\mathcal{G}^{(e)}_{AB}$, which encodes the single-particle
delocalization between $A$ and $B$, requires interfering the two channels
on a quantum point contact (QPC) and applying the same single-electron
tomography. 

This can be done either by placing the beam splitter before the
photo-assisted filters or by driving one channel before recombination at a beam
splitter. Both
routes, detailed in Appendix~\ref{appendix/GAB}, reduce to the single-electron
inverse problem. This stage is required in geometries such as
\cref{fig/geometries}-(c), where single-particle excitations are delocalized
between the channels. But more generally, it is required in any other geometry where one
expects tunneling between channels $A$ and $B$. The key point is that it
is part of the singular component of the inter-channel two-electron
coherence $\Gee_{AB}$ and its knowledge, as we shall see, is mandatory
to fully reconstruct this object.

\subsection{Two-electron tomography} 
\label{sec/protocols/G2}

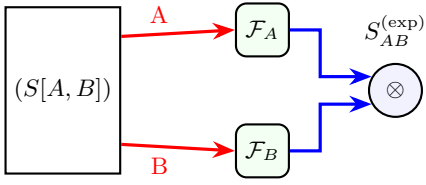
\begin{figure} 	
	\centering 	
	\begin{tikzpicture}[
	x=0.8cm,y=0.8cm,
	>=Stealth,
	thick,
	channel/.style={-Stealth, very thick},
	doublesrc/.style={draw, minimum width=1.1cm, minimum height=2.2cm},
	filter/.style={draw, rounded corners, fill=green!5,
		minimum width=0.7cm, minimum height=0.7cm},
	correlator/.style={draw, circle, fill=blue!5, minimum size=7mm}
    ]
	
	\node[doublesrc] (S) at (-2.,0.) {$(S[A,B])$};
	
	\node[filter] (FA) at (1.3,1.) {$\mathcal{F}_A$};
	\node[filter] (FB) at (1.3,-1.) {$\mathcal{F}_B$};
	
    \draw[channel,red] ([yshift=-4mm]S.north east) -- (FA.west)
    node[midway, above left, red] {A};
    
    \draw[channel,red] ([yshift=4mm]S.south east) -- (FB.west)
    node[midway, below left, red] {B};

    \node[correlator] (Corr) at (3.5, 0) {$\otimes$};
	\node[above=0.1cm] at (Corr.north) {$S_{AB}^{(\text{exp})}$};

    \draw[channel, blue] (FA.east) -- ++(0.5,0) |- (Corr.150);
    \draw[channel, blue] (FB.east) -- ++(0.5,0) |- (Corr.210);

\end{tikzpicture}%
 	
	\caption{\label{fig/tomo/G2AB-measurement} Two photo-assisted
	filters $\mathcal{F}_A$ and $\mathcal{F}_B$ of the type depicted on
	Fig.~\ref{fig/photo-assisted-filter} are placed on the two outgoing
	channels $A$ and $B$ from a two-channel blackbox emitter $(S[A,B])$.
	The low frequency time averaged current correlation
	$S_{AB}^{(\text{exp})}$ defined by \cref{eq/correlation-signal}
	between the outputs of the photo-assisted filters is then measured.}
\end{figure}

In this section, we discuss how to access the irreducible two-particle
part $\DWGee_{AB}$ of the
inter-channel two-electron coherence (see \cref{eq/basics/DWGee})
in the frequency domain using two ideal photo-assisted filters as depicted on
\cref{fig/tomo/G2AB-measurement}. This two-filter current-correlation scheme
generalizes, to photo-assisted filtering and the interacting regime, the
static-filter measurement of two-electron coherence introduced for
non-interacting sources in Ref.~\cite{Moskalets-2014-1}.

\subsubsection{General filtering equations} 
\label{sec/protocols/G2/introduction}

The starting point is the expression connecting the bare current
operator $\widehat{I}(t)=-ev_F\,(\psi^\dagger\psi)(t)$ to a formal
electronic Wigner operator $\widehat{W}(t,\omega)$: 
\begin{subequations}     
	\label{eq/tomo-g2/current-and-Wigner}
	\begin{align}         
		\widehat{I}_{\text{dc}}&=-e\int_{\mathbb{R}^2}     
		\mathcal{F}_{\Omega,\varphi}(t,\omega)\,\widehat{W}(t,\omega)\,\frac{\md\omega\,\md t}{2\pi}\,\\         
		\widehat{W}(t,\omega)&=v_F\int_{\mathrm{R}} 
		\me^{\mi\omega \tau} \psi^\dagger\left(t-\frac{\tau}{2}\right)\, \psi\left(t+\frac{\tau}{2}\right)\,\md\tau  
	\end{align} 
\end{subequations} 
in which $\widehat{I}_{\text{dc}}=\widehat{\tilde{I}}(\omega=0)$ denotes
the $\omega=0$ Fourier component of the current. When the photo-assisted filter is
irradiated with a classical radiation
at frequency $\Omega/2\pi$ with
phase $\varphi$ and its static filtering is tuned at energy $\omega_e$, we
have:
\begin{align}     
	\label{eq/tomo-g2/filter-expression}     
	\mathcal{F}&(t,\omega)= \mathcal{T}(\omega-\omega_e)\nonumber\\
	&+\varepsilon\,\sin(\Omega t-\varphi)\mathcal{T}^{(1)}(\omega-\omega_e;\Omega)\,. 
\end{align} 
As before, $\mathcal{T}^{(1)}(\omega-\omega_e;\Omega)$ is defined by
Eq.~\eqref{eq/tomo-g2/photoassisted/10}. The quantity of interest is the
time averaged low frequency current correlator
\begin{equation}   
   \label{eq/correlation-signal}   
   S_{AB}^{(\text{exp})}=\int_{\mathbb{R}} 
   \overline{\langle i_A(t_\Sigma+t_\Delta/2)\,i_B(t_\Sigma-t_\Delta/2)\rangle_c}^{t_\Sigma}\, 
   \md t_\Delta\, 
\end{equation}
where $\overline{X(t_\Sigma)}^{t_\Sigma}$ denotes the time average over
$t_\Sigma$. 
In the limit of a very large period $T$ ($f=1/T\to 0$) considered here
for simplicity, it
is related to the connected correlator of the zero frequency current
\begin{equation}
    S_{AB}^{(\text{exp})}\simeq f\langle
\widehat{I}_{B,\text{dc}}  \widehat{I}_{A,\text{dc}}
\rangle_c\,.
\end{equation}
\cref{eq/tomo-g2/current-and-Wigner} connects the non-connected correlator $C_{AB}=f\langle
\widehat{I}_{B,\text{dc}}  \widehat{I}_{A,\text{dc}}\rangle$ 
to the average value of the double Wigner operator
$\widehat{W}_B(t_B,\omega_B)\,\widehat{W}_A(t_A,\omega_A)$. Its average value is
nothing but the two-particle Wigner distribution function. When considering the
excess current operator $i_A(t_A)=-ev_F:(\psi_A^\dagger\psi_A)(t_A):$,
the Fermi sea contributions are removed and we 
obtain the intrinsic two-particle coherence $\DGee_{AB}$ (defined by
\eqref{eq/g2/excess/inter}). 
Introducing the compact notation $\mathbf{t}=(t_A,t_B)$ and
$\pmb{\omega}=(\omega_A,\omega_B)$, we have:
\begin{equation} 
	C_{AB}=e^2f\int_{\mathbb{R}^4} \mathcal{F}_{A,B}(\mathbf{t},\pmb{\omega})\, 
	\Delta W^{(2e)}_{AB}(\mathbf{t},\pmb{\omega}) \frac{\md^2\mathbf{t}\md^2\pmb{\omega}}{(2\pi)^2}\,. 
\end{equation} 
where the linear filter connecting two-electron excess correlation to $C_{AB}$ is 
\begin{equation} 
	\label{eq/tomo-g2/energy-filter-notation} 
	\mathcal{F}_{A,B}(\mathbf{t},\pmb{\omega})= \mathcal{F}_A(t_A,\omega_A)\,\mathcal{F}_B(t_B, \omega_B)\,. 
\end{equation} 
This then leads to the final two-electron tomography equation collecting the
connected current correlation
$S_{AB}=C_{AB}-f\langle i_{A,\text{dc}}\rangle\langle i_{B,\text{dc}}\rangle $ to
the connected inter-channel two-electron coherence $\DcGee_{AB}$: 
\begin{equation}     
	\label{eq/tomo/final}     
	S_{AB}^{(\text{exp})}=e^2f\int_{\mathbb{R}^4} \mathcal{F}_{A,B}(\mathbf{t},\pmb{\omega})\,
	\Delta_c W^{(2e)}(\mathbf{t},\pmb{\omega})
	\frac{\md^2\mathbf{t}\md^2\pmb{\omega}}{(2\pi)^2}\,. 
\end{equation} 
in which $\DcWee_{AB}$ is the Wigner representation of $\DcGee_{AB}$ introduced
in \eqref{eq/basics/DWGee}. Note that
\begin{subequations} 	
	\label{eq/tomo-g2/Delta-cW} 
	\begin{align}     
		\Delta_c W^{(2e)}_{AB}(\mathbf{t},\pmb{\omega})&=     
		\left(W^{(2e)}_{AB} -W^{(e)}_A\times W^{(e)}_B\right)(\mathbf{t},\pmb{\omega})\\     
		&=\left(\Delta W^{(2e)}_{AB} 
		-\Delta W^{(e)}_A \times \Delta W^{(e)}_B\right)(\mathbf{t},\pmb{\omega})\,. 
	\end{align} 
\end{subequations} 
Eq.~\eqref{eq/tomo/final} is the key equation for two-electron tomography: it
shows how the current correlations after the two photo-assisted filters is
related to the connected two-electron Wigner distribution between the two
channels $A$ and $B$. This is the crucial two-particle stage needed to implement the
nested tomography protocol sketched in \cref{sec/protocols/strategy}.

\subsubsection{The $T$-periodic case}
\label{sec/protocols/G2/T-periodic}

In practice, experiments are
performed on a $T$-periodic source to ensure a good signal-to-noise ratio.
This is
the regime we now describe from the experimentalist's standpoint, in the
perspective of the tomographic reconstruction of
$\Gee_{AB}$ presented in \cref{sec/protocols/strategy}.

\begin{figure} 	
	\begin{tikzpicture}
        
    \draw [-{Stealth}, thick] (7.1, 2.5) -- (10.9, 2.5) node[font=\large, right] {$\Omega_{A}$};
    \draw [-{Stealth}, thick] (8.5, 1.1) -- (8.5, 4.9) node[font=\large, above] {$\Omega_{B}$};

    \draw [-{Stealth}, gray, thick] (7.3, 1.3) -- (10, 4)
        node[font=\large, above right, color=black] {$\Omega_\Sigma$};

    \draw [-{Stealth}, gray, thick] (7.3, 3.7) -- (10, 1.0)
        node[font=\large, below right, color=black] {$\Omega_\Delta$};

    \foreach \s/\n in {-0.6/-1, 0/0, 0.6/1} {
        \draw [dashed, color=blue, opacity=1]
            (7.1 + \s, 3.9 + \s) -- (9.9 + \s, 1.1 + \s)
            node[pos=0.2, above, sloped, opacity=1, blue, font=\small] {$n=\n$};
    }
	
    \fill [black] (8.5, 2.5) circle (2.5pt);

	\fill [blue] (9.7, 2.5) circle (2.5pt);

\end{tikzpicture}%
 	
	\caption{\label{fig/tomo-2e} Two-electron coherence reconstruction
	for a $T$ periodic source in the off-diagonal coordinates
	$(\Omega_A,\Omega_B)$: the center of mass and relative particle
	$(\Omega_\Sigma,\Omega_\Delta)$ axis are indicated. The $(0,0)$
	point (black dot) selects the diagonal plane in the 4D frequency
	space. The points with either $\Omega_A=0$ or $\Omega_B=0$
	correspond to the measurements performed with only one rf drive
	active. Probing two-electron coherence for the generic points
	located on the $\Omega_\Sigma=2\pi nf$ 3D volumes (dashed blue
	lines) involves two rf drives.}
\end{figure}
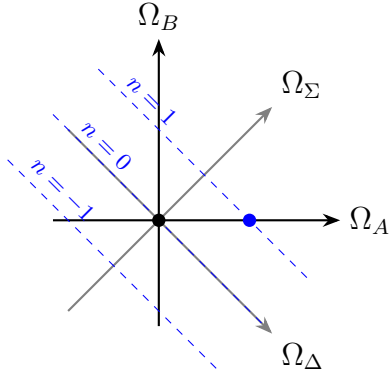

For a $T$-periodic source ($f=1/T$) the time integrals in
Eq.~\eqref{eq/tomo/final} are understood as an average over one period in the
centre-of-mass time $t_\Sigma$ and an integral over the relative time
$t_\Delta$, 
and the measured
signal is the time-averaged zero-frequency current correlation defined
by \cref{eq/correlation-signal}.
Accordingly, $\Wee_{AB}$ is
$T$-periodic in $t_\Sigma$ and decomposes into harmonics
\begin{align}  
	\begin{split}
   		\label{eq/tomo-g2/T-periodic/2}     
		W_{AB}^{(2e)}&(\mathbf{t},\pmb{\omega})=\\
   		&\sum_{n\in\mathbb{Z}} \int_{\mathbb{R}} 
   		\me^{-\mi(2\pi nft_\Sigma+\Omega_\Delta t_\Delta)} 
   		\Wee_{AB,n}(\pmb{\omega}|\Omega_\Delta)\, \frac{\md\Omega_{\Delta}}{2\pi}
	\end{split}
\end{align} 
so that the tomographic task is to reconstruct the harmonics
$\Wee_{AB,n}(\boldsymbol\omega\,|\,\Omega_\Delta)$. Here the
centre-of-mass modulation frequency $\Omega_\Sigma=\Omega_A+\Omega_B=2\pi n f$
is discrete (indexed by $n\in\mathbb{Z}$), whereas the relative
modulation frequency $\Omega_\Delta=(\Omega_A-\Omega_B)/2$, conjugate to
$t_\Delta$, is \emph{continuous}. Reconstructing a given harmonic thus
requires generating mutually coherent tones on the two filters at
$\Omega_{A/B}=\pi n f\pm\Omega_\Delta$. 
The corresponding sectors of the frequency space are those depicted in
Fig.~\ref{fig/tomo-2e}.

However, one also has to take into account the anatomy of two-electron
coherence discussed in \ref{sec/g2-and-entanglement/G2-anatomy}: 
the full two-electron coherence $\Gee_{AB}$ (or equivalently
$\Wee_{AB}$)
is a sum of a Wick's contribution that can be expressed in terms of
single-electron coherence (see \cref{eq/basics/Wick/1}) and 
an irreducible two-particle contribution $\DWGee_{AB}$. 
The Wick contribution is
a sum of two contributions: a correlation contribution involving $\Ge_A$
and $\Ge_B$ and
an exchange one involving $\Ge_{AB}$. 

As discussed in \cref{appendix/G2/T-periodic}, the correlation part
$\Ge_A\,\Ge_B$ gives $\delta$-singularities whose support consists of
planes
$\mathcal{C}_{n,k}$
corresponding to $\Omega_\Sigma=2\pi nf$ ($n\in\mathbb{Z}$) and
$\Omega_\Delta=2\pi (k-n/2) f$
($k\in\mathbb{Z}$).
When the inter-channel coherence $\Ge_{AB}$ is non zero, the
exchange contribution $\Ge_{AB}\Ge_{BA}$ gives $\delta$-singularities
whose support are the planes $\mathcal{E}_{n,k}$
corresponding to $\Omega_\Sigma=2\pi nf$ ($n\in\mathbb{Z}$) and
$\omega_\Delta=\pi (k-n/2) f$ ($k\in\mathbb{Z}$). The resulting decomposition of
$\Wee_{AB,n}(\pmb{\omega}|\Omega_\Delta)$ into a regular part 
$\DWWee_{AB,n}$ and singular parts located on these $\mathcal{C}$ and 
$\mathcal{E}$-planes is given by \cref{eq/G2/singularities} and
can be rewritten as a relation between $\DcWee_{AB}$, which is the
quantity connected to the current correlation signal as discussed in
\cref{sec/protocols/G2/introduction}, and $\DWWee_{AB}$:
\begin{subequations}
	\label{eq/G2/Dc-to-DW}
\begin{align}
	\DcWee_{AB,n}&(\pmb{\omega}|\Omega_\Delta)=
	\DWWee_{AB,n}(\pmb{\omega}|\Omega_\Delta)\\
	&-2\pi\sum_{k\in\mathbb{Z}}
	\delta(\omega_\Delta-\pi
	f(k-n/2))\,\We_{AB,k}\left(\frac{\omega_\Sigma+\Omega_\Delta}{2}\right)\nonumber\\
	&\times\We_{BA,n-k}\left(\frac{\omega_\Sigma-\Omega_\Delta}{2}\right)\,.
\end{align}
\end{subequations}
Once single-electron tomography has been
performed, the harmonics $\We_{AB,n}$ are known. The
challenge of the two-particle stage of two-electron tomography is to determine the
two-particle irreducible part of $\DWGee_{AB}$ in the frequency domain, {\it i.e.}
$\DWWee_{AB,n}(\pmb{\omega}|\Omega_\Delta)$. Since
the time-averaged low frequency current correlation is
directly related to the connected two-electron coherence $\DcGee_{AB}$,
the single and two-particle stages are nested as soon as $\Ge_{AB}\neq
0$.

As soon as $\Ge_{AB}\neq0$, the
connected coherence $\DcWee_{AB}$ is not a regular function: it carries the
singular contribution of \cref{eq/G2/Dc-to-DW}, pinned to the
exchange singularity planes
$\mathcal{E}_{n,k}$. Two important points follow: first, the measured signal stays finite
only because the static energy filters have a \emph{finite bandwidth}
and regularize divergences occurring each time the
operating point meets such a singularity. Second, isolating the
irreducible
two-body part $\DWWee_{AB}$ requires subtracting this singular contribution
entirely fixed by single-electron tomography from the experimental
signal before inversion of the linear tomography equation. 

Before explaining how to implement the reconstruction procedure in
practice, we will discuss how the necessary current correlation
measurements can be organized in three consistent packages, each of them
enabling the reconstruction of physically relevant information.

\subsubsection{Tomographic measurements}
\label{sec/protocols/G2/tomography-measurements}

The natural way to organize the two-particle stage of the 
tomography protocol is by the
degree of 
energy coherence: when are we probing the classical (diagonal) or quantum
(off-diagonal) part of two-electron coherence and in which energy
variable.
This yields three successive
layers, of increasing information content.

Let us now describe these three layers, mentioning briefly where 
these singularity issues matter
before we explain how to implement the inversion process underlying
two-electron coherence tomographic reconstruction.

\paragraph{Two-electron spectroscopy.}
With no drive applied, both filters are static energy filters and the
correlation is diagonal in both energies ($n=0$, $\Omega_\Delta=0$). Sweeping
the filtering energies $\pmb{\omega}_e=(\omega_{e,A},\omega_{e,B})$ probes the
diagonal $\DcWee_{AB,0}(\pmb{\omega}\,|\,0)$ through the finite-width kernel
$\mathcal{T}_A(-\omega_A)\,\mathcal{T}_B(-\omega_B)$ [row $(0,0)$ of
Table~\ref{table/2e-tomography}].

When the excitation is delocalized between the two channels ($\Ge_{AB}\neq0$),
this diagonal is not regular: the exchange singularity of
\cref{eq/G2/Dc-to-DW} contributes on the planes $\mathcal{E}_{0,k}$
($k\in\mathbb{Z}$) at $\omega_A-\omega_B=2\pi k f$, with weight
$\big|\Ge_{AB,k}(\omega_\Sigma/2)\big|^2$. Each resonance is the spectroscopic
fingerprint of a $k$-photon-assisted process through which the periodic source
builds the inter-channel coherence. 
It is only
because the static filters have a \emph{finite bandwidth} that these resonances
appear as finite dips rather than divergences.

Following the general principle stated above, as soon as $\Ge_{AB}\neq
0$, one does not invert directly for
the singular $\DcWee_{AB,0}$: the regularized exchange comb, entirely fixed by
the inter-channel single-electron coherence $\Ge_{AB}$, is moved to the experimental side, and
the linear inverse problem is solved for the regular part
$\DWWee_{AB,0}(\pmb{\omega}\,|\,0)$. The two-electron energy distribution is then
assembled from it,
\begin{equation}
  \Wee_{AB,0}(\omega_A,\omega_B\,|\,0)
  = \We_{A,0}(\omega_A)\,\We_{B,0}(\omega_B) + \DcWee_{AB,0}(\pmb{\omega}\,|\,0)\,,
\end{equation}
with $\DcWee_{AB,0}=\DWWee_{AB,0}-(\text{exchange comb})$ recombining the
reconstructed regular part and the known singular comb.
$\We_{A,0}$ and $\We_{B,0}$ are obtained via single-channel spectroscopies. 
The resulting two-electron distribution $\Wee_{AB,0}(\pmb{\omega}|0)$ is a purely
classical quantity, {\it i.e.} the probability of finding the two electrons at given
energies.

\paragraph{The $n=0$ slice: 
the minimal entanglement test.} 
Switching on a drive at $\Omega_\Delta\neq0$ while keeping $n=0$ leaves the
total energy classical ($\Omega_\Sigma=0$) but now probes the quantum coherence
in the relative energy,  {\it i.e.} the inter-channel entanglement of
\cref{sec/g2-and-entanglement} at fixed total energy. It is the natural stage
to test the witness \eqref{eq/gen-crit/G2} in the frequency domain
without a full tomography as will be discussed in
\cref{sec/2e-collision}.
Moreover, it is
experimentally the most accessible off-diagonal measurement, since $n=0$
imposes $\Omega_A=-\Omega_B=\Omega_\Delta$: a single microwave generator
feeding both filters is sufficient to perform the experiment.

As for two-electron spectroscopy, the reconstruction targets the regular part: the exchange
comb of \cref{eq/G2/Dc-to-DW}, known from single-electron tomography, is
joined to the current correlation signal
and the linear inverse problem is solved for the regular
$\DWWee_{AB,0}(\pmb{\omega}\,|\,\Omega_\Delta)$. The full inter-channel
coherence $\Gee_{AB}$ --- on which the witness \eqref{eq/gen-crit/G2} is
evaluated --- is then assembled from it and the single-electron data.

The reduction of the $T$-periodic witness, in the $f\to0$ limit, to the
single-collision frequency-domain test of \cref{sec/2e-collision}, together
with the condition (all probed modes above the Fermi sea) under which the
violation is preserved, is established in Appendix \ref{appendix/dilute-to-periodic}.

\paragraph{The $n\neq0$ slices: the complete second-order coherence.}
Driving the centre-of-mass modulation at $\Omega_\Sigma=2\pi n f$ probes quantum
coherence in the total energy as well. Sweeping $\Omega_\Delta$, the
single-drive points $\Omega_\Delta=\pm\pi n f$ ($\Omega_A=0$ or $\Omega_B=0$)
reduce to single-electron tomography (see rows $(1,0)$ and $(0,1)$ of
Table~\ref{table/2e-tomography}), while generic $\Omega_\Delta$ requires two
mutually coherent tones at $\pi nf\pm\Omega_\Delta$ (see row $(1,1)$).
It thus adds a supplementary experimental complication compared to the
$n=0$ slice discussed before.

In all cases,
the reconstruction proceeds as before: when present, the exchange singularities of
\cref{eq/G2/Dc-to-DW} are joined to the experimental signal and the linear inverse
problem is solved for the regular part $\DWWee_{AB,n}(\pmb{\omega}\,|\,
\Omega_\Delta)$ as discussed in the next Section.

Collecting all harmonics $n$ assembles the full inter-channel
two-electron coherence $\Gee_{AB}$, allowing the witness to be tested on
generic orthonormal mode families, beyond plane waves and involving the centre-of-mass
degrees of freedom as well. 
Further uses of the full access to inter-channel two-electron coherence
are discussed in the concluding
perspectives.

\subsubsection{The tomographic reconstruction algorithm}
\label{sec/protocols/G2/tomographic-reconstruction}

These three layers being described, we can now cast all of them as
instances of a single tomographic equation. Regardless
of which filters are driven, the ideal tomography equation 
takes the standard form of a linear convolution equation similar
to \cref{eq/tomo-g2/photoassisted/G1-tomography}, expressing the
ideal experimental signal  
$\mathbf{X}_{n,\Omega_\Delta}(\pmb{\omega}_e)$
as a convolution of a kernel
$\mathbf{H}_{n,\Omega_\Delta}$ with a Fourier component of the quantum
signal to be reconstructed $\DcWee_{AB}$:
\begin{align}
   \label{eq/tomo-g2/T-periodic/generic-form}
   \mathbf{X}_{n,\Omega_\Delta}&(\pmb{\omega}_e)=\nonumber \\
   &\left[\mathbf{H}_{n,\Omega_\Delta}
   \star\overline{\sin(2\pi nft_\Sigma+\Omega_\Delta t_\Delta-\varphi)
	\,\DcWee_{AB}}^{\mathbf{t}}\right](\pmb{\omega}_e)\,
\end{align}
In this equation, the off-diagonal coordinates $\Omega_{A/B}=\pi n f\pm\Omega_\Delta$ 
play the role of the single harmonic $n$ in the single-electron tomography
equation \eqref{eq/tomo-g2/photoassisted/G1-tomography}.
The experimental signal $\mathbf{X}_{n,\Omega_\Delta}(\boldsymbol\omega_e)$
and kernel $\mathbf{H}_{n,\Omega_\Delta}(\boldsymbol\omega)$ for each of the three
layers are collected in Table~\ref{table/2e-tomography}. The shape of the
filtering in the $(\omega_A,\omega_B)$ plane is presented in 
\cref{fig/3D-principle-of-tomography}, showing
in particular the four peaks
at the corners of a square of width $|\Omega_\Delta|$ for the driven
cases.
The explicit derivation of
these equations, order by order in the drive amplitudes, is given in
Appendix~\ref{appendix/tomo-g2}. 

But contrary to the single-particle case where the ideal tomography
equation \eqref{eq/tomo-g2/photoassisted/G1-tomography} 
directly involved well-behaved functions $\We_n(\omega)$ of $\omega$,
one has to account for the $\delta$-singularities of 
$\DcWee_{AB}$ in the frequency domain whenever $\Ge_{AB}\neq 0$. 
As explained before, the idea is to split $\DcWee_{AB}$ as a sum of $\DWWee_{AB}$ and of the
exchange contribution given by \eqref{eq/G2/Dc-to-DW} and join the
latter to the experimental signal thereby defining:
\begin{align}
	\label{eq/tomo-g2/T-periodic/Z-definition}
	&\mathbf{Z}_{n,\Omega_\Delta}(\pmb{\omega}_e)=\mathbf{X}_{n,\Omega_\Delta}(\pmb{\omega}_e)
	\nonumber\\
	&-\left[
        \mathbf{H}_{n,\Omega_{\Delta}}\star 
        \overline{\sin(2\pi nft_\Sigma+\Omega_\Delta
        t_\Delta-\varphi)\,\Wee_{AB,\text{exch}}}^{\mathbf{t}}\right]
	(\pmb{\omega}_e)
\end{align}
the final form of the ideal tomography equations is then
\begin{align}
	\label{eq/tomo-g2/T-periodic/final-tomography}
   &\mathbf{Z}_{n,\Omega_\Delta}(\pmb{\omega}_e)=\nonumber \\
   &\left[\mathbf{H}_{n,\Omega_\Delta}
   \star\overline{\sin(2\pi nft_\Sigma+\Omega_\Delta t_\Delta-\varphi)
    \,\DWWee_{AB}}^{\mathbf{t}}\right](\pmb{\omega}_e)\,
\end{align}
where each harmonic $\DWWee_{AB,n}(\pmb{\omega}_e|\Omega_\Delta)$ selected by
the time-average is a regular function of
$(\pmb{\omega}_e,\Omega_\Delta)$. One can then solve the proper noisy linear inverse problem
(see \cref{sec/protocols/G1/inverse-problem}) using for example the
Bayesian deconvolution method used in Ref.~\cite{Bisognin-2019-2}
to reconstruct 
the harmonics
$\DWGee_{AB,n}(\pmb{\omega}_e|\Omega_\Delta)$. Note that here
one
needs to know the error on the experimental signal but also the error on
the reconstructed single-electron coherence $\We_{AB}$ which now
appears within $\mathbf{Z}_{n,\Omega_\Delta}(\pmb{\omega}_e)$. Finally,
an operational set of positivity constraints on $\Gee_{AB}$ has to be
formulated on the quantity to be reconstructed, {\it i.e.} $\DWGee_{AB}$
as discussed in \cref{sec/protocols/strategy/positivity} and Appendix
\ref{appendix/positivity/G2}. 

\begin{table*} 	
	\renewcommand{\arraystretch}{1.8} 	
	\begin{tabular}{|c|c||c|c|c|} 
		\hline 		
		Order in $(\varepsilon_A,\varepsilon_B)$ & $n$ & $\Omega_\Delta$ & $\mathbf{X}_{n,\Omega_\Delta}(\pmb{\omega})$ & $\mathbf{H}_{n,\Omega_\Delta}(\pmb{\omega})$ \\ 
		\hline \hline 		
		$(0,0)$ & $n=0$ & $\Omega_\Delta=0$ & $\left(\Sigma_{AB}\right)_{\pmb{\varepsilon}=0}$ & $\mathcal{T}_{A}(-\omega_A)\mathcal{T}_{B}(-\omega_B)$\\ 		
		$(1,0)$ & $n\neq 0$ & $\Omega_\Delta=\pi nf$ & $\left(\frac{\partial \Sigma_{AB}}{\partial\varepsilon_A}\right)_{\pmb{\varepsilon}=0}$ & $\mathcal{T}^{(1)}_{A}(-\omega_A;2\pi nf)\mathcal{T}_{B}(-\omega_B)$ 		\\ $(0,1)$ & $n\neq 0$ & $\Omega_\Delta=-\pi nf$ &    $\left(\frac{\partial \Sigma_{AB}}{\partial\varepsilon_B} \right)_{\pmb{\varepsilon}=0}$ & $\mathcal{T}_{A}(-\omega_A) \,\mathcal{T}^{(1)}_{B}(-\omega_B;2\pi nf)$ \\ 
		$(1,1)$ & $n\in\mathbb{Z}$ & $\Omega_\Delta\neq \pm\pi nf$ & $\left(\frac{\partial^2\widetilde{\Sigma}_{AB}(\varphi)}{\partial\varepsilon_A\partial\varepsilon_B} 		\right)_{\pmb{\varepsilon}=0}$ & $\mathcal{T}^{(1)}_{A}(-\omega_A;\Omega_A) 		\,\mathcal{T}^{(1)}_{B}(-\omega_B;\Omega_B)$ \\
		[1mm] \hline 
	\end{tabular} 
	\caption{\label{table/2e-tomography} Experimental signal
	$\mathbf{X}_{n,\Omega_\Delta}(\pmb{\omega}_e)$ and kernel
	$\mathbf{H}_{n,\Omega_\Delta}(\pmb(\omega)$ for the two-electron
	tomography protocol equations
	\eqref{eq/tomo-g2/T-periodic/generic-form}, depending on the
	frequencies $\Omega_{A/B}=\pi nf\pm\Omega_\Delta$ of the tone
	applied to the photo-assisted filters $\mathcal{F}_A$ and
	$\mathcal{F}_B$. Here  $\Sigma_{AB}=S^{(\text{exp})}_{AB}/e^2f$
	denotes the dimensionless low frequency time averaged current
	correlation and  $\widetilde{\Sigma}_{AB}(\varphi)$ denotes the
	average of $\Sigma_{AB}$ over the two phase configurations given by
	\cref{eq/tomo-g2/T-periodic/phase-configuration}.} 
\end{table*}

\subsubsection{Order of magnitude of the experimental signal}
\label{sec/protocols/G2/signal-estimate}

\begin{figure} 	
	\resizebox{7.5cm}{!}{
\begin{tikzpicture}

    \draw[dashed, line width=1pt] (-3, 0, -3) -- (3, 0, -3) -- (3, 0, 3) -- (-3, 0, 3) node[left, scale=1.75, opacity=1, xshift=-3pt] {$\Omega_\Delta = 0\,$} -- cycle;
    \node[inner sep=0, cm={1.4, 0, 0.55, 0.55, (0, 0)}] at (0, 0, 0) {\includegraphics[width=4cm]{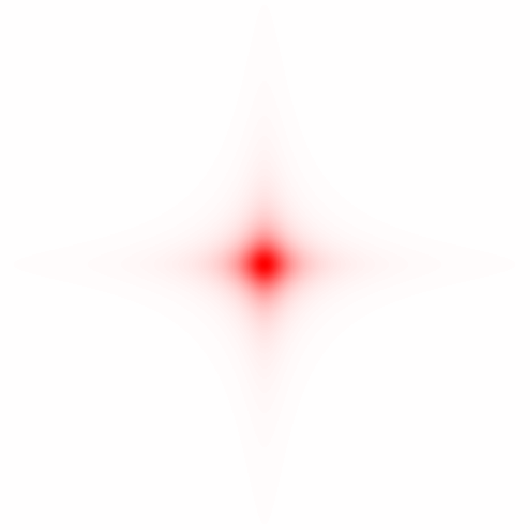}};
    \filldraw (0, 0, 0) circle (2pt);

    \draw[dashed, line width=1pt] (-3, 3, -3) -- (3, 3, -3) -- (3, 3, 3) -- (-3, 3, 3) node[left, scale=1.75, opacity=1, xshift=-3pt] {$\Omega_\Delta = 0.5\, \gamma$} -- cycle;
    \node[inner sep=0, cm={1.4, 0, 0.55, 0.55, (0, 0)}] at (0, 3, 0) {\includegraphics[width=4cm]{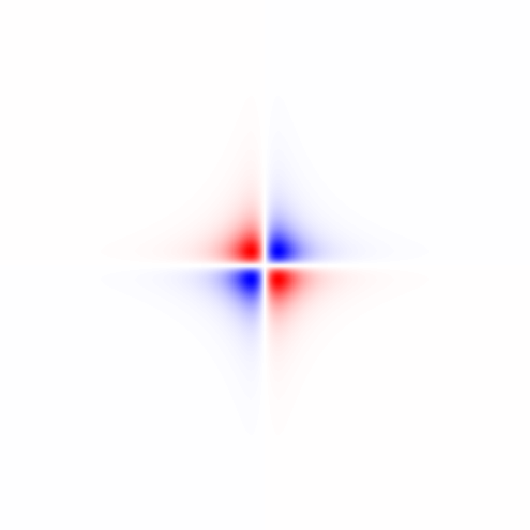}};
    \filldraw (0, 3, 0) circle (2pt);

    \draw[dashed, line width=1pt] (-3, 6, -3) -- (3, 6, -3) -- (3, 6, 3) -- (-3, 6, 3) node[left, scale=1.75, opacity=1, xshift=-3pt] {$\Omega_\Delta = 8\, \gamma\,$} -- cycle;
    \node[inner sep=0, cm={1.4, 0, 0.55, 0.55, (0, 0)}] at (0, 6, 0) {\includegraphics[width=4cm]{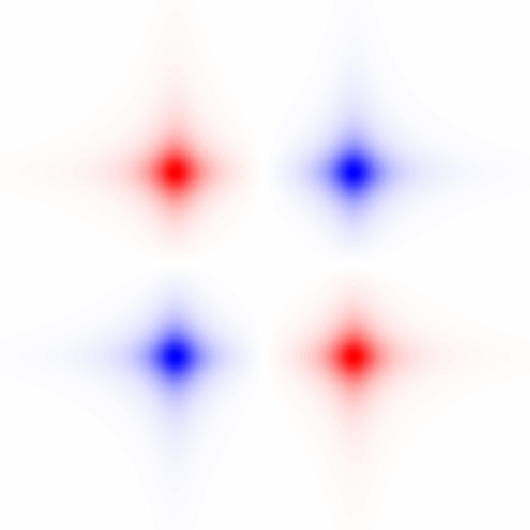}};
    \filldraw (0, 6, 0) circle (2pt);

    \draw[-{Stealth}, line width=1pt] (0, -2, 0) -- (0, 8, 0) node[above, scale=1.75] {$\Omega_\Delta$};
    \draw[-{Stealth}, line width=1pt] (-4, 0, 0) -- (4, 0, 0) node[right, scale=1.75] {$\omega_A$};
    \draw[-{Stealth}, line width=1pt] (0, 0, 4) -- (0, 0, -4) node[right, scale=1.75] {$\omega_B$};

    \node at (7, 4) {\includegraphics[width=2.25cm]{figures/colorbar.png}};

\end{tikzpicture}%
} 	
	\caption{\label{fig/3D-principle-of-tomography} Three dimensional
	view of the filtering used to relate
	$\DGee_{AB,n=0}(\pmb{\omega}|\Omega_\Delta)$ to the experimental
	tomography current correlation signal. For $\Omega_\Delta=0$, the filtering corresponds to a
	two-electron spectroscopy via the two static filters which, for
	convenience, are assumed here to have the same spectral width
	$\gamma$. For a given $\Omega_\Delta\neq 0$, the filtering is given
	by the product of two $\mathcal{T}^{(1)}$ filters (see
	Table~\ref{table/2e-tomography}).}
\end{figure}

The quantity reconstructed at the two-particle stage is the regular two-body part
$\DWWee_{AB}$. At fixed $(n,\Omega_\Delta)$, the maximum signal is expected when one peak of the filter
depicted on \cref{fig/3D-principle-of-tomography}
captures a local maximum of $\DWWee_{AB,n}(\pmb{\omega}|\Omega_\Delta)$,
which has the dimension of a time. Denoting by
$\Gamma_{\alpha}=\int_{\mathbb{R}}\mathcal{T}_{\alpha}(\omega)\,\md\omega$ the total
capture rate of each energy filter $\alpha=A$ or $B$ of spectral width
$\gamma_\alpha$ ($\Gamma_\alpha=\pi\gamma_\alpha/2$ for a Lorentzian filter
of full width at half maximum $\gamma_\alpha$ and unit maximum
transmission), the
signal associated with the two-body irreducible part $\DWGee_{AB}$ at operating point
$(\pmb{\omega}_e,\Omega_\Delta)$ is of order
\begin{equation}
	\label{eq/G2/signal-estimation/1}
	\frac{S_{AB}^{\text{2-body}}}{e^2f}\sim
	\frac{\Gamma_A\Gamma_B}{f^2}\,f\,\big|\DWWee_{AB,n}(\pmb{\omega}_e|\Omega_\Delta)\big|\,.
\end{equation}
This estimate holds for filters narrow compared to the scale over which
$\DWWee_{AB}$ varies. In the dilute regime, the harmonics are proportional
to the repetition rate, $\DWWee_{AB,n}=f\,\DWWee_{AB,(1)}|_{\Omega_\Sigma=2\pi nf}$,
where the subscript $(1)$ refers to a single two-electron collision
(\cref{appendix/dilute-to-periodic/limit/comb}), so that
\begin{equation}
	 \label{eq/G2/signal-estimation/2}
S^{\text{2-body}}_{AB}\sim
	e^2f\,\Gamma_A\Gamma_B\,|\DWWee_{AB,(1)}(\pmb{\omega}_e|2\pi
	nf,\Omega_\Delta)|\,
\end{equation}
scales as $e^2f$, as expected for one collision per period. One expects 
$|\DWWee_{AB,(1)}(\pmb{\omega}_e|2\pi
nf,\Omega_\Delta)|$ to be of the order of the product of two coherence
times, one relative to the center of mass and the other to the 
relative particle, times a dimensionless function
$\mathcal{K}(\pmb{\omega}_e|\Omega_\Sigma,\Omega_\Delta)$, thus leading
to an estimate of the order of
\begin{equation}
     \label{eq/G2/signal-estimation/3}
S^{\text{2-body}}_{AB}\sim
e^2f\,\Gamma_A\Gamma_B\tau_\Delta\tau_\Sigma\,
\mathcal{K}(\pmb{\omega}_e|2\pi nf,\Omega_\Delta)\,.
\end{equation}
Here $\tau_\Delta$ is of the order of the inverse of the energy scale
exchanged between the two electrons during the collision, whereas
$\tau_\Sigma$ is set by the spread of the total energy of the pair, which
can only be modified by exchanges with the outside (dissipation,
time-dependent drives).

The current correlation scales as the product of the two capture rates,
$\Gamma_A\Gamma_B$: the narrow bandwidth of the static filters reduces the
signal by the dimensionless factor $\Gamma_A\Gamma_B\tau_\Sigma\tau_\Delta$,
which compares the capture rates to the inverse coherence times. The ratios
$\Gamma_{A/B}/f$ entering \eqref{eq/G2/signal-estimation/1} compare instead
the capture rate of each filter to the repetition rate. For Lorentzian
filters with unit maximum probability of transmission, they can be expressed
in terms of the saturation currents
$I^{(\text{sat})}_{A/B}$ across the corresponding static energy filter 
\begin{equation} 	
	\label{eq/G2/signal-estimation/prefactors} 
	\frac{\Gamma_{A/B}}{f}=2\pi\,\frac{I^{(\text{sat})}_{A/B}}{ef}\,. 
\end{equation} 
A new feature appears, however, as soon as the source emits a delocalized
excitation ($\Ge_{AB}\neq0$). The connected coherence then carries the
singular exchange comb of \cref{eq/G2/Dc-to-DW}, which is joined to the
experimental signal in the tomography equation
\eqref{eq/tomo-g2/T-periodic/Z-definition}. Convolving a
$\delta$-singularity with the finite-bandwidth filters saturates one
of the two energy integrals, so that this exchange contribution scales as a
\emph{single} power $\Gamma$, rather than $\Gamma_A\Gamma_B$. Denoting by
$\Delta\omega_{\text{coh}}$ the energy scale over which $\DWWee_{AB}$ varies,
and assuming for simplicity that the two coherence times entering
\eqref{eq/G2/signal-estimation/3} are comparable,
$\tau_\Sigma\sim\tau_\Delta\sim1/\Delta\omega_{\text{coh}}$, the ratio of the singular exchange contribution to the genuine two-body
signal, evaluated on an exchange plane $\mathcal{E}$, is of order
\begin{equation}
	\label{eq/G2/singular-vs-regular}
	\frac{S_{AB}^{\text{exch}}}{S_{AB}^{\text{2-body}}}\sim
	\frac{\Delta\omega_{\text{coh}}}{\gamma}\,\big|\Ge_{AB}\big|^2\,.
\end{equation}
For narrow band filters ($\gamma\ll\Delta\omega_{\text{coh}}$) required to
resolve the coherence and for a well-delocalized excitation
($|\Ge_{AB}|^2\sim\mathcal{O}(1)$), this ratio is large: on and near the
exchange planes the singular contribution dominates the current
correlation. Consequently, the two-body signal is only a small residual on top of it.
This is precisely the regime in which an energy detector with too sharp a
resolution becomes counterproductive.

This has two consequences: First, the experimental noise is set by the total
correlation signal, itself dominated by the singular contribution. The
signal-to-noise ratio for $\DWWee_{AB}$ is accordingly degraded by a factor
$\sim\gamma/\Delta\omega_{\text{coh}}$. Second, because the exchange
contribution is computed from the reconstructed $\Ge_{AB}$ and subtracted
from the raw signal to form
\eqref{eq/tomo-g2/T-periodic/Z-definition}, one extracts a small residual
from a large known quantity: a relative error $\varepsilon$ on the reconstructed
$\Ge_{AB}$ produces an absolute error of order
$\varepsilon\,(\Delta\omega_{\text{coh}}/\gamma)$ on the two-body signal, which
must remain below it. This sets a stringent requirement on the precision of
the first, single-electron stage. This is a quantitative expression of the nested
character of the reconstruction.

Both effects are mitigated by operating away from the exchange planes
$\mathcal{E}_{n,k}$.
At a detuning $\delta_\omega\gg\gamma$ from a plane $\mathcal{E}$, the Lorentzian tail of
the filters suppresses the singular contribution as $(\gamma/\delta_\omega)^2$. Choosing
$\sqrt{\gamma\,\Delta\omega_{\text{coh}}}\ll\delta_\omega\ll\Delta\omega_{\text{coh}}$,
which is possible since $\gamma\ll\Delta\omega_{\text{coh}}$, reverses the ratio
\eqref{eq/G2/singular-vs-regular} so that the regular two-body signal now
dominates, while keeping the excluded neighborhoods narrow compared to the
scale $\Delta\omega_{\text{coh}}$ over which $\DWWee_{AB}$ varies.
Since $\DWWee_{AB}$ is assumed to carry no $\delta$-singularity
(\cref{appendix/G2/T-periodic}), it can be reconstructed at operating points
away from the planes and interpolated across them, provided it is also
continuous across the planes $\mathcal{E}_{n,k}$. The
filter width $\gamma$ is thus chosen as a compromise: narrow enough to
resolve the coherence but broad enough to keep the singular contribution
manageable.

Finally, this complication is specific to geometries with inter-channel
delocalization. In the Coulomb-coupler geometry of
\cref{fig/geometries}-(b), $\Ge_{AB}=0$: the exchange comb is absent,
$\DcGee_{AB}=\DWGee_{AB}$, and the whole current correlation is the two-body
signal itself. The estimates given for the collision experiment in
\cref{sec/2e-collision} therefore apply directly, with no singular
background to subtract.

Note that, since the ratio \eqref{eq/G2/singular-vs-regular} decreases
as $1/\gamma$, broader energy filters would not only increase the
collected signal but also dilute the singular exchange contribution. A
detailed analysis of such broadband detection schemes requires a more
realistic modeling of the photo-assisted filters and is left for future
work.

\subsection{Electronic decoherence effects} 
\label{sec/protocols/decoherence}

The photo-assisted filter considered in this work consists of two
successive stages. A side-gate driven by a weak monochromatic voltage
\begin{equation}
	V_g(t)=V_{\rm ac}\cos(\Omega t+\varphi),
\end{equation}
first generates photo-assisted transitions, while a downstream
quantum-dot filter selects electrons at energy $\hbar\omega_e$.
In the ideal description, the side-gate only provides the
time-dependent phase responsible for the photo-assisted amplitudes
$p_n$. In a realistic device, however, the gate is capacitively coupled
to the edge channel and to its electromagnetic environment which leads
to electronic decoherence and breaks the ideal picture discussed in
\cref{sec/protocols/principles}.

For example, in the
absence of external drive, {\it id est} when performing electronic spectroscopy, 
an electron
of energy $\hbar\omega$ propagating beneath the gate may 
either leave the interaction region with the same energy,
or lose energy in its external environment or by creating
electron-hole pairs. Since the quantum-dot detector
is energy selective, inelastic processes induce errors in the
spectroscopy. 

The key point is that this discussion remains valid even in the presence of
a coherent external drive. In the presence of an external drive, 
the amplitude for the electron to be scattered according to the
photo-assisted transition amplitudes \cref{eq/tomo-g2/photoassisted/6}
is indeed multiplied
by the elastic scattering amplitude 
$\mathcal{Z}_1(\omega)$ which is the elastic scattering amplitude for an
electron of energy $\hbar\omega$ crossing the gate in the absence of
the ac drive \cite{Degio-2009-1}. All other processes lead to inelastic errors
whose total probability is thus given by
\begin{equation}
	\eta(\omega)
	=
	1-|\mathcal{Z}_1(\omega)|^2 .
	\label{eq:eta_def}
\end{equation}
The problem is thus reduced
to estimating the elastic scattering probability given by the r.h.s. of
\cref{eq:eta_def}.
This quantity has been studied previously for a capacitively
coupled gate connected to an external electromagnetic impedance.

At low energy, the finite-frequency admittance of the gate-edge system
is characterized by an electrochemical capacitance $C_\mu$ and a charge
relaxation resistance $R_q$. The inelastic probability behaves as
\cite{Degio-2011-1}:
\begin{equation}
	\eta(\omega)
	\simeq
	\left(
	\frac{R_q}{R_K}
	-\frac12
	\right)
	\left(
	\omega R_KC_\mu
	\right)^2,
	\label{eq:eta_Rq}
\end{equation}
where $R_K=h/e^2$ is the resistance quantum and $R_q$ is the relaxation
resistance \cite{Buttiker-1993-1}. When the side-gate is connected to
a transmission line of impedance $R$, 
$R_q=R_K/2+R$, which leads to the error probability estimate:
\begin{equation}
	\eta(\omega)
	\simeq
	\frac{R}{R_K}
	\left(
	\omega R_KC_\mu
	\right)^2\,.
	\label{eq:eta_low}
\end{equation}
Equation~\eqref{eq:eta_low} shows that the error probability grows
quadratically in the electron energy. Requiring this error to remain below a prescribed
value $\eta$ constrains the energy range 
for the quantum-dot energy filter: $|\omega_e|$ must remain below
\begin{equation}
	\omega_e(\eta)
	\simeq
	\frac{1}{R_KC_\mu}
	\sqrt{\frac{\eta R_K}{R}},
	\label{eq:omegamax}
\end{equation}
which provides a practical criterion for choosing both the operating
energy and the gate geometry. 

The characteristic energy governing
electronic relaxation is
$E_c=\hbar/R_KC_\mu$
where $C_\mu$ denotes the electrochemical capacitance of the coupled
gate-edge system. Using the estimate
\begin{equation}
	\frac{1}{R_KC_\mu}
	\simeq
	\frac{v_F}{l}
	\left(
	1+\frac{C_q}{C_g}
	\right),
	\label{eq:Cmu_est}
\end{equation}
together with
\begin{equation}
	\frac{C_q}{C_g}
	\simeq
	\frac{\alpha_{\rm qed}}{\varepsilon_r}
	\frac{c}{v_F}\sim 1.76,
	\label{eq:Cratio}
\end{equation}
in AlGaAs/GaAs where $\varepsilon_r\sim 12.4$ and
$v_F\simeq\SI{e5}{\meter\per\second}$, one finds
that $E_c$ scales approximately as $1/l$. Hence, reducing the
gate length pushes the onset of inelastic relaxation towards higher
energies, thereby enlarging the energy window over which the
photo-assisted filtering stage remains essentially decoherence free.

For a $\SI{0.5}{\micro\meter}$ long side-gate,
a more detailed estimate (see \cref{appendix/collision-model/coupling-constant})
leads to
$E_c\sim 100$ to $\SI{200}{\micro\electronvolt}$ corresponding to
a characteristic frequency  $E_c/h$ of the order of
$24$ to $\SI{48}{\giga\hertz}$. 
For a
$\SI{50}{\ohm}$ transmission line and a 1~\% error probability, 
\cref{eq:omegamax} gives $\omega_e(\eta=0.01)\simeq 2.27/R_KC_\mu$
corresponding to a frequency
between $50$ and
$\SI{100}{\giga\hertz}$. This indicates that, for electron quantum
optics experiments performed with electronic sources such as the
mesoscopic capacitor \cite{Feve-2007-1} or state of the art Leviton
sources such as the ones used in Ref.~\cite{Bartolomei-2025-1}, 
electronic decoherence only provides a minor correction to the ideal
photo-assisted filtering process.

In practice, 
switching off the ac drive
($V_{\rm ac}=0$) and performing the energy spectroscopy of a
well-controlled nearby stationary source such as a biased QPC
\cite{Altimiras-2010-1} enables determining
the elastic
transmission probability across the photo-assisted filtering stage as a
function of $\omega$ and
therefore of the error probability $\eta(\omega)$ over the targeted range
of electronic energies.%

\section{Interaction-induced entanglement detection}
\label{sec/2e-collision}

In this section, we illustrate the evolution of the two-electron coherence and the emergence of electronic entanglement during a two-electron coherent collision. To do so, we consider a minimal model of two electronic excitations propagating along different chiral channels and interacting through a finite-length capacitive Coulomb coupler (see \cref{fig/geometries}-(b)). The resulting energy entanglement can be detected using the witness introduced in \cref{eq/gen-crit/G2}.

\subsection{A simple physical picture of energy exchange}
\label{sec/2e-collision/picture}

Before entering into the details of the model, it is useful to sketch
the simple physical picture that underlies the entanglement generation
discussed throughout this section. We are considering the collision of
two particles
in a setup depicted on \cref{fig/geometries}-(b). In such a
process, Coulomb interactions within the coupler
conserves the total energy $\omega_\Sigma=\omega_A+\omega_B$ of
the two incoming electrons but coherently redistributes energy
between them. In the familiar two-body problem of mechanics, this
redistribution changes the energy of the \emph{relative particle}, here given
by the imbalance 
$\omega_\Delta=(\omega_A-\omega_B)/2$.

Consider first the idealized situation of two electrons with perfectly
well-defined energies, entering the coupler exactly at the same time.
Coulomb interaction then coherently exchanges a characteristic energy
quantum $\hbar\omega_{AB}$ between them, and does so with equal
probability amplitude in both directions: electron $A$ can hand over
$\hbar\omega_{AB}$ to electron $B$ or receive it from $B$. Because
these two outcomes originate from the very same coherent scattering
event, the relative particle is left in a coherent {superposition} of an
energy gain and an energy loss, i.e.\ of
$\omega_\Delta\to\omega_\Delta+\omega_{AB}$ and
$\omega_\Delta\to\omega_\Delta-\omega_{AB}$, rather than in a
statistical mixture of the two. It is precisely the type of
superposition considered in
\cref{sec/g2-and-entanglement/examples-setups}: a superposition of two
Slater determinants which is not separable between the modes of $A$ and
$B$ but entangled.

In practice, the incoming electrons are finite-duration wave packets
rather than monochromatic plane waves, so this elementary picture gets
dressed by their finite spectral width. It nonetheless remains the
common thread behind all the structures discussed below: the side peaks
of the diagonal coherence
(\cref{sec/2e-collision/outgoing-G2/diagonal}), the off-diagonal lobes
of the frequency-domain coherence
(\cref{sec/2e-collision/outgoing-G2/off-diagonal}), and the interference
fringes of the relative-particle Wigner function
(\cref{sec/2e-collision/entanglement/Wigner}) are all different
manifestations of the same coherent superposition between two opposite
energy-exchange processes.

\subsection{Capacitive model for two-electron scattering} \label{sec/2e-collision/model}

\begin{figure}[h!]
	\centering
	\resizebox{0.4\textwidth}{!}{\begin{tikzpicture}[line cap=round, line join=round]
	
	\path[use as bounding box] (-6, -4) rectangle (6, 4);
		
		\coordinate (Lt1) at (-5, 3);
		\coordinate (Lt2) at (-3, 1);
		\coordinate (Rt1) at (3, 1);
		\coordinate (Rt2) at (5, 3);
		
		\coordinate (Lb1) at (-5, -3);
		\coordinate (Lb2) at (-3, -1);
		\coordinate (Rb1) at (3, -1);
		\coordinate (Rb2) at (5, -3);
		
		\coordinate (Cmid) at (0, 0);
	
		\draw[gray!80, line width=2pt] ($(Cmid)+(0, 0.2)$) -- ++(0, 0.8);
		\draw[gray!80, line width=2pt] ($(Cmid)+(-0.5, 0.2)$) -- ++(1, 0);
		\draw[gray!80, line width=2pt] ($(Cmid)+(0, -0.2)$) -- ++(0, -0.8);
		\draw[gray!80, line width=2pt] ($(Cmid)+(-0.5, -0.2)$) -- ++(1, 0);
		
		\node[gray!80, scale=1.5] at (-1, 0) {$C$};

		\coordinate (CgA) at (1.5, 1);
		
		\draw[gray!80, line width=2pt] ($(CgA)$) -- ++(0, 0.7);
		\draw[gray!80, line width=2pt] ($(CgA)+(-0.3, 0.7)$) -- ++(0.6, 0);
		\draw[gray!80, line width=2pt] ($(CgA)+(-0.3, 0.9)$) -- ++(0.6, 0);
		\draw[gray!80, line width=2pt] ($(CgA)+(0, 0.9)$) -- ++(0, 0.5);
	
		\draw[gray!80, line width=2pt] ($(CgA)+(-0.3, 1.4)$) -- ++(0.6, 0);
		\draw[gray!80, line width=2pt] ($(CgA)+(-0.2, 1.55)$) -- ++(0.4, 0);
		\draw[gray!80, line width=2pt] ($(CgA)+(-0.1, 1.7)$) -- ++(0.2, 0);
	
		\node[gray!80, scale=1.5] at (0.5, 2.1) {$c_{gA}$};
		
		\coordinate (CgB) at (1.5, -1);
		
		\draw[gray!80, line width=2pt] ($(CgB)$) -- ++(0, -0.7);
		\draw[gray!80, line width=2pt] ($(CgB)+(-0.3, -0.7)$) -- ++(0.6, 0);
		\draw[gray!80, line width=2pt] ($(CgB)+(-0.3, -0.9)$) -- ++(0.6, 0);
		\draw[gray!80, line width=2pt] ($(CgB)+(0, -0.9)$) -- ++(0, -0.5);

		\draw[gray!80, line width=2pt] ($(CgB)+(-0.3, -1.4)$) -- ++(0.6, 0);		
		\draw[gray!80, line width=2pt] ($(CgB)+(-0.2, -1.55)$) -- ++(0.4, 0);
		\draw[gray!80, line width=2pt] ($(CgB)+(-0.1, -1.7)$) -- ++(0.2, 0);

		\node[gray!80, scale=1.5] at (0.5, -2.1) {$c_{gB}$};

		\draw[red, dashed, line width=2pt] ($(Lt2)+(0, 0.3)$) -- ($(Lb2)+(0, -0.3)$) -- ($(Rb1)+(0, -0.3)$) -- ($(Rt1)+(0, 0.3)$) -- cycle;
		
		\draw[blue, line width=2pt, fill, fill opacity=0.3] ($(Lt1)!0.3!(Lt2)$) .. controls (-2.8, 3.4) .. ($(Lt1)!0.65!(Lt2)$) -- cycle;
		\node[blue, scale=1.5] at (-2.8, 3.4) {$\omega_{A}$};
		
		\draw[red, line width=2pt, fill, fill opacity=0.3] ($(Lb1)!0.2!(Lb2)$) .. controls (-5, -1) .. ($(Lb1)!0.75!(Lb2)$) -- cycle;
		\node[red, scale=1.5] at (-5.1, -0.9) {$\omega_{B}$};

		\node[left=5pt, scale=1.5]  at (Lt1) {$A_{\rm in}$};
		\draw[line width=2.3pt] (Lt1) -- (Lt2) -- (Rt1) -- (Rt2);
		\draw[-{Triangle[length=9pt,width=9pt]}, line width=2pt] ($(Lt1)!0.5!(Lt2)$) -- ($(Lt1)!0.55!(Lt2)$);
		\draw[-{Triangle[length=9pt,width=9pt]}, line width=2pt] ($(Lt2)!0.3!(Rt1)$) -- ($(Lt2)!0.35!(Rt1)$);
		\draw[-{Triangle[length=9pt,width=9pt]}, line width=2pt] ($(Rt1)!0.5!(Rt2)$) -- ($(Rt1)!0.55!(Rt2)$);
		\node[right=5pt, scale=1.5] at (Rt2) {$A_{\rm out}$};

		\node[left=5pt, scale=1.5]  at (Lb1) {$B_{\rm in}$};
		\draw[line width=2.3pt] (Lb1) -- (Lb2) -- (Rb1) -- (Rb2);
		\draw[-{Triangle[length=9pt,width=9pt]}, line width=2pt] ($(Lb1)!0.5!(Lb2)$) -- ($(Lb1)!0.55!(Lb2)$);
		\draw[-{Triangle[length=9pt,width=9pt]}, line width=2pt] ($(Lb2)!0.3!(Rb1)$) -- ($(Lb2)!0.35!(Rb1)$);	
		\draw[-{Triangle[length=9pt,width=9pt]}, line width=2pt] ($(Rb1)!0.5!(Rb2)$) -- ($(Rb1)!0.55!(Rb2)$);
		\node[right=5pt, scale=1.5] at (Rb2) {$B_{\rm out}$};

		\draw[red, {Stealth}-{Stealth}, line width=1.5pt] ($(Lb2)+(0, -2)$) -- ($(Rb1)+(0, -2)$) node[midway, below, red, scale=1.5] {$L$};
	
	\end{tikzpicture}%
}
	\caption{\label{cap_model} Schematic of the Coulomb coupler between two chiral edge channels, $A$ and $B$, carrying single-electron excitations with respective energies $\omega_A$ and $\omega_B$. The red dashed contour highlights the finite interaction region of length $L$, where the channels are coupled through an inter-edge capacitance $C$. Each channel is also connected to the ground through gate capacitances $c_{gA}$ and $c_{gB}$.}
\end{figure}
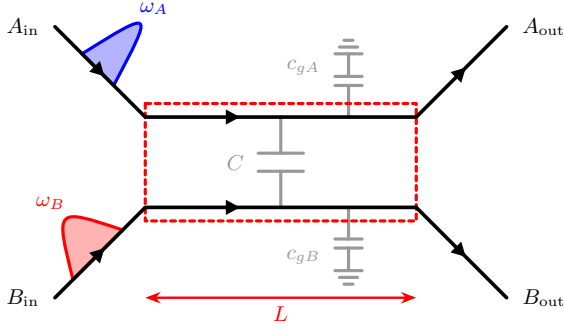

The system studied here is composed of two co-propagating chiral channels labeled $A$ and $B$, and schematically shown in \cref{cap_model}. A single-electron wave packet is injected in each channel, propagating ballistically at velocity $v_F$.

The coupling between the two channels originates from Coulomb interactions. It is accounted for in our system by using an effective capacitive coupling, based on a discrete element model over a finite region of length $L$ introduced in Refs.~\cite{Bocquillon-2013-1,Grenier-2013-1}. Inside the Coulomb coupler, the electrons experience within each channel a potential $U_\alpha(t)$ ($\alpha=A$ or $B$) depending on the total charge present within the coupler. The capacitance matrix describing these two conductors involves the mutual capacitance $C$ coupling the two channels, in addition to capacitances to the ground $c_{gA}$ and $c_{gB}$. The characteristic interaction time is set by the ballistic traversal time $\tau_L=L/v_F$ across the radiation coupler.

This model has already been used to study electronic decoherence and fractionalization \cite{Cabart-2018-1,Grenier-2013-1} within quantum Hall edge channels. Here however, we consider a much simpler situation in which only one electron propagates within each chiral channel, thereby realizing a coherent two-particle scattering without any quantum exchange effect. In such a collision, the total energy is conserved and energy is exchanged between the two particles - this is precisely the coherent exchange described in \cref{sec/2e-collision/picture}, and the cause of the resulting energy entanglement.

A key ingredient of the model is the electrostatic phase $\phi_{A/B}(t)=\frac{i
e}{\hbar} \int_{-\infty}^tU_{A/B}(t')\, dt'$ accumulated by each particle as it
flies across the Coulomb coupler \cite{Souquet-Basiege-2024-1,Frigerio-2024-1}. 
Due to the mutual capacitance $C$, it contains
a coupling contribution $\phi_{AB}(t_\Delta^0)$, function of the time delay
$t_\Delta^0 = t_A^0 - t_B^0$ between the injections of the two electrons in the
coupler. In the present geometry and model, $\phi_{AB}(t_\Delta^0)$ is non-zero
only when the two electrons overlap within the Coulomb coupler, and varies
linearly with their time delay (see
\cref{eq/collision-model/interaction-phase}). The magnitude of this
interaction-induced phase determines the characteristic energy scale
$\hbar\omega_{AB}$ governing coherent energy exchange between the two electrons.
The product $\alpha_{\text{eff}}=\omega_{AB}\tau_L$ is the maximum phase shift
arising from Coulomb interactions (see Eq.~\eqref{eq/collision-model/9}). This
dimensionless coupling constant can also be expressed as
\begin{equation}
	\alpha_{\text{eff}}=4\pi\,\frac{\alpha_{\text{QED}}}{\varepsilon_R}\,\frac{c}{v_F} \,\zeta^{-1} \,.
\end{equation}
Under this form, it appears as an effective electromagnetic coupling constant, renormalizing the fine structure constant $\alpha_{\text{QED}}$ by the relative permittivity $\varepsilon_R$ of the underlying material, the ratio of velocities $c/v_F$, and a dimensionless factor $\zeta$ encoding the geometry of the Coulomb coupler. As in previous works \cite{Grenier-2013-1,Cabart-2018-1,Delgard-2021-1}, the assumption that capacitances are proportional to lengths implies that $\alpha_{\text{eff}}$ only depends on the geometry through the aspect ratio factor $\zeta$ given by \cref{eq/collision-model/8}; the choice of $L$ mainly fixes the frequency scale $\tau_L^{-1}=v_F/L$ to which $\omega_{AB}$ is proportional.

For typical AlGaAs/GaAs heterostructures, $v_F \simeq \SI{e5}{\meter\second^{-1}}$ and $\varepsilon_R=12.4$. Besides, for two $\SI{100}{\nano\meter}$-wide edge channels separated by $\SI{1}{\micro\meter}$, and co-propagating over a distance $L=\SI{50}{\micro\meter}$, one obtains $\zeta=2.47$ (see Appendix \ref{appendix/collision-model}). These parameters finally give $\alpha_{\text{eff}} \sim 9$, corresponding to $\omega_{AB}/2\pi\sim \SI{2.9}{\giga\hertz}$ and $\tau_L\sim \SI{0.5}{\nano\second}$. The outgoing two-electron coherence can now be computed from the incoming two-electron wave function.

\subsection{Energy-domain structure of $\mathcal{G}_{AB}^{(2e)}$} \label{sec/2e-collision/frequency-space-discussion}

Because Coulomb interaction induces a coherent exchange of energy
between the two electrons, one expects entanglement to emerge in the
energy sector. This motivates the study of two-electron coherence in the
frequency domain. 
As expected from the toy example considered in
\cref{sec/g2-and-entanglement/examples-setups}, the inter-channel two-electron
coherence in the frequency domain then contains off-diagonal components,
reflecting coherent superpositions of processes involving different
energy exchanges.

In the frequency domain, the two-electron coherence is naturally
expressed in terms of the variables $(\omega_{A}^{+}, \omega_{B}^{+},
\omega_{A}^{-}, \omega_{B}^{-})$ associated with each particle. We can
therefore work with the corresponding classical and quantum variables
$(\omega_{A}, \omega_{B}; \Omega_A, \Omega_B)$, namely the Fourier
conjugates of $(\tau_A,\tau_B;t_{A}, t_{B})$.

We are dealing here with a two-body problem in which each electron is
assigned to a given chiral channel. 
Hence, we naturally introduce the corresponding
center-of-mass and relative particle coordinates already anticipated in
\cref{sec/2e-collision/picture}. In the frequency domain, the classical
variables are the Fourier conjugates to $(\tau_\Sigma, \tau_\Delta)$:
$\omega_\Sigma = \omega_A + \omega_B$ is associated with the total
energy of the two-particle system, and $\omega_\Delta = \frac{1}{2}\,
(\omega_A - \omega_B)$ represents the energy imbalance between them,
i.e.\ the energy of the relative particle. We then have $\omega_{A/B} =
\frac{\omega_\Sigma}{2} \pm \omega_\Delta$. The corresponding quantum
variables are $\Omega_\Sigma = \Omega_A + \Omega_B$ and $\Omega_\Delta =
\frac{1}{2} ( \Omega_A - \Omega_B)$, the Fourier conjugate to
$(t_\Sigma, t_\Delta)$. All the variables are summarized in Table
\ref{table/conjugate-variables} in Appendix \ref{appendix/G2}.

\subsection{Post-collision two-electron coherence} 
\label{sec/2e-collision/outgoing-G2}

We consider the case of Landau excitations, for instance generated by a mesoscopic capacitor in the single-electron regime of operation \cite{Feve-2007-1}. In the time domain, a Landau wave packet emitted at time $ t_{\xi}^{0}$ around the energy $\hbar\omega_{\xi}^{0}$, with a duration $\gamma_{\xi}^{-1}$ is given by
\begin{equation}
	\label{landau_wp}
	\psi_{\xi}(t) = -i \sqrt{\gamma_{\xi}}\, \exp\bigg[\Big( \mi\, \omega_{\xi}^{0} - \frac{\gamma_{\xi}}{2} \Big) \big( t - t_{\xi}^{0} \big) \bigg]\, \Theta \big( t - t_{\xi}^{0} \big),
\end{equation}
where $\xi\in\{A,B\}$ denotes the injection channel. Typical operating conditions for mesoscopic capacitors correspond to $\omega^0_\xi/2\pi\sim \SI{15}{\giga\hertz}$ and $\gamma_{\xi}^{-1}\sim \SI{0.3}{\nano\second}$, yielding $\omega^0_\xi/\gamma_\xi\sim 28$. This large ratio indicates that the emitted wave packet is well localized in energy far above the Fermi sea, thereby justifying the neglect of Fermi sea excitations in the present description. Moreover, our estimate of $\omega_{AB}/2\pi$ shows that the interaction-induced exchange remains much smaller than the central injection energy of the incoming electrons. As a consequence, the two electrons remain well above the Fermi sea even after propagation across the Coulomb coupler.

The outgoing two-electron coherence can then be computed explicitly from the time-domain scattering amplitude, as explained in Appendix \ref{appendix/collision-model}. All the results presented here have been obtained with a Python numerical code able to generate results for generic normalized wave packets \footnote{This code has been checked against the closed-form analytical expressions that can be obtained for the case of Landau excitations.}.

\subsubsection{Diagonal coherence}
\label{sec/2e-collision/outgoing-G2/diagonal}

The diagonal part of two-electron coherence is obtained for $\Omega_\Sigma=\Omega_\Delta=0$. It corresponds to the two-electron probability distribution in energy.

\begin{figure}[h!]
	\includegraphics[width=\columnwidth]{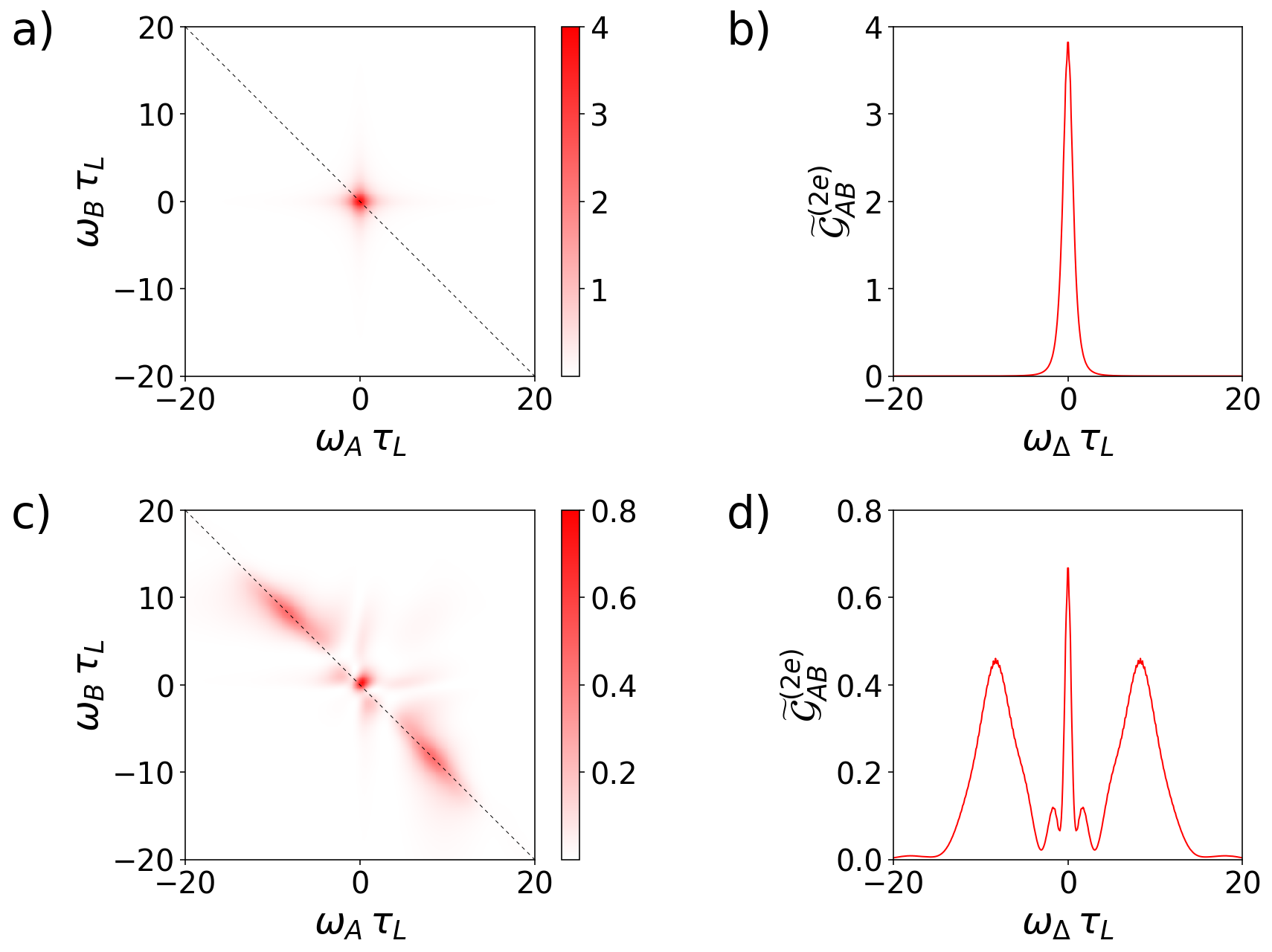}
	\caption{Incoming (top row) and outgoing (bottom row) diagonal two-electron
	coherences for two Landau wave packets injected simultaneously into the
	Coulomb coupler. Both wave packets have identical injection energies, times,
	and durations $(\omega_{A}^{0}, t_{A}^{0}, \gamma_A^{-1}) =(\omega_{B}^{0},
	t_{B}^{0}, \gamma_B^{-1}) = (0, 0, \tau_L / 2)$. The interaction strength is
	given by the coupling parameter $\alpha_{\text{eff}} = 9$. {\bf (a) and (c)}
	Incoming and outgoing coherences in the $(\omega_{A}, \omega_{B})$ plane (in
	units of $1 / \tau_{L}$). {\bf (b) and (d)} Incoming and outgoing coherences
	along the $\omega_{\Delta} = \omega_{A} - \omega_{B}$ direction (in units of
	$1/\tau_{L}$), corresponding to cuts of the left panels along the grey
	dashed lines.}
	\label{fig_g2_top}
\end{figure}

For two Landau wave packets, the incoming diagonal two-electron coherence exhibits a four-branch star-shaped structure centered at $(\omega_{A}^{0}, \omega_{B}^{0})$ (see \cref{fig_g2_top}-(a) and (b)). It reflects the Lorentzian energy distributions of the incoming states. By contrast, \cref{fig_g2_top}-(c) and (d) show the outgoing diagonal two-electron coherence after propagation through the Coulomb coupler.

The incoming cross-like structure, centered at $(\omega_A^0,\omega_B^0)$, is split during the collision into three peaks spread along the direction of constant total energy, as expected. The dominant central peak at $(\omega_{A}^{0}, \omega_{B}^{0})$ corresponds to the absence of energy transfer. Two smaller and broader side peaks appear along the $\omega_{\Delta}$ direction: they are the diagonal-coherence signature of the two branches of the superposition introduced in \cref{sec/2e-collision/picture} - the top-left peak results from electron $A$ losing energy to electron $B$, while the bottom-right one corresponds to the opposite process. The most probable energy transfer is $\pm\hbar\omega_{AB}$. Here, we focus on the strong coupling regime $\alpha_{\text{eff}} \gg 1$, that underlies the structure observed in \cref{fig_g2_top}. The weak coupling regime is discussed in Appendix \ref{appendix/collision-model/coupling-constant}, where the corresponding version of \cref{fig_g2_top} is shown.

\subsubsection{Off-diagonal coherence}
\label{sec/2e-collision/outgoing-G2/off-diagonal}

\begin{figure}[h!]
	\includegraphics[width=0.8\columnwidth]{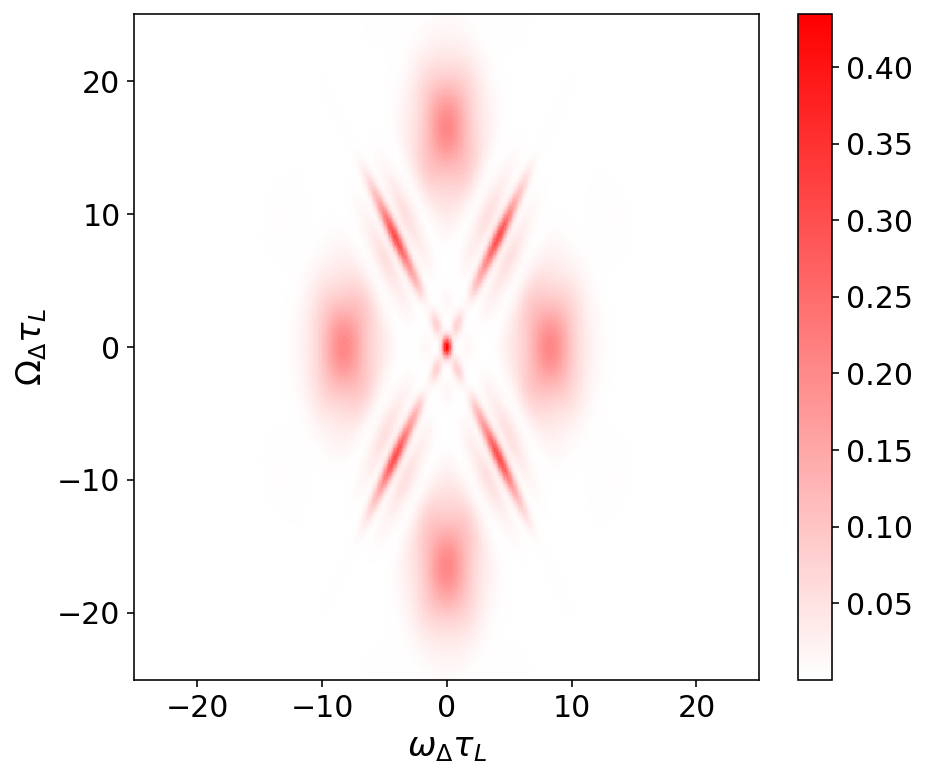}
	\caption{\label{fig_g2_face} Square modulus of the outgoing second-order coherence of two identical Landau wave packets injected simultaneously into the Coulomb coupler, in the $(\omega_{\Delta}, \Omega_{\Delta})$ plane (in units of $1 / \tau_{L}$) for $\omega_\Sigma=\omega_A^0+\omega_B^0$ and $\Omega_\Sigma=0$. The wave packets have the same parameters as in \cref{fig_g2_top}. As before, the coupling parameter is $\alpha_{\text{eff}} = 9$.}
\end{figure}

The coherent superposition described in \cref{sec/2e-collision/picture} leaves its imprint in the off-diagonal two-electron coherence. It is most conveniently visualized in the section of the four-dimensional frequency domain $(\omega_\Sigma,\omega_\Delta;\Omega_\Sigma,\Omega_\Delta)$ defined by $\Omega_\Sigma=0$ and $\omega_\Sigma=\omega_A^0+\omega_B^0$. \cref{fig_g2_face} shows the square modulus of $\widetilde{\mathcal{G}}^{(2e)}_{AB}$ in this plane, as a function of $(\omega_\Delta,\Omega_\Delta)$, for two identical Landau wave packets entering the Coulomb coupler simultaneously.

Its restriction to the line $\Omega_{\Delta} = 0$ corresponds to
\cref{fig_g2_top}-(d), and the two side-lobes visible there are seen to extend
slightly beyond this diagonal. Further away from it, two purely off-diagonal
lobes appear around $\omega_\Delta\sim 0$ and $\Omega_{\Delta} \tau_L \sim \pm 2
\alpha_{\text{eff}}$: they are the direct frequency-domain signature of the
interference between the two branches of the superposition - electron $A$ losing
an energy $\sim\omega_{AB}$ to $B$, and the reverse process. Thinner fringes
visible between the four main lobes result from the interference of either
lateral lobe with the central, no-exchange peak.

\subsection{Entanglement generation}
\label{sec/2e-collision/entanglement}

Motivated by the discussion of
\cref{sec/g2-and-entanglement/examples-setups}, 
we can now use the entanglement witness introduced in \cref{eq/gen-crit}
to probe two-electron energy entanglement generated by the coherent
scattering processes within the Coulomb coupler.

\subsubsection{Entanglement witness}
\label{sec/2e-collision/entanglement/witness}

\cref{eq/gen-crit/G2} can be rewritten in the frequency domain by
\begin{align}
	\begin{split}
		\vert \widetilde{\mathcal{G}}&^{(2e)}_{AB}(\omega_{A}^{+}, \omega_{B}^{+}\, \vert\, \omega_{A}^{-}, \omega_{B}^{-}) \vert^{2}\\ &\leq \widetilde{\mathcal{G}}^{(2e)}_{AB}(\omega_{A}^{+}, \omega_{B}^{-}\, \vert\, \omega_{A}^{+}, \omega_{B}^{-})\,  \widetilde{\mathcal{G}}^{(2e)}_{AB}(\omega_{A}^{-}, \omega_{B}^{+}\, \vert\, \omega_{A}^{-}, \omega_{B}^{+}).
		\label{crit_g2}
	\end{split}
\end{align}
As expected from \cref{sec/g2-and-entanglement/examples-setups}, violation 
of this inequality signals entanglement between 
the frequency
modes of $A$ and $B$. It is a non-linear entanglement witness well
suited to probe an entangled state generated by interactions in the setup
of \cref{fig/geometries}-(b).

For finite-duration wave packets,
a key parameter is the time delay $t_{\Delta}^{0} = t_{A}^{0} -
t_{B}^{0}$ between the two excitations. When the two wave packets are
sufficiently delayed such that they do not overlap within the radiation
coupler, entanglement is expected to be negligible. In contrast, when
they arrive simultaneously, a strong violation of \cref{crit_g2} is
expected.

This suggests analyzing the difference between the two sides of
\cref{crit_g2} as a function of the time delay $t_\Delta^0$ between the
two wave packets and of $\Omega_\Delta$, at fixed $\Omega_\Sigma=0$ and
$(\omega_A,\omega_B)=(\omega_A^0,\omega_B^0)$. Equivalently, this
corresponds to fixing $\omega_\Sigma=\omega_A^0+\omega_B^0$ and
$\omega_\Delta=0$. As discussed in \cref{sec/protocols/G2}, the
condition $\Omega_\Sigma=0$ corresponds to selecting the $n=0$ harmonic
in a $T$-periodic experiment, while $\Omega_\Delta/2\pi$ is set by the
frequency of the r.f.\ tone applied to the two photo-assisted filters.
The parameters $\omega_\Sigma$ and $\omega_\Delta$ can then be tuned
independently through the static filters via gate voltages, following
the procedure of \cite{Altimiras-2010-1}. Whether this single-collision
violation survives on the actual $T$-periodic source is not obvious:
\cref{appendix/dilute-to-periodic} shows that \cref{crit_g2} is the
$f\to0$ limit of the periodic witness, and that the periodic test inherits
the violation provided all four probed modes lie above the Fermi sea.

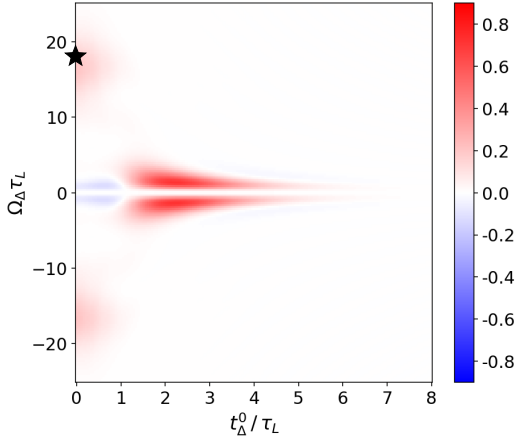
\begin{figure}[h!]
	\includegraphics[width=0.8\columnwidth]{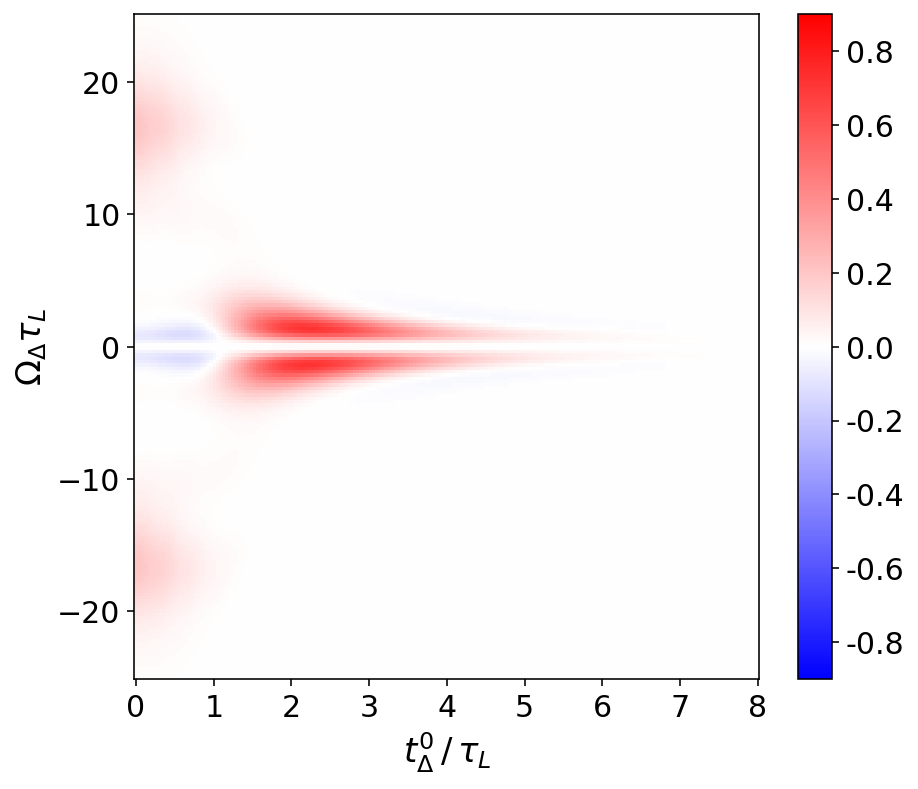}
	\caption{\label{fig_witness} Evolution of the Cauchy--Schwarz entanglement witness along the $\Omega_{\Delta}$-axis as a function of the injection time delay $t_{\Delta}^{0} = t_{A}^{0} - t_{B}^{0}$ between the two wave packets. Exactly as for \cref{fig_g2_top}, $\omega_{\Sigma} = \omega_{\Delta} = \Omega_{\Sigma} = 0$, $(\omega_{A}^{0}, t_{A}^{0}, \gamma_{A}^{-1}) = (\omega_{B}^{0}, t_{B}^{0}, \gamma_{B}^{-1}) = (0, 0, \tau_L / 2)$, and $\alpha_{\text{eff}} = 9$. The plotted quantity is the difference between the left- and right-hand sides of \cref{crit_g2}. Red regions indicate violation of the witness, and thus entanglement.}
	\begin{tikzpicture}[overlay, remember picture]
		\node[scale=0.4, star, star points=5, star point height=2mm, draw=black, fill=black] at (-2.45, 8.88) {};
	\end{tikzpicture}
\end{figure}

Figure \ref{fig_witness} exhibits two regions where the witness is violated. The first consists of two red areas located at $t_{\Delta}^{0} \lesssim \tau_L$ and $\Omega_\Delta\sim 2\omega_{AB}$, which correspond to the upper and lower structures visible in \cref{fig_g2_face}. This violation directly comes from the coherent energy-transfer processes taking place when both wave packets are sent simultaneously, and constitutes a signature of the entanglement generated by coherent two-electron scattering in the strong-coupling regime. The behavior of the witness for different wave packet durations is discussed in \cref{appendix/dependence-witness}.

\begin{figure}
	\center
	\resizebox{0.48\textwidth}{!}{\begin{tikzpicture}

    \draw[blue, very thick, fill, fill opacity=0.25] (0.8, 0) -- (0.8, 3.5) .. controls(1.3, 0.25) .. (3.8, 0) -- cycle;
    \draw[very thick] (0.8, 0.2) -- (0.8, -0.2) node[blue, below, xshift=5pt, scale=2] {$t_A^0$};
    
    \draw[red, very thick, fill, fill opacity=0.25] (0.5, 0) -- (0.5, 3.5) .. controls(1, 0.25) .. (3.5, 0) -- cycle;
    \draw[very thick] (0.5, 0.2) -- (0.5, -0.2) node[red, below, xshift=-5pt, scale=2] {$t_B^0$};
    
    \draw[-{Stealth}, very thick] (-0.5, 0) -- (5.5, 0) node[above left, scale=2] {$t$};
    \draw[-{Stealth}, very thick] (0, -0.5) -- (0, 4) node[above, scale=2] {$\lvert \psi \rvert^2$};

    \begin{scope}
        \clip (0.5, 0) -- (0.5, 3.5) .. controls(1, 0.25) .. (3.5, 0) -- cycle;
        \fill[pattern=north east lines] (0.8, 0) -- (0.8, 3.5) .. controls(1.3, 0.25) .. (3.8, 0) -- cycle;
    \end{scope}

    \draw[blue, very thick, fill, fill opacity=0.25] (8.5, 0) -- (8.5, 3.5) .. controls(9, 0.25) .. (11.5, 0) -- cycle;
    \draw[very thick] (8.5, 0.2) -- (8.5, -0.2) node[blue, below, scale=2] {$t_A^0$};

    \draw[red, very thick, fill, fill opacity=0.25] (7, 0) -- (7, 3.5) .. controls(7.5, 0.25) .. (10, 0) -- cycle;
    \draw[very thick] (7, 0.2) -- (7, -0.2) node[red, below, scale=2] {$t_B^0$};

    \draw[-{Stealth}, very thick] (6, 0) -- (12, 0) node[above left, scale=2] {$t$};
    \draw[-{Stealth}, very thick] (6.5, -0.5) -- (6.5, 4) node[above, scale=2] {$\lvert \psi \rvert^2$};

    \begin{scope}
        \clip (7, 0) -- (7, 3.5) .. controls(7.5, 0.25) .. (10, 0) -- cycle;
        \fill[pattern=north east lines] (8.5, 0) -- (8.5, 3.5) .. controls(9, 0.25) .. (11.5, 0) -- cycle;
    \end{scope}
\end{tikzpicture}%
}
	\caption{Schematics of the temporal overlap between two identical Landau
	wave packets for two distinct time delays $t_\Delta^0 = t_A^0 - t_B^0$. Due
	to their particular temporal shape - a decreasing exponential -, the overlap
	between such wave packets is non negligible only when $t_\Delta^0 \ll 1$
	(left panel). This is the reason why the energy transferred between the two
	electronic excitations decreases drastically when $t_\Delta^0$ increases
	(right panel).}
	\label{fig_overlap}
\end{figure}
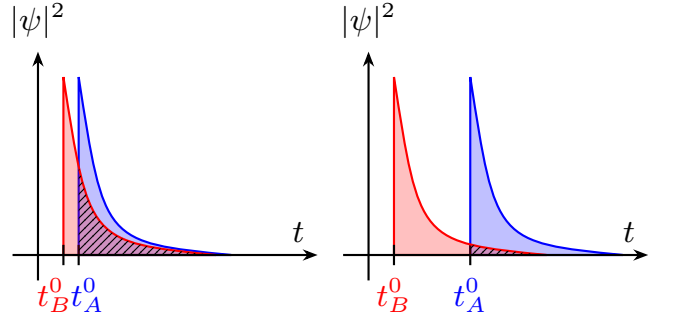

A second violation occurs for time delays in the range $\tau_{L} \lesssim t_{\Delta}^{0} \lesssim 6\tau_{L}$, in a region lying close to the diagonal axis. When the two wave packets arrive with a slight time delay, they still influence each other, but the coherent energy transfer induced by Coulomb interaction is much smaller, as shown in \cref{fig_overlap}.

Finally, for sufficiently large time delays, the difference between the two sides of \cref{crit_g2} vanishes: the wave packets are then sufficiently separated in time for the first electron to exit the interaction region before the second enters it, the interaction between the two electrons disappears, and so does any entanglement signature.

\subsubsection{Wigner representation}
\label{sec/2e-collision/entanglement/Wigner}

The superposition between energy-gain and energy-loss processes introduced in \cref{sec/2e-collision/picture} can also be visualized in a time-frequency representation of the relative particle, obtained by integrating the full two-electron Wigner function $\Wee_{AB}(t_\Sigma,t_\Delta;\omega_\Sigma,\omega_\Delta)$ over the center-of-mass variables $t_\Sigma$ and $\omega_\Sigma$:
\begin{equation}
	W_{\Delta}(t_\Delta,\omega_\Delta)= 	\int_{\mathbb{R}^2}\Wee_{AB}(t_\Sigma,t_\Delta;\omega_\Sigma,\omega_\Delta)\, \frac{\md t_\Sigma\,\md\omega_\Sigma}{2\pi}.
\end{equation}

\begin{figure}[h!]
	\includegraphics[width=0.8\columnwidth]{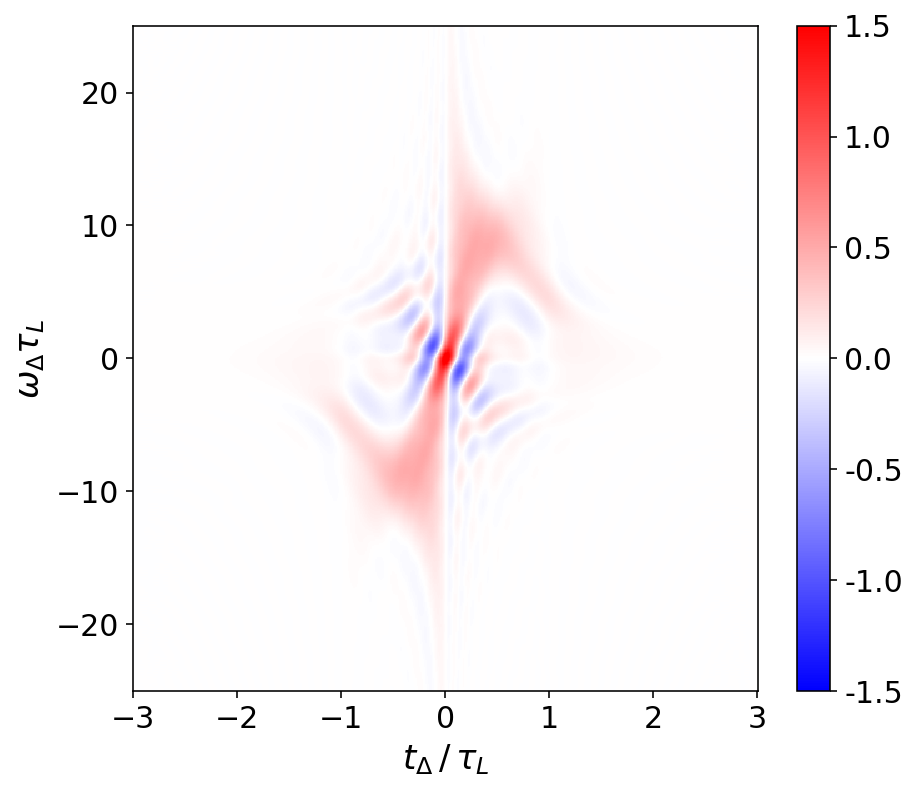}
	\caption{\label{fig/Wigner} Relative-particle Wigner representation
	of the post-collision coherence, obtained by integrating the
	two-electron Wigner function $\Wee_{AB}(\omega_\Sigma,
	\omega_\Delta; t_\Sigma, t_\Delta)$ over $t_\Sigma$ and
	$\omega_\Sigma$. The result, a function of the variables
	$(t_{\Delta}, \omega_{\Delta})$, is depicted in units of
	$t_\Delta/\tau_L$ and $\omega_\Delta\tau_L$. The parameters used to
	plot are identical to the ones used in \cref{fig_g2_top}:
	$(\omega_{A}^{0}, \gamma_{A}^{-1}) = (\omega_{B}^{0},
	\gamma_{B}^{-1}) =(0, \tau_L / 2)$, $\alpha_{\text{eff}} = 9$ and
	$t_\Delta^0=0$ (synchronous arrival of the incoming wave packets).}
\end{figure}

\cref{fig/Wigner} shows $W_\Delta$ for two identical wave packets
entering the coupler synchronously ($t_\Delta^0=0$). The two red areas
centered around $\omega_\Delta\sim\pm\omega_{AB}$, occurring at finite
$t_\Delta$, are the direct time-domain signature of the two branches of
the superposition described in \cref{sec/2e-collision/picture}, each
spread over a time window set by the wave packet duration $1/\gamma$.
Between these two lobes, the distribution develops negative fringes:
because the two branches interfere coherently rather than adding as
classical probabilities, $W_\Delta$ cannot remain positive everywhere.
Such negativities are a well-known signature of quantum interference
\cite{Wigner-1932-1}.

Physically, the two tilted lobes are those of a \emph{separable} (Wick)
state built from the outgoing marginals: they merely encode the
snowplough energy exchange discussed in \cref{appendix/collision-model/snowplough}, 
slice by slice. 
The genuinely non-classical
content is carried by the negative fringes between them. They are absent from
any separable state: they stem from the Wick deviation $\DWGee_{AB}$ as
expected from the fact that the outgoing
two-particle state
realizes an energy domain analogue of 
the non-Gaussian superposition of Slater
determinants $\SumOfSl$ of
\cref{sec/g2-and-entanglement/G2-anatomy}. Since the inter-channel
geometry enforces $\Ge_{AB}=0$ (hence $\DWGee_{AB}=\DcGee_{AB}$), the
Wick deviation $\DWGee_{AB}$ 
cannot be reduced to zero: this component is responsible for the
violation of the entanglement witness as well as of the fringes
appearing on \cref{fig/Wigner}. 

This is precisely the point made by Bell in his analysis of EPR
correlations through Wigner distributions~\cite{Bell-EPR-Wigner}: a
positive Wigner function admits a classical, local statistical model, and
a Gaussian entangled state has a Wigner
function that remains everywhere positive. It is therefore the negativity
of $W_\Delta$ that certifies the non-Gaussian character of the
entanglement generated by the collision.

Experimentally, only "slices" of $W_\Delta$ at fixed $\omega_\Sigma$ are
directly accessible, each obtained from the reconstruction protocol of
\cref{sec/protocols/G2} at $\Omega_\Sigma=0$. The most informative
slice, at $\omega_\Sigma=\omega_A^0+\omega_B^0$, is precisely the Wigner
transform of the coherence shown in \cref{fig_g2_face}; it displays the
same lobes and negativities as the fully integrated $W_\Delta$ in
\cref{fig/Wigner}, only somewhat sharper, since integrating over
$\omega_\Sigma$ mildly smooths the pattern as the spectral width of the
wave packets increases (see Appendix~\ref{appendix/collision-model} for the
quantitative reconstruction procedure and its dependence on the wave
packets' spectral width).

\subsection{Realistic estimation of the tomography signal} \label{sec/2e-collision/entanglement/exp-signal}

We finally estimate the experimental signal expected from the photo-assisted
two-electron tomography protocol of \cref{sec/protocols/G2}. We focus on the
off-diagonal sector $\Omega_\Sigma=0$, finite $\Omega_\Delta$, using the
filtering equations of Appendix~\ref{appendix/tomo-g2} with Lorentzian static
energy filters of widths $J_A$ and $J_B$. Although computed for a single collision, this estimate 
is relevant for the realistic
$T$-periodic case (see \cref{appendix/dilute-to-periodic/limit/comb} and \cref{sec/protocols/G2/signal-estimate}).

\begin{figure}[t]
	\centering
	\includegraphics[width=0.78\linewidth]{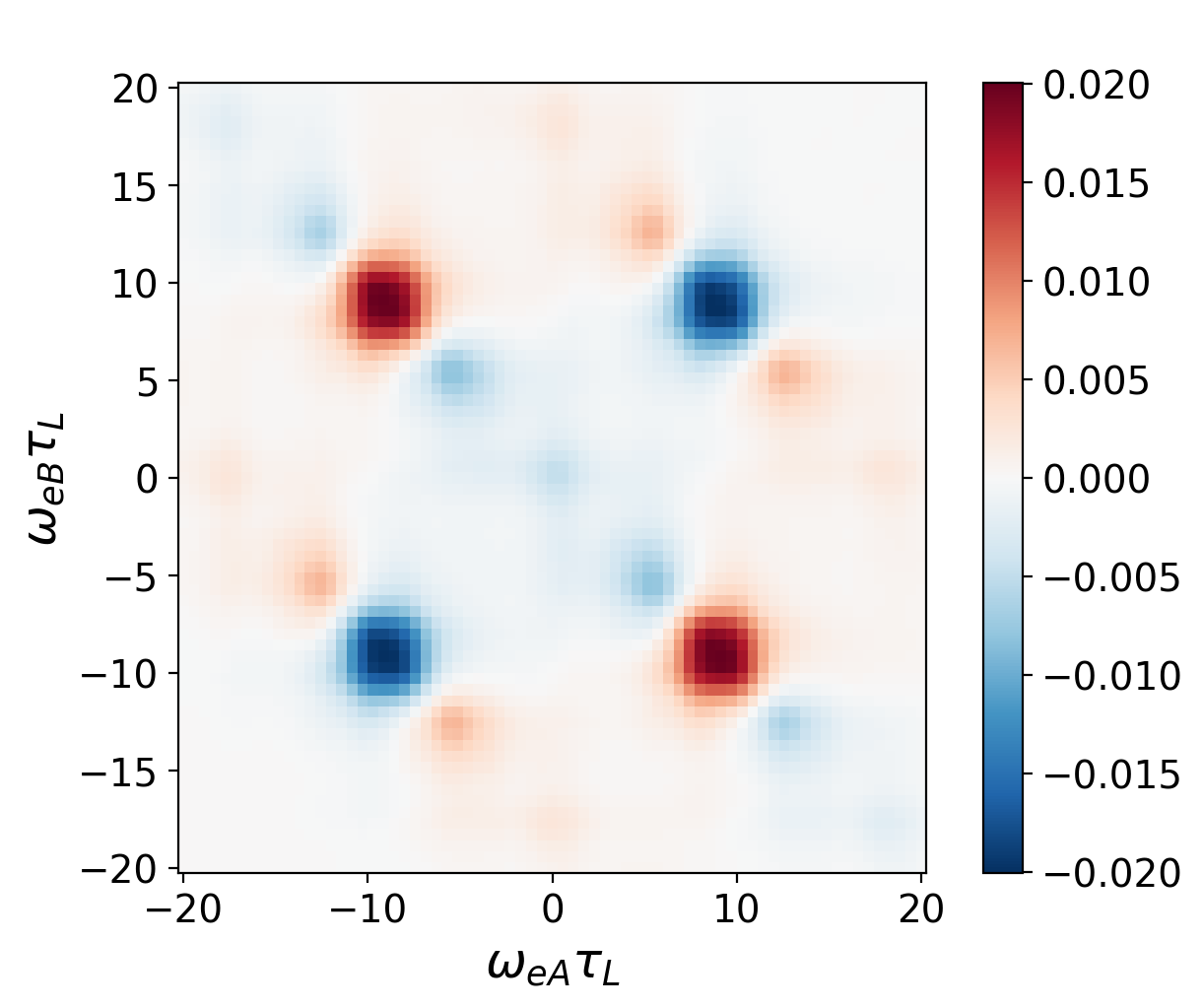}
	\caption{Dimensionless experimental signal $\mathcal{C}_{\mathrm{exp}}/(e^2f)$ as a function of $(\omega_{eA}\tau_L,\omega_{eB}\tau_L)$ associated with the stationary off-diagonal sector $\Omega_\Sigma=0$ and $\Omega_\Delta\tau_L=18$ of the two-electron coherence. The calculation is performed in the strong-coupling regime $\alpha_{\text{eff}} = 9$ for Landau wave packets with $\gamma_A=\gamma_B=2/\tau_L$ and Lorentzian static energy filters of widths $J_A=J_B=1/\tau_L$.}
	\label{fig:exp-signal-realistic}
\end{figure}

\cref{fig:exp-signal-realistic} shows the resulting dimensionless signal $\mathcal{C}_\mathrm{exp}/(e^2f)$ as a function of the filtering energies $(\omega_{eA}\tau_L, \omega_{eB}\tau_L)$, for the strongly off-diagonal configuration $\Omega_\Delta\tau_L=18$ marked by the black star in \cref{fig_witness}. The dominant lobes near $(\omega_{eA}\tau_L,\omega_{eB}\tau_L)\simeq(\pm\alpha_{\text{eff}},\mp\alpha_{\text{eff}})$ reflect the interaction-induced energy exchange, consistent with the fourfold filtering geometry discussed in \cref{fig/3D-principle-of-tomography}. Since the signal is evaluated at $\Omega_\Delta\tau_L=18$, well off-diagonal, these structures constitute a genuine signature of two-electron quantum coherence beyond diagonal energy spectroscopy.

For the realistic parameters of \cref{sec/2e-collision/model}, the interaction-induced energy scale $\omega_{AB}$ lies in the few-\si{\giga\hertz} range, comfortably above the thermal scale $k_BT_\mathrm{el}/h\sim\SI{2}{\giga\hertz}$ at $T_\mathrm{el}\sim\SI{100}{\milli\kelvin}$. Resolving the interaction-induced sidebands requires a filter tuning range of order $2\hbar\omega_{AB}$, which is accessible with sub-micrometric quantum dots \cite{Altimiras-2010-1}. The structures displayed in \cref{fig:exp-signal-realistic} therefore remain well resolved despite thermal broadening and finite filter resolution.%

\section{Photo-assisted Franson interferometry}
\label{sec/Franson}

In this section, we examine how the photo-assisted tomography for
inter-channel two-electron coherence presented in \cref{sec/protocols} can be
used to probe intra-channel two-electron coherence. 

\subsection{Measurement of intra-channel two-electron coherence}
\label{sec/Franson/intra-channel-G2-tomography}

The photo-assisted filtering protocol described in \cref{sec/protocols} 
reconstructs the Wick deviation 
$\DWGee_{AB}$ of the outgoing inter-channel two-electron coherence. We now show that the same
protocol gives access to the \emph{intra}-channel two-electron coherence of a
single source. 

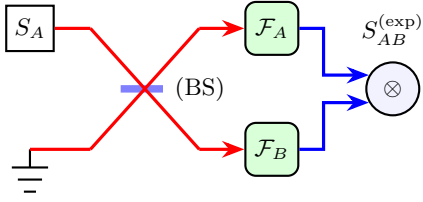
\begin{figure}
	\begin{tikzpicture}[
        x=0.8cm,y=0.8cm,
        >=Stealth,
        thick,
        channel/.style={-Stealth, very thick},
        src/.style={draw, minimum width=0.6cm, minimum height=0.6cm},
        filter2/.style={draw, rounded corners, fill=green!15,
                minimum width=0.7cm, minimum height=0.7cm},
        correlator/.style={draw, circle, fill=blue!5, minimum size=7mm},
        ]
        
        \node[src] (S1) at (-2., 1.) {$S_A$};
        
		\coordinate (G) at (-2.,-1.);
        \draw[thick] (G) -- ++(0,-0.3);
        \draw[thick] ($(G)+(-0.3,-0.3)$) -- ++(0.6,0);
        \draw[thick] ($(G)+(-0.2,-0.5)$) -- ++(0.4,0);
        \draw[thick] ($(G)+(-0.1,-0.7)$) -- ++(0.2,0);
        
        \draw[channel,-,blue, line width=2.8pt, color=blue!50]
            (-0.5,0) -- (0.2,0) node[right,black] {$(\text{BS})$};
        
        \draw[channel,-,red] (S1.east) -- (-1., 1.);
        \draw[channel,-,red] (G)       -- (-1.,-1.);
        \draw[channel,-,red] (-1., 1.) -- (0.8,-1.);
        \draw[channel,-,red] (-1.,-1.) -- (0.8, 1.);
        
        \node[filter2] (FA) at (2., 1.) {$\mathcal{F}_A$};
        \node[filter2] (FB) at (2.,-1.) {$\mathcal{F}_B$};
        
        \draw[channel,red] (0.8, 1.) -- (FA.west) node[midway, above, red] {};
        \draw[channel,red] (0.8,-1.) -- (FB.west) node[midway, below, red] {};
        
        \node[correlator] (Corr) at (4., 0.) {$\otimes$};
        \node[above=0.1cm] at (Corr.north) {$S_{AB}^{(\text{exp})}$};
        
        \draw[channel,blue] (FA.east) -- ++(0.4,0) |- (Corr.150);
        \draw[channel,blue] (FB.east) -- ++(0.4,0) |- (Corr.210);

\end{tikzpicture}%

	\caption{\label{fig/Franson-photoassisted} Generalized Franson
	interferometer for intra-channel two-electron tomography via
	photoassisted filtering. It is a generalization of the Franson
	interferometer \cite{Franson-1989-1}, in which the MZIs
	are replaced by two photo-assisted filters presented in
	\cref{sec/protocols/principles}, and where the signal is the low-frequency
	time-averaged current correlator $S_{AB}^{\text{(exp)}}$ defined by
	\cref{eq/correlation-signal}.}
\end{figure}

The idea is a generalized Franson interferometer: the Franson
setup, originally proposed in optics~\cite{Franson-1989-1}, pipelines a
Hanbury-Brown--Twiss (HBT) interferometer with two Mach-Zehnder
interferometers and coincidence detection. Its electron quantum optics
version replaces single-shot photodetection by low-frequency current-correlation
measurements~\cite{Thibierge-2016-1}. Here we replace the two Mach-Zehnder
interferometers by two photo-assisted filters (\cref{sec/protocols/principles}),
so that the measured signal is the time-averaged current correlator
$S_{AB}^{(\text{exp})}$ of \cref{eq/correlation-signal}
(\cref{fig/Franson-photoassisted}). Note that the presence of the QPC defines it
as a \cref{fig/geometries}-(c) type setup. 

In this setup, a single source $S_A$ is fed into one input of QPC while
the other one is connected to the ground. 
The ideal electronic beam splitter turns the incoming
intrinsic two-electron
intra-channel coherence $\DGee_{S_A}$ 
into the outgoing inter-channel
one \cite{Thibierge-2016-1}: 
\begin{equation}
\label{eq/franson/basic-conversion}
\DGee_{AB,\text{out}}(1,2|1',2')=RT\,\DGee_{S_A}(1,2|1',2')\,;
\end{equation}
while the single-electron coherences transform according to
\begin{subequations}
	\label{eq/Franson/G1-propagation}
\begin{align}
	\label{eq/Franson/G1-propagation/1}
\DGe_{A,\text{out}}(1|1')&=T\DGe_{S_A}(1|1')\\
	\label{eq/Franson/G1-propagation/2}
\DGe_{B,\text{out}}(2|2')&=R\DGe_{S_A}(2|2')\\
	\label{eq/Franson/G1-propagation/3}
	\Ge_{AB,\text{out}}(1|2')&=
	-\mi\sqrt{RT}\,\DGe_{S_A}(1|2')\\
	\label{eq/Franson/G1-propagation/4}
	\Ge_{BA,\text{out}}(2|1')&=\mi\sqrt{RT}\,\DGe_{S_A}(2|1')\,.
\end{align}
\end{subequations}
Consequently, the protocol of
\cref{sec/protocols} reconstructs
\begin{equation}
	\label{eq/Franson/DGee/HBT-propagation}
\DWGee_{AB,\text{out}}(1,2|1',2')=RT\,\DWGee_{S_A}(1,2|1',2'),
\end{equation}
the \emph{intra-channel} Wick deviation of two-electron coherence of
$S_A$ up to the factor $RT$. 
Since $S_B$ is switched off, this is a genuine HBT geometry: there is no
two-particle interference between excitations emitted by two distinct
sources. Unlike the collision of \cref{sec/2e-collision}, the reconstructed
coherence is that of the source alone.

Note that, due to \cref{eq/Franson/G1-propagation/3}, we are necessarily in the situation where
$\DGe_{AB}\neq 0$ since $\DGe_{S_A}$ is generically non
vanishing when $S_A$ is active.
The reconstruction procedure then proceeds along the line of
\cref{sec/protocols}: $\Ge_{AB,\text{out}}$ at the output of
the QPC is reconstructed following the guidelines of \cref{sec/protocols/G1/inter-channel}.
Then one proceeds by reconstructing $\DWGee_{AB,\text{out}}$ following the
procedure detailed in \cref{sec/protocols/G2/T-periodic} and
\cref{sec/protocols/G2/tomography-measurements}. This leads to
$\DWGee_{AB,\text{out}}$ which is proportional to $\DWGee_{S_A}$ by
\cref{eq/Franson/DGee/HBT-propagation}. In the end, the
reconstruction of the outgoing
$\Ge_{AB,\text{out}}$ is equivalent to the one of $\DGe_{S_A}$ and thus of
$\Ge_{S_A}$. This is how $\Gee_{S_A}$ as well as its anatomy can be
reconstructed. 

We now turn to the structure of this intra-channel coherence, and to the
entanglement it can carry between mode subspaces of the channel.

\subsection{Intra-channel two-electron coherence and mode entanglement}
\label{sec/Franson/anatomy}

Having reconstructed the intra-channel $\Gee_{S_A}$ and its anatomy,
we can ask what entanglement it carries. A single channel has no {\it a priori}
bipartite structure. Following Ref.~\cite{Benatti-2020-1}, we introduce one by hand, splitting 
its single-particle
space into two orthogonal mode subspaces, $A=A_1\oplus A_2$. The
Cauchy--Schwarz witness of \cref{sec/g2-and-entanglement/witness} can
then be rewritten to provide a witness for mode entanglement between 
the $(A_1,A_2)$ sectors:
\begin{equation}
	\label{eq/Franson/A1-A2-witness}
	\left| \Gee_{S_A}(1,2|1',2')\right|^2\leq \Gee_{S_A}(1,2'|1,2')\,
	\Gee_{S_A}(1',2|1',2).
\end{equation}
A violation of this inequality for $(1,1')\in A_1^2$ and 
$(2,2')\in A_2^2$ certifies entanglement between
the $A_1$ and $A_2$ modes. The essential difference with
\cref{sec/g2-and-entanglement} is that $A_1|A_2$ is now a \emph{choice} made on
the reconstructed coherence. It is not associated with two different
physical channels.

The anatomy of the two-electron coherence remains unchanged:
\begin{align}
	\Gee_{S_A}(1,2|1',2')&=\Ge_{S_A}(1|1')\Ge_{S_A}(2|2')\nonumber \\
	&-\Ge_{S_A}(1|2')\Ge_{S_A}(2|1')\nonumber \\
	&+\DWGee_{S_A}(1,2|1',2').
\end{align}
The violation of \cref{eq/Franson/A1-A2-witness} can still come from the
two cases discussed in \cref{sec/g2-and-entanglement/G2-anatomy}: the
exchange term $\Ge_{S_A}(1|2')\Ge_{S_A}(2|1')$ - one-body, Gaussian,
non-zero only when the two subspaces $A_1$ and $A_2$ share single-electron
coherence -, and the connected deviation $\DWGee_{S_A}$ - genuine two-body,
non-Gaussian. The exchange term is nothing but the Pauli correlation
between two electrons propagating within the same channel.

The two examples of \cref{sec/g2-and-entanglement} have direct intra-channel
counterparts. Considering mutually-orthogonal normalized single-particle states
$(\varphi_1,\varphi_{1'}) \in A_1^2$ and
$(\varphi_2,\varphi_{2'}) \in A_2^2$, we can choose states mirroring the
ones considered in \cref{sec/g2-and-entanglement/examples-setups}. 

The superposition of two Slater determinants,
$\alpha\,\psi_A^\dagger[\varphi_1]\psi_A^\dagger[\varphi_{2}]\ket0
+\beta\,\psi_A^\dagger[\varphi_{1'}]\psi_A^\dagger[\varphi_{2'}]\ket0$,
with $|\alpha|^2+|\beta|^2=1$
is non-Gaussian
($\DWGee_{S_A}(\varphi_1,\varphi_2|\varphi_{1'},\varphi_{2'})\neq0$ and
$\Ge_{S_A}(\varphi_1|\varphi_{2'})=\Ge_{A}(\varphi_2|\varphi_{1'})=0$) and violates
\cref{eq/Franson/A1-A2-witness} via $\DWGee_{S_A}$: its l.h.s. is
$|\alpha\beta|^2$ while its r.h.s. vanishes.

A two-electron Slater determinant built from normalized \emph{delocalized} modes,
$\psi_A^\dagger[\alpha_1\varphi_1+\alpha_{2'}\varphi_{2'}]\,\psi_A^\dagger
[\alpha_{1'}\varphi_{1'}+\alpha_2\varphi_{2}]\ket0$, is Gaussian ($\DWGee_{S_A}
(\varphi_1,\varphi_2|\varphi_{1'},\varphi_{2'})=0$) and violates the witness through 
the exchange term since $\Ge_{S_A}(\varphi_1|\varphi_{2'})\neq0$ and
$\Ge_{S_A}(\varphi_2|\varphi_{1'})\neq 0$: the l.h.s. is
$|\alpha_1\alpha_2\alpha_{1'}\alpha_{2'}|^2$ while the r.h.s. vanishes
due to the Pauli principle: for example, the diagonal coherence
$\Gee_{S_A}(\varphi_1,\varphi_{2'}|\varphi_1,\varphi_{2'})$ vanishes
because it probes the modes $\varphi_1$ and $\varphi_{2'}$ belonging to
the same filled orbital $\alpha_1\varphi_1+\alpha_{2'}\varphi_{2'}$. 

These two examples also point to a subtlety proper to the intra-channel
case: the bipartition $A_1|A_2$ is conventional, and both the exchange
term and the Wick deviation depend on it, so the value of the witness at
given arguments is not an intrinsic quantity. Three statements of
decreasing scope should therefore be distinguished. A violation of
\cref{eq/Franson/A1-A2-witness} certifies mode entanglement between $A_1$
and $A_2$, for \emph{any} state, pure or mixed. A non-vanishing
$\DWGee_{S_A}$ certifies that the emitted state is non-Gaussian. This is 
a weaker property, since a classical statistical mixture of Slater
determinants is already non-Gaussian while remaining separable. Only for
a \emph{pure two-electron state} do the two notions coincide: $\DWGee_{S_A}\neq0$ is
then equivalent to the state not being a single Slater determinant in any
single-particle basis, a basis-independent obstruction that is the
genuine, interaction-induced content.
Under this hypothesis, one may bring the
reconstructed coherence to its canonical form, a sum of Slater
determinants built on orthogonal pairs of natural orbitals: a single
determinant signals an uncorrelated pair, two or more genuine
two-electron correlations. This is relevant for experiments in which we are sure that the
two-electron pure state physics is realized, as in the discussion of
\cref{sec/2e-collision}. In the case of a $T$-periodic driven system,
this assumes that we are dealing with two ideal single-electron sources
or with a single ideal two-electron source~\cite{Moskalets-2020-1}, emitting exactly one pair
of excitations in a pure state per period. We also need interactions to
have
no influence between adjacent periods. Testing this
requires working with a
discrete, normalizable basis of wavepackets rather than in the extended
Floquet-Bloch modes: the single-particle Floquet-Wannier basis being presumably the best
suited.

Reading the entanglement content directly off $\DWGee_{S_A}$ in the
general, mixed or interacting case is, by contrast, an open problem: the
two-electron coherence is a two-body reduced quantity, and knowing it
does not by itself classify many-body entanglement. Dedicated
quantifiers such as the fermionic partial-transpose negativity 
\cite{Shapourian-2017-1,Shapourian-2019-1}
or
lower bounds on the entanglement of formation ~\cite{Giovannetti-2006-2} 
are then required. These lie beyond the scope of the present paper and
are deferred to a companion study. 

What the present protocol establishes
is access to the regime where the effects of Pauli statistics and of
interactions coexist \emph{together}:
two electrons
emitted into a single channel that have interacted, so that 
$\Gee_{S_A}$ 
carries both an exchange term and a non-vanishing 
$\DWGee_{S_A}$, 
beyond the inter-channel collision of \cref{sec/2e-collision} 
where $\DWGee_{AB}\neq 0$ but
the two electrons occupy distinct channels and exchange is absent.
This is the regime of a two-electron source
in the presence of
interactions~\cite{Schulenborg-2024-1}.
This regime is also realized when
exchange (created by the beam-splitter of \cref{fig/geometries}-(c))
coexists with interactions either by an interacting beam-splitter or by Coulomb
interactions right after it. %

\section{Conclusions and perspectives}
\label{sec/conclusion}

In this work, we have introduced a photo-assisted tomography protocol
that grants access to single- and two-electron coherences through
dc-current and time-averaged, low-frequency current correlation measurements,
which are already routinely performed in state-of-the-art electronic
interferometry experiments.

The protocol leverages photo-assisted transitions to recombine
coherences at different energies into a detectable signal, effectively
mapping off-diagonal elements of the electronic density matrix onto
experimentally accessible observables. For single-electron tomography,
the signal is a dc-current linear response, a significant practical
advantage over HOM tomography \cite{Bisognin-2019-2} which requires
high-sensitivity, time-consuming low-frequency noise measurements.
For two-electron tomography, we have shown that, combined with
single-electron tomography, current correlations
after two photo-assisted filters reconstruct the irreducible
two-particle part of the
two-electron coherence $\DWGee_{AB}$, thus providing
direct access to the full two-electron coherence and its anatomy. The resulting
linear inverse problem, summarized in Table~\ref{table/2e-tomography} 
and described in \cref{sec/protocols/G2/tomographic-reconstruction},
covers the full four-dimensional frequency space by systematically
combining zero, one, or two rf drives. From a quantum signal processing
perspective, this realizes a generalization of Franson interferometry
\cite{Franson-1989-1} to the energy domain in the spirit of
Ref.~\cite{Thibierge-2016-1}. Moreover, electronic decoherence within
the photo-assisted filters only provides a minor correction over the
energy range relevant for current single-electron sources, and can be
calibrated independently by energy spectroscopy
(\cref{sec/protocols/decoherence}).

To illustrate the power of this approach, we have applied it to a
minimal yet physically relevant model: two co-propagating chiral edge
channels coupled via a finite-length Coulomb coupler. Coulomb
interactions generate energy-domain entanglement between the two
electrons via coherent superpositions of states with the
same total energy but different energy imbalances $\omega_\Delta$. This
entanglement is detectable via an adaptation 
of the Cauchy-Schwarz-based entanglement witness formulated in
\cite{Wolk-2014-1} which is violated over a range of injection time
delays and off-diagonal frequencies $\Omega_\Delta$. The Wigner
function of the relative particle further exhibits regions of
negativity, a hallmark of non-classical correlations. Numerical
simulations for experimentally realistic parameters 
confirm that
the interaction-induced sidebands remain well resolved above the
thermal background, and we have estimated the expected correlation
signal in terms of directly measurable saturation currents through the
static energy filters. We stress that the tomography protocol itself is
model-independent: the specific example discussed here 
serves as a transparent illustration, but
the framework applies to any microscopic description of the
interacting system, provided $\DWGee_{AB}$ is regular
(\cref{appendix/G2/T-periodic}), including electronic decoherence effects as well as more realistic
geometries.

We have also shown that combining an electronic
Hanbury-Brown--Twiss interferometer with two photo-assisted filters
realizes a generalized Franson interferometer that probes
intra-channel two-electron coherences emitted by a single source. This
setup unifies the description of interaction-induced entanglement and
single-source quantum statistics within the same framework, and clarifies
the connection between HOM interferometry and two-electron tomography by
separating Wick's theorem contributions from genuine irreducible
two-particle correlations.

The proposals presented here are grounded in statistical observables
such as
average currents, time-averaged low-frequency current noise
\cite{Parmentier-2010-1} and current correlations that have been
routinely used in electronic interferometry
\cite{Ji-2003-1,Roulleau-2008-2} as well as in electronic HBT and HOM
experiments \cite{Bocquillon-2012-1,Bocquillon-2013-1} and, more
recently, in experiments with anyons \cite{Bartolomei-2020-1}. This
grounding in well-established measurement techniques ensures direct
experimental relevance. A practical limitation of the present scheme is
the narrow bandwidth of the quantum-dot energy filters, which reduces
the overall signal level. Moving to broader-band filtering devices such 
as directly ac-driven quantum dots or top-gated integrated
designs \cite{Locane-2019-1} could improve the
signal-to-noise ratio, at the cost of energy resolution, and open an appealing direction for further
investigation.

Looking further ahead, the forthcoming development of single-shot
electron detection via direct analogues of photodetection
\cite{Freise-2020-1}, coupling to a spin qubit
\cite{Thalineau-2014-1} or ultrafast capacitive sensing
\cite{Glattli-2020-1} may open a new era of electron quantum optics
with access to full counting statistics. As a recent example, the
partitioning statistics of few-electron droplets at a Y-junction has
been studied \cite{Shaju-2025-1} in a regime where coherence effects
average due to disorder. Future experiments in the coherent regime will 
require accessing higher-order electronic coherences. With single-shot
electron detectors, the photo-assisted tomography protocol introduced
here could be generalized toward architectures closer to those employed
in attosecond physics \cite{Laurell-2025-1,Laurell-2025-2} while
exploiting the additional versatility of nanolithography: electrons
could be partitioned at multi-terminal ballistic conductors and
directed onto several photo-assisted energy filters equipped with
single-shot energy-resolved detectors, potentially granting access to
higher-order electronic coherences.

Beyond the specific model discussed here, the two-electron coherence tomography
developed in this work offers a characterization of
two-particle entanglement in electron quantum optics, complementary to the many
proposals for entanglement generation and detection in mesoscopic conductors, 
from superconducting entanglers \cite{Chtchelkatchev-2002-1,
Lesovik-2001-1,Recher-2001-1,Samuelsson-2004-2} to orbital and
electron–hole entanglement in the quantum Hall regime
\cite{Samuelsson-2003-1,Beenakker-2006-1} and single-electron sources
\cite{Hofer-2016-2,Dasenbrook-2016-1,Dasenbrook-2015-1}. Rather than testing a single Bell-type
inequality, our approach reconstructs the full coherence and disentangles its
single-particle (Wick) and irreducible two-body (Wick-deviation) contributions. A
detailed discussion of how this tomographic viewpoint connects to, and could
serve, this body of work goes beyond the scope of the present paper, which is a
first step in this direction.
More broadly, the Floquet--Bloch structure underlying our analysis---conservation
of the quasienergy modulo $\hbar\omega$, and the attendant distinction between
elastic and inelastic processes---is generic to periodically driven systems, and
governs, for instance, the scattering of interacting particles in driven cold-atom
lattices~\cite{Bilitewski-2015-1}.

Beyond witnessing entanglement, turning the reconstructed single and two-electron
coherences into a \emph{quantitative} measure of it, such as the 
lower-bound of
entanglement of formation from beam-splitter correlations
in Refs.~\cite{Giovannetti-2006-2,Giovannetti-2007-1}
is a natural direction that we shall pursue elsewhere.

Finally, these experimental perspectives as well as the content of the present work
point to a question of a broader scope. 
Reconstructing a coherence from noisy data is an inverse problem whose
regularization requires knowing which coherences are physically admissible, that
is arising from a physically acceptable reduced density operator at the many
body level. This
is the
many-body analogue of the constraint that single-particle occupation numbers lie
in $[0,1]$.
Already at the two-electron level, this admissibility is governed by the
positivity of the two-particle density matrix and, more restrictively, by its
$N$-representability \cite{Coleman-1963-1}, a problem known to be computationally hard
\cite{Liu-2007-1}.
Characterizing the admissible domain of the coherence hierarchy
$\mathcal{G}^{(n)}$
from practical Cauchy–Schwarz constraints to the full 
$N$-representability conditions is thus both a foundational question and an
important issue for further developments in electron quantum optics as well as
in many-body physics.
Explicit necessary conditions at the two-electron level are derived in
\cref{appendix/positivity/G2/Wannier}. A systematic study of these
constraints, common to all orders of the hierarchy, is left for future work.%

\acknowledgements{
This work was funded by the e-qubit-fly grant (ANR-22-PETQ-0012) from the France 2030 program 
managed by the Agence Nationale
de la Recherche as well as by the ANR Grant ANR grant ``QuSig4QuSense''
(ANR-21-CE47-0012) and the ElQuRes project from the EU~H2020 Research \&
Innovation program under the grant agreement 101017733 and the ANR funding
(ANR-24-QUA2-0002). We thank
M. Brune and J.-M. Raimond for their hospitality 
at Collège de France 
where this work was initiated as well as the University of Latvia where
we had the opportunity to present it.
We warmly thank
C.~Bauerle, C.~Altimiras and A.~Peugeot 
for useful discussions, V.~Kashcheyevs and N.~Ubbelohde for feedback and
discussions as well as F.~Lépine and A.~Lhuillier for informing us
on similar techniques used within the attosecond community.
The authors used Claude (Anthropic; Opus~4.8, 5.0 and 5.5, July--October 2026) to
assist with consistency checks, rewording, verification of derivations and
conventions, and literature search during manuscript preparation. All
suggestions were reviewed paragraph by paragraph and accepted, modified or
rejected by the authors, who verified all content and take full
responsibility for it.} %

\appendix
\crefalias{section}{appendix}
\crefalias{subsection}{appendix}
\crefalias{subsubsection}{appendix}

\section{Single-electron coherence}
\label{appendix/G1}

In this Appendix, we summarize the definition of single-electron
coherences as well as their various representations that are used in the
present work. Most of these definitions and formulae are known from the litterature
but summarized here for self-suficiency.

\subsection{Definition}
\label{appendix/G1/definitions}

Single-electron or first-order electronic 
coherence is usually defined \cite{Degio-2011-1,Haack-2012-2} 
by
\begin{equation}
	\label{eq/G1/definition}
	\Ge_{\rho}(1|1')=\mathrm{Tr}\left(\psi(1)\,\rho\,\psi^\dagger(1')\right),
\end{equation}
in which $1$ and $1'$ caracterize a position within the electrical
circuit and a time of measurement: $1=(\alpha,x,t)$ and
$1'=(\alpha',x',t')$ where $\alpha$ denotes a quantum Hall edge channel (for example), $x$ the
position and $t$ the time. As measurements are performed at a given position in quantum 
coherent electronics, the position index is
often removed and the lead indexes $(\alpha,\alpha')$ are sometimes
forgotten, as well as $\rho$ for the sake of simplicity. 

The case where $\alpha=\alpha'$ corresponds to electronic
coherence within a given channel and $\alpha\neq\alpha'$ to
inter-channel electronic coherences. 

\subsection{Excess single-electron coherence}
\label{appendix/G1/excess}

In a conductor, the reference state is not the true vacuum but a Fermi
sea at a given chemical potential. What is probed experimentally are
deviations relative to this reference state when a source is turned
on. These relative deviations are defined recursively from electronic coherences by removing
Fermi sea contributions of lower order. 

At the single electron
coherence level, this leads to defining the excess or intrinsic single electron
coherence $\DGe$ by~\cite{Degio-2011-1,Haack-2012-2}:
\begin{align}
\Ge_\rho(1|1') = \Ge_F(1|1') + \DGe_\rho(1|1')
\,
\label{eq/G1/excess}
\end{align}
in which $\Ge_F(I|1')$ denotes the single-electron coherence of the
reference Fermi sea. Note that, when considering two different channels
$A$ and $B$
\begin{equation}
	\Ge_{AB}(1|1')=\DGe_{AB}(1|1')
	\label{eq/G1/excess/inter-channel}
\end{equation}
since the interchannel coherence $\Ge_{AB}$ vanishes in the reference
two-channel Fermi sea $\ket{F_A}\wedge \ket{F_B}$.

\subsection{Representations of single electron coherence}
\label{appendix/G1/representations}

Using Fourier transformation, four different representations of single-electron 
coherences can be defined. Two of them are related by double
Fourier transforms and correspond to the time and frequency domains
representations of single-electron coherences:
\begin{equation}
	\label{eq/G1/representations}
		\widetilde{\Ge}(\omega_+|\omega_-)=\int_{\mathbb{R}^2}
		\Ge(t_+ \vert t_-)\,\me^{\mi(\omega_+t_+-\omega_-t_-)}\md^2\mathbf{t}\,.
\end{equation}
The other two are time-frequency representations. One of them
is the electronic Wigner function \cite{Ferraro-2013-1}, defined as:
\begin{equation}
	\label{eq/G1/Wigner}
	\We(t,\omega)=
	\int_{\mathbb{R}}
	v_F\,\Ge\left(t+\frac{\tau}{2}\Big|t-\frac{\tau}{2}\right)\,\me^{\mi\omega\tau}\md\tau.
\end{equation}
Note that the Wigner distribution is obtained by taking a Fourier
transform with respect to $\tau=t_+-t_-$, the quantum (or off-diagonal)
part of the time domain $(t_+,t_-)$. 

The last one is the electronic ambiguity function
\cite{Souquet-Basiege-2026-1}, which
charterizes the resolving power of a single electron Mach-Zehnder
interferometer used to probe classical or quantum electromagnetic
radiation \cite{Souquet-Basiege-2024-1}. It is not used in the present
work.

\subsection{Properties}
\label{appendix/G1/properties}

The electronic Wigner function is also related to the frequency domain
representation of single-electron coherence by:
\begin{equation}
	\We(t,\omega)=\int_{\mathbb{R}}
	v_F\,\widetilde{\Ge}\left(\omega+\frac{\Omega}{2}\Big|\omega-\frac{\Omega}{2}\right)\,
	\me^{-\mi\Omega t}\frac{\md\Omega}{2\pi}
\end{equation}
Here again, the Wigner distribution is obtained by taking an inverse Fourier
transform with respect to $\Omega=\omega_+-\omega_-$, the quantum
(or off-diagonal) part of the frequency domain $(\omega_+,\omega_-)$.
It thus depends only
on the diagonal coordinates of the time and frequency domains:
\begin{equation}
	t=\frac{t_++t_-}{2},\quad \omega=\frac{\omega_++\omega_-}{2}.
\end{equation}
Let us call $\Omega$ the Fourier conjugate variable relative to $t$. 
For a time periodic system with period $T$, $\Omega$ is quantized as a multiple of $2\pi f$
($f=1/T$). This leads to the Fourier series expansion of the electronic
Wigner function, directly relevant for describing time-repeated
experiments:
\begin{equation}
	\label{eq/G1/Fourier-series}
	\We(t,\omega)=\sum_{n\in\mathbb{Z}}\We_n(\omega)\,\me^{-2\pi \mi
	nft},
\end{equation}
in which $\We_0(\omega)$ is the electronic distribution function,
measured through electronic spectroscopy \cite{Altimiras-2010-1}.
The recently implemented  Hong-Ou-Mandel tomography protocol
\cite{Bisognin-2019-2} consists in reconstructing the various harmonics
$\We_n(\omega)$, and thereby access the time dependance of the
electronic Wigner distribution function.

\subsection{Single particle coherence operators}
\label{appendix/G1/operators}

As explained in Ref.~\cite{Roussel-2020-1}, it is useful to unify the
time and frequency representations of single electron coherence by
introducing the single-electron coherence operator defined by
\begin{equation}
	\label{eq/G1/operators/definition}
	\OGe=v_F^2\int_{\mathbb{R}^2}\ket{t}\,\mathcal{G}^{(e)}(t|t')\,\bra{t'}\,\md
	t\,\md t'
\end{equation}
Using the energy basis $(\ket{\omega})_\omega$ related to the time basis
$(\ket{t})_t$ by
\begin{subequations}
	\label{eq/G1/operators/bases}
	\begin{align}
		\ket{\omega} &= \sqrt{\frac{v_F}{2\pi}} \int_{\mathbb{R}}
	\me^{-\mi\omega t}\ket{t}\,\md t \\
		\ket{t} &= \frac{1}{\sqrt{2\pi v_F}}
		\int_{\mathbb{R}} \me^{\mi\omega t}\ket{\omega}\,\md\omega
	\end{align}
\end{subequations}
so that $\langle \omega|\omega'\rangle=\delta(\omega-\omega')$ and $\langle
t|t'\rangle=\delta(t-t')/v_F$, one has
\begin{equation}
		\langle \omega|\OGe|\omega'\rangle
		=\frac{v_F}{2\pi}\,\widetilde{\mathcal{G}}^{(e)}(\omega|\omega')\,.
\end{equation}
We introduce the charge conjugation
operator $\OC$ defined on the space of single particle states by
\begin{equation}
	\label{eq/G1/operators/C-definition}
	\langle t|\OC\varphi\rangle = \langle t|\varphi\rangle^*\,.
\end{equation}
Note that it is an anti-linear involution and $\OC\ket{t}=\ket{t}$ but
$\OC\ket{\omega}=\ket{-\omega}$. 
Its extension to the full Fock space satisfies
\begin{equation}
	\label{eq/G1/operators/1}
	\OC\,\psi(t)\,\OC^\dagger =
	\psi^\dagger(t)\,.
\end{equation}
It is then natural to identify $\psi^\dagger(t)$ with $\psi_h(t)$, the
operator destroying a hole at time $t$. Having defined the hole creation
and destruction operators, 
the
anti-linearity of $\OC$ as well as Eqs. \eqref{eq/G1/operators/1} and
\eqref{eq/G1/operators/C-definition} imply that, for any single particle
state $\varphi$:
\begin{equation}
	\label{eq/G1/operators/2}
	\OC\,\psi^\dagger[\varphi]\,\OC^\dagger =
	\psi_h^\dagger[\OC\varphi]\,.
\end{equation}
The charge conjugation operator $\OC$ 
can then be used to define the hole coherence by
\begin{equation}
	\label{eq/G1/operators/3}
	\mathcal{G}^{(h)}(1|1')=\mathrm{Tr}(\psi_h(1)\,\rho\,\psi_h^\dagger(1'))=
	\mathrm{Tr}(\psi^\dagger(\OC\,1)\,\rho\,\psi(\OC\,1'))\,.
\end{equation}
Then, using the canonical anti-commutation relation
$\{\psi(t),\psi^\dagger(t')\}=v_F^{-1}\delta(t-t')$ and
Eq.~\eqref{eq/G1/operators/C-definition}, we can derive the
operatorial relation between single-electron and single-hole operators
\begin{equation}
	\label{eq/G1/operators/Ge-Gh-relation}
	\OGh=\OId-\OC\,\OGe\,\OC^\dagger\,.
\end{equation}

\section{Two-electron coherence}
\label{appendix/G2}

In this Appendix, we summarize the definition of two-electron
coherences as well as their various representations that are used in the
present work. some of these definitions and formulae are known from the litterature
\cite{Moskalets-2014-1,Thibierge-2016-1}
but summarized here for self-suficiency.

\subsection{Definition}
\label{appendix/G2/definition}

Let us now consider two electron coherence which is defined as
\cite{Moskalets-2014-1,Thibierge-2016-1}:
\begin{equation}
	\label{eq/G2/definition}
	\Ge_{\rho}(1,2|1',2')=\mathrm{Tr}\left(\psi(2)\psi(1)\,\rho\,\psi^\dagger(1')\psi^\dagger(2')\right)
\end{equation}
The two-electron coherence satisfies symmetry properties that reflect
the anti-commutation of two $\psi$ (or two $\psi^\dagger$) operators:
\begin{subequations}
	\label{eq/G2/symmetries}
	\begin{align}
		\label{eq/G2/symmetries/anti-symmetry/a}
	\Ge_{\rho}(1,2|1',2')&=-\Ge_{\rho}(2,1|1',2')\\
		&=-\Ge_{\rho}(1,2|2',1')\label{eq/G2/symmetries/anti-symmetry/b}\\
	&=\Ge_{\rho}(2,1|2',1')\,.
		\label{eq/G2/symmetries/classical-exchange}
	\end{align}
\end{subequations}
However, 
\cref{eq/G2/symmetries}
list
equalities for the same two-electron coherence
only for intra-channel two-electron coherence. Reintroducing the channel
indices,
\cref{eq/G2/symmetries/anti-symmetry/a,eq/G2/symmetries/anti-symmetry/b}
imply that
$\Gee_{AA}(1,2|1',2')=0$ as soon as $1=2$ or $1'=2'$ which is a direct
expression of Pauli exclusion
principle. It also shows that for the intra-channel two-electron coherence, all the
physical 
information is contained in a fourth of the 4D-space of arguments of
$\Gee$, a direct consequence of indiscernability.

By contrast, when considering the inter-channel two-electron 
coherence $\Gee_{AB}$, \cref{eq/G2/symmetries/classical-exchange}
connects $\Gee_{AB}$ to $\Gee_{BA}$. Note that $\Gee_{AB}(1,2|1',2')$ does not
vanish when $1=2$ nor when $1'=2'$ and the full range of parameters
$(1,2|1'2')$ is required to specify $\Gee_{AB}$, as the channel index
breaks the indiscernability.

\subsection{Intrinsic two-electron coherence}
\label{appendix/G2/intrinsic}

At the-two electron level, the intrinsic two electron coherence is
defined from the full two electron coherence by removing the Fermi
sea contributions as well as those involving a contribution of the excess single
electron coherence to two-electron detection. They involve processes
where the origin of detected electrons can formally be traced back to the Fermi
sea and to the excess single electron coherence but also processes
involving quantum exchange thereby leading to \cite{Thibierge-2016-1}
\begin{subequations}
\label{eq/g2/excess/intra}
\begin{align}
    \Gee_{\rho}(1,2|1',2') = & \; \Gee_{F} (1,2|1',2') \\
    &+ \Ge_F(1|1') \, \DGe_\rho(2|2')\nonumber \\
    &+ \Ge_F(2|2') \DGe_\rho(1|1') \label{eq:g2:corr}\\
    &- \Ge_F(1|2') \, \DGe_\rho(2|1')\nonumber \\
    &- \Ge_F(2|1') \, \DGe_\rho(1|2') \label{eq:g2:hbt}\\
    &+ \DGee_\rho(1,2|1',2')
\end{align}
\end{subequations}
which defines the intra-channel intrinsic two electron coherence
$\DGee_\rho(1,2|1',2')$.

Because there is no
inter-channel
single electron coherence in the reference two-channel Fermi sea
$\ket{F_{AB}}=\ket{F_A}\wedge\ket{F_B}$ ($\Ge_{AB,\ket{F_{AB}}}(1|1')=0$),
the quantum exchange contributions corresponding to term
\eqref{eq:g2:hbt} in Eq.~\eqref{eq/g2/excess/intra} are not present. The intrinsic
inter-channel two electron coherence $\DGee_{AB,\rho}$ is therefore defined by:
\begin{subequations}
\label{eq/g2/excess/inter}
\begin{align}
    &\Gee_{AB,\rho}(1,2|1',2') = \Gee_{AB,F_{AB}} (1,2|1',2') \\
    &+ \Ge_{F_A}(1|1') \, \DGe_{B,\rho}(2|2')
    + \Ge_{F_B,\rho}(2|2') \DGe_{A,\rho}(1|1')
\label{eq:g2:inter:corr}\\
&+ \DGee_{AB,\rho}(1,2|1',2')\,.
\end{align}
\end{subequations}
Note that in \cref{eq:g2:inter:corr}, the excess single electron
coherences within each channel can be evaluated using the reduced density operators $\rho_A$
and $\rho_B$ obtained by tracing out the other channels electronic
degrees of freedom from $\rho$.

\subsection{Representations of two-electron coherence}
\label{appendix/G2/representations}

Exactly as before, two-electron coherence admits several representations
that are related via Fourier transform. Transposing
\cref{eq/G1/representations,eq/G1/Wigner}:
\begin{subequations}
	\label{eq/G2/representations}
	\begin{align}
		\widetilde{\Gee}(\pmb{\omega}_+|\pmb{\omega}_-) &=
		\int_{\mathbb{R}^4}
		\Gee(\mathbf{t}_+|\mathbf{t}_-)
		\,\me^{\mi(\pmb{\omega}_+\mathbf{t}_+-\pmb{\omega}_-\cdot\mathbf{t}_-)}
		\md^4\mathbf{t}
	\label{eq/G2/representations/frequency}	\\
		\Wee(\mathbf{t},\pmb{\omega})&=
		\int_{\mathbb{R}^2}
		v_F^2\Gee\left(\mathbf{t}+\frac{\pmb{\tau}}{2}\Big|\mathbf{t}-\frac{\pmb{\tau}}{2}\right)\,
		\me^{\mi\pmb{\omega}\cdot\pmb{\tau}}
		\md^2\pmb{\tau}
		\label{eq/G2/representations/Wigner}
	\end{align}
\end{subequations}
in which we use a vector notation as a shorthand to describe the times
or frequency domain coordinates for the two particles:
\begin{equation}
	\mathbf{t}=\begin{pmatrix}
		t_1\\
		t_2
	\end{pmatrix}
	\quad
	\pmb{\omega}=\begin{pmatrix}
		\omega_1\\
		\omega_2
	\end{pmatrix}.
\end{equation}
Note that both the inter-channel $\Wee_{AB}$ and intra-channel
$\Wee_{AA}$ two-electron Wigner
distributions are real. 

\subsection{Two-particle coherence operators}
\label{appendix/G2/operators}

The purpose of this section is to define two particle coherence
operators, thereby generalizing what has been done
at the single-particle level. 
There are four of them, reflecting the four quadrant structure
$(ee)$, $(eh)$, $(he)$ and $(hh)$ of the diagonal plane for
two-electron coherence in the frequency domain:
\begin{subequations}
\label{eq/eh-note/19}
    \begin{align}
        \Gee(1,2|1',2')
        &=\mathrm{Tr}(\psi(2)\psi(1)\,\rho\,\psi^\dagger(1')\psi^\dagger(2'))\\
        \Geh(1,2|1',2')&=\mathrm{Tr}(\psi_h(2)\psi(1)\,\rho\,\psi^\dagger(1')\psi_h^\dagger(2'))\nonumber\\
		&=\mathrm{Tr}(\psi^\dagger(\OC\,2)\psi(1)\,\rho\,\psi^\dagger(1')\psi(\OC\,2'))\\
        \Ghe(1,2|1',2')&=\mathrm{Tr}(\psi(2)\psi_h(1)\,\rho\,\psi_h^\dagger(1')\psi^\dagger(2'))\nonumber\\
		&=\mathrm{Tr}(\psi(2)\psi^\dagger(\OC\,1)\,\rho\,\psi(\OC\,1')\psi^\dagger(2'))\\
        \Ghh(1,2|1',2')&=\mathrm{Tr}(\psi_h(2)\psi_h(1)\,\rho\,\psi_h^\dagger(1')\psi_h^\dagger(2'))\nonumber\\
		&=\mathrm{Tr}(\psi^\dagger(\OC\,2)\psi^\dagger(\OC\,1)\,\rho\,\psi(\OC\,1')\psi(\OC\,2'))\,.
    \end{align}
\end{subequations}

\subsubsection{Two-particle spaces and coherence operators}
\label{appendix/G2/operators/spaces}

The four coherences of \cref{eq/eh-note/19} do not all live on the same
space. Two excitations \emph{of the same nature} --- two electrons, or
two holes --- are indistinguishable, so the corresponding operators act
on the exterior square
$\mathcal{H}_{\text{2p}}=\mathcal{H}_{\text{1p}}\wedge\mathcal{H}_{\text{1p}}$
of the single-particle space. Two excitations \emph{of different
nature} --- one electron and one hole --- are distinguishable, and the
corresponding operators act on the ordinary tensor product
$\mathcal{H}_{\text{1p}}\otimes\mathcal{H}_{\text{1p}}$: there is no
antisymmetrisation between electronic and hole excitations
since $\psi_h(t)=\psi^\dagger(t)$ does not anticommute with $\psi(t')$
for all $(t,t')$. The
$(2e)$ and $(2h)$ sectors therefore obey a different algebra from the
mixed $(eh)$ and $(he)$ ones and we present both of them here.

\paragraph*{Labels.}
We keep here the full single-particle label $1=(\alpha,t)$ of
\cref{appendix/G1/definitions}, the position being fixed as usual, so
that the channel index travels with the particle rather than being
carried separately. The scalar product and the integration measure then
read
\begin{equation}
	\label{eq/G2/operators/labels}
	\langle 1|1'\rangle=\frac{\delta_{\alpha\alpha'}\,\delta(t-t')}{v_F}\,,
	\qquad
	\int\md 1=\sum_\alpha\int_{\mathbb{R}}\md t\,.
\end{equation}
This is the only place where the nature of the label matters: everything
below uses the single-particle space and
\cref{eq/G2/operators/labels} alone. Keeping the channel index is what
makes intra-channel ($\alpha=\alpha'$) and inter-channel
($\alpha\neq\alpha'$) coherences two specialisations of one and the
same operator instead of two separate constructions.

\paragraph*{The exterior square.}
We use the normalised exterior product
\begin{equation}
	\label{eq/G2/operators/wedge-convention}
	\ket{a}\wedge\ket{b}=\frac{1}{\sqrt{2}}
	\left(\ket{a}\otimes\ket{b}-\ket{b}\otimes\ket{a}\right)\,,
\end{equation}
for which the scalar product carries no extra factor,
\begin{equation}
	\label{eq/G2/operators/wedge-scalar-product}
	\left(\bra{a}\wedge\bra{b}\right)\left(\ket{c}\wedge\ket{d}\right)
	=\langle a|c\rangle\langle b|d\rangle-\langle a|d\rangle\langle b|c\rangle\,.
\end{equation}
Two-particle states are written $\ket{1}\wedge\ket{2}$, in the order of
the arguments of \cref{eq/eh-note/19}.
The resolution of the identity on $\mathcal{H}_{\text{2p}}$ reads
\begin{equation}
	\label{eq/G2/operators/identity}
	\OId\wedge\OId=\frac{v_F^2}{2!}\int
	\ket{1}\wedge\ket{2}\;\bra{1}\wedge\bra{2}\;\md 1\,\md 2\,,
\end{equation}
the $1/2!$ compensating the fact that the integration runs over all
pairs rather than over ordered ones.

\paragraph*{The $(2e)$ and $(2h)$ coherence operators.}
Sandwiching the coherence between two such resolutions of the identity
gives the two-particle analogue of
\cref{eq/G1/operators/definition}:
\begin{align}
	\label{eq/G2/operators/definition}
	\OGee=\frac{v_F^4}{(2!)^2}&\int
	\ket{1}\wedge\ket{2}\Gee(1,2|1',2')\bra{1'}\wedge\bra{2'}\nonumber
	\\
	&\md 1\,\md 2\,\md 1'\,\md 2'\,,
\end{align}
and likewise for $\OGhh$ from $\Ghh$. As at the single
particle level, the prefactor is precisely what makes the kernel of the
operator the coherence itself:
\begin{equation}
	\label{eq/G2/operators/kernel}
	\bra{1}\wedge\bra{2}\OGee
	\ket{1'}\wedge\ket{2'}=\Gee(1,2|1',2')\,.
\end{equation}
Note that this is also fully compatible with the antisymmetry
properties of two-electron coherences (see \cref{eq/G2/symmetries}) 
and avoids
the corresponding over-counting. 
Both operators $\OGee$ and $\OGhh$ 
are Hermitian on $\mathcal{H}_{\text{2p}}$.

The exterior product of two single-particle operators is defined by its
kernel,
\begin{equation}
	\label{eq/G2/operators/wedge-of-operators}
	\bra{1}\wedge\bra{2}\mathbf{A}\wedge\mathbf{B}
	\ket{1'}\wedge\ket{2'}
	=A(1|1')\,B(2|2')-A(1|2')\,B(2|1')\,,
\end{equation}
which is bilinear in $(\mathbf{A},\mathbf{B})$ and consistent with
\cref{eq/G2/operators/identity}. For
$\mathbf{A}=\mathbf{B}=\OGe$, the right hand side of
\cref{eq/G2/operators/wedge-of-operators} is nothing but Wick's
expression for a two-particle kernel built from a single-particle one,
which is why the anatomy of two-electron coherence takes the compact
operatorial form:
\begin{equation}
	\label{eq/G2/operators/G2e-anatomy}
	\OGee=\OGe\wedge \OGe+
	\ODWGee\,
\end{equation}
in which $\ODWGee$ is defined by replacing
$\mathcal{G}^{(2e)}$ by 
$\Delta_W\mathcal{G}^{(2e)}$ in the r.h.s. of
Eq.~\eqref{eq/G2/operators/definition}. 

\paragraph*{The mixed sectors.}
For $(eh)$ and $(he)$ the two slots are distinguishable, so neither
antisymmetrisation nor combinatorial factor occurs: the identity on
$\mathcal{H}_{\text{1p}}\otimes\mathcal{H}_{\text{1p}}$ is simply
$v_F^2\int\ket{1}\otimes\ket{2}\,\bra{1}\otimes\bra{2}\,\md 1\,\md 2$,
and the corresponding operators are
\begin{align}
	\label{eq/G2/operators/definition-mixed}
	\OGeh=v_F^4&\int
	\ket{1}\otimes\ket{2}\;\Geh(1,2|1',2')\;\bra{1'}\otimes\bra{2'}
	\nonumber \\
	&\md 1\,\md 2\,\md 1'\,\md 2'\,,
\end{align}
and likewise for $\OGhe$ from $\Ghe$, with the same kernel
property as \cref{eq/G2/operators/kernel}.

Although only $n=2$ is needed in the present work, nothing in this construction is specific to two particles: the same
three ingredients, the exterior power of the single-particle space,
the normalisation \cref{eq/G2/operators/wedge-convention} and one
factor $v_F$ per particle and per insertion of the identity,
define the $n$-particle coherence operators on
$\mathcal{H}_{\text{1p}}^{\wedge n}$. 

As in the single-particle case discussed in
\cref{appendix/G1/operators}, these four operators are not independant. Using the canonical
anti-commutation relation, one can write down the relations between
them. 

\subsubsection{The $(2e)$ and $(2h)$ sectors}
\label{appendix/G2/operators/2e-2h}

The use of CAR on $\Ghh$ translates into:
\begin{subequations}
    \label{eq/eh-note/23}
    \begin{align}
    \OGhh&=\left(\OC\wedge\OC\right)
    \,\left[
        \OId^{\wedge 2}-\left(\OId\wedge
        \OGe+\OGe\wedge\OId\right)
		\right.\nonumber\\
		&+\left.
        \OGee\right]
    \,\left(\OC\wedge\OC\right)^\dagger\\
    &=\left(\OC\wedge\OC\right)
    \left[(\OId-\OGe)^{\wedge
    2}+\OGee\right.\nonumber\\
		&-\left.\OGe\wedge
        \OGe\right]
    \,\left(\OC\wedge\OC\right)^\dagger\\
    &=\left(\OC\wedge\OC\right)
        \left[\left(\OId-\OGe\right)^{\wedge 2}
    +\ODWGee
        \right]
    \,\left(\OC\wedge\OC\right)^\dagger
    \end{align}
\end{subequations}
We then use \cref{eq/G1/operators/Ge-Gh-relation} to rewrite
\begin{equation}
    \label{eq/eh-note/24}
\left(\OC\wedge\OC\right)
    \,\left[
(\OId-\OGe)^{\wedge
    2}\right]\,\left(\OC\wedge\OC\right)^\dagger=\OGh\wedge
\OGh\,.
\end{equation}
thereby leading to
\begin{equation}
	\OGhh=\OGh\wedge \OGh+
	(\OC\wedge\OC)\,\ODWGee(\OC\wedge
	\OC)^\dagger\,.
\end{equation}
This immediately leads to the relation between the deviations to Wick
for two-electron and two-hole coherences at the operatorial level:
\begin{equation}
    \ODWGhh=
    (\OC\wedge\OC)\,\ODWGee(\OC\wedge
    \OC)^\dagger\,.
\end{equation}

\subsubsection{The $(eh)$ and $(he)$ sectors}
\label{appendix/G2/operators/eh-he}

In the mixed sectors the two slots carry excitations of different
nature, so that charge conjugation acts on a \emph{single} slot:
through $\OId\otimes\OC$ for $(eh)$, the hole sitting on the
second slot, and through $\OC\otimes\OId$ for $(he)$.

\paragraph*{Reduction to two-electron coherences.}
Using $\psi_h=\psi^\dagger$ together with the canonical
anticommutation relations, the $(eh)$ coherence reduces to
\begin{equation}
	\label{eq/G2/operators/eh-reduction}
	\Geh(1,2|1',2')=\delta_{22'}\,\Ge(1|1')-\Gee(1,2'|1',2)\,.
\end{equation}
The $(he)$ sector is not its mirror image. There the hole operators
\emph{enclose} the electron slot, so reducing the expression requires
moving the electron operators across them, which generates additional
\emph{contact} terms:
\begin{align}
	\label{eq/G2/operators/he-reduction}
	\Ghe(1,2|1',2')&=\delta_{11'}\,\Ge(2|2')-\Gee(2,1'|2',1)\nonumber \\
	&+\underbrace{\delta_{12}\,\delta_{1'2'}
	-\delta_{12}\,\Ge(1'|2')-\delta_{1'2'}\,\Ge(2|1)}_{\text{contact terms,
	absent from }(eh)}\,.
\end{align}
Together with \cref{appendix/G2/operators/2e-2h}, this shows that the four
quadrants are four readings of one and the same object: each of them is
expressed in terms of $\Gee$ and of the single-particle coherence
alone.

\paragraph*{Anatomy.}
The single-particle contribution is now an \emph{ordinary} product,
$\OGe\otimes\OGh$ for $(eh)$ and $\OGh\otimes\OGe$ for $(he)$. Isolating
it in \cref{eq/G2/operators/eh-reduction,eq/G2/operators/he-reduction}
gives
\begin{subequations}
	\label{eq/G2/operators/mixed-anatomy}
	\begin{align}
	\Geh&=\Ge\otimes\Gh
	+\underbrace{\Ge(1|2)\,\Ge(2'|1')}_{e\text{-}h\
		\text{exchange}}\nonumber \\
		&-\DWGee(1,2'|1',2)\,,\\
	\Ghe&=\Gh\otimes\Ge
	+\underbrace{\Ge(2|1)\,\Ge(1'|2')}_{e\text{-}h\
		\text{exchange}}\nonumber \\
		&-\DWGee(2,1'|2',1)\nonumber\\
	&\quad+\delta_{12}\,\delta_{1'2'}-\delta_{12}\,\Ge(1'|2')
	-\delta_{1'2'}\,\Ge(2|1)\,.
	\end{align}
\end{subequations}
Contrary to the $(2e)$ and $(2h)$ case, the deviation from the
single-particle product is therefore not purely $\OC$-covariant: it
mixes the irreducible two-particle part $\DWGee$ with a single-particle
$e$-$h$ exchange, and, for $(he)$, with contact terms.

\paragraph*{Inter-channel versus intra-channel content.}
Let us write the slot channels explicitly, $\alpha$ for the first slot
and $\beta$ for the second. The two kinds of extra terms then behave
differently. The contact terms $\delta_{12}$ and $\delta_{1'2'}$ carry a
factor $\delta_{\alpha\beta}$, since the elementary contraction is
$\langle 1|1'\rangle=\delta_{\alpha\alpha'}\delta(t-t')/v_F$: they are
present in the intra-channel case only. The $e$-$h$ exchange, by
contrast, is a product of \emph{inter-channel} single-electron
coherences and carries no such factor. Inter-channel mixed coherences
are therefore given by
\begin{subequations}
	\label{eq/G2/operators/mixed-interchannel}
\begin{align}
	\label{eq/G2/operators/mixed-interchannel/eh}
	\Geh_{\alpha\beta}&=\Ge_\alpha\otimes\Gh_\beta
	+\Ge(1|2)\,\Ge(2'|1')-\DWGee_{\alpha\beta}\,,
	\\
	\label{eq/G2/operators/mixed-interchannel/he}
	\Ghe_{\alpha\beta}&=\Gh_\alpha\otimes\Ge_\beta
	+\Ge(2|1)\,\Ge(1'|2')-\DWGee_{\beta\alpha}\,,
\end{align}
\end{subequations}
the exchange terms involving $\Ge_{\alpha\beta}$ and
$\Ge_{\beta\alpha}$, whereas the contact singularities ($(he)$ only)
survive in the intra-channel case together with the exchange.

When the inter-channel single-electron coherence vanishes,
$\Ge_{\alpha\beta}=0$, the exchange contribution drops as well and
\cref{eq/G2/operators/mixed-interchannel} reduces to the clean products
\begin{subequations}
	\label{eq/G2/operators/mixed-interchannel-noG1}
\begin{align}
	\Geh_{\alpha\beta}&=\Ge_\alpha\otimes\Gh_\beta-\DWGee_{\alpha\beta}\,,\\
	\Ghe_{\alpha\beta}&=\Gh_\alpha\otimes\Ge_\beta-\DWGee_{\beta\alpha}\,.
\end{align}
\end{subequations}
In that situation, and in that situation only, the mixed sectors give
access to the intrinsic two-electron coherence without any exchange
contribution.

\paragraph*{Why the mixed sectors are richer.}
The ordinary tensor product, contrary to the exterior one, does not
reabsorb the exchange between the two excitations:
$\{\psi_h(t),\psi(t')\}=\delta (t-t')/v_F$. In other terms, an
electron and a hole living on the same channel are not supported on
orthogonal single-particle subspaces, so that their exchange
contribution does not vanish. The clean product structure
$\OGe\otimes\OGh$ is recovered only when the electron and the hole
occupy \emph{orthogonal} single-particle subspaces: distinct channels,
as above, or disjoint energy supports, the electron lying above and the
hole below the Fermi level.

\subsection{Two-body variables}
\label{appendix/G2/two-body}

Two-electron coherence encodes the coherence properties for two-electron
physics in a quantum circuit. 
In classical physics, the two-body problem
can be conveniently described using the center of
mass and relative coordinates. 

\subsubsection{Two-body variables in the time domain}
\label{appendix/G2/two-body/time-domain}

In the time domain, this translates into
the following definitions for the center of mass and relative times:
\begin{equation}
	\label{eq/variables/2body-definition}
t_{\Sigma} = \frac{t_A + t_B}{2} 
\,,\quad
t_{\Delta} = t_A - t_B
\,.
\end{equation}
Since here, we are dealing with quantum coherences, these coordinates
have to be
introduced for the $+$ as well as for the $-$ time coordinates. We can
then move to the center of mass and relative particle classical
(diagonal) variables:
\begin{subequations}
	\label{eq/variables/classical-variables-expressions}
	\begin{align}
	t_\Sigma &=\frac{t_{\Sigma,+}+t_{\Sigma,-}}{2}
	=\frac{t_{A,+}+t_{B,+}+t_{A,-}+t_{B,-}}{4}\\
	t_\Delta &=\frac{t_{\Delta,+}+t_{\Delta,-}}{2}
	=\frac{t_{A,+}-t_{B,+}+t_{A,-}-t_{B,-}}{2}
\end{align}
\end{subequations}
as well as the quantum (off-diagonal) ones:
\begin{subequations}
	\label{eq/variables/quantum-variables-expressions}
	\begin{align}
	\tau_\Sigma &= t_{\Sigma,+}-t_{\Sigma,-}
	= \frac{t_{A,+}+t_{B,+}-t_{A,-}-t_{B,-}}{2}\\
	\tau_{\Delta} &= t_{\Delta,+}-t_{\Delta,-}
		=t_{A,+}-t_{B,+}-t_{A,-}+t_{B,-}.
\end{align}
\end{subequations}
Note that the quantum (off-diagonal) parts of the center of mass and
relative particles can be expressed in terms of $(\tau_A,\tau_B)$ as
\begin{equation}
	\label{eq/variables/2body/quantum-expressions}
	\tau_{\Sigma} = \frac{\tau_A + \tau_B}{2} 
\quad \text{and}\quad
\tau_{\Delta} = \tau_B - \tau_A,
\end{equation} 
thus mirroring the expressions for the classical variables (see 
\cref{eq/variables/2body-definition}). 

\begin{figure}
	\usetikzlibrary{decorations.pathreplacing}
\usetikzlibrary{shadows.blur}

\begin{tikzpicture}

    \begin{scope}
         \shade[shading=radial, inner color=blue!60] (0, 5.75) ellipse (0.25 and 2);
    \end{scope}

    \begin{scope}
         \shade[shading=radial, inner color=blue!60] (0, 1.25) ellipse (0.25 and 1.5);
    \end{scope}

    \draw[-{Stealth}, thick] (0, -0.5) -- (0, 8) node[above] {$t_{A/B}$};

    \filldraw[blue] (0, 7) circle (2pt) node[right] {$t_{A}^{+} = t_{A} + \frac{\tau_{A}}{2}$};
    \filldraw[black] (0, 5.75) circle (2pt) node[right] {$t_{A} = \frac{1}{2} (t_{A}^{+} + t_{A}^{-})$};
    \filldraw[blue] (0, 4.5) circle (2pt) node[right] {$t_{A}^{-} = t_{A} - \frac{\tau_{A}}{2}$};
    \draw[-{Stealth}, thick] (-0.25, 4.5) -- (-0.25, 7) node[midway, left, sloped, xshift=30pt, yshift=10pt] {$\tau_{A} = t_{A}^{+} - t_{A}^{-}$};

    \filldraw[blue] (0, 2) circle (2pt) node[right] {$t_{B}^{+} = t_{B} + \frac{\tau_{B}}{2}$};
    \filldraw[black] (0, 1.25) circle (2pt) node[right] {$t_{B} = \frac{1}{2} (t_{B}^{+} + t_{B}^{-})$};
    \filldraw[blue] (0, 0.5) circle (2pt) node[right] {$t_{B}^{-} = t_{A} - \frac{\tau_{B}}{2}$};
    \draw[-{Stealth}, thick] (-0.25, 0.5) -- (-0.25, 2) node[midway, left, sloped, xshift=30pt, yshift=10pt] {$\tau_{B} = t_{B}^{+} - t_{B}^{-}$};

    \begin{scope}
         \shade[shading=radial, inner color=green!60] (4, 5.75) ellipse (0.25 and 0.5);
    \end{scope}

    \begin{scope}
         \shade[shading=radial, inner color=red!60] (4, 3.5) ellipse (0.25 and 1.75);
    \end{scope}

    \draw[-{Stealth}, thick] (4, -0.5) -- (4, 8) node[above] {$t_{\Sigma/\Delta}$};

    \filldraw[red] (4, 3.5) circle (2pt) node[below right] {$t_{\Sigma} = \frac{1}{2} (t_{A} + t_{B})$};
    \draw[red, -{Stealth}, thick] (4.25, 3.5) -- (4.25, 5) node[midway, right] {$\tau_{\Sigma} = \frac{1}{2} (\tau_{A} + \tau_{B})$};

    \draw[green, thick, dashed] (2.5, 1.25) -- (3.95, 1.25);
    \draw[green, thick, dashed] (2.5, 5.75) -- (3.95, 5.75);
    
    \draw[green, -{Stealth}, thick] (3.5, 1.25) -- (3.5, 5.75) node[midway, left, sloped, xshift=30pt, yshift=10pt] {$t_{\Delta} = t_{A} - t_{B}$};

    \draw[green, {Stealth}-, thick] (4.25, 6.15) -- (4.25, 5.35) node[midway, right] {$\tau_{\Delta} = \tau_{A} - \tau_{B}$};

\end{tikzpicture}%

	\caption{\label{fig/coordinates} Time variables for the quantum
	two-particle problem. {\emph Left panel}: the four Keldysh time
	variables
	$(t_{A}^{+}, t_{B}^{+}, t_{A}^{-},
	t_{B}^{-})$ can be expressed in terms of classical and quantum
	variables $(t_A,t_B)$ and $(\tau_A,\tau_B)$ defined by
	$t_\alpha^\pm=t_\alpha\pm\tau_\alpha/2$ for $\alpha=A$ or $B$. 
	The blue blurring represents the scale of quantum fluctuations for
	$\tau_A$ and $\tau_B$. 
	{\emph Right panel}: the
	center of mass and relative time variables are defined by
	$t^\varepsilon_\Sigma=(t_A^\varepsilon+t_B^\varepsilon)/2$ and 
	$t_\Delta^\varepsilon=t_A^\varepsilon-t_B^\varepsilon$ for $\varepsilon=\pm$.
	This leads to the center of mass and relative classical and quantum
	time variables:
	$(t_{\Sigma}, t_{\Delta})$ describe the system's center of mass
	and relative classical time variables, while $(\tau_{\Sigma}, \tau_{\Delta})$
	correspond to their quantum fluctuations which are associated
	respectively with the red and green blurrings.}
	\label{fig/coordinates_tikz}
\end{figure}
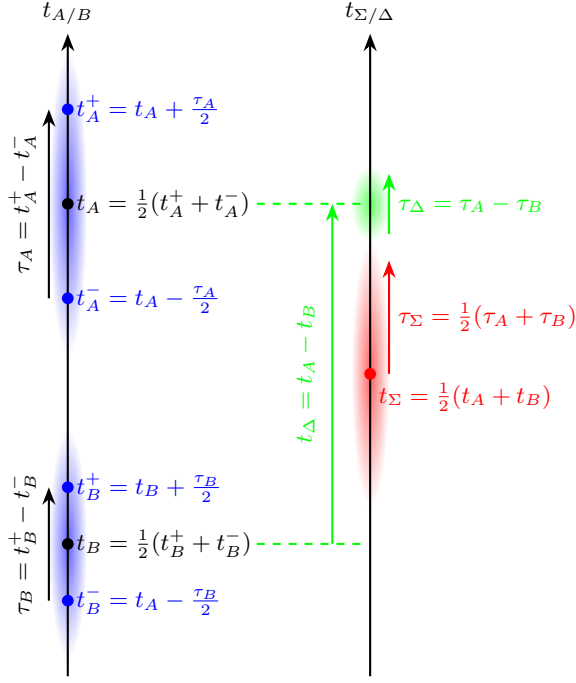

\Cref{fig/coordinates} graphically sums up these definitions. Depending on the situation, 
we can now use any of these choices of time coordinates
to represent two-electron coherence. In the next-subsection, we discuss
the frequency domain version of this representation, which enables
to properly identify the two-electron excitations, and provides the
appropriate coordinates to discuss the simple model of two-electron
collisions of \cref{sec/2e-collision}.

\subsubsection{Two-body frequency domain variables}
\label{appendix/G2/two-body/frequency}

For a single particle, the variables conjugated to $t$ and $\tau$
respectively are $\Omega$ and $\omega$ (see \cref{eq/G1/Wigner}). At the two-particle 
level, we have $\Omega_j$ conjugated to $t_j$ and $\omega_j$
conjugated to $\tau_j$ for $j=A$ or $B$. It is then easy to find the
conjugated variables to $t_\Sigma$ and $\tau_\Sigma$, as well as to
$t_\Delta$ and $\tau_\Delta$. As expected,
$\hbar\omega_\Sigma=\hbar\omega_A+\hbar\omega_B$ is the total energy of
the two particles and $\hbar\omega_\Delta=\hbar(\omega_A-\omega_B)/2$ is
the energy imbalance between the two particles.

Table \ref{table/conjugate-variables} summarizes these results including the
Fourier conjugates to the classical variables $t_\Sigma$ and $t_\Delta$. 

\begin{table} 
\renewcommand{\arraystretch}{1.5}
\begin{tabular}{|c|c|}\hline 
Time domain & Frequency domain \\ \hline\hline 
$(t_A,\tau_A)$ & $(\Omega_A,\omega_A)$ \\ 
$(t_B,\tau_B)$ & $(\Omega_B,\omega_B)$ \\ \hline\hline 
$(t_\Sigma,\tau_\Sigma)$ & 
$(\Omega_\Sigma=\Omega_A+\Omega_B,\omega_\Sigma=\omega_A+\omega_B)$\\ 
$(t_\Delta,\tau_\Delta)$ & 
    $(\Omega_\Delta=(\Omega_A-\Omega_B)/2,\omega_\Delta=(\omega_A-\omega_B)/2)$ \\ \hline 
\end{tabular} 
    \caption{\label{table/conjugate-variables} Couples of 
	classical/quantum variables and their Fourier 
conjugates. The first two lines correspond to each particle, 
individually whereas the last two lines correspond to the decomposition 
into ``center of mass'' and ``relative particle'' time coordinates for
	the two electrons.} 
\end{table} 

\subsection{The $T$-periodic case}
\label{appendix/G2/T-periodic}

At the two-electron level, $T$-periodicity leads to the quantization $\Omega_\Sigma=2\pi nf$ ($f=1/T$) since
translating all times simultaneously by $T$ leaves $(t_\Delta,\tau_\Sigma,\tau_\Delta)$ invariant and changes $t_\Sigma\mapsto 
t_\Sigma+T$. The two electron coherence is therefore localized within the codimension
one subspaces $\Omega_\Sigma=2\pi nf$ in the frequency domain:
\begin{equation}
\label{eq/G2/singularities/T-periodic}
v_F^2\widetilde{\Gee}(\pmb{\omega};\pmb{\Omega})=2\pi \sum_{n\in\mathbb{Z}}
\delta(\Omega_\Sigma-2\pi nf)\,\Wee_n(\pmb{\omega}|\Omega_\Delta)
\end{equation}
This generalizes the decomposition given by \cref{eq/G1/singularities} at the single electron
level. But there are additional singularities in the two-electron case. For
example, the full inter-channel two-electron coherence can be decomposed as
\begin{subequations}
\label{eq/G2/full-decomposition}
\begin{align}
\Gee_{AB}(1,2|1',2')&=\Ge_A(1|1')\,\Ge_B(2|2')
\label{eq/G2/full-decomposition/1}\\
&-\Ge_{AB}(1|2')\,\Ge_{BA}(2|1')
\label{eq/G2/full-decomposition/2}\\
&+\DWGee_{AB}(1,2|1',2')
\label{eq/G2/full-decomposition/3}
\end{align}
\end{subequations}
in which $\DWGee_{AB}(1,2|1',2')$ is the Wick deviation to $\Gee_{AB}$. But
$\Ge_A$, $\Ge_B$ and $\Ge_{AB}$ and $\Ge_{BA}$ are also time periodic in the
translation of all their time arguments. In the frequency domain, they are
singular distributions of the following form ($\xi=A$, $B$, $AB$ or $BA$):
\begin{equation}
\label{eq/G1/singularities}
v_F\widetilde{\Ge_\xi}(\omega;\Omega)=2\pi\sum_{n\in\mathbb{Z}}
\delta(\Omega-2\pi nf)\,\We_{\xi,n}(\omega)\,.
\end{equation}
Consequently terms given by Eqs.~\eqref{eq/G2/full-decomposition/1} and
\eqref{eq/G2/full-decomposition/2} contain two-delta singularities and thus
define a singular part of $\Gee$ as a distribution with support on a codimension
2 affine subspace of the frequency domain. As explained in \cref{sec/g2-and-entanglement/G2-anatomy}, the
Wick deviation \cref{eq/G2/full-decomposition/3} represents the effect of
interactions at the two particle level that are not arising from single particle
kinematics. Based on the experience from single electron decoherence
\cite{Ferraro-2013-1} as well as from explorations of its generalization to
higher orders, it is reasonable to assume that $\DWGee$ does not include
extra $\delta$-singularities in the frequency domain and corresponds to a
regular distribution associated with a function with discontinuities.

The singular parts associated with terms \eqref{eq/G2/full-decomposition/1}
and \eqref{eq/G2/full-decomposition/2} can be obtained from the decomposition
given by \cref{eq/G1/singularities}. Combined with \cref{eq/G1/singularities}, 
this finally leads to the complete
decomposition of $\Gee$ into singular contributions with supports of codimension
$2$ and $1$:
\begin{widetext}
	\begin{subequations}
	\label{eq/G2/singularities}
		\begin{align}
			\Wee_{AB,n}(\pmb{\omega}|\Omega_\Delta)&=2\pi\sum_{k\in\mathbb{Z}}
			\delta(\Omega_\Delta-\pi f(2k-n))\,\Ge_{A,k}(\omega_A)\,\Ge_{B,n-k}(\omega_B)
			\label{eq/G2/singularities/correlations}\\
			&-2\pi\sum_{k\in\mathbb{Z}}
			\delta\left(\omega_\Delta-\pi f\left(k-\frac{n}{2}\right)\right)\,
			\Ge_{AB,k}\left(\frac{\omega_\Sigma+\Omega_\Delta}{2}\right)\,
			\Ge_{BA,n-k}\left(\frac{\omega_\Sigma-\Omega_\Delta}{2}\right)
			\label{eq/G2/singularities/exchange}\\
			&+\DWGee_{AB,n}(\pmb{\omega}|\Omega_\Delta)
			\label{eq/G2/singularities/regular}
		\end{align}
	\end{subequations}
\end{widetext}
The singular support of $v_F^2\widetilde{\Gee}$
is thus confined to an arrangement of affine hyperplanes in the 4D frequency
domain forming a hierarchical comb structure. 
At the outer level, we have 
a pure Dirac comb defined by $\Omega_\Sigma=2\pi nf$ ($n\in\mathbb{Z}$). 
Within each leaf $\Omega_\Sigma=2\pi nf$,
the singularity structure
further decomposes into two transverse codimension-1 combs in $\Omega_\Delta$
and
$\omega_\Delta$ which lead to codimension-2 singularities in the 4D frequency
domain. These carry the
contributions given by 
Eqs.~\eqref{eq/G2/singularities/correlations} and
\eqref{eq/G2/singularities/exchange}. 
The first ones are always present since, in the presence
of a Fermi sea, it will always contribute to $\Gee$. The second ones are only
present when there is non-zero inter-channel single electron coherence.
The intersections of these singularity
carrying codimension-2 spaces are of codimension 3 but do not carry a specific
singularity.  
Finally, \cref{eq/G2/singularities/regular} is the 
codimension one singularity of
two-electron coherence in the frequency domain. 
As we shall see now, this singularity structure has deep consequences when
deriving the Cauchy-Schwarz constraints at the two particle level.

\section{Positivity constraints}
\label{appendix/positivity}

In this Appendix, we provide a review of the positivity constraints on
single and two-electron coherences. After a rapid review of known
results in Sec.~\ref{appendix/positivity/review}, we discuss the
Cauchy-Schwarz constraints on single and two-electron coherence. 

\subsection{General discussion}
\label{appendix/positivity/review}

A Hermitian operator $\mathbf{M}$ is called positive semi-definite when
\begin{equation}
	\langle \varphi |\mathbf{M}| \varphi\rangle\geq 0 
\end{equation}
for all $\ket{\varphi}$. Among positive semi-definite (PSD) operators, 
$\mathbf{M}$ is called positive definite if
and only if $\langle \varphi |\mathbf{M}| \varphi\rangle=0$ implies that
$\ket{\varphi}=0$, or equivalently that $\langle \varphi |\mathbf{M}|
\varphi\rangle>0$ for any non-zero $\ket{\varphi}$. 
Based
on the discussion of \cref{appendix/G1/operators}, the single particle coherences
$\OGe$ and $\OGh$ are expected to be PSD (but
not necessarily positive definite)
operators as well as the two-particle coherences
$\mathbf{G}^{(\alpha\beta)}$ with $\alpha$ and $\beta$ being either
$e$ or $h$. 

Positive semi-definiteness and positive definiteness are equivalent to conditions on
the spectrum of $\mathbf{M}$: positive semi-definiteness (resp. positive
definiteness) is
equivalent to $\text{Spec}(\mathbf{M})\subset \mathbb{R}^+=[0,+\infty)$ (resp.
$\text{Spec}(\mathbf{M})\subset \mathbb{R}^{+\,\star}$). 

Positivity can also be recast as a condition on two-dimensional
subspaces: $\mathbf{M}$ is positive semi-definite if and only if, for
every pair $(\ket{\varphi},\ket{\chi})$, its restriction
$\Pi\,\mathbf{M}\,\Pi$ with $\Pi$ being the orthogonal projector onto 
the plane
they span, is positive semi-definite. The off-diagonal part of this
condition is the Cauchy-Schwarz inequality
\begin{equation}
    \label{eq/positivity/CS-abstract}
    \left|\langle\varphi|\mathbf{M}|\chi\rangle\right|^2\leq
    \langle\varphi|\mathbf{M}|\varphi\rangle\,
    \langle\chi|\mathbf{M}|\chi\rangle \,,
\end{equation}
and its diagonal part is $\langle\varphi|\mathbf{M}|\varphi\rangle\geq0$.
Note that neither half suffices on its own: $\mathbf{M}=-\OId$ satisfies the
Cauchy-Schwarz inequality for every pair of states while being negative
definite.

None of these criteria is practical for implementation within
a tomography protocol which reconstructs the matrix elements of the 
electronic coherence in a specific basis. We are thus looking for criteria that can be
expressed within a given basis well-suited to the specific
protocol that is considered. Let $(\ket{\psi_i})_i$ denote such a
basis, then, Sylvester's criterion provides us with necessary and
sufficient conditions for positive semi-definiteness (resp. positive
definiteness) 
using only as data the matrix elements $M_{i,j}$ of $\mathbf{M}$ in this
basis. With $M$ the matrix
formed by these elements:
\begin{itemize}
	\item The operator $\mathbf{M}$ is positive definite if and only if all the
leading principal minors of $M$ are strictly positive.
\item The operator $\mathbf{M}$ is positive semi-definite if and only if
all the principal minors of $M$ are non-negative.
\end{itemize}
The Cauchy-Schwarz inequality within a given basis
\begin{equation}
	\forall (i,j)\quad \left|M_{i,j}\right|^2\leq M_{i,i}\,M_{j,j}
\end{equation}
as well as the positivity of diagonal elements
\begin{equation}
	\forall i,\quad M_{i,i}\geq 0
\end{equation}
only form a set of necessary conditions for positive semi-definiteness but
do not suffice to enforce it as it misses all the constraints on $M$
coming from order three and above principal minors. 

\subsection{The single-particle Cauchy-Schwarz constraints} 
\label{appendix/positivity/G1}

Let us now revisit the single-particle 
Cauchy-Schwarz inequalities 
first mentioned in \cite{Ferraro-2013-1} and
more generally in \cite{Wolk-2014-1}. These are necessary conditions ensuring that any
detection probability computed from single-electron coherence is 
non-negative.
Another set of inequalities is obtained by ensuring
that the hole coherence
$\mathcal{G}^{(h)}_\rho(1|1')=\mathrm{Tr}(\psi^\dagger(1)\,\rho\,\psi(1'))$
also leads to positive hole detection probabilities. 
Using the relation between electronic and hole coherence
leads to a complete set of Cauchy-Schwarz inequalities satisfied by
single particle coherences.

For a $T$-periodic source, these are expressed on the Wigner
distribution harmonics $\We_n(\omega)$ (see \cref{eq/G1/Fourier-series}) as:
\begin{subequations}
	\label{eq/G1/CS}
	\begin{align}
		\label{eq/G1/CS/electron}
		\left|\We_n(\omega)\right|^2 & \leq
		\We_0(\omega-\pi nf)\We_0(\omega+\pi nf)\\
		\label{eq/G1/CS/hole}
		\left|\We_n(\omega)\right|^2 & \leq \left(1-\We_0(\omega-\pi
		nf)\right)\left(1-\We_0(\omega+\pi nf)\right)
			\\
		\label{eq/G1/CS/electron/inter-channel}
		\left|\We_{AB,n}(\omega)\right|^2 & \leq
		\We_{AA,0}(\omega+\pi nf)\,\We_{BB,0}(\omega-\pi nf)
\end{align}
\end{subequations}
\Cref{eq/G1/CS/electron/inter-channel} involves the harmonics
$\We_{AB,n}$ of the inter-channel single-electron Wigner distribution,
whose reconstruction is discussed in
\cref{sec/protocols/G1/inter-channel}. It follows from the positivity of
$\OGe$ on $\mathcal{H}_A\oplus\mathcal{H}_B$, and not from the
positivity of each channel taken separately.

The derivation of these inequalities relies on the use of localizable
probes which are electronic wavepackets able to regularize the
$\delta$-singularies of two-electron frequencies in the frequency
domain. These are
the Martin Landauer wavepackets (also known as Shannon wavelets in
classical signal processing)
associated with energy ranges $[2\pi f_Nk,2\pi f_N(k+1)]$ with
$f_N=f/N$ ($N\in\mathbb{N}^*$) and
centered at times $t_l=NlT$ for $(k,l)\in\mathbb{Z}^2$. 
In practice, for each value of $N$, it is sufficient to consider the $l=0$
family
\begin{equation}
\text{ML}^{(N)}_{k,0}(t)=\frac{1}{\sqrt{v_FNT}}\frac{\sin(\pi
ft/N)}{\pi ft/N}\,\me^{-2\pi\mi f(k+1/2)t/N}\,
\end{equation}
with $k\in\mathbb{Z}$.
Sending $N$ to $+\infty$ localizes each wavepacket in the frequency domain
since their width is $2\pi f_N\rightarrow 0$. It is thus possible to
pin-point a plane wave of specific pulsation $\omega$ with a proper family 
of probe single electron $\text{ML}^{(N)}_{k_N(\omega),0}$ with $2\pi
f_N(k_N(\omega)+1/2)\rightarrow \omega$
as $N\rightarrow +\infty$. Given a pair of Martin Landauer wavepackets
enables us to pin-point a point $(\omega_+,\omega_-)$ within the
frequency domain of single electron coherence. 

The keypoint is that, using these wavepackets enables us to regularize
the singularity of the single electron coherence within the frequency
domain associated with $T$-periodicity. This is why, for single electron
coherence, 
this {\it local analysis} enables to
derive the Cauchy-Schwarz
inequality \cref{eq/G1/CS/electron}: it regularizes the
$\delta$-singularities present in the frequency domain. 

\subsection{The two-particle Cauchy-Schwarz constraints}
\label{appendix/positivity/G2}

We now discuss these inequalities at the two-particle level, starting
with two-electron coherence $\OGee$. Two points should be kept in mind.
First, positivity bears on the \emph{complete} two-particle coherence,
that is on the operator acting on the two-particle space built from the
single-particle space $\mathcal{H}_A\oplus\mathcal{H}_B$ of the two
channels, and not on each sector separately: constraints involving the
inter-channel sector $\Gee_{AB}$ therefore arise from probe pairs that
mix the two channels --- exactly as, at the single-particle level, the
bound $\OId\succeq\OGe\succeq0$ bears on the complete coherence
including its inter-channel block (see
\cref{sec/protocols/strategy/positivity}). Second, the Cauchy-Schwarz
inequality is only the off-diagonal half of positivity; the diagonal
half, whose two-particle version is the coincidence bound
\cref{eq/tomo/G2-positivity}, carries constraints of its own.

We will first adapt the local analysis used to derive Cauchy-Schwarz
constraints on single-particle coherences and see that it does not
constrain the Wick deviation to two-electron coherence $\ODWGee$, in
any of the four sectors $(ee)$, $(hh)$, $(eh)$ and $(he)$. Nevertheless, we will discuss the resulting constraints on
$\OGe$, which turn out to go beyond the pairwise Cauchy-Schwarz
inequalities \cref{eq/G1/CS}. Using single-particle atoms of signal, in
an approach adapted from Ref.~\cite{Roussel-2020-1}, we will then derive
explicit necessary constraints on $\DWGee_{AB}$ for the geometries of
\cref{fig/geometries}-(b) and (c).

\subsubsection{Local analysis}
\label{appendix/positivity/G2/local}

In order to write them down explicitly, one has to take into account the singularity structure of
$\Gee_{AB}$ unraveled in \cref{appendix/G2/T-periodic}. The less
singular contribution $\DWGee_{AB}$ lives within the $\Omega_\Sigma=2\pi nf$
codimension-1 subspaces. Then, we have
up to two families of singularities of codimension two
associated with planes
defined by
\begin{subequations}
\begin{align}
\mathcal{C}_{n,k}:\ \Omega_\Sigma&=2\pi nf\quad \&\ \Omega_\Delta=2\pi f(k-n/2)\\
\mathcal{E}_{n,k}:\ \Omega_\Sigma&=2\pi nf\quad \&\ \omega_\Delta=\pi f(k-n/2)
\end{align}
\end{subequations}
with $(k,n)\in\mathbb{Z}^2$. The diagonal corresponds to the
$n=k=0$ member of the first family. It is denoted by $\mathcal{C}$ to remind
that it arises from the correlations terms between channels $A$ and $B$. Due to
the presence of the Fermi seas, this first family is always non-empty. The
second family, denoted by $\mathcal{E}$ to emphasize that it arises from
exchange terms, is only present in the presence of non-vanishing inter-channel
single electron coherence.

Deriving the Cauchy-Schwarz constraints thus requires using two-particle
wave-functions coming from an orthonormal basis of single particle
wave-functions that provide a regularization for these singularities and which can
be localized around a given point of the two-particle frequency domain. This is
the two-particle version of the {\it local analysis} technique used to derive
\cref{eq/G1/CS/electron}. 

But at the two-electron level, the nested singularity structure described in the
previous subsection has to be taken into account. It introduces different scaling in $N$ 
of the codimension one versus
codimension two singularity contributions. 

A careful investigation
shows that the {\it local analysis} technique sketched at the end of
\cref{appendix/positivity/G1}
does not lead to any
Cauchy-Schwarz (CS) constraint on the $\DWGee_{AB}$ part of two-electron
coherence. This comes from the fact that, in the l.h.s. of the
Cauchy-Schwarz inequality for two-electron coherences, 
the contribution stemming from $\Ge$ via
Wick's formula has a $N^0$ scaling while $\DWGee_{AB}$ has a $N^{-1}$
scaling. Importantly, the contribution stemming from $\Ge$ via 
Wick's formula is present in the l.h.s. of the CS inequality when the 
off-diagonal target point belongs to codimension $2$
singularities supports. But since the diagonal is a codimension $2$
singularity support, the r.h.s. of the CS inequality always has a $N^0$
scaling. In the end, for a generic off-diagonal target point, that is 
a point within the co-dimension one singularities $\Omega_\Sigma=2\pi
nf$ for $n\in\mathbb{Z}$ that does not belong to
codimension-$2$ singularities, the $\DWGee_{AB}$ contribution disappears
in the $N\rightarrow +\infty$ limit compared to the r.h.s., 
thereby making the local analysis
unable to constrain it from Cauchy-Schwarz inequalities.

The other case is when the off-diagonal target point belongs to one of
the codimension-two singularities $\mathcal{C}_{n,k}$ with
$(n,k)\neq(0,0)$, or $\mathcal{E}_{n,k}$ with
$(n,k)\in\mathbb{Z}^2$. For such points, the local analysis does lead to
$N^0$ scaling constraints. On $\mathcal{C}_{n,k}$, only the direct term
survives and the constraint factorizes into two intra-channel
inequalities, which are exactly \cref{eq/G1/CS/electron}. On
$\mathcal{E}_{n,k}$, only the exchange term survives and one obtains
instead the inter-channel inequality
\cref{eq/G1/CS/electron/inter-channel}. In both cases, nothing is
obtained beyond the single-particle inequalities \cref{eq/G1/CS}.

The situation is different for off-diagonal points belonging to the
intersection of two codimension-two singularities. There is first a
codimension locking result: the off-diagonal point belongs to an
intersection $\mathcal{C}_{n,k}\cap\mathcal{E}_{n,k'}$ if and only if
both of its diagonal descendants belong to an intersection of two
codimension-two singularities, one of which is the diagonal
$\mathcal{C}_{0,0}$. We denote these two intersections by
$\mathcal{C}_{0,0}\cap\mathcal{E}_{0,k_\pm}$, with $k_+=k+k'-n$ and
$k_-=k'-k$.

When the two probe pairs entering the Cauchy-Schwarz inequality share a
single-particle mode, the resulting constraint is a principal minor of
order three of the single-electron coherence (see
\cref{appendix/positivity/review}) --- one mode in one channel, two
modes in the other. Therefore, the resulting cubic constraint on its matrix elements
is \emph{not} contained in the pairwise inequalities
\cref{eq/G1/CS}. At the remaining intersections, the two probe pairs
involve four distinct modes and the constraint becomes a
\emph{quartic} constraint on the matrix elements of $\OGe$. We have
determined that it is
neither a principal minor of order four, nor even a polynomial function
of the principal minors. It nevertheless still follows from
$\OGe\succeq0$. Explicit expressions can be written down but are not
needed here.

In all cases, the constraint bears on $\OGe$ alone and leaves $\ODWGee$
unconstrained.

The constraints obtained at the intersections of codimension-two
singularities are of a different nature. Although implied by
$\OGe\succeq0$, they couple several harmonics at once. Their systematic
use, like that of the full operator condition
$\mathbf{0}\preceq\OGe\preceq\OId$, goes beyond the scope of the
present work.

\subsubsection{Constraints from the hole and mixed sectors}
\label{appendix/positivity/G2/sectors}

Now, exactly as for single-electron coherence, the two-electron
coherence can be related to one electron, one hole coherence as well as
two hole coherences (see \cref{appendix/G2/operators/eh-he}). Imposing
the positivity of the probability for detecting two hole excitations or
one electron and one hole excitation also leads to some constraints on
two-electron coherences.

These four coherences are not independent: as shown in
\cref{appendix/G2/operators}, each of them is expressed in terms of
$\Gee$ and of the single-particle coherence alone. In the frequency
domain, these relations amount to linear changes of the frequency
variables --- charge conjugation reverses the frequencies attached to
the hole slots, and the mixed sectors further exchange the ket and bra
arguments of one slot --- supplemented by single-particle
contributions. A linear change of variables preserves the codimension
of the singular supports: in each sector, the Wick contributions remain
supported on codimension-two subspaces whereas $\DWGee$ remains a
regular function on the codimension-one subspaces
$\Omega_\Sigma=2\pi nf$. The scaling argument of
\cref{appendix/positivity/G2/local} therefore applies to each sector,
and the local analysis does not constrain $\DWGee$ in any of them. The
contact terms of the intra-channel $(he)$ sector do not involve
$\DWGee$ and do not alter this conclusion. The hole and mixed sectors
nevertheless carry constraints of their own on $\DWGee_{AB}$, which
become explicit in the basis of atoms of signal
(\cref{appendix/positivity/G2/Wannier}).

\subsubsection{Single particle atoms of signal}
\label{appendix/positivity/G2/Wannier}

The local analysis fails to constrain $\ODWGee$ when considering 
Martin-Landauer probes of width $2\pi f/N$ for $N\rightarrow +\infty$: the Wick
contributions, supported on codimension-two subspaces, then dominate
the codimension-one contribution of $\DWGee$ by a factor $N$. Probes of
finite width do not suffer from this suppression. A natural choice
for $T$-periodic sources is the basis of single-particle atoms of
signal built along a construction similar to the one presented in 
Ref.~\cite{Roussel-2020-1}. They are normalizable wavepackets
localized over roughly one period, built from a full
Floquet-Brillouin zone $2\pi f$. In this basis, all matrix elements are
finite, and the Wick and irreducible contributions are comparable. This
is the basis in which explicit constraints on $\DWGee_{AB}$ can be
written.

\paragraph*{Definitions.}
For each channel $\alpha\in\{A,B\}$, the $T$-periodicity of the complete
single-electron coherence $\OGe_\alpha$, Fermi sea included, implies
that it commutes with time translations by $T$. One can then follow the
road of Ref.~\cite{Roussel-2020-1}, with a key difference: in that
work, electron and hole atoms of signal were constructed separately by
projecting onto the electron and hole quadrants, whereas here we start
by diagonalizing the complete coherence of each channel, which provides
a Floquet-Bloch eigenbasis of the full single-electron coherence
$\OGe_\alpha$. 

When Fourier transformed with respect to the quasi-energy,
it defines single-particle atoms of signal $\varphi^\alpha_{a,l}$, labelled by a band
index $a$ and a period index $l\in\mathbb{Z}$, which are orthonormal,
time translated from each other ($\mathbf{T}_T\varphi^\alpha_{a,l}=
\varphi^\alpha_{a,l+1}$),
and in which
\begin{equation}
	\label{eq/G2/Wannier/G1}
	\langle\varphi^\alpha_{a,l}|\OGe_\alpha|\varphi^\alpha_{a',l'}\rangle
	=\delta_{a,a'}\,g^\alpha_a(l-l')\,.
\end{equation}
The diagonal element $n^\alpha_a\equiv g^\alpha_a(0)\in[0,1]$ is the
probability for the atom $\varphi^\alpha_{a,l}$ 
to be occupied and $g^\alpha_a(l\neq0)$
its coherence with the same atom $l$ periods apart. Because
single-particle atoms of signal give 
the single-electron coherence the
band-diagonal form \cref{eq/G2/Wannier/G1}, they simplify Wick
contributions below.

\paragraph*{The geometry of \cref{fig/geometries}-(b).}
In the Coulomb-coupler geometry depicted on \cref{fig/geometries}-(b), 
the two sources are independent and
the coupler does not transfer electrons between the two channels. 
Consequently, the
numbers of electrons $N_A$ and $N_B$ in each channel are separately
conserved. All matrix elements of $\OGee$ that change $(N_A,N_B)$
therefore vanish and $\OGee$ is block diagonal with respect to the
decomposition of the two-particle space into its $AA$, $AB$ and $BB$
sectors. Its positivity is equivalent to the positivity of each block,
and $\ODWGee_{AB}$ only appears in the $AB$ block, which acts on
$\mathcal{H}_A\otimes\mathcal{H}_B$. Since separate conservation also
implies $\Ge_{AB}=0$, the
constraints on $\ODWGee_{AB}$ only involve $\OGe_A$ and $\OGe_B$. Using
\cref{eq/eh-note/23,eq/G2/operators/eh-reduction,eq/G2/operators/he-reduction},
the positivity of the four sectors $(ee)$, $(hh)$, $(eh)$ and $(he)$ of
this block reads
\begin{subequations}
	\label{eq/G2/Wannier/four-sectors}
	\begin{align}
		\ODWGee_{AB} &\succeq -\,\OGe_A\otimes\OGe_B\,,
		\label{eq/G2/Wannier/four-sectors/ee}\\
		\ODWGee_{AB} &\succeq -\,(\OId-\OGe_A)\otimes(\OId-\OGe_B)\,,
		\label{eq/G2/Wannier/four-sectors/hh}\\
		\big(\ODWGee_{AB}\big)^{\mathsf{T}_B} &\preceq
		\OGe_A\otimes(\OId-\OGe_B)^{\mathsf{T}}\,,
		\label{eq/G2/Wannier/four-sectors/eh}\\
		\big(\ODWGee_{AB}\big)^{\mathsf{T}_B} &\preceq
		(\OId-\OGe_A)\otimes\big(\OGe_B\big)^{\mathsf{T}}\,,
		\label{eq/G2/Wannier/four-sectors/he}
	\end{align}
\end{subequations}
where $\mathsf{T}_B$ denotes the partial transposition with respect to
channel $B$, taken in a fixed basis of $\mathcal{H}_B$ (here, the atoms
of signal of channel $B$) in which the transposition $\mathsf{T}$ is
also evaluated. The
$(ee)$ and $(hh)$ sectors bound $\ODWGee_{AB}$ from below, whereas the
mixed sectors bound its partial transpose from above. There is no
contact term, since we are considering inter-channel two-electron
coherence.

\paragraph*{Diagonal elements: Fr\'echet-Hoeffding bounds.}
Consider a pair of atoms $P=(\varphi^A_{a,l},\varphi^B_{b,m})$ and the
diagonal element
$W_{PP}=\DWGee_{AB}(\varphi^A_{a,l},\varphi^B_{b,m}|\varphi^A_{a,l},\varphi^B_{b,m})$.
Since partial transposition leaves it unchanged, taking the diagonal
element on $P$ of each inequality of \cref{eq/G2/Wannier/four-sectors}
gives
\begin{align}
	\label{eq/G2/Wannier/Frechet}
	\max\big[-n^A_a n^B_b,\,-(1-n^A_a)(1-n^B_b)\big]
	&\leq W_{PP} \nonumber\\
	\leq \min\big[n^A_a(1-n^B_b),&\,(1-n^A_a)\,n^B_b\big]\,.
\end{align}
This result can also be derived from a physical argument:
Consider an ideal detector measuring the
occupations of the two atoms $\varphi^A_{a,l}$ and $\varphi^B_{b,m}$,
described by the occupation operators $\hat{n}^A_{a,l}$ and
$\hat{n}^B_{b,m}$, whose averages are
$\langle\hat{n}^A_{a,l}\rangle=n^A_a$ and
$\langle\hat{n}^B_{b,m}\rangle=n^B_b$ (independent of $l$ and $m$ by
$T$-periodicity, see \cref{eq/G2/Wannier/G1}). Since these operators
commute, they can be measured jointly, each outcome being $0$ or $1$.
The probabilities $p_{x,y}$ of the four possible outcomes ($x$,
$y\in\{0,1\}$ being the occupations of $\varphi^A_{a,l}$ and
$\varphi^B_{b,m}$) are
$p_{1,1}=\langle\hat{n}^A_{a,l}\hat{n}^B_{b,m}\rangle$,
$p_{0,0}=\langle(1-\hat{n}^A_{a,l})(1-\hat{n}^B_{b,m})\rangle$,
$p_{1,0}=\langle\hat{n}^A_{a,l}(1-\hat{n}^B_{b,m})\rangle$ and
$p_{0,1}=\langle(1-\hat{n}^A_{a,l})\hat{n}^B_{b,m}\rangle$, which are precisely
the diagonal elements on $P$ of the $(ee)$, $(hh)$, $(eh)$ and $(he)$
sectors. \Cref{eq/G2/Wannier/Frechet} thus expresses the non-negativity
of these four probabilities: it is the argument of
\cref{sec/protocols/strategy/positivity} applied to this particular
detection.
Equivalently, the probability
$p_{1,1}=n^A_an^B_b+W_{PP}$ that both single particle atoms of signal are occupied lies within the
Fr\'echet-Hoeffding bounds~\cite{Hoeffding-1940-1,Frechet-1951-1}:
\begin{equation}
\max(0,n^A_a+n^B_b-1)\leq p_{1,1}\leq\min(n^A_a,n^B_b)\,. 
\end{equation}
Since $p_{1,1}=n^A_an^B_b+W_{PP}$, the covariance of the two occupations
is $\mathrm{Cov}(\hat{n}^A_{a,l},\hat{n}^B_{b,m})=W_{PP}$: this is
\cref{eq/basics/DWGee/2} projected onto a pair of atoms, which relates
the connected coherence accessed by current correlations to the Wick
deviation, here without exchange contribution since $\Ge_{AB}=0$. The
coherence between different pairs of atoms is carried by the
off-diagonal elements $W_{PQ}$, constrained below.
The lower bound coming from the $(ee)$ sector is the
coincidence bound \cref{eq/tomo/G2-positivity} evaluated on atoms of
signal. Note finally that for an atom with a definite occupation
($n=0$ or $1$), \cref{eq/G2/Wannier/Frechet} forces $W_{PP}=0$: the
constraints only give information on partially occupied single particle atoms of
signal.

\paragraph*{Off-diagonal elements.}
For two pairs of atoms $P=(\varphi^A_{a,l},\varphi^B_{b,m})$ and
$Q=(\varphi^A_{a',l'},\varphi^B_{b',m'})$, the Cauchy-Schwarz
inequality associated with \cref{eq/G2/Wannier/four-sectors/ee} reads,
using \cref{eq/G2/Wannier/G1},
\begin{align}
	\label{eq/G2/Wannier/CS}
	\big|\,\delta_{a,a'}\delta_{b,b'}\,g^A_a(l-l')\,g^B_b(m-m')
	+W_{PQ}\big|^2 \leq&\nonumber\\
	\big(n^A_a n^B_b+W_{PP}\big)\,&\big(n^A_{a'}n^B_{b'}+W_{QQ}\big)\,,
\end{align}
with $W_{PQ}=\langle P|\ODWGee_{AB}|Q\rangle$. When $a\neq a'$ or
$b\neq b'$, the Wick contribution to the left-hand side vanishes and
the inequality bounds the irreducible part alone. The other three
sectors yield analogous inequalities.

\Cref{eq/G2/Wannier/Frechet,eq/G2/Wannier/CS} are explicit necessary,
but not sufficient, conditions for the positivity of $\OGee_{AB}$.
\Cref{eq/G2/Wannier/Frechet} is a box constraint on each diagonal
element $W_{PP}$, whose bounds only involve single-electron data
reconstructed at the first stage of the protocol, whereas the
right-hand side of \cref{eq/G2/Wannier/CS} also involves the diagonal
elements $W_{PP}$ and $W_{QQ}$. In Ref.~\cite{Bisognin-2019-2}, the
analogous single-electron constraints could be enforced sequentially,
the stationary part of the coherence being fixed beforehand by an
independent spectroscopy. Whether a similar ordering is possible here
depends on the structure of the tomographic equations in the basis of
atoms of signal, and is left for future work.

\paragraph*{The geometry of \cref{fig/geometries}-(c).}
When two independent sources $S_A$ and $S_B$, each conserving its
charge, feed an ideal electronic beam splitter described by the
single-particle unitary $\mathbf{U}$, the outgoing fields are linear in the
incoming ones. The Wick deviation then transforms covariantly,
$\ODWGee_{\text{out}}=(\mathbf{U}\wedge
\mathbf{U})\,\ODWGee_{\text{in}}\,(\mathbf{U}\wedge \mathbf{U})^\dagger$,
without any Gaussianity assumption on the sources. Independence and
charge conservation imply that $\ODWGee_{\text{in}}$ only lives in the
$AA$ and $BB$ sectors, where it reduces to the intra-channel Wick
deviations $\ODWGee_{S_A}$ and $\ODWGee_{S_B}$ of the two sources. The
positivity of the outgoing two-electron coherence is therefore
equivalent, together with that of the single-electron coherences, to
that of $\OGee_{S_A}$ and $\OGee_{S_B}$, while the
inter-channel Wick deviation accessed by the protocol is
\begin{align}
	\label{eq/G2/Wannier/BS-sum}
	\DWGee_{AB,\text{out}}&(1,2|1',2')=\nonumber\\
	&RT\,\big[\DWGee_{S_A}(1,2|1',2')+\DWGee_{S_B}(1,2|1',2')\big]\,,
\end{align}
which reduces
to \cref{eq/Franson/DGee/HBT-propagation} when $S_B$ is switched off.
The constraints thus bear on each source separately, each
$\DWGee_{S}$ being accessible by switching the other source off, and are
written in the atoms of signal of $\OGe_{S_A}$ (resp. $\OGe_{S_B}$). Two atoms of
the same channel being indistinguishable, the Wick contribution now
includes the exchange term. For two orthogonal atoms $i\neq j$ of the
same source, with $G_{ij}=\langle\varphi_i|\OGe_{S_A}|\varphi_j\rangle$
and $W_{ij}$ the corresponding diagonal element of $\ODWGee_{S_A}$, the
joint occupation probability is $n_in_j-|G_{ij}|^2+W_{ij}$ and the same
argument gives
\begin{widetext}
\begin{align}
	\label{eq/G2/Wannier/Frechet-intra}
	\max\big[-(n_in_j-|G_{ij}|^2),\,-((1-n_i)(1-n_j)-|G_{ij}|^2)\big]
	&\leq W_{ij}\nonumber\\
	\leq \min\big[n_i(1-n_j)+|G_{ij}|^2,&\,(1-n_i)n_j+|G_{ij}|^2\big]\,.
\end{align}
\end{widetext}
By \cref{eq/G2/Wannier/G1}, $G_{ij}$ vanishes for atoms of different
bands, and the bounds take the same form as
\cref{eq/G2/Wannier/Frechet}; for atoms of the same band, the inter-period coherence $g_a(l-l')$ shifts the bounds.

\paragraph*{Beyond these geometries.}
When the inter-channel coherence $\Ge_{AB}$ is generated within the
source itself, for instance by an interacting beam splitter, none of
these simplifications holds. The compression of $\OGee$ onto
$\mathcal{H}_A\otimes\mathcal{H}_B$, written in channel-resolved atoms
of signal, still provides necessary conditions on $\DWGee_{AB}$, with
the exchange contribution included in its Wick part. The complete
positivity condition, however, involves sectors that the present
protocol does not reconstruct. Finally, the atoms of signal are only
defined up to a quasi-energy dependent gauge transformation, under
which the Wick contributions are invariant whereas $\DWGee$ is not:
each gauge choice thus provides a different set of valid inequalities
on the same $\DWGee$.

\section{From a single collision to the \texorpdfstring{$T$}{T}-periodic source}
\label{appendix/dilute-to-periodic}

Section~\ref{sec/2e-collision} computes the two-electron coherence produced by
a \emph{single} collision of two energy-resolved wave packets in the
Coulomb-coupler geometry of \cref{fig/geometries}-(b), where
$\Ge_{AB}=0$ so that no exchange term is present. 

In practice, experiments are performed using
a $T$-periodic source ($T=1/f$). In this Appendix, we discuss how to
connect the single collision discussion to the $T$-periodic framework
presented in \cref{sec/protocols/G2} for two-electron tomography.

In particular, we show 
that the single-collision result
is the $f\to0$ limit of the periodic one, that periodization amounts to
sampling the single-shot envelope on a frequency comb and that the measured
$T$-periodic signal follows from the single collision through a simple
dictionary. We also discuss how the proper $T$-periodic entanglement
witness reduces to the frequency domain one discussed in
\cref{sec/2e-collision/entanglement/witness} thus clarifying its safe
application domain beyond the ideal two-electron collision.

Throughout this Appendix, a subscript $(1)$ denotes a single-shot
(one collision) quantity and a subscript $T$ its periodized counterpart.
$\Delta$ denotes the excess with respect to the stationary reference (Fermi
seas).

\subsection{Dilute construction}
\label{appendix/dilute-to-periodic/construction}

We assume that in the \emph{dilute limit} the wave packets of successive periods do not
overlap: their duration $\tau_{e}$ is $\ll T$. We also assume that
successive collisions are independent: no inter-period correlations are
generated beyond those internal to a single collision.
The excess coherences of the $T$-periodic
train are then obtained by superposing the single-shot event, rigidly
translated by integer numbers of periods (all time arguments shifted
together):
\begin{subequations}
\begin{align}
	\label{eq/append/dilute/G1}
	\Delta\Ge_{\alpha,T}(t\,|\,t')
	&=\sum_{m\in\mathbb Z}\Delta\Ge_{\alpha,(1)}(t-mT\,|\,t'-mT)\\
	\label{eq/append/dilute/G2}
	\Delta\Gee_{AB,T}&(t_1,t_2|t_1',t_2')=\nonumber\\
	&\sum_{m\in\mathbb Z}\Delta\Gee_{AB,(1)}(t_1-mT,t_2-mT\,|\,t_1'-mT,t_2'-mT)\,
\end{align}
\end{subequations}
in which $\alpha=A$ or $B$ in \cref{eq/append/dilute/G1}.
A key point is that there are no inter-period terms
$\Delta\Gee_{AB,(1)}$ evaluated at $(t_1-mT,t_2-mT\,|\,t_1'-m'T,t_2'-m'T)$ with $m\neq m'$ 
for the two-body Wick deviation $\DWGee_{AB}$: this quantity is built
from the interactions between electronic excitations belonging to the
same time period: we assume that the pairs of electrons belonging to different time
period do not see each-other.
The bridge between
the single collision and the periodic source is therefore cleanest precisely
on the object that carries the entanglement, $\DWGee_{AB}$, which remains a sum
of \emph{single-collision} contributions.

By contrast, the Wick (single-particle) part
$\Gee_{AB,W,T}=\Ge_{A,T}\,\Ge_{B,T}$ (no exchange terms are present since $\Ge_{AB}=0$)
receives inter-period cross terms, in which the electron detected in $A$
originates from one period and that detected in $B$ from another. As shown
below, these cross terms generate the codimension-two singularity grid
$\mathcal{C}_{n,k}$ of \cref{appendix/G2/T-periodic}.

\subsection{The \texorpdfstring{$f\to0$}{f->0} limit}
\label{appendix/dilute-to-periodic/limit}

\subsubsection{The comb sampling}
	\label{appendix/dilute-to-periodic/limit/comb}

Because \cref{eq/append/dilute/G1,eq/append/dilute/G2} are invariant under a
\emph{global} shift $t_i\to t_i+T$, the corresponding centre-of-mass
frequency
is locked onto a Dirac comb. Using
$\sum_{m}\me^{\mi\Omega mT}=2\pi f\sum_{n}\delta(\Omega-2\pi nf)$, the
periodized single-particle coherence reads (with $\Omega_\alpha$ the
centre-of-mass frequency of channel $\alpha$)
\begin{equation}
	\label{eq/append/comb/G1}
	\Delta\widetilde\Ge_{\alpha,T}(\omega|\omega')
	=2\pi f\sum_{n}\delta(\Omega_\alpha-2\pi nf)\;
	\Delta\widetilde\Ge_{\alpha,(1)}\big|_{\Omega_\alpha=2\pi nf}\,.
\end{equation}
For the irreducible two-particle Wick deviation, only
$\Omega_\Sigma$ is quantized:
\begin{equation}
	\label{eq/append/comb/G2}
	\Delta_W\widetilde\Gee_{AB,T}
	=2\pi f\sum_{n}\delta(\Omega_\Sigma-2\pi nf)\;
	\Delta_W\widetilde\Gee_{AB,(1)}\big|_{\Omega_\Sigma=2\pi nf}\,.
\end{equation}
The single-collision calculation thus provides the \emph{continuous
envelope} sampled by
the periodic train at $\Omega_\Sigma=2\pi nf$ with weight
$2\pi f$. The ``singular combs'' of the $T$-periodic description are exactly
these Dirac contributions. In the limit $f\to0$,
$$2\pi f\sum_n\delta(\,\cdot\,)\to\int\md\Omega_\Sigma\,(\cdot)\,.$$
The comb thus
becomes continuous and one recovers the single-collision object of
\cref{sec/2e-collision}. This is the precise sense in which
\cref{sec/2e-collision} is the $f\to0$ limit of the periodic protocol.

\subsubsection{The correlation grid $\mathcal{C}_{n,k}$.}
\label{appendix/dilute-to-periodic/limit/C-singularities}

The Wick part $\Gee_{AB,W,T}=\Ge_{A,T}\Ge_{B,T}$ is a product of two
independently combed single-particle coherences, and thus enforces the two
quantization
conditions simultaneously, $\Omega_A=2\pi n_Af$ and
$\Omega_B=2\pi n_Bf$ with $(n_A,n_B)\in\mathbb{Z}^2$. In the collective variables of
\cref{table/conjugate-variables} ($\Omega_\Sigma=\Omega_A+\Omega_B$,
$\Omega_\Delta=\tfrac12(\Omega_A-\Omega_B)$), this is
$\Omega_\Sigma=2\pi nf$ and $\Omega_\Delta=2\pi f(k-n/2)$ with
$(n,k)=(n_A+n_B,n_A)$. This gives the family $\mathcal{C}_{n,k}$ derived
top-down in \cref{eq/G2/singularities/correlations}. The present dilute
construction gives its bottom-up reading, and identifies where the Fermi seas
sit within it: since a sea coherence is stationary, hence diagonal in energy
($\Ge_F(\omega|\omega')\propto\delta(\omega-\omega')$) and pinned to
$\Omega=0$, the four Wick terms distribute as
\begin{align}
	\Gee_{AB,W}&=\underbrace{\Ge_{A,F}\Ge_{B,F}}_{\mathcal{C}_{0,0}}\\
	&+\underbrace{\Ge_{A,F}\,\Delta\Ge_B}_{\mathcal{C}_{n,0}}
	+\underbrace{\Delta\Ge_A\,\Ge_{B,F}}_{\mathcal{C}_{n,n}}\\
	&+\underbrace{\Delta\Ge_A\,\Delta\Ge_B}_{\text{all other}\
	\mathcal{C}_{n,k}\mathrm{s}}\,.
\end{align}
The Fermi seas occupy the corner $\mathcal{C}_{0,0}$ and the two edges
$k\in\{0,n\}$. The interior of the grid ($k\neq 0,n$) is the classically
correlated periodic source content. For a bare Fermi sea only $\mathcal{C}_{0,0}$
survives, which is why this family is always present. In the $f\to0$ limit the
grid carried by the $\Delta\Ge$ factors densifies into the continuous Wick
background of \cref{sec/2e-collision}, while the Fermi sea contribution $\mathcal{C}_{0,0}$
stays pinned at the origin.

\subsection{Entanglement witness}
\label{appendix/dilute-to-periodic/witness}

We now discuss the entanglement witness by bridging the gap between the
single two-electron collision case of \cref{sec/2e-collision} and the $T$-periodic
case of experiments.

\paragraph{The witness survives periodization.}
The witness \eqref{eq/gen-crit/G2} is quadratic in $\Gee_{AB}$. For a
$T$-periodic source it is naturally evaluated on Martin--Landauer wave
packets \cite{Martin-1992-1}, an orthonormal basis of normalized modes each
spanning one energy bin of width $2\pi f$ over a period. Their spectral
width is set by $f$, and as $f\to0$ they contract onto energy eigenstates:
this single parameter interpolates between the periodic source and the
single collision.

Above the Fermi sea, where the excitations of \cref{sec/2e-collision} live,
all matrix elements share a common $f$-scaling. For four such packets
centred at $(\omega_A,\omega_B\,|\,\omega_A',\omega_B')$, to leading order,
\begin{equation}
	\label{eq/append/ML-block}
	\Gee_{AB,T}(\varphi_A,\varphi_B\,|\,\varphi_A',\varphi_B')
	= f^2\,\widetilde\Gee_{AB,(1)}(\omega_A,\omega_B\,|\,\omega_A',\omega_B')
\end{equation}
with $\widetilde\Gee_{AB,(1)}=\widetilde\Ge_{A,(1)}\,\widetilde\Ge_{B,(1)}
+\Delta_W\widetilde\Gee_{AB,(1)}$ the single-collision coherence in the
frequency domain. The prefactor $f^2$ is common to the disconnected and the
irreducible parts, although they are singular in different ways: the
disconnected part carries two frequency combs and leaves two free frequency
integrations, the irreducible part a single comb and three integrations,
both yielding the same net power $f^2$.

Because \eqref{eq/append/ML-block} holds uniformly in the four arguments,
both sides of the witness carry the same factor $f^4$. At the operating
point of \cref{sec/2e-collision} the left-hand side reduces to
$$f^4\,\big|\widetilde\Gee_{AB,(1)}(\omega_A^+,\omega_B^+|\omega_A^-,\omega_B^-)\big|^2$$
and the right-hand side to
$$f^4\,\widetilde\Gee_{AB,(1)}(\omega_A^+,\omega_B^-|\omega_A^+,\omega_B^-)\,
\widetilde\Gee_{AB,(1)}(\omega_A^-,\omega_B^+|\omega_A^-,\omega_B^+)\,.$$
Eliminating the common $f^4$, the $T$-periodic witness reduces to the
single-collision frequency-domain witness of
\cref{sec/2e-collision/entanglement/witness}. The violation established
there therefore survives periodization, the $f\to0$ limit recovering it at
sharp energies.

\paragraph{Validity: keeping the modes above the Fermi sea.}
The reduction above rests on the uniform building block
\eqref{eq/append/ML-block}, which holds only for modes lying above the
Fermi sea. Being stationary, the Fermi sea has an intra-channel
single-electron coherence that is diagonal in energy: whenever one of the
four modes falls into the Fermi sea, an excitation amplitude
$\Delta\Ge_\alpha=\mathcal{O}(f)$ is replaced by a Fermi sea occupation
$n_{\alpha,F}=\mathcal{O}(1)$. This substitution can occur on either side
of the witness.

Its effect is most damaging on the right-hand side of the entanglement
witness. There, the two diagonal
coincidences carry the product of occupations $n_A\,n_B$: when both modes
reach the Fermi sea one recovers the corner $\mathcal{C}_{0,0}$, contributing at
$\mathcal{O}(1)$, and a single Fermi sea factor (the $\mathcal{C}_{n,0}$ and
$\mathcal{C}_{n,n}$ planes) contributes at $\mathcal{O}(f)$. On the
left-hand side, a Fermi sea factor in one channel together with an off-diagonal
excitation in the other gives $n_{\alpha,F}\,\Delta\Ge_\beta=\mathcal{O}(f)$,
hence $\mathcal{O}(f^2)$ once squared. In every case the Fermi-sea
contribution overwhelms the two-particule irreducible signal, which is only
$\mathcal{O}(f^4)$ on both sides. The $f^4$ homogeneity underlying the
reduction to the single-two particle collision is lost. 
This is the $f\to0$ counterpart of the local-analysis
obstruction discussed in \cref{appendix/positivity/G2/local}.

The remedy is simply to keep all four modes above the Fermi sea. Then, no
Fermi sea occupation factor appears, both sides of the witness are genuinely
$\mathcal{O}(f^4)$, and \eqref{eq/append/ML-block} applies uniformly. Under
this single condition, valid here because $\Ge_{AB}=0$ leaves no exchange
singularities, the $T$-periodic witness reduces to, and inherits the
violation of, the single-collision frequency-domain witness of
\cref{sec/2e-collision/entanglement/witness}.%

\section{HOM tomography}
\label{appendix/HOM-tomography}

The excess outgoing current noise in an HOM interferometer when both
sources placed on the incoming channels are switched ``on'' can be
written as \cite{Degio-2010-4}:
\begin{equation}
	\label{eq/HOM/A1}
	\Delta S_{11}^{(\text{on}/\text{on})}=RT\,\Delta
	\mathcal{Q}^{(\text{HOM})}+\Delta S_{11}^{(\text{on}/\text{off})}
	+\Delta S_{11}^{(\text{off}/\text{on})}
\end{equation}
where
\begin{equation}
	\label{eq/HOM/A2}
	\mathcal{Q}^{(\text{HOM})}=-2e^2\int_{\mathbb{R}}\,\overline{
		\DWe_1(t,\omega)\,\DWe_2(t,\omega)}^t\frac{\md\omega}{2\pi}
\end{equation}
is nothing but the overlap of the two incoming excess electronic Wigner
distributions. This is the quantity of interest that contains the effect
of two-particle interferences between the electron and hole
excitations emitted by both sources $S_1$ and $S_2$. 

An explicit computation of the excess Wigner function for a small ac-drive
at frequency $V_{\text{ac}}(t)=\cos(2\pi nft+\varphi_n)$ on top of
a dc-bias $-eV_{\text{dc}}=\hbar\omega_{\text{dc}}$ leads to
\begin{align}
	\DWe_{2}(t,\omega)&=\frac{eV_{\text{ac}}}{nhf}\,\sin(2\pi
	nft+\varphi_n)\,\left[f_{eq}(\omega-\omega_{\text{dc}}+\pi
	nf)\right.\nonumber\\ 
	&\left.-f_{\text{eq}}(\omega-\pi nf-\omega_{\text{dc}})\right]
	\label{eq/HOM/A3}
\end{align}
at first order in $\varepsilon_n=eV_{\text{ac}}/nhf$. In this expression, 
\begin{equation}
	\label{eq/HOM/A4}
	f_{\text{eq}}(\omega)=\frac{1}{\me^{\frac{\hbar\omega}{k_BT_{\text{el}}}}+1}
\end{equation}
is the Fermi-Dirac equilibrium distribution at temperature
$T_{\text{el}}$. Substituting \eqref{eq/HOM/A3} into Eq.~\eqref{eq/HOM/A2} leads
to the derivative of the HOM contribution to
the experimental noise signal with respect to the dimensionless ac
voltage drive $\varepsilon_n$:
\begin{widetext}
	\begin{equation}
		\label{eq/HOM/A5}
		\frac{\partial\,\mathcal{Q}^{(\text{HOM})}}{\partial
		\omega_{\text{dc}}}=-2e^2\,\varepsilon_n\,\int_{\mathbb{R}}\,
		\mathcal{F}_{\text{th}}(\omega)\,
		\overline{\left[
			\DWe_1(t,\omega+\omega_{\text{dc}}+\pi
			nf)-\DWe_1(t,\omega+\omega_{\text{dc}}-\pi nf)\right]\,
		\sin(2\pi nft+\varphi_n)}^t\,
		\frac{\md\omega}{2\pi},
	\end{equation}
\end{widetext}
in which the thermal smearing is accounted for by:
\begin{equation}
	\mathcal{F}_{\text{th}}(\omega)=\frac{\hbar}{4k_BT_{\text{el}}}\,\frac{1}{\cosh^2\left(
        \frac{\hbar\omega}{2k_BT_{\text{el}}}\right)}\,.
\end{equation}

\section{Reconstruction of the inter-channel single-electron coherence}
\label{appendix/GAB}

The single-electron tomography of \cref{sec/protocols/G1/general}
reconstructs the intra-channel coherences $\mathcal{G}^{(e)}_A$ and
$\mathcal{G}^{(e)}_B$ directly. Accessing the inter-channel coherence
$\mathcal{G}^{(e)}_{AB}$, which encodes single-particle delocalization
between the channels, requires interfering $A$ and $B$. Both protocols below
rely on an ideal QPC of transmission $T$ and reflection $R$ recombining the
incoming channels into two outgoing ones
$A_{\text{out}},{B_{\text{out}}}$,
\begin{subequations}
\begin{align}
	\psi_{A_{\text{out}}} &= \sqrt{T}\,\psi_A + \mi\sqrt{R}\,\psi_B\\
	\psi_{B_{\text{out}}} &= \mi\sqrt{R}\,\psi_A + \sqrt{T}\,\psi_B ,
\end{align}
\end{subequations}
and reduce the reconstruction of $\mathcal{G}^{(e)}_{AB}$ to the
single-electron tomography of \cref{sec/protocols/G1/general}.%

\subsection{Beam splitter followed by photo-assisted filtering}
\label{appendix/GAB/recombine-and-filter}

Here the QPC is placed first and a photo-assisted filter is applied to each
outgoing channel. The output intra-channel coherences read
\begin{align}
	\mathcal{G}^{(e)}_{A_{\text{out}}} &= T\,\mathcal{G}^{(e)}_A + R\,\mathcal{G}^{(e)}_B
   + \mi\sqrt{TR}\,\bigl(\mathcal{G}^{(e)}_{AB}-\mathcal{G}^{(e)}_{BA}\bigr),
\\
	\mathcal{G}^{(e)}_{B_{\text{out}}} &= R\,\mathcal{G}^{(e)}_A + T\,\mathcal{G}^{(e)}_B
   - \mi\sqrt{TR}\,\bigl(\mathcal{G}^{(e)}_{AB}-\mathcal{G}^{(e)}_{BA}\bigr).
\end{align}
Reconstructing $\mathcal{G}^{(e)}_{A_{\text{out}}}$ and
$\mathcal{G}^{(e)}_{B_{\text{out}}}$ by the protocol
of Sec.~\ref{sec/protocols/G1} and taking their difference at a balanced
QPC ($T=R=1/2$) eliminates the intra-channel contributions:
\begin{equation}
	\mathcal{G}^{(e)}_{A_{\text{out}}} -
	\mathcal{G}^{(e)}_{B_{\text{out}}}
   = \mi\,\bigl(\mathcal{G}^{(e)}_{AB}-\mathcal{G}^{(e)}_{BA}\bigr),
\label{eq:GAB_diff}
\end{equation}
so that no subtraction of the separately measured $\mathcal{G}^{(e)}_A$,
$\mathcal{G}^{(e)}_B$ is required. Equation~\eqref{eq:GAB_diff} yields the
antisymmetric combination $\mathcal{G}^{(e)}_{AB}-\mathcal{G}^{(e)}_{BA}$;
inserting a tunable phase $\varphi$ in one input arm
($\psi_B\to e^{\mi\varphi}\psi_B$) rotates
$\mathcal{G}^{(e)}_{AB}\to e^{\mi\varphi}\mathcal{G}^{(e)}_{AB}$. Combining
the results at $\varphi=0$ and $\varphi=\pi/2$ yields the full complex
$\mathcal{G}^{(e)}_{AB}$. This variant is a close adaptation of the
intra-channel  single-electron tomography protocol of
\cref{sec/protocols/G1/general} obtained at
at the cost of an added interferometer phase and of filtering both
outputs.

\subsection{Photo-assisted drive followed by a beam splitter}
\label{appendix/GAB/filter-and-recombine}

Here the photo-assisted drive acts on the incoming channels before the QPC,
and only a static energy filter (at energy $\hbar\omega_e$) sits on each
output. 
Retropropagating the average outcoing current across the static energy
filter, the QPC and the photo-assisted stage leads to:
\begin{align}
	\left.\frac{\partial \langle \mi_{A_{\text{out}}}\rangle}{\partial \varepsilon_A}
	\right|_{\mathrm{cross}}
	&\propto \mi\sqrt{TR}\,
\left[\, e^{\mi\varphi_A}\,\mathcal{G}^{(e)}_{AB}(\omega_e|\omega_e-2\pi
	f)\right.\nonumber \\
	   &+ \left.
	   e^{-\mi\varphi_A}\,\mathcal{G}^{(e)}_{AB}(\omega_e|\omega_e+2\pi
	   f)\,\right]
+ \mathrm{c.c.}\,
\label{eq:GAB_drive}
\end{align}
in which $\varphi_A$ denotes the phase of the drive applied to channel
$A$ and $\varepsilon_A$.. 
The inter-channel coherence at off-diagonal parameter $\Omega=\pm 2\pi f$ is
thus accessed at first order by driving the channel $A$, with the drive
phase $\varphi_A$ playing the role of an interferometer phase. Measurements
at $\varphi_A$ and $\varphi_A+\pi/2$ separate the two quadratures. The
intra-channel contribution $\propto\mathcal{G}^{(e)}_A$ is removed by the
balanced difference of the two outputs,
\begin{equation}
\frac{\partial}{\partial \varepsilon_A}
	\bigl(\langle i_{A_{\text{out}}}\rangle - \langle i_{B_{\text{out}}}\rangle\bigr)
\propto \mathcal{G}^{(e)}_{AB},
\end{equation}
and the resulting inverse problem is identical in form to
the one obtained for intra-channel single electron tomography
\cref{eq/tomo-g2/photoassisted/12}
with the inter-channel Wigner distribution
$\Delta W^{(e)}_{AB}$ in place of the intra-channel one. This variant needs no
additional interferometer phase and only
static filters on the outputs. Driving the $n$-th harmonic gives access to
single electron coherences along the codimension one singularities at
$\Omega=\pm 2\pi nf$.

Both variants access the full inter-channel coherence and reduce to the
single-electron inverse problem of
\ref{sec/protocols/G1/inverse-problem}, the
single-electron counterpart of Hong--Ou--Mandel
tomography~\cite{Degio-2010-4}.

\section{Two-electron tomography equations}
\label{appendix/tomo-g2}

In this Appendix, one derives the explicit tomography equations for 
reconstructing $\DcWee_{AB}$. The appendix is organized into three
subsections corresponding to the presence or absence of an rf drive
on each of the photo-assisted filters $\mathcal{F}_{A/B}$.

\subsection{Contribution at order $(\varepsilon_A^0,\varepsilon_B^0)$}
\label{appendix/tomo-g2/00}

In this case, no rf drive is applied. The filters $\mathcal{F}_A$ and
$\mathcal{F}_B$ consist of static energy filters such as the one
considered in Ref.~\cite{Altimiras-2010-1}. 
This enables reconstructing
the diagonal part of two-electron coherence:
\begin{equation}
    \label{eq/g2-tomo/final/00-equation}
    \left(S_{AB}\right)_{\pmb{\varepsilon}=0}
    =e^2\int_{\mathbb{R}^2}\mathcal{T}_{\pmb{\omega}_e}(\pmb{\omega})
    \,\Delta_cW_{AB,0}^{(2e)}(\pmb{\omega}|\Omega_\Delta=0)\,
    \frac{\md^2\pmb{\omega}}{(2\pi)^2}\,
\end{equation}
where
$\mathcal{T}_{\pmb{\omega}_e}(\pmb{\omega})=
    \mathcal{T}_{\omega_{e,A}}(\omega_A)
    \mathcal{T}_{\omega_{e,B}}(\omega_B)$.
Sweeping the static filter energies enables sweeping the diagonal plane
$\Omega_A=\Omega_B=0$. This is the two-channel, two-electron version of
the spectroscopy presented in Ref.~\cite{Altimiras-2010-1}.

\subsection{Contribution at order $(\varepsilon_A,\varepsilon_B^0)$}
\label{appendix/tomo-g2/10}

In this case, no rf drive is applied to $\mathcal{F}_B$. We thus select $\Omega_B=0$ 
which corresponds, for the $n$-th
harmonic $\DcWee_{AB,n}$ to $\Omega_A=2\pi nf$. 
The linear response of the current
correlation to $\varepsilon_A$ is then
\begin{widetext}
\begin{equation}
    \label{eq/g2-tomo/final/01-equation}
    \left(\frac{\partial
    S_{AB}}{\partial\varepsilon_A}\right)_{\pmb{\varepsilon}=0}=
    e^2\int_{\mathbb{R}^2}
    \mathcal{T}_{\omega_{e,B}}(\omega_B)\,\left[
        \mathcal{T}_{\omega_{e,A}}\left(\omega_A+\pi nf\right)
        -\mathcal{T}_{\omega_{e,A}}\left(\omega_A-\pi nf\right)
        \right]\,
\overline{\sin(2\pi
    nft_A-\varphi_A)\,\Delta_cW^{(2e)}_{AB}(\mathbf{t},\pmb{\omega})}^{\mathbf{t}}\,
        \frac{\md^2\pmb{\omega}}{(2\pi)^2}\,.
\end{equation}
\end{widetext}
in which $\overline{\ }^{\mathbf{t}}$ denotes averaging over one period on
$t_\Sigma$ and integrating over $\mathbb{R}$ for $t_\Delta$.
Exchanging the role of $A$ and $B$ in the above relation gives the
expression for the contribution at order
$(\varepsilon_A^0,\varepsilon_B)$.

\subsection{Contribution at order $(\varepsilon_A,\varepsilon_B)$}
\label{appendix/tomo-g2/11}

Finally, let us consider the case where both photoassisted filters
$\mathcal{F}_A$ and $\mathcal{F}_B$ are driven. 
The
time dependence of the
filtering $\mathcal{F}_{A,B}(\mathbf{t},\pmb{\omega})$ then involves
\begin{align}
    \sin(\Omega_At_A &-\varphi_A)\sin(\Omega_Bt_B-\varphi_B)=\nonumber \\
    &\frac{1}{2}\left[\cos(2\Omega_\Delta t_\Sigma+\Omega_\Sigma
    t_\Delta/2-(\varphi_A-\varphi_B))\right.\nonumber\\
    &\left. -
    \cos(\Omega_\Sigma t_\Sigma+\Omega_\Delta
    t_\Delta-(\varphi_A+\varphi_B))
    \right]
    \label{eq/g2-tomo/final/11-intermediate}
\end{align}
with $\Omega_\Sigma=\Omega_A+\Omega_B$ and
$\Omega_\Delta=(\Omega_A-\Omega_B)/2$. The
$T$-periodicity implies that one has to fix
\begin{equation}
    \Omega_{A/B}=\pi nf\pm\Omega_\Delta
\end{equation}
to capture specifically the $n$-th harmonic
$\Delta_cW^{(2e)}_{AB,n}(\pmb{\omega}|\Omega_\Delta)$ at fixed $n\in\mathbb{Z}^*$
and $\Omega_\Delta$.

Selecting the second term in the r.h.s of
Eq.~\eqref{eq/g2-tomo/final/11-intermediate} is achieved
by adding the experimental data for two configurations of the phases
$(\varphi_A,\varphi_B)$ such that $\varphi_A+\varphi_B=\varphi$ and
$\varphi_A-\varphi_B=0$ or $\pi$. This leads to the two phase
configurations:
\begin{equation}
	\label{eq/tomo-g2/T-periodic/phase-configuration}
    \left(\frac{\varphi}{2},\frac{\varphi}{2}\right)\quad
    \left(\frac{\pi}{2}+\frac{\varphi}{2},-\frac{\pi}{2}+\frac{\varphi}{2}\right)\,.
\end{equation}
Denoting by $\Sigma_{AB}(\varphi)$ the corresponding linear combination
of experimental results, which now depends on the phase $\varphi$, we
have
\begin{widetext}
    \begin{equation}
        \label{eq/g2-tomo/final/11-equation}
        \left(
        \frac{\partial^2\Sigma_{AB}(\varphi)}{\partial\varepsilon_A\,\partial\varepsilon_B}
        \right)_{\pmb{\varepsilon}=0}
        =e^2\int_{\mathbb{R}^2}\prod_{\alpha\in\{A,B\}}
        \left[\mathcal{T}_{\omega_{e,\alpha}}
        \left(\omega_\alpha+\frac{\Omega_\alpha}{2}\right)
        -\mathcal{T}_{\omega_{e,\alpha}}\left(\omega_{\alpha}-\frac{\Omega_{\alpha}}{2}\right)
        \right]\,
    \overline{\cos(2\pi nft_\Sigma +\Omega_\Delta t_\Delta-\varphi)
        \,\Delta_cW^{(2e)}_{AB}(\mathbf{t},\pmb{\omega})}^{\mathbf{t}}\,
        \frac{\md^2\pmb{\omega}}{(2\pi)^2}\,.
    \end{equation}
\end{widetext}
with $\Omega_{A/B}=\pi nf\pm\Omega_\Delta$. Shifting $\varphi\mapsto
\pi/2+\varphi$ then ensures that we are capturing
\begin{equation}
\overline{\sin(2\pi nft_\Sigma +\Omega_\Delta t_\Delta-\varphi)
		\,\Delta_cW^{(2e)}_{AB}(\mathbf{t},\pmb{\omega})}^{\mathbf{t}}
\end{equation}
which is consistent with the common form given by
Eq.~\eqref{eq/tomo-g2/T-periodic/generic-form}.

\section{Capacitive model for $e+e$ collisions}
\label{appendix/collision-model}

In this Appendix, we present the capacitive model used in \cref{sec/2e-collision}
and derive the two-particle scattering amplitude for two-electron collision
within the Coulomb coupler.

\subsection{The model}
\label{appendix/collision-model/model}

We consider two copropagating edge channels that are capacitively coupled to the
ground as well as between them, as depicted on \cref{cap_model}. The capacitance matrix
\begin{equation}
	\label{eq/collision-model/1}
	\OC =
	\begin{pmatrix}
		c_{gA} + C & -C \\
		-C & c_{gB} + C
	\end{pmatrix}
\end{equation}
determines the uniform time-dependent potential within each chiral channel
as a function of the charge present on them. Note that due to the coupling to
the ground plane, this capacitance matrix is invertible
with determinant $D=C(c_{gA}+c_{gB})+c_{gA}c_{gB}$. 

As it propagates across the Coulomb coupler at velocity $v_F$, 
an electron within channel $\xi \in \{A,B\}$ acquires an electrical phase term
\begin{equation}
	\label{eq/collision-model/2}
	\exp\left(\frac{\mi e}{\hbar}\int_{t_{\xi,\text{in}}}^{t_{\xi,\text{out}}}
	U_\xi(t')\,\md t'\right)
\end{equation}
where $t_{\xi,\text{in}/\text{out}}$ denotes the entrance (resp. exit) time into
(resp. from) the Coulomb coupler and $U_{\xi}(t')$ denotes the time-dependant 
uniform potential within channel $\xi$ 
which depends on the time-dependent charges $Q_A(t)$ and $Q_B(t)$
during its transit. In a two-electron collision, because one electron is present all along its travel,
this phase contains 
a constant phase shift $\varphi_\xi$ due to its electrostatic self-interaction. 
This phase reads $\varphi_\xi= -(\pi \tau_L)/(R_K C^{(\xi)}_{\mathrm{eff}})$ with $R_K=h/e^2$ 
the Von Klitzing constant and $C^{(\xi)}_{\mathrm{eff}}=D/(C+c_{g\xi})$.

But the potential $U_{\xi}$ is also function of the charge on the other edge
channel. When 
both electrons are simultaneously present in the interaction region, an additional two-body phase is accumulated, 
which depends on their relative injection times. 
Assuming that the interaction is effective only when the particles overlap, this phase reads
\begin{subequations}
\label{eq/collision-model/interaction-phase}
\begin{align}
	\varphi_{AB}(t_A - t_B) &= 
	-\omega_{AB}\tau_L\,\mathcal{F}
	\left(\frac{|t_A - t_B|}{\tau_L}\right)\\
	\mathcal{F}(x)&=\left(1 - |x|\right)\,\Theta(1-|x|)
\end{align}
\end{subequations}
where $\omega_{AB} = 2\pi/(R_K C^{(AB)}_{\mathrm{eff}})$. The
effective capacitance
with $ C^{(AB)}_{\mathrm{eff}}=D/C$ sets the 
Coulomb interaction energy scale and $\Theta(x)$ is the Heaviside distribution. 

\subsection{Interaction parameter estimates}
\label{appendix/collision-model/coupling-constant}

Moreover, using experimentally relevant parameters for quantum Hall edge
channels, we estimate the dimensionless interaction strength $\alpha_{\text{eff}}$ as the maximum
phase acquired during the two-electron collision:
\begin{equation}
	\label{eq/collision-model/6}
	\alpha_{\text{eff}} = \omega_{AB}\tau_L = \frac{e^2 \tau_L}{C_{AB} \hbar}\, ,
\end{equation}
where $C_{AB}$ denotes the effective capacitance of the interaction region.
Approximating the interaction region as two capacitively coupled conductors, the
capacitance is of generic form
\begin{equation}
	\label{eq/collision-model/7}
	C_{AB} = \varepsilon_0 \varepsilon_R L\, \zeta,
\end{equation}
where $L$ denotes the length of the Coulomb coupler and $\zeta$ is a geometric factor that
depends on the specific layout of the edge channels. In the case of two parallel
channels separated by a distance $d$ and of width $w$, one has
\begin{equation}
	\label{eq/collision-model/8}
	\zeta^{-1} = \ln\left(\frac{d+2w}{8w}\right) \, .
\end{equation}
The interaction strength can be conveniently expressed in terms of the
fine-structure constant $\alpha_{\mathrm{QED}} = e^2/4\pi \varepsilon_0 c
\hbar\simeq 1/137$, leading to
\begin{equation}
	\label{eq/collision-model/9}
	\alpha_{\text{eff}} = 4\pi\,
	\frac{\alpha_{\mathrm{QED}}}{\varepsilon_R}\,\frac{c}{v_F}\,\zeta^{-1}\, .
\end{equation}
For typical experimental values $v_F=\SI{e5}{\meter\second^{-1}}$,
$\varepsilon_R=12.4$, $d = \SI{1}{\micro\meter}$, and $w=\SI{100}{\nano\meter}$, we obtain
\begin{equation}
	\label{eq/collision-model/10}
	\alpha_{\text{eff}} \simeq 9 \, .
\end{equation}
We emphasize that these parameters correspond to a strong coupling regime, 
which underlies the structure of the spectra discussed in the main text. 

\subsection{The two-electron scattering matrix}
\label{appendix/collision-model/scattering-matrix}

The two-electron propagator across the Coulomb coupler can now be computed. 
Denoting $(t_A,t_B)$ the two times at which the wavepackets in A and B enter the
interaction region and $(t'_A,t'_B)$ the times at which they exit, the full
propagator describing our model is then
\begin{equation}
	\label{eq/collision-model/time-propagator}
	K(t'_A,t'_B; t_A,t_B) = e^{i \varphi_{AB} (t_A-t_B)} 
	\prod_{\xi \in \{A,B\}} \delta(t'_\xi-t_\xi -\tau_L) \, e^{i\varphi_\xi} \, ,
\end{equation}
where the Dirac distributions account for ballistic propagation at constant
velocity and the phase factors encode the effect of Coulomb interactions,
including the Coulomb interaction inter-channel phase $\varphi_{AB} (t_A-t_B)$.

Note that this scattering matrix can also be expressed in terms of the
center of mass and relative particle coordinates $(t_\Sigma,t_\Delta$
defined in \cref{eq/variables/2body-definition}:
\begin{align}
    \label{eq/collision-model/time-propagator/CM-RP}
	K(t'_\Sigma,t'_\Delta; t_\Sigma,t_\Delta) &= \me^{\mi \varphi_{AB}
	(t_\Delta)}\me^{\mi(\varphi_A+\varphi_B)}\nonumber \\
	& \times
	\delta(t_\Sigma-t'_\Sigma-\tau_L)\,\delta(t'_\Delta-t_\Delta)\,.
\end{align}
which shows the expected factorization between free propagation for the
center of mass particle and a time dependent phase for the relative
particle.

For a two-electron collision with incoming state $\ket{\psi_{\text{in}}}$, 
the two-particle scattering amplitude in the energy domain is obtained by projecting the propagated state onto 
outgoing energy eigenstates
$|\omega_A,\omega_B\rangle=\ket{\omega_A}_A\wedge\ket{\omega_B}_B$:
\begin{equation}
	\label{eq/collision-model/3}
	S(\omega_A,\omega_B|\psi_{\text{in}}) = \langle \omega_A,\omega_B|K|
	\psi_{\text{in}} \rangle\,
\end{equation}
This leads to the two-electron coherence in the frequency domain:
\begin{equation}
	\label{eq/collision-model/4}
	\Gee_{AB}(\omega_A^+,\omega_B^+|\omega_A^-,\omega_B^-)=
	S(\omega_A^+,\omega_B^+|\psi_{\text{in}})\,S(\omega_A^-,\omega_B^-|\psi_{\text{in}})^*\,.
\end{equation}
In \cref{sec/2e-collision} of the main text, we consider incoming electrons described by Landau wave packets. 
Denoting by $\psi_\xi$ the wave packet injected in channel $\xi \in \{A,B\}$, we take
\begin{equation}
	\label{eq/collision-model/5}
	\psi_\xi(t) = -i \sqrt{\gamma_\xi}\,
	\exp\!\left[i\omega_{0\xi} t - \frac{\gamma_\xi}{2}(t - t_{0\xi})\right]\,
	\Theta(t - t_{0\xi}),
\end{equation}
where $\gamma_\xi$ sets the energy width of the wave packet, $\omega_{0\xi}$ its
central frequency, and $t_{0\xi}$ its injection time. The incoming two-electron
wavepackets is then $\psi_{\text{in}}(t_A,t_B)=\psi_A(t_A)\,\psi_B(t_B)$.

For this choice of wave packets, the time integrals defining the scattering
amplitude can be evaluated analytically, leading to closed-form expressions for
the two-electron coherence.  In practice, however, the results presented in the
main text are obtained using numerical evaluation, which allows for a more
general treatment of different wave packets and a direct visualization of
energy-domain correlations.

\subsection{Effect of the coupling strength}
\label{appendix/collision-model/coupling-strength-effect}

\begin{figure}[h!]
	\includegraphics[width=\columnwidth]{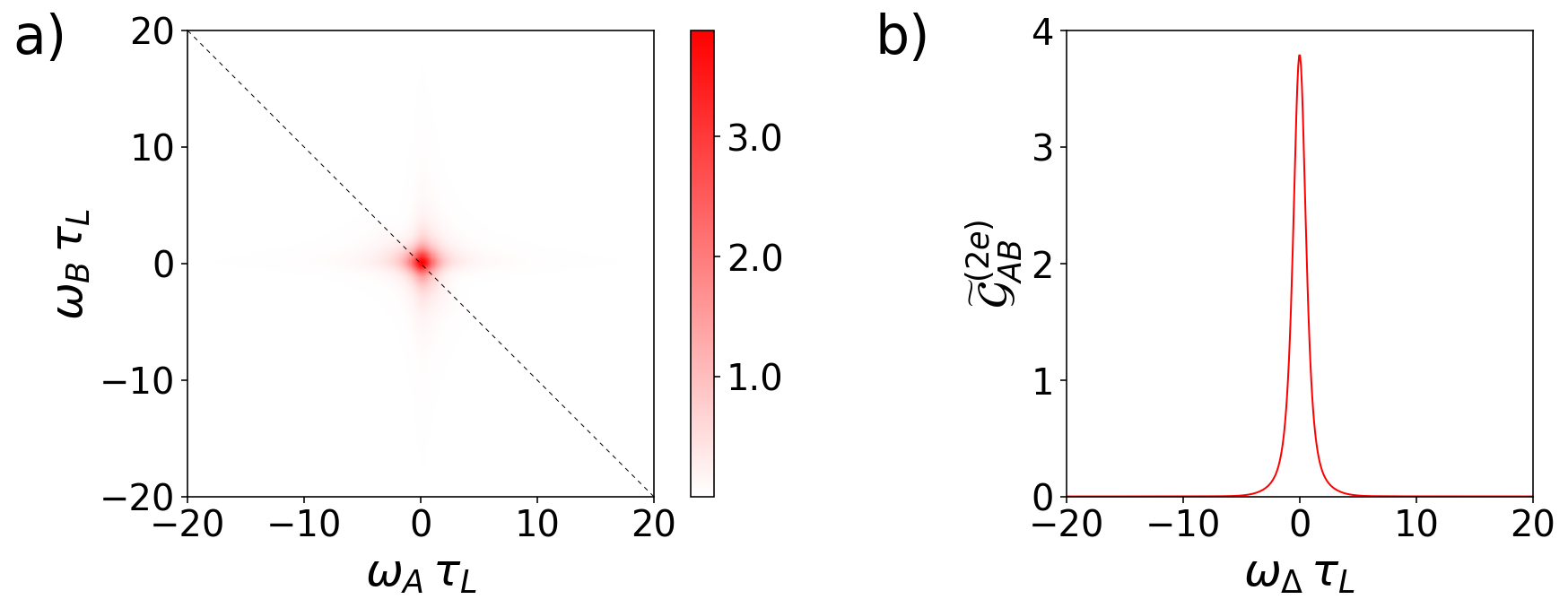}
	\caption{Outgoing diagonal two-electron coherence for two Landau
		wavepackets injected simultaneously into the Coulomb coupler. Both
		wavepackets have identical parameters $(\omega_{A}^{0}, t_{A}^{0},
		\gamma_{A}) = (\omega_{B}^{0}, t_{B}^{0}, \gamma_{B}) = (0, 0, 2 /
		\tau_{L})$. The interaction strength is given by the coupling
		parameter $\alpha_{\text{eff}} = 0.5$. {(a)} Coherence in the $(\omega_{A},
		\omega_{B})$ plane (in units of $1 / \tau_{L}$). { (b)} Coherence
		along the $\omega_{\Delta} = \omega_{A} - \omega_{B}$ direction (in
		units of $1/\tau_{L}$), corresponding to a cut of the left panel
		along the grey dashed line.}
	\label{g2_top_05}
\end{figure}

\begin{figure}[h!]
	\includegraphics[width=\columnwidth]{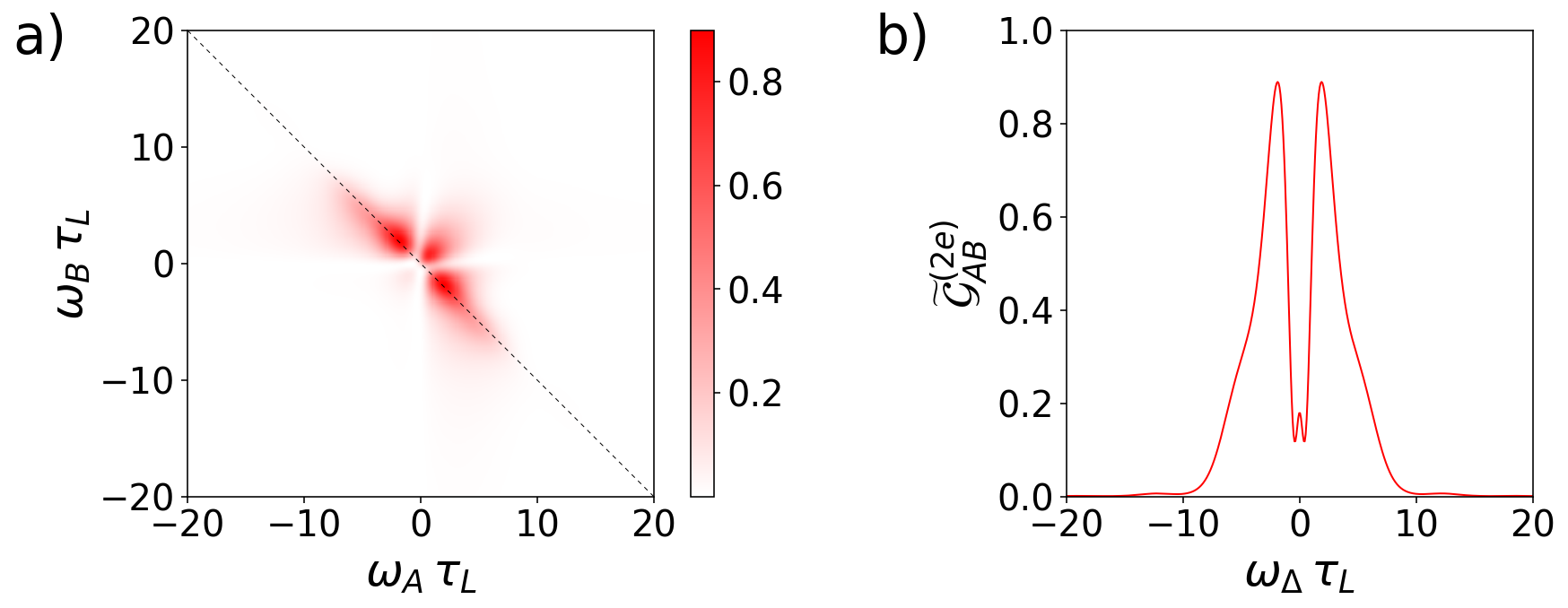}
	\caption{Outgoing diagonal two-electron coherence for two Landau
		wavepackets injected simultaneously into the Coulomb coupler.
		Same parameters as in \cref{g2_top_05}, but with $\alpha_{\text{eff}} = 4$ . {(a)} Coherence in the $(\omega_{A},
		\omega_{B})$ plane (in units of $1 / \tau_{L}$). { (b)} Coherence
		along the $\omega_{\Delta} = \omega_{A} - \omega_{B}$ direction (in
		units of $1/\tau_{L}$), corresponding to a cut of the left panel
		along the grey dashed line.}
	\label{g2_top_4}
\end{figure}

For completeness, we also provide the corresponding results in the weak
coupling regime, illustrating how the spectral features evolve as the
interaction strength is reduced. The spectra for weaker coupling regimes
are shown in \cref{g2_top_05}, where the coupling strength is set to
$\alpha_{\text{eff}} = 0.5$, and in \cref{g2_top_4}, where $\alpha_{\text{eff}} = 4$, allowing for a direct comparison with \cref{fig_g2_top} of the
main text.

As the interaction strength is reduced, the characteristic multi-peak
structure observed in the strong coupling regime progressively collapses
into a single narrow peak centered at $(\omega_A^0,\omega_B^0)$,
indicating the suppression of energy exchange processes between the two
electrons. In the weak coupling limit, the spectra are therefore
dominated by elastic processes, with negligible spectral weight along
the $\omega_\Delta$ direction.

\subsection{Effect of the injection times delay}
\label{appendix/collision-model/snowplough}

\begin{figure}[h!]
	\centering
	\begin{minipage}[c]{0.5\textwidth}
	\includegraphics[width=0.45\textwidth]{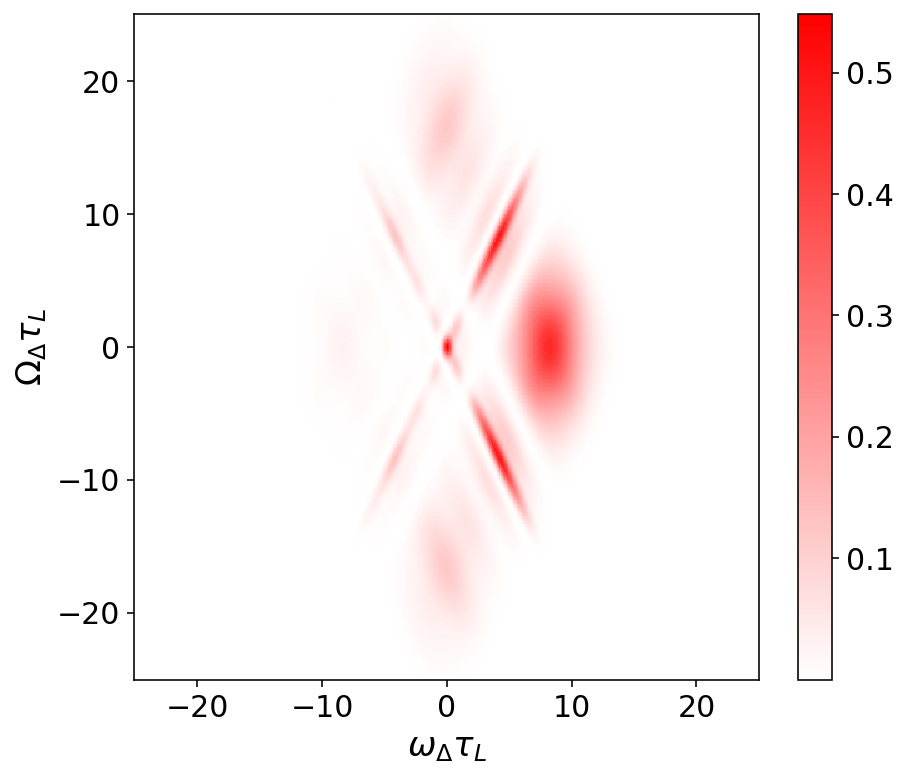}
	\hspace{0.5em}
	\includegraphics[width=0.45\textwidth]{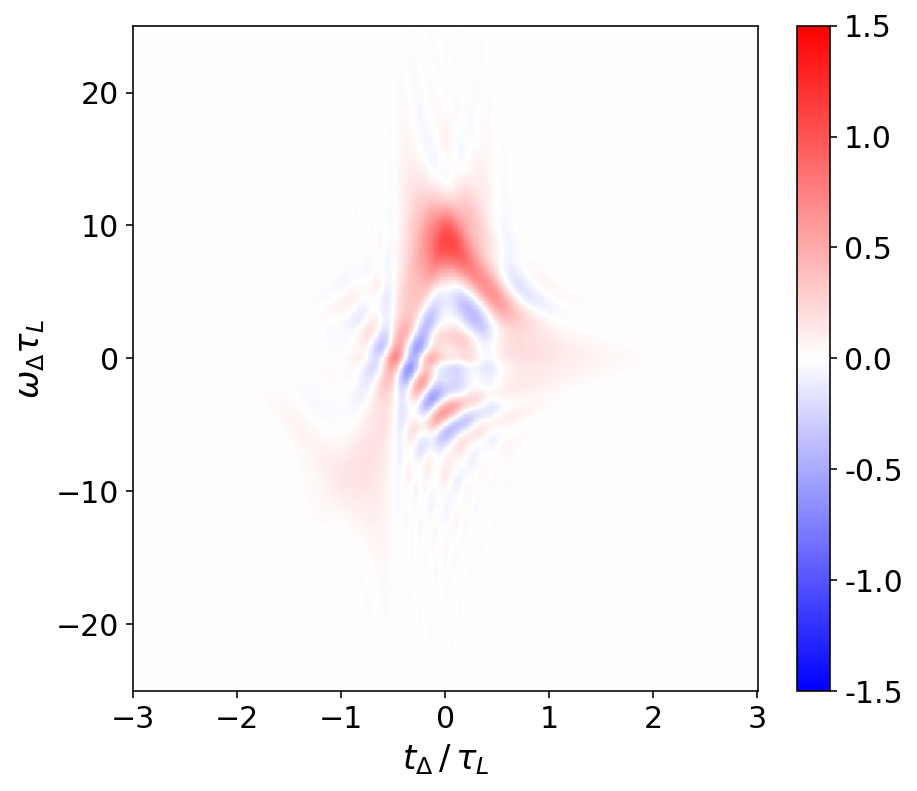}
	\end{minipage}
	\caption{\label{fig/snowplough} Post-interaction square modulus of the second-order coherence (left panel) and Wigner function
	of the relative particle (right panel) for $t_\Delta^0 = 0.5\, \tau_L$. The two wave
	packets are identical, $(\omega_A^0, \gamma_A) = (\omega_B^0, \gamma_B) =
	(0, 2/\tau_L)$ but are injected at different times: $t_A^0 = 0.5\, \tau_L$
	and $t_B^0 = 0$. Electron $B$ is therefore injected first, followed by
	electron $A$ at $t = 0.5\, \tau_L$. The coupling parameter is $\alpha_{\text{eff}} = 9$.}
\end{figure}

This section extends a remark made in
\cref{sec/2e-collision/entanglement/Wigner}, concerning the injection-time
offset between the two injected wavepackets. We analyze the post-collision dynamics when
the electrons are not injected simultaneously in the collision model ($t_\Delta^0
\neq 0$). The resulting two-electron coherence and the Wigner representation of
the relative particle for $t_\Delta^0 = 0.5\,\tau_L$ are shown in
\cref{fig/snowplough}.

In both representations, one of the two lobes associated with inter-channel
energy exchange is significantly more pronounced than the other. This indicates
that energy transfer between electrons is directionally biased rather than
symmetric. In particular, a positive injection delay $t_\Delta^0 > 0$ enhances
the lobe at $\omega_\Delta > 0$, implying that $t_A^0 > t_B^0$ correlates with
$\omega_A > \omega_B$: electron $A$ gains energy from electron $B$ rather than
the reverse.

This behavior is counterintuitive, as it suggests that the electron injected
first tends to lose energy while the later electron gains energy, effectively
resulting in a transfer that resembles an “acceleration of the second electron”
at the expense of the first. This appears to contrast with the discussion of
Sec.~V.B of
Ref.~\cite{Silvestrov-2025-1}
in which the first-injected electron instead gains energy from the second.

The discrepancy arises from a key difference in geometry: in the present
situation, electrons propagate in separated channels. They do not co-propagate within a single channel 
while they interact. As a result, direct Coulomb repulsion within the same
channel is absent. Because the Coulomb coupler involves an inter-channel
capacitance, 
the first electron induces a slight positive charge polarization in the opposite channel. 
The second electron then benefits from the corresponding electric field
generated by this charge polarization and is accelerated. When it comes out, the
first electron has already exited the coupling region and thus, the deceleration
experienced upon exiting the Coulomb coupler
does not compensate the acceleration upon arrival.
This mechanism is analogous to a snowplough effect: 
the leading particle reshape the potenial landscape for the trailing electron,
thereby facilitating its propagation.

\subsection{Dependence of the witness on the wave-packet duration}
\label{appendix/dependence-witness}

In \cref{sec/2e-collision/entanglement/witness}, we focused on the case of incoming Landau
wave packets with a duration $\gamma^{-1}$ significantly smaller than
the time of flight
across the Coulomb coupler
$\gamma^{-1}=\tau_L/2$.
It is instructive to examine how the violation of the entanglement
witness evolves when varying the wave-packet duration.

\begin{figure}[t]
	\includegraphics[width=\columnwidth]{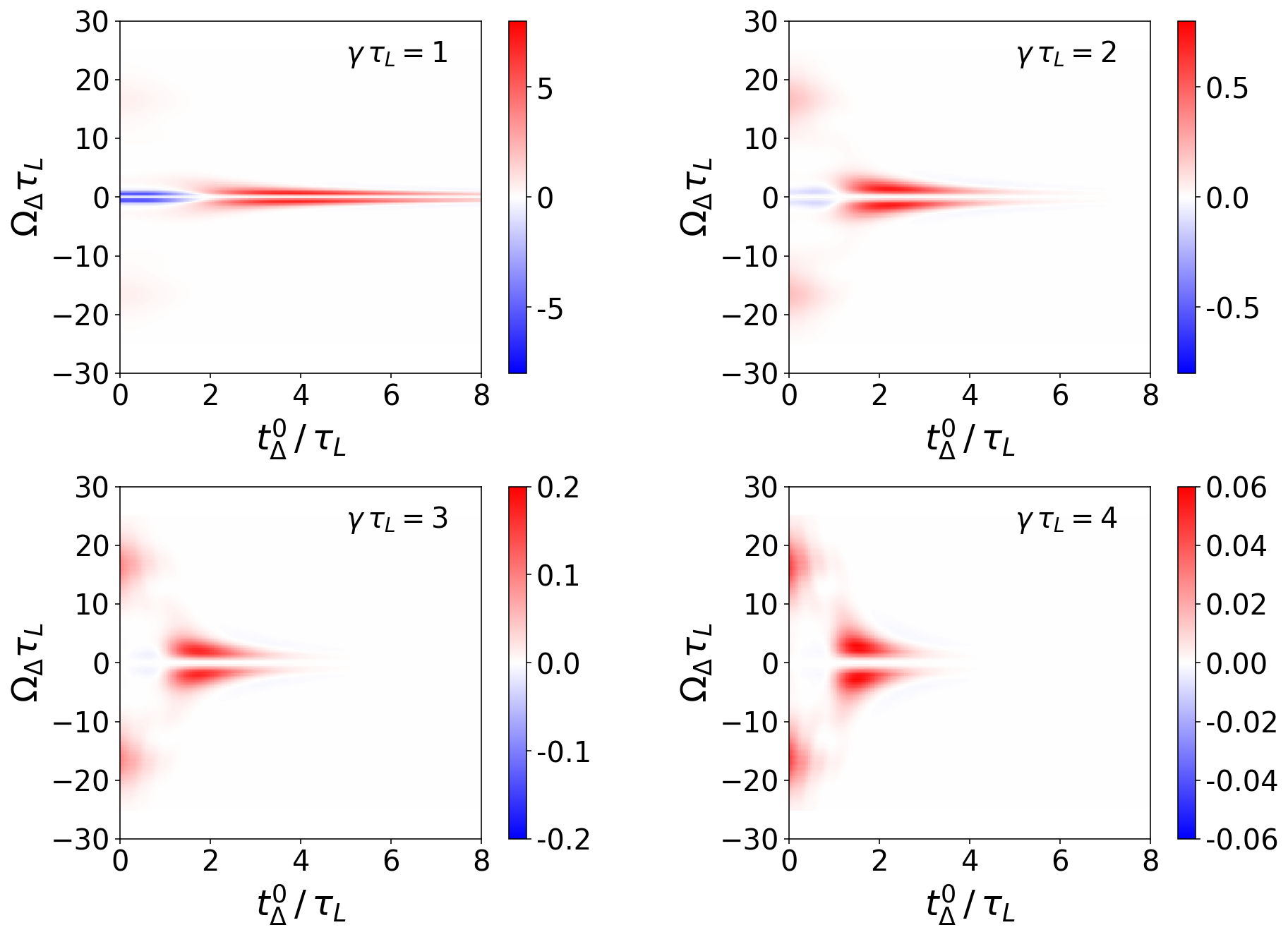}
	\caption{
	Interaction-induced contribution to the entanglement witness
	for increasing values of $\gamma\tau_L$ (from left to right and
	top to bottom: $\gamma\tau_L=1$, $2$, $3$, and $4$).
	Small values of $\gamma\tau_L$ correspond to long incoming
	wave packets, leading to broad violations in relative injection
	time, whereas larger values correspond to shorter wave packets
	for which the violation becomes stronger and more localized
	within the interaction time window $|t_\Delta^0|\lesssim\tau_L$.
	The interaction strength is fixed to $\alpha_{\text{eff}}=9$.}
	\label{fig:witness_gamma}
\end{figure}

Figure~\ref{fig:witness_gamma} shows the interaction-induced
contribution to the entanglement witness for several values of
$\gamma \tau_L$, ranging from long to short incoming
wave packets.

For long wave-packets ($\gamma\tau_L\lesssim 1$), the violation remains concentrated
around small values of $\Omega_\Delta$ but becomes weaker and spread over a borader range of relative injection times. 
This regime corresponds to a partial averaging of the interaction
phase over the temporal extent of the incoming excitations.

By contrast, shorter wave packets
($\gamma\tau_L\gg 1$)
lead to stronger violations of the entanglement witness.
In this regime, the interaction occurs over a narrower temporal
window, thereby enhancing the coherence of the energy exchange
processes responsible for entanglement generation.

The violation remains confined to the region $|t_\Delta^0|\lesssim\tau_L+2\gamma^{-1}$
which tends towards the interaction time $\tau_L$ in the limit
of very short wave packets.

\begin{thebibliography}{89}%
\makeatletter
\providecommand \@ifxundefined [1]{%
 \@ifx{#1\undefined}
}%
\providecommand \@ifnum [1]{%
 \ifnum #1\expandafter \@firstoftwo
 \else \expandafter \@secondoftwo
 \fi
}%
\providecommand \@ifx [1]{%
 \ifx #1\expandafter \@firstoftwo
 \else \expandafter \@secondoftwo
 \fi
}%
\providecommand \natexlab [1]{#1}%
\providecommand \enquote  [1]{``#1''}%
\providecommand \bibnamefont  [1]{#1}%
\providecommand \bibfnamefont [1]{#1}%
\providecommand \citenamefont [1]{#1}%
\providecommand \href@noop [0]{\@secondoftwo}%
\providecommand \href [0]{\begingroup \@sanitize@url \@href}%
\providecommand \@href[1]{\@@startlink{#1}\@@href}%
\providecommand \@@href[1]{\endgroup#1\@@endlink}%
\providecommand \@sanitize@url [0]{\catcode `\\12\catcode `\$12\catcode
  `\&12\catcode `\#12\catcode `\^12\catcode `\_12\catcode `\%12\relax}%
\providecommand \@@startlink[1]{}%
\providecommand \@@endlink[0]{}%
\providecommand \url  [0]{\begingroup\@sanitize@url \@url }%
\providecommand \@url [1]{\endgroup\@href {#1}{\urlprefix }}%
\providecommand \urlprefix  [0]{URL }%
\providecommand \Eprint [0]{\href }%
\providecommand \doibase [0]{https://doi.org/}%
\providecommand \selectlanguage [0]{\@gobble}%
\providecommand \bibinfo  [0]{\@secondoftwo}%
\providecommand \bibfield  [0]{\@secondoftwo}%
\providecommand \translation [1]{[#1]}%
\providecommand \BibitemOpen [0]{}%
\providecommand \bibitemStop [0]{}%
\providecommand \bibitemNoStop [0]{.\EOS\space}%
\providecommand \EOS [0]{\spacefactor3000\relax}%
\providecommand \BibitemShut  [1]{\csname bibitem#1\endcsname}%
\let\auto@bib@innerbib\@empty
\bibitem [{\citenamefont {Slussarenko}\ and\ \citenamefont
  {Pryde}(2019)}]{Slussarenko-2019-1}%
  \BibitemOpen
  \bibfield  {author} {\bibinfo {author} {\bibfnamefont {S.}~\bibnamefont
  {Slussarenko}}\ and\ \bibinfo {author} {\bibfnamefont {G.}~\bibnamefont
  {Pryde}},\ }\bibfield  {title} {\bibinfo {title} {Photonic quantum
  information processing: a concise review},\ }\href@noop {} {\bibfield
  {journal} {\bibinfo  {journal} {Applied Physics Review}\ }\textbf {\bibinfo
  {volume} {6}},\ \bibinfo {pages} {041303} (\bibinfo {year}
  {2019})}\BibitemShut {NoStop}%
\bibitem [{\citenamefont {Bertoni}\ \emph {et~al.}(2000)\citenamefont
  {Bertoni}, \citenamefont {Bordone}, \citenamefont {Brunetti}, \citenamefont
  {Jacoboni},\ and\ \citenamefont {Reggiani}}]{Bertoni-2000-1}%
  \BibitemOpen
  \bibfield  {author} {\bibinfo {author} {\bibfnamefont {A.}~\bibnamefont
  {Bertoni}}, \bibinfo {author} {\bibfnamefont {P.}~\bibnamefont {Bordone}},
  \bibinfo {author} {\bibfnamefont {R.}~\bibnamefont {Brunetti}}, \bibinfo
  {author} {\bibfnamefont {C.}~\bibnamefont {Jacoboni}},\ and\ \bibinfo
  {author} {\bibfnamefont {S.}~\bibnamefont {Reggiani}},\ }\bibfield  {title}
  {\bibinfo {title} {Quantum logic gates based on coherent electron transport
  in quantum wires},\ }\href {https://doi.org/10.1103/PhysRevLett.84.5912}
  {\bibfield  {journal} {\bibinfo  {journal} {Phys. Rev. Lett.}\ }\textbf
  {\bibinfo {volume} {84}},\ \bibinfo {pages} {5912} (\bibinfo {year}
  {2000})}\BibitemShut {NoStop}%
\bibitem [{\citenamefont {Ionicioiu}\ \emph {et~al.}(2001)\citenamefont
  {Ionicioiu}, \citenamefont {Amaratunga},\ and\ \citenamefont
  {Udrea}}]{Ionicioiu-2001-1}%
  \BibitemOpen
  \bibfield  {author} {\bibinfo {author} {\bibfnamefont {R.}~\bibnamefont
  {Ionicioiu}}, \bibinfo {author} {\bibfnamefont {G.}~\bibnamefont
  {Amaratunga}},\ and\ \bibinfo {author} {\bibfnamefont {F.}~\bibnamefont
  {Udrea}},\ }\bibfield  {title} {\bibinfo {title} {Quantum computation with
  ballistic electrons},\ }\href@noop {} {\bibfield  {journal} {\bibinfo
  {journal} {Int. J. Mod. Phys. B}\ }\textbf {\bibinfo {volume} {15}},\
  \bibinfo {pages} {125} (\bibinfo {year} {2001})}\BibitemShut {NoStop}%
\bibitem [{\citenamefont {Bertoni}(2007)}]{Bertoni-2007-1}%
  \BibitemOpen
  \bibfield  {author} {\bibinfo {author} {\bibfnamefont {A.}~\bibnamefont
  {Bertoni}},\ }\bibfield  {title} {\bibinfo {title} {Perspectives on
  solid-state flying qubits},\ }\href@noop {} {\bibfield  {journal} {\bibinfo
  {journal} {J. Comput. Electron.}\ }\textbf {\bibinfo {volume} {6}},\ \bibinfo
  {pages} {67} (\bibinfo {year} {2007})}\BibitemShut {NoStop}%
\bibitem [{\citenamefont {Bisognin}\ \emph {et~al.}(2019)\citenamefont
  {Bisognin}, \citenamefont {Marguerite}, \citenamefont {Roussel},
  \citenamefont {Kumar}, \citenamefont {Cabart}, \citenamefont {Chapdelaine},
  \citenamefont {Mohammad-Djafari}, \citenamefont {Berroir}, \citenamefont
  {Bocquillon}, \citenamefont {Pla{\c c}ais}, \citenamefont {Cavanna},
  \citenamefont {Gennser}, \citenamefont {Jin}, \citenamefont {Degiovanni},\
  and\ \citenamefont {F{\`e}ve}}]{Bisognin-2019-2}%
  \BibitemOpen
  \bibfield  {author} {\bibinfo {author} {\bibfnamefont {R.}~\bibnamefont
  {Bisognin}}, \bibinfo {author} {\bibfnamefont {A.}~\bibnamefont
  {Marguerite}}, \bibinfo {author} {\bibfnamefont {B.}~\bibnamefont {Roussel}},
  \bibinfo {author} {\bibfnamefont {M.}~\bibnamefont {Kumar}}, \bibinfo
  {author} {\bibfnamefont {C.}~\bibnamefont {Cabart}}, \bibinfo {author}
  {\bibfnamefont {C.}~\bibnamefont {Chapdelaine}}, \bibinfo {author}
  {\bibfnamefont {A.}~\bibnamefont {Mohammad-Djafari}}, \bibinfo {author}
  {\bibfnamefont {J.~M.}\ \bibnamefont {Berroir}}, \bibinfo {author}
  {\bibfnamefont {E.}~\bibnamefont {Bocquillon}}, \bibinfo {author}
  {\bibfnamefont {B.}~\bibnamefont {Pla{\c c}ais}}, \bibinfo {author}
  {\bibfnamefont {A.}~\bibnamefont {Cavanna}}, \bibinfo {author} {\bibfnamefont
  {U.}~\bibnamefont {Gennser}}, \bibinfo {author} {\bibfnamefont
  {Y.}~\bibnamefont {Jin}}, \bibinfo {author} {\bibfnamefont {P.}~\bibnamefont
  {Degiovanni}},\ and\ \bibinfo {author} {\bibfnamefont {G.}~\bibnamefont
  {F{\`e}ve}},\ }\bibfield  {title} {\bibinfo {title} {Quantum tomography of
  electrical currents},\ }\href {https://doi.org/10.1038/s41467-019-11369-5}
  {\bibfield  {journal} {\bibinfo  {journal} {Nature Communications}\ }\textbf
  {\bibinfo {volume} {10}},\ \bibinfo {pages} {3379} (\bibinfo {year}
  {2019})}\BibitemShut {NoStop}%
\bibitem [{\citenamefont {Fletcher}\ \emph {et~al.}(2019)\citenamefont
  {Fletcher}, \citenamefont {Johnson}, \citenamefont {Locane}, \citenamefont
  {See}, \citenamefont {Griffiths}, \citenamefont {Farrer}, \citenamefont
  {Ritchie}, \citenamefont {Brouwer}, \citenamefont {Kashcheyevs},\ and\
  \citenamefont {Kataoka}}]{Fletcher-2019-1}%
  \BibitemOpen
  \bibfield  {author} {\bibinfo {author} {\bibfnamefont {J.~D.}\ \bibnamefont
  {Fletcher}}, \bibinfo {author} {\bibfnamefont {N.}~\bibnamefont {Johnson}},
  \bibinfo {author} {\bibfnamefont {E.}~\bibnamefont {Locane}}, \bibinfo
  {author} {\bibfnamefont {P.}~\bibnamefont {See}}, \bibinfo {author}
  {\bibfnamefont {J.~P.}\ \bibnamefont {Griffiths}}, \bibinfo {author}
  {\bibfnamefont {I.}~\bibnamefont {Farrer}}, \bibinfo {author} {\bibfnamefont
  {D.~A.}\ \bibnamefont {Ritchie}}, \bibinfo {author} {\bibfnamefont {P.~W.}\
  \bibnamefont {Brouwer}}, \bibinfo {author} {\bibfnamefont {V.}~\bibnamefont
  {Kashcheyevs}},\ and\ \bibinfo {author} {\bibfnamefont {M.}~\bibnamefont
  {Kataoka}},\ }\bibfield  {title} {\bibinfo {title} {Continuous-variable
  tomography of solitary electrons},\ }\href
  {https://doi.org/10.1038/s41467-019-13222-1} {\bibfield  {journal} {\bibinfo
  {journal} {Nature Communications}\ }\textbf {\bibinfo {volume} {10}},\
  \bibinfo {pages} {5298} (\bibinfo {year} {2019})}\BibitemShut {NoStop}%
\bibitem [{\citenamefont {Duprez}\ \emph {et~al.}(2019)\citenamefont {Duprez},
  \citenamefont {Sivre}, \citenamefont {Anthore}, \citenamefont {Aassime},
  \citenamefont {Cavanna}, \citenamefont {Ouerghi}, \citenamefont {Gennser},\
  and\ \citenamefont {Pierre}}]{Duprez-2019-1}%
  \BibitemOpen
  \bibfield  {author} {\bibinfo {author} {\bibfnamefont {H.}~\bibnamefont
  {Duprez}}, \bibinfo {author} {\bibfnamefont {E.}~\bibnamefont {Sivre}},
  \bibinfo {author} {\bibfnamefont {A.}~\bibnamefont {Anthore}}, \bibinfo
  {author} {\bibfnamefont {A.}~\bibnamefont {Aassime}}, \bibinfo {author}
  {\bibfnamefont {A.}~\bibnamefont {Cavanna}}, \bibinfo {author} {\bibfnamefont
  {A.}~\bibnamefont {Ouerghi}}, \bibinfo {author} {\bibfnamefont
  {U.}~\bibnamefont {Gennser}},\ and\ \bibinfo {author} {\bibfnamefont
  {F.}~\bibnamefont {Pierre}},\ }\bibfield  {title} {\bibinfo {title}
  {Macroscopic electron quantum coherence in a solid-state circuit},\ }\href
  {https://doi.org/10.1103/PhysRevX.9.021030} {\bibfield  {journal} {\bibinfo
  {journal} {Phys. Rev. X}\ }\textbf {\bibinfo {volume} {9}},\ \bibinfo {pages}
  {021030} (\bibinfo {year} {2019})}\BibitemShut {NoStop}%
\bibitem [{\citenamefont {Wei}\ \emph {et~al.}(2017)\citenamefont {Wei},
  \citenamefont {van~der Sar}, \citenamefont {Sanchez-Yamagishi}, \citenamefont
  {Watanabe}, \citenamefont {Taniguchi}, \citenamefont {Jarillo-Herrero},
  \citenamefont {Halperin},\ and\ \citenamefont {Yacoby}}]{Wei-2017-1}%
  \BibitemOpen
  \bibfield  {author} {\bibinfo {author} {\bibfnamefont {D.~S.}\ \bibnamefont
  {Wei}}, \bibinfo {author} {\bibfnamefont {T.}~\bibnamefont {van~der Sar}},
  \bibinfo {author} {\bibfnamefont {J.~D.}\ \bibnamefont {Sanchez-Yamagishi}},
  \bibinfo {author} {\bibfnamefont {K.}~\bibnamefont {Watanabe}}, \bibinfo
  {author} {\bibfnamefont {T.}~\bibnamefont {Taniguchi}}, \bibinfo {author}
  {\bibfnamefont {P.}~\bibnamefont {Jarillo-Herrero}}, \bibinfo {author}
  {\bibfnamefont {B.~I.}\ \bibnamefont {Halperin}},\ and\ \bibinfo {author}
  {\bibfnamefont {A.}~\bibnamefont {Yacoby}},\ }\bibfield  {title} {\bibinfo
  {title} {Mach-{Z}ehnder interferometry using spin- and valley-polarized
  quantum {Hall} edge states in graphene},\ }\href
  {https://doi.org/10.1126/sciadv.1700600} {\bibfield  {journal} {\bibinfo
  {journal} {Science advances}\ }\textbf {\bibinfo {volume} {3}},\ \bibinfo
  {pages} {e1700600} (\bibinfo {year} {2017})}\BibitemShut {NoStop}%
\bibitem [{\citenamefont {Jo}\ \emph {et~al.}(2022)\citenamefont {Jo},
  \citenamefont {Lee}, \citenamefont {Assouline}, \citenamefont {Brasseur},
  \citenamefont {Watanabe}, \citenamefont {Taniguchi}, \citenamefont {Roche},
  \citenamefont {Glattli}, \citenamefont {Kumada}, \citenamefont {Parmentier},
  \citenamefont {Sim},\ and\ \citenamefont {Roulleau}}]{Jo-2022-1}%
  \BibitemOpen
  \bibfield  {author} {\bibinfo {author} {\bibfnamefont {M.}~\bibnamefont
  {Jo}}, \bibinfo {author} {\bibfnamefont {J.-Y.~M.}\ \bibnamefont {Lee}},
  \bibinfo {author} {\bibfnamefont {A.}~\bibnamefont {Assouline}}, \bibinfo
  {author} {\bibfnamefont {P.}~\bibnamefont {Brasseur}}, \bibinfo {author}
  {\bibfnamefont {K.}~\bibnamefont {Watanabe}}, \bibinfo {author}
  {\bibfnamefont {T.}~\bibnamefont {Taniguchi}}, \bibinfo {author}
  {\bibfnamefont {P.}~\bibnamefont {Roche}}, \bibinfo {author} {\bibfnamefont
  {D.~C.}\ \bibnamefont {Glattli}}, \bibinfo {author} {\bibfnamefont
  {N.}~\bibnamefont {Kumada}}, \bibinfo {author} {\bibfnamefont {F.~D.}\
  \bibnamefont {Parmentier}}, \bibinfo {author} {\bibfnamefont {H.~S.}\
  \bibnamefont {Sim}},\ and\ \bibinfo {author} {\bibfnamefont {P.}~\bibnamefont
  {Roulleau}},\ }\bibfield  {title} {\bibinfo {title} {Scaling behavior of
  electron decoherence in a graphene mach-zehnder interferometer},\ }\href
  {https://doi.org/10.1038/s41467-022-33078-2} {\bibfield  {journal} {\bibinfo
  {journal} {Nature Communications}\ }\textbf {\bibinfo {volume} {13}},\
  \bibinfo {pages} {5473} (\bibinfo {year} {2022})}\BibitemShut {NoStop}%
\bibitem [{\citenamefont {D{\'e}prez}\ \emph {et~al.}(2021)\citenamefont
  {D{\'e}prez}, \citenamefont {Veyrat}, \citenamefont {Vignaud}, \citenamefont
  {Nayak}, \citenamefont {Watanabe}, \citenamefont {Taniguchi}, \citenamefont
  {Gay}, \citenamefont {Sellier},\ and\ \citenamefont
  {Sac{\'e}p{\'e}}}]{Deprez-2021-2}%
  \BibitemOpen
  \bibfield  {author} {\bibinfo {author} {\bibfnamefont {C.}~\bibnamefont
  {D{\'e}prez}}, \bibinfo {author} {\bibfnamefont {L.}~\bibnamefont {Veyrat}},
  \bibinfo {author} {\bibfnamefont {H.}~\bibnamefont {Vignaud}}, \bibinfo
  {author} {\bibfnamefont {G.}~\bibnamefont {Nayak}}, \bibinfo {author}
  {\bibfnamefont {K.}~\bibnamefont {Watanabe}}, \bibinfo {author}
  {\bibfnamefont {T.}~\bibnamefont {Taniguchi}}, \bibinfo {author}
  {\bibfnamefont {F.}~\bibnamefont {Gay}}, \bibinfo {author} {\bibfnamefont
  {H.}~\bibnamefont {Sellier}},\ and\ \bibinfo {author} {\bibfnamefont
  {B.}~\bibnamefont {Sac{\'e}p{\'e}}},\ }\bibfield  {title} {\bibinfo {title}
  {A tunable {Fabry--P{\'e}rot} quantum {Hall} interferometer in graphene},\
  }\href {https://doi.org/10.1038/s41565-021-00847-x} {\bibfield  {journal}
  {\bibinfo  {journal} {Nature Nanotechnology}\ }\textbf {\bibinfo {volume}
  {16}},\ \bibinfo {pages} {555} (\bibinfo {year} {2021})}\BibitemShut
  {NoStop}%
\bibitem [{\citenamefont {Edlbauer}\ \emph {et~al.}(2022)\citenamefont
  {Edlbauer}, \citenamefont {Wang}, \citenamefont {Crozes}, \citenamefont
  {Perrier}, \citenamefont {Ouacel}, \citenamefont {Geffroy}, \citenamefont
  {Georgiou}, \citenamefont {Chatzikyriakou}, \citenamefont {Lacerda-Santos},
  \citenamefont {Waintal}, \citenamefont {Glattli}, \citenamefont {Roulleau},
  \citenamefont {Nath}, \citenamefont {Kataoka}, \citenamefont
  {Splettstoesser}, \citenamefont {Acciai}, \citenamefont {da~Silva~Figueira},
  \citenamefont {{\"O}ztas}, \citenamefont {Trellakis}, \citenamefont {Grange},
  \citenamefont {Yevtushenko}, \citenamefont {Birner},\ and\ \citenamefont
  {B{\"a}uerle}}]{Edlbauer-2022-1}%
  \BibitemOpen
  \bibfield  {author} {\bibinfo {author} {\bibfnamefont {H.}~\bibnamefont
  {Edlbauer}}, \bibinfo {author} {\bibfnamefont {J.}~\bibnamefont {Wang}},
  \bibinfo {author} {\bibfnamefont {T.}~\bibnamefont {Crozes}}, \bibinfo
  {author} {\bibfnamefont {P.}~\bibnamefont {Perrier}}, \bibinfo {author}
  {\bibfnamefont {S.}~\bibnamefont {Ouacel}}, \bibinfo {author} {\bibfnamefont
  {C.}~\bibnamefont {Geffroy}}, \bibinfo {author} {\bibfnamefont
  {G.}~\bibnamefont {Georgiou}}, \bibinfo {author} {\bibfnamefont
  {E.}~\bibnamefont {Chatzikyriakou}}, \bibinfo {author} {\bibfnamefont
  {A.}~\bibnamefont {Lacerda-Santos}}, \bibinfo {author} {\bibfnamefont
  {X.}~\bibnamefont {Waintal}}, \bibinfo {author} {\bibfnamefont {D.~C.}\
  \bibnamefont {Glattli}}, \bibinfo {author} {\bibfnamefont {P.}~\bibnamefont
  {Roulleau}}, \bibinfo {author} {\bibfnamefont {J.}~\bibnamefont {Nath}},
  \bibinfo {author} {\bibfnamefont {M.}~\bibnamefont {Kataoka}}, \bibinfo
  {author} {\bibfnamefont {J.}~\bibnamefont {Splettstoesser}}, \bibinfo
  {author} {\bibfnamefont {M.}~\bibnamefont {Acciai}}, \bibinfo {author}
  {\bibfnamefont {M.~C.}\ \bibnamefont {da~Silva~Figueira}}, \bibinfo {author}
  {\bibfnamefont {K.}~\bibnamefont {{\"O}ztas}}, \bibinfo {author}
  {\bibfnamefont {A.}~\bibnamefont {Trellakis}}, \bibinfo {author}
  {\bibfnamefont {T.}~\bibnamefont {Grange}}, \bibinfo {author} {\bibfnamefont
  {O.~M.}\ \bibnamefont {Yevtushenko}}, \bibinfo {author} {\bibfnamefont
  {S.}~\bibnamefont {Birner}},\ and\ \bibinfo {author} {\bibfnamefont
  {C.}~\bibnamefont {B{\"a}uerle}},\ }\bibfield  {title} {\bibinfo {title}
  {Semiconductor-based electron flying qubits: review on recent progress
  accelerated by numerical modelling},\ }\href
  {https://doi.org/10.1140/epjqt/s40507-022-00139-w} {\bibfield  {journal}
  {\bibinfo  {journal} {EPJ Quantum Technology}\ }\textbf {\bibinfo {volume}
  {9}},\ \bibinfo {pages} {21} (\bibinfo {year} {2022})}\BibitemShut {NoStop}%
\bibitem [{\citenamefont {Assouline}\ \emph {et~al.}(2023)\citenamefont
  {Assouline}, \citenamefont {Pugliese}, \citenamefont {Chakraborti},
  \citenamefont {Lee}, \citenamefont {Bernabeu}, \citenamefont {Jo},
  \citenamefont {Watanabe}, \citenamefont {Taniguchi}, \citenamefont {Glattli},
  \citenamefont {Kumada}, \citenamefont {Sim}, \citenamefont {Parmentier},\
  and\ \citenamefont {Roulleau}}]{Assouline-2023-1}%
  \BibitemOpen
  \bibfield  {author} {\bibinfo {author} {\bibfnamefont {A.}~\bibnamefont
  {Assouline}}, \bibinfo {author} {\bibfnamefont {L.}~\bibnamefont {Pugliese}},
  \bibinfo {author} {\bibfnamefont {H.}~\bibnamefont {Chakraborti}}, \bibinfo
  {author} {\bibfnamefont {S.}~\bibnamefont {Lee}}, \bibinfo {author}
  {\bibfnamefont {L.}~\bibnamefont {Bernabeu}}, \bibinfo {author}
  {\bibfnamefont {M.}~\bibnamefont {Jo}}, \bibinfo {author} {\bibfnamefont
  {K.}~\bibnamefont {Watanabe}}, \bibinfo {author} {\bibfnamefont
  {T.}~\bibnamefont {Taniguchi}}, \bibinfo {author} {\bibfnamefont {D.~C.}\
  \bibnamefont {Glattli}}, \bibinfo {author} {\bibfnamefont {N.}~\bibnamefont
  {Kumada}}, \bibinfo {author} {\bibfnamefont {H.-S.}\ \bibnamefont {Sim}},
  \bibinfo {author} {\bibfnamefont {F.~D.}\ \bibnamefont {Parmentier}},\ and\
  \bibinfo {author} {\bibfnamefont {P.}~\bibnamefont {Roulleau}},\ }\bibfield
  {title} {\bibinfo {title} {Emission and coherent control of levitons in
  graphene},\ }\href {https://doi.org/10.1126/science.adf9887} {\bibfield
  {journal} {\bibinfo  {journal} {Science}\ }\textbf {\bibinfo {volume}
  {382}},\ \bibinfo {pages} {1260} (\bibinfo {year} {2023})}\BibitemShut
  {NoStop}%
\bibitem [{\citenamefont {Dubois}\ \emph {et~al.}(2013)\citenamefont {Dubois},
  \citenamefont {Jullien}, \citenamefont {Grenier}, \citenamefont {Degiovanni},
  \citenamefont {Roulleau},\ and\ \citenamefont {Glattli}}]{Dubois-2013-1}%
  \BibitemOpen
  \bibfield  {author} {\bibinfo {author} {\bibfnamefont {J.}~\bibnamefont
  {Dubois}}, \bibinfo {author} {\bibfnamefont {T.}~\bibnamefont {Jullien}},
  \bibinfo {author} {\bibfnamefont {C.}~\bibnamefont {Grenier}}, \bibinfo
  {author} {\bibfnamefont {P.}~\bibnamefont {Degiovanni}}, \bibinfo {author}
  {\bibfnamefont {P.}~\bibnamefont {Roulleau}},\ and\ \bibinfo {author}
  {\bibfnamefont {D.~C.}\ \bibnamefont {Glattli}},\ }\bibfield  {title}
  {\bibinfo {title} {Integer and fractional charge lorentzian voltage pulses
  analyzed in the framework of photon-assisted shot noise},\ }\href
  {https://doi.org/10.1103/PhysRevB.88.085301} {\bibfield  {journal} {\bibinfo
  {journal} {Phys. Rev. B}\ }\textbf {\bibinfo {volume} {88}},\ \bibinfo
  {pages} {085301} (\bibinfo {year} {2013})}\BibitemShut {NoStop}%
\bibitem [{\citenamefont {Glauber}(1963)}]{Glauber-1963-1}%
  \BibitemOpen
  \bibfield  {author} {\bibinfo {author} {\bibfnamefont {R.}~\bibnamefont
  {Glauber}},\ }\bibfield  {title} {\bibinfo {title} {The quantum theory of
  optical coherence},\ }\href {https://doi.org/10.1103/PhysRev.130.2529}
  {\bibfield  {journal} {\bibinfo  {journal} {Phys. Rev.}\ }\textbf {\bibinfo
  {volume} {130}},\ \bibinfo {pages} {2529} (\bibinfo {year}
  {1963})}\BibitemShut {NoStop}%
\bibitem [{\citenamefont {W\"olk}\ \emph {et~al.}(2014)\citenamefont {W\"olk},
  \citenamefont {Huber},\ and\ \citenamefont {G\"uhne}}]{Wolk-2014-1}%
  \BibitemOpen
  \bibfield  {author} {\bibinfo {author} {\bibfnamefont {S.}~\bibnamefont
  {W\"olk}}, \bibinfo {author} {\bibfnamefont {M.}~\bibnamefont {Huber}},\ and\
  \bibinfo {author} {\bibfnamefont {O.}~\bibnamefont {G\"uhne}},\ }\bibfield
  {title} {\bibinfo {title} {Unified approach to entanglement criteria using
  the {Cauchy-Schwarz} and {H}\"older inequalities},\ }\href
  {https://doi.org/10.1103/PhysRevA.90.022315} {\bibfield  {journal} {\bibinfo
  {journal} {Phys. Rev. A}\ }\textbf {\bibinfo {volume} {90}},\ \bibinfo
  {pages} {022315} (\bibinfo {year} {2014})}\BibitemShut {NoStop}%
\bibitem [{\citenamefont {Peugeot}\ \emph {et~al.}(2021)\citenamefont
  {Peugeot}, \citenamefont {M\'enard}, \citenamefont {Dambach}, \citenamefont
  {Westig}, \citenamefont {Kubala}, \citenamefont {Mukharsky}, \citenamefont
  {Altimiras}, \citenamefont {Joyez}, \citenamefont {Vion}, \citenamefont
  {Roche}, \citenamefont {Esteve}, \citenamefont {Milman}, \citenamefont
  {Lepp\"akangas}, \citenamefont {Johansson}, \citenamefont {Hofheinz},
  \citenamefont {Ankerhold},\ and\ \citenamefont {Portier}}]{Peugeot-2021-1}%
  \BibitemOpen
  \bibfield  {author} {\bibinfo {author} {\bibfnamefont {A.}~\bibnamefont
  {Peugeot}}, \bibinfo {author} {\bibfnamefont {G.}~\bibnamefont {M\'enard}},
  \bibinfo {author} {\bibfnamefont {S.}~\bibnamefont {Dambach}}, \bibinfo
  {author} {\bibfnamefont {M.}~\bibnamefont {Westig}}, \bibinfo {author}
  {\bibfnamefont {B.}~\bibnamefont {Kubala}}, \bibinfo {author} {\bibfnamefont
  {Y.}~\bibnamefont {Mukharsky}}, \bibinfo {author} {\bibfnamefont
  {C.}~\bibnamefont {Altimiras}}, \bibinfo {author} {\bibfnamefont
  {P.}~\bibnamefont {Joyez}}, \bibinfo {author} {\bibfnamefont
  {D.}~\bibnamefont {Vion}}, \bibinfo {author} {\bibfnamefont {P.}~\bibnamefont
  {Roche}}, \bibinfo {author} {\bibfnamefont {D.}~\bibnamefont {Esteve}},
  \bibinfo {author} {\bibfnamefont {P.}~\bibnamefont {Milman}}, \bibinfo
  {author} {\bibfnamefont {J.}~\bibnamefont {Lepp\"akangas}}, \bibinfo {author}
  {\bibfnamefont {G.}~\bibnamefont {Johansson}}, \bibinfo {author}
  {\bibfnamefont {M.}~\bibnamefont {Hofheinz}}, \bibinfo {author}
  {\bibfnamefont {J.}~\bibnamefont {Ankerhold}},\ and\ \bibinfo {author}
  {\bibfnamefont {F.}~\bibnamefont {Portier}},\ }\bibfield  {title} {\bibinfo
  {title} {Generating two continuous entangled microwave beams using a
  dc-biased {Josephson} junction},\ }\href
  {https://doi.org/10.1103/PhysRevX.11.031008} {\bibfield  {journal} {\bibinfo
  {journal} {Phys. Rev. X}\ }\textbf {\bibinfo {volume} {11}},\ \bibinfo
  {pages} {031008} (\bibinfo {year} {2021})}\BibitemShut {NoStop}%
\bibitem [{\citenamefont {{The ATLAS Collaboration}}(2024)}]{Atlas-2024-1}%
  \BibitemOpen
  \bibfield  {author} {\bibinfo {author} {\bibnamefont {{The ATLAS
  Collaboration}}},\ }\bibfield  {title} {\bibinfo {title} {Observation of
  quantum entanglement with top quarks at the {ATLAS} detector},\ }\href
  {https://doi.org/10.1038/s41586-024-07824-z} {\bibfield  {journal} {\bibinfo
  {journal} {Nature}\ }\textbf {\bibinfo {volume} {633}},\ \bibinfo {pages}
  {542} (\bibinfo {year} {2024})}\BibitemShut {NoStop}%
\bibitem [{\citenamefont {Burkard}\ and\ \citenamefont
  {Loss}(2003)}]{Burkard-2003-1}%
  \BibitemOpen
  \bibfield  {author} {\bibinfo {author} {\bibfnamefont {G.}~\bibnamefont
  {Burkard}}\ and\ \bibinfo {author} {\bibfnamefont {D.}~\bibnamefont {Loss}},\
  }\bibfield  {title} {\bibinfo {title} {Lower bound for electron spin
  entanglement from beam splitter current correlations},\ }\href
  {https://doi.org/10.1103/PhysRevLett.91.087903} {\bibfield  {journal}
  {\bibinfo  {journal} {Phys. Rev. Lett.}\ }\textbf {\bibinfo {volume} {91}},\
  \bibinfo {pages} {087903} (\bibinfo {year} {2003})}\BibitemShut {NoStop}%
\bibitem [{\citenamefont {Giovannetti}\ \emph {et~al.}(2006)\citenamefont
  {Giovannetti}, \citenamefont {Frustaglia}, \citenamefont {Taddei},\ and\
  \citenamefont {Fazio}}]{Giovannetti-2006-2}%
  \BibitemOpen
  \bibfield  {author} {\bibinfo {author} {\bibfnamefont {V.}~\bibnamefont
  {Giovannetti}}, \bibinfo {author} {\bibfnamefont {D.}~\bibnamefont
  {Frustaglia}}, \bibinfo {author} {\bibfnamefont {F.}~\bibnamefont {Taddei}},\
  and\ \bibinfo {author} {\bibfnamefont {R.}~\bibnamefont {Fazio}},\ }\bibfield
   {title} {\bibinfo {title} {Electronic hong-ou-mandel interferometer for
  multimode entanglement detection},\ }\href
  {https://doi.org/10.1103/PhysRevB.74.115315} {\bibfield  {journal} {\bibinfo
  {journal} {Phys. Rev. B}\ }\textbf {\bibinfo {volume} {74}},\ \bibinfo
  {pages} {115315} (\bibinfo {year} {2006})}\BibitemShut {NoStop}%
\bibitem [{\citenamefont {Benatti}\ \emph {et~al.}(2020)\citenamefont
  {Benatti}, \citenamefont {Floreanini}, \citenamefont {Franchini},\ and\
  \citenamefont {Marzolino}}]{Benatti-2020-1}%
  \BibitemOpen
  \bibfield  {author} {\bibinfo {author} {\bibfnamefont {F.}~\bibnamefont
  {Benatti}}, \bibinfo {author} {\bibfnamefont {R.}~\bibnamefont {Floreanini}},
  \bibinfo {author} {\bibfnamefont {F.}~\bibnamefont {Franchini}},\ and\
  \bibinfo {author} {\bibfnamefont {U.}~\bibnamefont {Marzolino}},\ }\bibfield
  {title} {\bibinfo {title} {Entanglement in indistinguishable particle
  systems},\ }\href {https://doi.org/10.1016/j.physrep.2020.07.003} {\bibfield
  {journal} {\bibinfo  {journal} {Physics Reports}\ }\textbf {\bibinfo {volume}
  {878}},\ \bibinfo {pages} {1} (\bibinfo {year} {2020})}\BibitemShut {NoStop}%
\bibitem [{\citenamefont {Thibierge}\ \emph {et~al.}(2016)\citenamefont
  {Thibierge}, \citenamefont {Ferraro}, \citenamefont {Roussel}, \citenamefont
  {Cabart}, \citenamefont {Marguerite}, \citenamefont {F{\`e}ve},\ and\
  \citenamefont {Degiovanni}}]{Thibierge-2016-1}%
  \BibitemOpen
  \bibfield  {author} {\bibinfo {author} {\bibfnamefont {E.}~\bibnamefont
  {Thibierge}}, \bibinfo {author} {\bibfnamefont {D.}~\bibnamefont {Ferraro}},
  \bibinfo {author} {\bibfnamefont {B.}~\bibnamefont {Roussel}}, \bibinfo
  {author} {\bibfnamefont {C.}~\bibnamefont {Cabart}}, \bibinfo {author}
  {\bibfnamefont {A.}~\bibnamefont {Marguerite}}, \bibinfo {author}
  {\bibfnamefont {G.}~\bibnamefont {F{\`e}ve}},\ and\ \bibinfo {author}
  {\bibfnamefont {P.}~\bibnamefont {Degiovanni}},\ }\bibfield  {title}
  {\bibinfo {title} {Two-electron coherence and its measurement in electron
  quantum optics},\ }\href {https://doi.org/10.1103/PhysRevB.93.081302}
  {\bibfield  {journal} {\bibinfo  {journal} {Phys. Rev. B}\ }\textbf {\bibinfo
  {volume} {93}},\ \bibinfo {pages} {081302(R)} (\bibinfo {year}
  {2016})}\BibitemShut {NoStop}%
\bibitem [{\citenamefont {Roussel}\ \emph {et~al.}(2021)\citenamefont
  {Roussel}, \citenamefont {Cabart}, \citenamefont {F{\`e}ve},\ and\
  \citenamefont {Degiovanni}}]{Roussel-2020-1}%
  \BibitemOpen
  \bibfield  {author} {\bibinfo {author} {\bibfnamefont {B.}~\bibnamefont
  {Roussel}}, \bibinfo {author} {\bibfnamefont {C.}~\bibnamefont {Cabart}},
  \bibinfo {author} {\bibfnamefont {G.}~\bibnamefont {F{\`e}ve}},\ and\
  \bibinfo {author} {\bibfnamefont {P.}~\bibnamefont {Degiovanni}},\ }\bibfield
   {title} {\bibinfo {title} {Processing quantum signals carried by electrical
  currents},\ }\href@noop {} {\bibfield  {journal} {\bibinfo  {journal}
  {Physical Review X Quantum}\ }\textbf {\bibinfo {volume} {2}},\ \bibinfo
  {pages} {020314} (\bibinfo {year} {2021})}\BibitemShut {NoStop}%
\bibitem [{\citenamefont {Laurell}\ \emph {et~al.}(2022)\citenamefont
  {Laurell}, \citenamefont {Finkelstein-Shapiro}, \citenamefont {Dittel},
  \citenamefont {Guo}, \citenamefont {Demjaha}, \citenamefont {Ammitzb\"oll},
  \citenamefont {Weissenbilder}, \citenamefont {Neori\ifmmode \check{c}\else
  \v{c}\fi{}i\ifmmode~\acute{c}\else \'{c}\fi{}}, \citenamefont {Luo},
  \citenamefont {Gisselbrecht}, \citenamefont {Arnold}, \citenamefont
  {Buchleitner}, \citenamefont {Pullerits}, \citenamefont {L'Huillier},\ and\
  \citenamefont {Busto}}]{Laurell-2022-1}%
  \BibitemOpen
  \bibfield  {author} {\bibinfo {author} {\bibfnamefont {H.}~\bibnamefont
  {Laurell}}, \bibinfo {author} {\bibfnamefont {D.}~\bibnamefont
  {Finkelstein-Shapiro}}, \bibinfo {author} {\bibfnamefont {C.}~\bibnamefont
  {Dittel}}, \bibinfo {author} {\bibfnamefont {C.}~\bibnamefont {Guo}},
  \bibinfo {author} {\bibfnamefont {R.}~\bibnamefont {Demjaha}}, \bibinfo
  {author} {\bibfnamefont {M.}~\bibnamefont {Ammitzb\"oll}}, \bibinfo {author}
  {\bibfnamefont {R.}~\bibnamefont {Weissenbilder}}, \bibinfo {author}
  {\bibfnamefont {L.}~\bibnamefont {Neori\ifmmode \check{c}\else
  \v{c}\fi{}i\ifmmode~\acute{c}\else \'{c}\fi{}}}, \bibinfo {author}
  {\bibfnamefont {S.}~\bibnamefont {Luo}}, \bibinfo {author} {\bibfnamefont
  {M.}~\bibnamefont {Gisselbrecht}}, \bibinfo {author} {\bibfnamefont {C.~L.}\
  \bibnamefont {Arnold}}, \bibinfo {author} {\bibfnamefont {A.}~\bibnamefont
  {Buchleitner}}, \bibinfo {author} {\bibfnamefont {T.}~\bibnamefont
  {Pullerits}}, \bibinfo {author} {\bibfnamefont {A.}~\bibnamefont
  {L'Huillier}},\ and\ \bibinfo {author} {\bibfnamefont {D.}~\bibnamefont
  {Busto}},\ }\bibfield  {title} {\bibinfo {title} {Continuous-variable quantum
  state tomography of photoelectrons},\ }\href
  {https://doi.org/10.1103/PhysRevResearch.4.033220} {\bibfield  {journal}
  {\bibinfo  {journal} {Phys. Rev. Res.}\ }\textbf {\bibinfo {volume} {4}},\
  \bibinfo {pages} {033220} (\bibinfo {year} {2022})}\BibitemShut {NoStop}%
\bibitem [{\citenamefont {Laurell}\ \emph
  {et~al.}(2025{\natexlab{a}})\citenamefont {Laurell}, \citenamefont
  {Ba{\~n}os-Guti{\'e}rrez}, \citenamefont {L'Huillier}, \citenamefont
  {Busto},\ and\ \citenamefont {Finkelstein-Shapiro}}]{Laurell-2025-1}%
  \BibitemOpen
  \bibfield  {author} {\bibinfo {author} {\bibfnamefont {H.}~\bibnamefont
  {Laurell}}, \bibinfo {author} {\bibfnamefont {J.}~\bibnamefont
  {Ba{\~n}os-Guti{\'e}rrez}}, \bibinfo {author} {\bibfnamefont
  {A.}~\bibnamefont {L'Huillier}}, \bibinfo {author} {\bibfnamefont
  {D.}~\bibnamefont {Busto}},\ and\ \bibinfo {author} {\bibfnamefont
  {D.}~\bibnamefont {Finkelstein-Shapiro}},\ }\bibfield  {title} {\bibinfo
  {title} {A multidimensional approach to quantum state tomography of
  photoelectron wavepackets},\ }\href
  {https://doi.org/10.1038/s41598-025-86701-9} {\bibfield  {journal} {\bibinfo
  {journal} {Scientific Reports}\ }\textbf {\bibinfo {volume} {15}},\ \bibinfo
  {pages} {3937} (\bibinfo {year} {2025}{\natexlab{a}})}\BibitemShut {NoStop}%
\bibitem [{\citenamefont {Laurell}\ \emph
  {et~al.}(2025{\natexlab{b}})\citenamefont {Laurell}, \citenamefont {Luo},
  \citenamefont {Weissenbilder}, \citenamefont {Ammitzb{\"o}ll}, \citenamefont
  {Ahmed}, \citenamefont {S{\"o}derberg}, \citenamefont {Petersson},
  \citenamefont {Poulain}, \citenamefont {Guo}, \citenamefont {Dittel},
  \citenamefont {Finkelstein-Shapiro}, \citenamefont {Squibb}, \citenamefont
  {Feifel}, \citenamefont {Gisselbrecht}, \citenamefont {Arnold}, \citenamefont
  {Buchleitner}, \citenamefont {Lindroth}, \citenamefont {Frisk~Kockum},
  \citenamefont {L'Huillier},\ and\ \citenamefont {Busto}}]{Laurell-2025-2}%
  \BibitemOpen
  \bibfield  {author} {\bibinfo {author} {\bibfnamefont {H.}~\bibnamefont
  {Laurell}}, \bibinfo {author} {\bibfnamefont {S.}~\bibnamefont {Luo}},
  \bibinfo {author} {\bibfnamefont {R.}~\bibnamefont {Weissenbilder}}, \bibinfo
  {author} {\bibfnamefont {M.}~\bibnamefont {Ammitzb{\"o}ll}}, \bibinfo
  {author} {\bibfnamefont {S.}~\bibnamefont {Ahmed}}, \bibinfo {author}
  {\bibfnamefont {H.}~\bibnamefont {S{\"o}derberg}}, \bibinfo {author}
  {\bibfnamefont {C.~L.~M.}\ \bibnamefont {Petersson}}, \bibinfo {author}
  {\bibfnamefont {V.}~\bibnamefont {Poulain}}, \bibinfo {author} {\bibfnamefont
  {C.}~\bibnamefont {Guo}}, \bibinfo {author} {\bibfnamefont {C.}~\bibnamefont
  {Dittel}}, \bibinfo {author} {\bibfnamefont {D.}~\bibnamefont
  {Finkelstein-Shapiro}}, \bibinfo {author} {\bibfnamefont {R.~J.}\
  \bibnamefont {Squibb}}, \bibinfo {author} {\bibfnamefont {R.}~\bibnamefont
  {Feifel}}, \bibinfo {author} {\bibfnamefont {M.}~\bibnamefont
  {Gisselbrecht}}, \bibinfo {author} {\bibfnamefont {C.~L.}\ \bibnamefont
  {Arnold}}, \bibinfo {author} {\bibfnamefont {A.}~\bibnamefont {Buchleitner}},
  \bibinfo {author} {\bibfnamefont {E.}~\bibnamefont {Lindroth}}, \bibinfo
  {author} {\bibfnamefont {A.}~\bibnamefont {Frisk~Kockum}}, \bibinfo {author}
  {\bibfnamefont {A.}~\bibnamefont {L'Huillier}},\ and\ \bibinfo {author}
  {\bibfnamefont {D.}~\bibnamefont {Busto}},\ }\bibfield  {title} {\bibinfo
  {title} {Measuring the quantum state of photoelectrons},\ }\href
  {https://doi.org/10.1038/s41566-024-01607-8} {\bibfield  {journal} {\bibinfo
  {journal} {Nature Photonics}\ }\textbf {\bibinfo {volume} {19}},\ \bibinfo
  {pages} {352} (\bibinfo {year} {2025}{\natexlab{b}})}\BibitemShut {NoStop}%
\bibitem [{\citenamefont {Roussel}\ \emph {et~al.}(2017)\citenamefont
  {Roussel}, \citenamefont {Cabart}, \citenamefont {F{\`e}ve}, \citenamefont
  {Thibierge},\ and\ \citenamefont {Degiovanni}}]{Roussel-2016-2}%
  \BibitemOpen
  \bibfield  {author} {\bibinfo {author} {\bibfnamefont {B.}~\bibnamefont
  {Roussel}}, \bibinfo {author} {\bibfnamefont {C.}~\bibnamefont {Cabart}},
  \bibinfo {author} {\bibfnamefont {G.}~\bibnamefont {F{\`e}ve}}, \bibinfo
  {author} {\bibfnamefont {E.}~\bibnamefont {Thibierge}},\ and\ \bibinfo
  {author} {\bibfnamefont {P.}~\bibnamefont {Degiovanni}},\ }\bibfield  {title}
  {\bibinfo {title} {Electron quantum optics as quantum signal processing},\
  }\href {https://doi.org/10.1002/pssb.201600621} {\bibfield  {journal}
  {\bibinfo  {journal} {Physica Status Solidi B}\ }\textbf {\bibinfo {volume}
  {254}},\ \bibinfo {pages} {16000621} (\bibinfo {year} {2017})}\BibitemShut
  {NoStop}%
\bibitem [{\citenamefont {Roussel}\ \emph {et~al.}(2016)\citenamefont
  {Roussel}, \citenamefont {Degiovanni},\ and\ \citenamefont
  {Safi}}]{Roussel-2016-1}%
  \BibitemOpen
  \bibfield  {author} {\bibinfo {author} {\bibfnamefont {B.}~\bibnamefont
  {Roussel}}, \bibinfo {author} {\bibfnamefont {P.}~\bibnamefont
  {Degiovanni}},\ and\ \bibinfo {author} {\bibfnamefont {I.}~\bibnamefont
  {Safi}},\ }\bibfield  {title} {\bibinfo {title} {Perturbative fluctuation
  dissipation relation for nonequilibrium finite-frequency noise in quantum
  circuits},\ }\href {https://doi.org/10.1103/PhysRevB.93.045102} {\bibfield
  {journal} {\bibinfo  {journal} {Phys. Rev. B}\ }\textbf {\bibinfo {volume}
  {93}},\ \bibinfo {pages} {045102} (\bibinfo {year} {2016})}\BibitemShut
  {NoStop}%
\bibitem [{\citenamefont {Franson}(1989)}]{Franson-1989-1}%
  \BibitemOpen
  \bibfield  {author} {\bibinfo {author} {\bibfnamefont {J.~D.}\ \bibnamefont
  {Franson}},\ }\bibfield  {title} {\bibinfo {title} {Bell inequality for
  position and time},\ }\href {https://doi.org/10.1103/PhysRevLett.62.2205}
  {\bibfield  {journal} {\bibinfo  {journal} {Phys. Rev. Lett.}\ }\textbf
  {\bibinfo {volume} {62}},\ \bibinfo {pages} {2205} (\bibinfo {year}
  {1989})}\BibitemShut {NoStop}%
\bibitem [{\citenamefont {Parmentier}\ \emph {et~al.}(2011)\citenamefont
  {Parmentier}, \citenamefont {Mah{\'e}}, \citenamefont {Denis}, \citenamefont
  {Berroir}, \citenamefont {Glattli}, \citenamefont {Pla{\c c}ais},\ and\
  \citenamefont {F{\`e}ve}}]{Parmentier-2010-1}%
  \BibitemOpen
  \bibfield  {author} {\bibinfo {author} {\bibfnamefont {F.}~\bibnamefont
  {Parmentier}}, \bibinfo {author} {\bibfnamefont {A.}~\bibnamefont
  {Mah{\'e}}}, \bibinfo {author} {\bibfnamefont {A.}~\bibnamefont {Denis}},
  \bibinfo {author} {\bibfnamefont {J.-M.}\ \bibnamefont {Berroir}}, \bibinfo
  {author} {\bibfnamefont {D.}~\bibnamefont {Glattli}}, \bibinfo {author}
  {\bibfnamefont {B.}~\bibnamefont {Pla{\c c}ais}},\ and\ \bibinfo {author}
  {\bibfnamefont {G.}~\bibnamefont {F{\`e}ve}},\ }\bibfield  {title} {\bibinfo
  {title} {A high sensitivity ultra-low temperature rf conductance and noise
  measurement setup},\ }\href@noop {} {\bibfield  {journal} {\bibinfo
  {journal} {Rev. Sci. Instrum.}\ }\textbf {\bibinfo {volume} {82}},\ \bibinfo
  {pages} {013904} (\bibinfo {year} {2011})}\BibitemShut {NoStop}%
\bibitem [{\citenamefont {Ji}\ \emph {et~al.}(2003)\citenamefont {Ji},
  \citenamefont {Chung}, \citenamefont {Sprinzak}, \citenamefont {Heiblum},
  \citenamefont {Mahalu},\ and\ \citenamefont {Shtrikman}}]{Ji-2003-1}%
  \BibitemOpen
  \bibfield  {author} {\bibinfo {author} {\bibfnamefont {Y.}~\bibnamefont
  {Ji}}, \bibinfo {author} {\bibfnamefont {Y.}~\bibnamefont {Chung}}, \bibinfo
  {author} {\bibfnamefont {D.}~\bibnamefont {Sprinzak}}, \bibinfo {author}
  {\bibfnamefont {M.}~\bibnamefont {Heiblum}}, \bibinfo {author} {\bibfnamefont
  {D.}~\bibnamefont {Mahalu}},\ and\ \bibinfo {author} {\bibfnamefont
  {H.}~\bibnamefont {Shtrikman}},\ }\bibfield  {title} {\bibinfo {title} {An
  electronic {Mach-Zehnder} interferometer},\ }\href
  {https://doi.org/doi:10.1038/nature01503} {\bibfield  {journal} {\bibinfo
  {journal} {Nature}\ }\textbf {\bibinfo {volume} {422}},\ \bibinfo {pages}
  {415} (\bibinfo {year} {2003})}\BibitemShut {NoStop}%
\bibitem [{\citenamefont {Roulleau}\ \emph {et~al.}(2008)\citenamefont
  {Roulleau}, \citenamefont {Portier}, \citenamefont {Roche}, \citenamefont
  {Cavanna}, \citenamefont {Faini}, \citenamefont {Gennser},\ and\
  \citenamefont {Mailly}}]{Roulleau-2008-2}%
  \BibitemOpen
  \bibfield  {author} {\bibinfo {author} {\bibfnamefont {P.}~\bibnamefont
  {Roulleau}}, \bibinfo {author} {\bibfnamefont {F.}~\bibnamefont {Portier}},
  \bibinfo {author} {\bibfnamefont {P.}~\bibnamefont {Roche}}, \bibinfo
  {author} {\bibfnamefont {A.}~\bibnamefont {Cavanna}}, \bibinfo {author}
  {\bibfnamefont {G.}~\bibnamefont {Faini}}, \bibinfo {author} {\bibfnamefont
  {U.}~\bibnamefont {Gennser}},\ and\ \bibinfo {author} {\bibfnamefont
  {D.}~\bibnamefont {Mailly}},\ }\bibfield  {title} {\bibinfo {title} {Direct
  measurement of the coherence length of edge states in the integer quantum
  {Hall} regime},\ }\href {https://doi.org/10.1103/PhysRevLett.100.126802}
  {\bibfield  {journal} {\bibinfo  {journal} {Phys. Rev. Lett.}\ }\textbf
  {\bibinfo {volume} {100}},\ \bibinfo {pages} {126802} (\bibinfo {year}
  {2008})}\BibitemShut {NoStop}%
\bibitem [{\citenamefont {Bocquillon}\ \emph {et~al.}(2012)\citenamefont
  {Bocquillon}, \citenamefont {Parmentier}, \citenamefont {Grenier},
  \citenamefont {Berroir}, \citenamefont {Degiovanni}, \citenamefont {Glattli},
  \citenamefont {Pla{\c c}ais}, \citenamefont {Cavanna}, \citenamefont {Jin},\
  and\ \citenamefont {F{\`e}ve}}]{Bocquillon-2012-1}%
  \BibitemOpen
  \bibfield  {author} {\bibinfo {author} {\bibfnamefont {E.}~\bibnamefont
  {Bocquillon}}, \bibinfo {author} {\bibfnamefont {F.~D.}\ \bibnamefont
  {Parmentier}}, \bibinfo {author} {\bibfnamefont {C.}~\bibnamefont {Grenier}},
  \bibinfo {author} {\bibfnamefont {J.-M.}\ \bibnamefont {Berroir}}, \bibinfo
  {author} {\bibfnamefont {P.}~\bibnamefont {Degiovanni}}, \bibinfo {author}
  {\bibfnamefont {D.~C.}\ \bibnamefont {Glattli}}, \bibinfo {author}
  {\bibfnamefont {B.}~\bibnamefont {Pla{\c c}ais}}, \bibinfo {author}
  {\bibfnamefont {A.}~\bibnamefont {Cavanna}}, \bibinfo {author} {\bibfnamefont
  {Y.}~\bibnamefont {Jin}},\ and\ \bibinfo {author} {\bibfnamefont
  {G.}~\bibnamefont {F{\`e}ve}},\ }\bibfield  {title} {\bibinfo {title}
  {Electron quantum optics: partitioning electrons one by one},\ }\href@noop {}
  {\bibfield  {journal} {\bibinfo  {journal} {Phys. Rev. Lett.}\ }\textbf
  {\bibinfo {volume} {108}},\ \bibinfo {pages} {196803} (\bibinfo {year}
  {2012})}\BibitemShut {NoStop}%
\bibitem [{\citenamefont {Bocquillon}\ \emph {et~al.}(2013)\citenamefont
  {Bocquillon}, \citenamefont {Freulon}, \citenamefont {Berroir}, \citenamefont
  {Degiovanni}, \citenamefont {Pla{\c c}ais}, \citenamefont {Cavanna},
  \citenamefont {Jin},\ and\ \citenamefont {F{\`e}ve}}]{Bocquillon-2013-1}%
  \BibitemOpen
  \bibfield  {author} {\bibinfo {author} {\bibfnamefont {E.}~\bibnamefont
  {Bocquillon}}, \bibinfo {author} {\bibfnamefont {V.}~\bibnamefont {Freulon}},
  \bibinfo {author} {\bibfnamefont {J.-M.}\ \bibnamefont {Berroir}}, \bibinfo
  {author} {\bibfnamefont {P.}~\bibnamefont {Degiovanni}}, \bibinfo {author}
  {\bibfnamefont {B.}~\bibnamefont {Pla{\c c}ais}}, \bibinfo {author}
  {\bibfnamefont {A.}~\bibnamefont {Cavanna}}, \bibinfo {author} {\bibfnamefont
  {Y.}~\bibnamefont {Jin}},\ and\ \bibinfo {author} {\bibfnamefont
  {G.}~\bibnamefont {F{\`e}ve}},\ }\bibfield  {title} {\bibinfo {title}
  {Coherence and indistinguability of single electrons emitted by independent
  sources},\ }\href {https://doi.org/10.1126/science.1232572} {\bibfield
  {journal} {\bibinfo  {journal} {Science}\ }\textbf {\bibinfo {volume}
  {339}},\ \bibinfo {pages} {1054} (\bibinfo {year} {2013})}\BibitemShut
  {NoStop}%
\bibitem [{\citenamefont {Bartolomei}\ \emph {et~al.}(2020)\citenamefont
  {Bartolomei}, \citenamefont {Kumar}, \citenamefont {Bisognin}, \citenamefont
  {Marguerite}, \citenamefont {Berroir}, \citenamefont {Bocquillon},
  \citenamefont {Pla{\c c}ais}, \citenamefont {Cavanna}, \citenamefont {Dong},
  \citenamefont {Gennser}, \citenamefont {Jin},\ and\ \citenamefont
  {F{\`e}ve}}]{Bartolomei-2020-1}%
  \BibitemOpen
  \bibfield  {author} {\bibinfo {author} {\bibfnamefont {H.}~\bibnamefont
  {Bartolomei}}, \bibinfo {author} {\bibfnamefont {M.}~\bibnamefont {Kumar}},
  \bibinfo {author} {\bibfnamefont {R.}~\bibnamefont {Bisognin}}, \bibinfo
  {author} {\bibfnamefont {A.}~\bibnamefont {Marguerite}}, \bibinfo {author}
  {\bibfnamefont {J.-M.}\ \bibnamefont {Berroir}}, \bibinfo {author}
  {\bibfnamefont {E.}~\bibnamefont {Bocquillon}}, \bibinfo {author}
  {\bibfnamefont {B.}~\bibnamefont {Pla{\c c}ais}}, \bibinfo {author}
  {\bibfnamefont {A.}~\bibnamefont {Cavanna}}, \bibinfo {author} {\bibfnamefont
  {Q.}~\bibnamefont {Dong}}, \bibinfo {author} {\bibfnamefont {U.}~\bibnamefont
  {Gennser}}, \bibinfo {author} {\bibfnamefont {Y.}~\bibnamefont {Jin}},\ and\
  \bibinfo {author} {\bibfnamefont {G.}~\bibnamefont {F{\`e}ve}},\ }\bibfield
  {title} {\bibinfo {title} {Fractional statistics in anyon collisions},\
  }\href {https://doi.org/10.1126/science.aaz5601} {\bibfield  {journal}
  {\bibinfo  {journal} {Science}\ }\textbf {\bibinfo {volume} {368}},\ \bibinfo
  {pages} {173} (\bibinfo {year} {2020})}\BibitemShut {NoStop}%
\bibitem [{\citenamefont {G{\"u}hne}\ and\ \citenamefont
  {L{\"u}tkenhaus}(2006)}]{Guhne-2006-1}%
  \BibitemOpen
  \bibfield  {author} {\bibinfo {author} {\bibfnamefont {O.}~\bibnamefont
  {G{\"u}hne}}\ and\ \bibinfo {author} {\bibfnamefont {N.}~\bibnamefont
  {L{\"u}tkenhaus}},\ }\bibfield  {title} {\bibinfo {title} {Nonlinear
  entanglement witnesses},\ }\href
  {https://doi.org/10.1103/PhysRevLett.96.170502} {\bibfield  {journal}
  {\bibinfo  {journal} {Phys. Rev. Lett.}\ }\textbf {\bibinfo {volume} {96}},\
  \bibinfo {pages} {170502} (\bibinfo {year} {2006})}\BibitemShut {NoStop}%
\bibitem [{\citenamefont {Terhal}(2000)}]{Terhal-2000-1}%
  \BibitemOpen
  \bibfield  {author} {\bibinfo {author} {\bibfnamefont {B.~M.}\ \bibnamefont
  {Terhal}},\ }\bibfield  {title} {\bibinfo {title} {Bell inequalities and the
  separability criterion},\ }\href
  {https://doi.org/https://doi.org/10.1016/S0375-9601(00)00401-1} {\bibfield
  {journal} {\bibinfo  {journal} {Physics Letters A}\ }\textbf {\bibinfo
  {volume} {271}},\ \bibinfo {pages} {319} (\bibinfo {year}
  {2000})}\BibitemShut {NoStop}%
\bibitem [{\citenamefont {Moskalets}(2014)}]{Moskalets-2014-1}%
  \BibitemOpen
  \bibfield  {author} {\bibinfo {author} {\bibfnamefont {M.}~\bibnamefont
  {Moskalets}},\ }\bibfield  {title} {\bibinfo {title} {Two-electron state from
  the {F}loquet scattering matrix perspective},\ }\href
  {https://doi.org/10.1103/PhysRevB.89.045402} {\bibfield  {journal} {\bibinfo
  {journal} {Phys. Rev. B}\ }\textbf {\bibinfo {volume} {89}},\ \bibinfo
  {pages} {045402} (\bibinfo {year} {2014})}\BibitemShut {NoStop}%
\bibitem [{\citenamefont {Ferraro}\ \emph {et~al.}(2013)\citenamefont
  {Ferraro}, \citenamefont {Feller}, \citenamefont {Ghibaudo}, \citenamefont
  {Thibierge}, \citenamefont {Bocquillon}, \citenamefont {F{\`e}ve},
  \citenamefont {Grenier},\ and\ \citenamefont {Degiovanni}}]{Ferraro-2013-1}%
  \BibitemOpen
  \bibfield  {author} {\bibinfo {author} {\bibfnamefont {D.}~\bibnamefont
  {Ferraro}}, \bibinfo {author} {\bibfnamefont {A.}~\bibnamefont {Feller}},
  \bibinfo {author} {\bibfnamefont {A.}~\bibnamefont {Ghibaudo}}, \bibinfo
  {author} {\bibfnamefont {E.}~\bibnamefont {Thibierge}}, \bibinfo {author}
  {\bibfnamefont {E.}~\bibnamefont {Bocquillon}}, \bibinfo {author}
  {\bibfnamefont {G.}~\bibnamefont {F{\`e}ve}}, \bibinfo {author}
  {\bibfnamefont {C.}~\bibnamefont {Grenier}},\ and\ \bibinfo {author}
  {\bibfnamefont {P.}~\bibnamefont {Degiovanni}},\ }\bibfield  {title}
  {\bibinfo {title} {Wigner function approach to single electron coherence in
  quantum {Hall} edge channels},\ }\href
  {https://doi.org/10.1103/PhysRevB.88.205303} {\bibfield  {journal} {\bibinfo
  {journal} {Phys. Rev. B}\ }\textbf {\bibinfo {volume} {88}},\ \bibinfo
  {pages} {205303} (\bibinfo {year} {2013})}\BibitemShut {NoStop}%
\bibitem [{\citenamefont {Ferraro}\ \emph {et~al.}(2014)\citenamefont
  {Ferraro}, \citenamefont {Wahl}, \citenamefont {Rech}, \citenamefont
  {Jonckheere},\ and\ \citenamefont {Martin}}]{Ferraro-2014-1}%
  \BibitemOpen
  \bibfield  {author} {\bibinfo {author} {\bibfnamefont {D.}~\bibnamefont
  {Ferraro}}, \bibinfo {author} {\bibfnamefont {C.}~\bibnamefont {Wahl}},
  \bibinfo {author} {\bibfnamefont {J.}~\bibnamefont {Rech}}, \bibinfo {author}
  {\bibfnamefont {T.}~\bibnamefont {Jonckheere}},\ and\ \bibinfo {author}
  {\bibfnamefont {T.}~\bibnamefont {Martin}},\ }\bibfield  {title} {\bibinfo
  {title} {Electronic {H}ong-{O}u-{M}andel interferometry in two-dimensional
  topological insulators},\ }\href {https://doi.org/10.1103/PhysRevB.89.075407}
  {\bibfield  {journal} {\bibinfo  {journal} {Phys. Rev. B}\ }\textbf {\bibinfo
  {volume} {89}},\ \bibinfo {pages} {075407} (\bibinfo {year}
  {2014})}\BibitemShut {NoStop}%
\bibitem [{\citenamefont {Moskalets}(2018)}]{Moskalets-2018-3}%
  \BibitemOpen
  \bibfield  {author} {\bibinfo {author} {\bibfnamefont {M.}~\bibnamefont
  {Moskalets}},\ }\bibfield  {title} {\bibinfo {title} {Single-electron
  second-order correlation function ${G}^{(2)}$ at nonzero temperatures},\
  }\href {https://doi.org/10.1103/PhysRevB.98.115421} {\bibfield  {journal}
  {\bibinfo  {journal} {Phys. Rev. B}\ }\textbf {\bibinfo {volume} {98}},\
  \bibinfo {pages} {115421} (\bibinfo {year} {2018})}\BibitemShut {NoStop}%
\bibitem [{\citenamefont {Mohammad-Djafari}(2010)}]{Book-Inverse-Problems}%
  \BibitemOpen
  \bibinfo {editor} {\bibfnamefont {A.}~\bibnamefont {Mohammad-Djafari}},\
  ed.,\ \href@noop {} {\emph {\bibinfo {title} {Inverse problems in vision and
  {3D} tomography}}}\ (\bibinfo  {publisher} {Wiley-ISTE},\ \bibinfo {year}
  {2010})\BibitemShut {NoStop}%
\bibitem [{\citenamefont {Blume-Kohout}(2010)}]{Blume-Kohout-2010-1}%
  \BibitemOpen
  \bibfield  {author} {\bibinfo {author} {\bibfnamefont {R.}~\bibnamefont
  {Blume-Kohout}},\ }\bibfield  {title} {\bibinfo {title} {Optimal, reliable
  estimation of quantum states},\ }\href
  {https://doi.org/10.1088/1367-2630/12/4/043034} {\bibfield  {journal}
  {\bibinfo  {journal} {New Journal of Physics}\ }\textbf {\bibinfo {volume}
  {12}},\ \bibinfo {pages} {043034} (\bibinfo {year} {2010})},\ \Eprint
  {https://arxiv.org/abs/quant-ph/0611080} {quant-ph/0611080} \BibitemShut
  {NoStop}%
\bibitem [{\citenamefont {Ubbelohde}\ \emph {et~al.}(2026)\citenamefont
  {Ubbelohde}, \citenamefont {Kokainis},\ and\ \citenamefont
  {Kashcheyevs}}]{Ubbelohde-Kokainis-Kashcheyevs-PC}%
  \BibitemOpen
  \bibfield  {author} {\bibinfo {author} {\bibfnamefont {N.}~\bibnamefont
  {Ubbelohde}}, \bibinfo {author} {\bibfnamefont {M.}~\bibnamefont
  {Kokainis}},\ and\ \bibinfo {author} {\bibfnamefont {V.}~\bibnamefont
  {Kashcheyevs}}} (\bibinfo {year} {2026}),\ \bibinfo {note} {private
  communication}\BibitemShut {NoStop}%
\bibitem [{\citenamefont {Souquet-Basi{\`e}ge}\ \emph
  {et~al.}(2025)\citenamefont {Souquet-Basi{\`e}ge}, \citenamefont {Roussel},
  \citenamefont {Rebora}, \citenamefont {M{\'e}nard}, \citenamefont {Safi},
  \citenamefont {F{\`e}ve},\ and\ \citenamefont
  {Degiovanni}}]{Souquet-Basiege-2024-1}%
  \BibitemOpen
  \bibfield  {author} {\bibinfo {author} {\bibfnamefont {H.}~\bibnamefont
  {Souquet-Basi{\`e}ge}}, \bibinfo {author} {\bibfnamefont {B.}~\bibnamefont
  {Roussel}}, \bibinfo {author} {\bibfnamefont {G.}~\bibnamefont {Rebora}},
  \bibinfo {author} {\bibfnamefont {G.}~\bibnamefont {M{\'e}nard}}, \bibinfo
  {author} {\bibfnamefont {I.}~\bibnamefont {Safi}}, \bibinfo {author}
  {\bibfnamefont {G.}~\bibnamefont {F{\`e}ve}},\ and\ \bibinfo {author}
  {\bibfnamefont {P.}~\bibnamefont {Degiovanni}},\ }\bibfield  {title}
  {\bibinfo {title} {Quantum sensing of time dependent electromagnetic fields
  with single electron excitations},\ }\href
  {https://doi.org/10.1103/1nfc-stxp} {\bibfield  {journal} {\bibinfo
  {journal} {Physical Review X}\ }\textbf {\bibinfo {volume} {15}},\ \bibinfo
  {pages} {021043} (\bibinfo {year} {2025})}\BibitemShut {NoStop}%
\bibitem [{\citenamefont {Altimiras}\ \emph
  {et~al.}(2010{\natexlab{a}})\citenamefont {Altimiras}, \citenamefont
  {le~Sueur}, \citenamefont {Gennser}, \citenamefont {Cavanna}, \citenamefont
  {Mailly},\ and\ \citenamefont {Pierre}}]{Altimiras-2010-1}%
  \BibitemOpen
  \bibfield  {author} {\bibinfo {author} {\bibfnamefont {C.}~\bibnamefont
  {Altimiras}}, \bibinfo {author} {\bibfnamefont {H.}~\bibnamefont {le~Sueur}},
  \bibinfo {author} {\bibfnamefont {U.}~\bibnamefont {Gennser}}, \bibinfo
  {author} {\bibfnamefont {A.}~\bibnamefont {Cavanna}}, \bibinfo {author}
  {\bibfnamefont {D.}~\bibnamefont {Mailly}},\ and\ \bibinfo {author}
  {\bibfnamefont {F.}~\bibnamefont {Pierre}},\ }\bibfield  {title} {\bibinfo
  {title} {Non-equilibrium edge channel spectroscopy in the integer quantum
  {H}all regime},\ }\href@noop {} {\bibfield  {journal} {\bibinfo  {journal}
  {Nature Physics}\ }\textbf {\bibinfo {volume} {6}},\ \bibinfo {pages} {34}
  (\bibinfo {year} {2010}{\natexlab{a}})}\BibitemShut {NoStop}%
\bibitem [{\citenamefont {Altimiras}\ \emph
  {et~al.}(2010{\natexlab{b}})\citenamefont {Altimiras}, \citenamefont
  {Le~Sueur}, \citenamefont {Gennser}, \citenamefont {Cavanna}, \citenamefont
  {Mailly},\ and\ \citenamefont {Pierre}}]{Altimiras-2010-2}%
  \BibitemOpen
  \bibfield  {author} {\bibinfo {author} {\bibfnamefont {C.}~\bibnamefont
  {Altimiras}}, \bibinfo {author} {\bibfnamefont {H.}~\bibnamefont {Le~Sueur}},
  \bibinfo {author} {\bibfnamefont {U.}~\bibnamefont {Gennser}}, \bibinfo
  {author} {\bibfnamefont {A.}~\bibnamefont {Cavanna}}, \bibinfo {author}
  {\bibfnamefont {D.}~\bibnamefont {Mailly}},\ and\ \bibinfo {author}
  {\bibfnamefont {F.}~\bibnamefont {Pierre}},\ }\bibfield  {title} {\bibinfo
  {title} {Tuning energy relaxation along quantum hall edge channels},\ }\href
  {https://doi.org/https://doi.org/10.1103/PhysRevLett.105.226804} {\bibfield
  {journal} {\bibinfo  {journal} {Phys. Rev. Lett.}\ }\textbf {\bibinfo
  {volume} {105}},\ \bibinfo {pages} {226804} (\bibinfo {year}
  {2010}{\natexlab{b}})}\BibitemShut {NoStop}%
\bibitem [{\citenamefont {Le~Sueur}\ \emph {et~al.}(2010)\citenamefont
  {Le~Sueur}, \citenamefont {Altimiras}, \citenamefont {Gennser}, \citenamefont
  {Cavanna}, \citenamefont {Mailly},\ and\ \citenamefont
  {Pierre}}]{LeSueur-2010-1}%
  \BibitemOpen
  \bibfield  {author} {\bibinfo {author} {\bibfnamefont {H.}~\bibnamefont
  {Le~Sueur}}, \bibinfo {author} {\bibfnamefont {C.}~\bibnamefont {Altimiras}},
  \bibinfo {author} {\bibfnamefont {U.}~\bibnamefont {Gennser}}, \bibinfo
  {author} {\bibfnamefont {A.}~\bibnamefont {Cavanna}}, \bibinfo {author}
  {\bibfnamefont {D.}~\bibnamefont {Mailly}},\ and\ \bibinfo {author}
  {\bibfnamefont {F.}~\bibnamefont {Pierre}},\ }\bibfield  {title} {\bibinfo
  {title} {Energy relaxation in the integer quantum hall regime},\ }\href
  {https://doi.org/10.1103/PhysRevLett.105.056803} {\bibfield  {journal}
  {\bibinfo  {journal} {Phys. Rev. Lett.}\ }\textbf {\bibinfo {volume} {105}},\
  \bibinfo {pages} {056803} (\bibinfo {year} {2010})}\BibitemShut {NoStop}%
\bibitem [{\citenamefont {Rodriguez}\ \emph {et~al.}(2020)\citenamefont
  {Rodriguez}, \citenamefont {Parmentier}, \citenamefont {Ferraro},
  \citenamefont {Roulleau}, \citenamefont {Gensser}, \citenamefont {Cavanna},
  \citenamefont {Sassetti}, \citenamefont {Portier}, \citenamefont {Mailly},\
  and\ \citenamefont {Roche}}]{Rodriguez-2020-1}%
  \BibitemOpen
  \bibfield  {author} {\bibinfo {author} {\bibfnamefont {R.}~\bibnamefont
  {Rodriguez}}, \bibinfo {author} {\bibfnamefont {F.}~\bibnamefont
  {Parmentier}}, \bibinfo {author} {\bibfnamefont {D.}~\bibnamefont {Ferraro}},
  \bibinfo {author} {\bibfnamefont {P.}~\bibnamefont {Roulleau}}, \bibinfo
  {author} {\bibfnamefont {U.}~\bibnamefont {Gensser}}, \bibinfo {author}
  {\bibfnamefont {A.}~\bibnamefont {Cavanna}}, \bibinfo {author} {\bibfnamefont
  {M.}~\bibnamefont {Sassetti}}, \bibinfo {author} {\bibfnamefont
  {F.}~\bibnamefont {Portier}}, \bibinfo {author} {\bibfnamefont
  {D.}~\bibnamefont {Mailly}},\ and\ \bibinfo {author} {\bibfnamefont
  {P.}~\bibnamefont {Roche}},\ }\bibfield  {title} {\bibinfo {title}
  {Relaxation and revival of quasiparticles injected in an interacting quantum
  {H}all liquid},\ }\href {https://doi.org/10.1038/s41467-020-16331-4}
  {\bibfield  {journal} {\bibinfo  {journal} {Nature Communications}\ }\textbf
  {\bibinfo {volume} {11}},\ \bibinfo {pages} {2426} (\bibinfo {year}
  {2020})}\BibitemShut {NoStop}%
\bibitem [{\citenamefont {Degiovanni}\ \emph {et~al.}(2009)\citenamefont
  {Degiovanni}, \citenamefont {Grenier},\ and\ \citenamefont
  {F{\`e}ve}}]{Degio-2009-1}%
  \BibitemOpen
  \bibfield  {author} {\bibinfo {author} {\bibfnamefont {P.}~\bibnamefont
  {Degiovanni}}, \bibinfo {author} {\bibfnamefont {C.}~\bibnamefont
  {Grenier}},\ and\ \bibinfo {author} {\bibfnamefont {G.}~\bibnamefont
  {F{\`e}ve}},\ }\bibfield  {title} {\bibinfo {title} {Decoherence and
  relaxation of single electron excitations in quantum {H}all edge channels},\
  }\href@noop {} {\bibfield  {journal} {\bibinfo  {journal} {Phys. Rev. B}\
  }\textbf {\bibinfo {volume} {80}},\ \bibinfo {pages} {241307(R)} (\bibinfo
  {year} {2009})}\BibitemShut {NoStop}%
\bibitem [{\citenamefont {Grenier}\ \emph
  {et~al.}(2011{\natexlab{a}})\citenamefont {Grenier}, \citenamefont
  {Herv{\'e}}, \citenamefont {F{\`e}ve},\ and\ \citenamefont
  {Degiovanni}}]{Degio-2011-1}%
  \BibitemOpen
  \bibfield  {author} {\bibinfo {author} {\bibfnamefont {C.}~\bibnamefont
  {Grenier}}, \bibinfo {author} {\bibfnamefont {R.}~\bibnamefont {Herv{\'e}}},
  \bibinfo {author} {\bibfnamefont {G.}~\bibnamefont {F{\`e}ve}},\ and\
  \bibinfo {author} {\bibfnamefont {P.}~\bibnamefont {Degiovanni}},\ }\bibfield
   {title} {\bibinfo {title} {Electron quantum optics in quantum {H}all edge
  channels},\ }\href@noop {} {\bibfield  {journal} {\bibinfo  {journal} {Mod.
  Phys. Lett. B}\ }\textbf {\bibinfo {volume} {25}},\ \bibinfo {pages} {1053 }
  (\bibinfo {year} {2011}{\natexlab{a}})},\ \bibinfo {note} {proceedings of the
  Statphys 24 satellite meeting "International Conference on Frustrated Spin
  Systems, Cold Atoms and Nanomaterials" (Hanoi, 14-16 July 2010)}\BibitemShut
  {NoStop}%
\bibitem [{\citenamefont {B{\"u}ttiker}\ \emph {et~al.}(1993)\citenamefont
  {B{\"u}ttiker}, \citenamefont {Pr{\^e}tre},\ and\ \citenamefont
  {Thomas}}]{Buttiker-1993-1}%
  \BibitemOpen
  \bibfield  {author} {\bibinfo {author} {\bibfnamefont {M.}~\bibnamefont
  {B{\"u}ttiker}}, \bibinfo {author} {\bibfnamefont {A.}~\bibnamefont
  {Pr{\^e}tre}},\ and\ \bibinfo {author} {\bibfnamefont {H.}~\bibnamefont
  {Thomas}},\ }\bibfield  {title} {\bibinfo {title} {Dynamic conductance and
  the scattering matrix of small conductors},\ }\href
  {https://doi.org/10.1103/PhysRevLett.70.4114} {\bibfield  {journal} {\bibinfo
   {journal} {Phys. Rev. Lett.}\ }\textbf {\bibinfo {volume} {70}},\ \bibinfo
  {pages} {4114} (\bibinfo {year} {1993})}\BibitemShut {NoStop}%
\bibitem [{\citenamefont {F{\`e}ve}\ \emph {et~al.}(2007)\citenamefont
  {F{\`e}ve}, \citenamefont {Mah{\'e}}, \citenamefont {Berroir}, \citenamefont
  {Kontos}, \citenamefont {Pla{\c c}ais}, \citenamefont {Glattli},
  \citenamefont {Cavanna}, \citenamefont {Etienne},\ and\ \citenamefont
  {Jin}}]{Feve-2007-1}%
  \BibitemOpen
  \bibfield  {author} {\bibinfo {author} {\bibfnamefont {G.}~\bibnamefont
  {F{\`e}ve}}, \bibinfo {author} {\bibfnamefont {A.}~\bibnamefont {Mah{\'e}}},
  \bibinfo {author} {\bibfnamefont {J.}~\bibnamefont {Berroir}}, \bibinfo
  {author} {\bibfnamefont {T.}~\bibnamefont {Kontos}}, \bibinfo {author}
  {\bibfnamefont {B.}~\bibnamefont {Pla{\c c}ais}}, \bibinfo {author}
  {\bibfnamefont {D.~C.}\ \bibnamefont {Glattli}}, \bibinfo {author}
  {\bibfnamefont {A.}~\bibnamefont {Cavanna}}, \bibinfo {author} {\bibfnamefont
  {B.}~\bibnamefont {Etienne}},\ and\ \bibinfo {author} {\bibfnamefont
  {Y.}~\bibnamefont {Jin}},\ }\bibfield  {title} {\bibinfo {title} {An
  on-demand single electron source},\ }\href
  {https://doi.org/10.1126/science.1141243} {\bibfield  {journal} {\bibinfo
  {journal} {Science}\ }\textbf {\bibinfo {volume} {316}},\ \bibinfo {pages}
  {1169} (\bibinfo {year} {2007})}\BibitemShut {NoStop}%
\bibitem [{\citenamefont {Bartolomei}\ \emph {et~al.}(2025)\citenamefont
  {Bartolomei}, \citenamefont {Frigerio}, \citenamefont {Ruelle}, \citenamefont
  {Rebora}, \citenamefont {Jin}, \citenamefont {Gennser}, \citenamefont
  {Cavanna}, \citenamefont {Baudin}, \citenamefont {Berroir}, \citenamefont
  {Safi}, \citenamefont {Degiovanni}, \citenamefont {M{\'e}nard},\ and\
  \citenamefont {F{\`e}ve}}]{Bartolomei-2025-1}%
  \BibitemOpen
  \bibfield  {author} {\bibinfo {author} {\bibfnamefont {H.}~\bibnamefont
  {Bartolomei}}, \bibinfo {author} {\bibfnamefont {E.}~\bibnamefont
  {Frigerio}}, \bibinfo {author} {\bibfnamefont {M.}~\bibnamefont {Ruelle}},
  \bibinfo {author} {\bibfnamefont {G.}~\bibnamefont {Rebora}}, \bibinfo
  {author} {\bibfnamefont {Y.}~\bibnamefont {Jin}}, \bibinfo {author}
  {\bibfnamefont {U.}~\bibnamefont {Gennser}}, \bibinfo {author} {\bibfnamefont
  {A.}~\bibnamefont {Cavanna}}, \bibinfo {author} {\bibfnamefont
  {E.}~\bibnamefont {Baudin}}, \bibinfo {author} {\bibfnamefont {J.-M.}\
  \bibnamefont {Berroir}}, \bibinfo {author} {\bibfnamefont {I.}~\bibnamefont
  {Safi}}, \bibinfo {author} {\bibfnamefont {P.}~\bibnamefont {Degiovanni}},
  \bibinfo {author} {\bibfnamefont {G.~C.}\ \bibnamefont {M{\'e}nard}},\ and\
  \bibinfo {author} {\bibfnamefont {G.}~\bibnamefont {F{\`e}ve}},\ }\bibfield
  {title} {\bibinfo {title} {Time-resolved sensing of electromagnetic fields
  with single-electron interferometry},\ }\href
  {https://doi.org/10.1038/s41565-025-01888-2} {\bibfield  {journal} {\bibinfo
  {journal} {Nature Nanotechnology}\ }\textbf {\bibinfo {volume} {20}},\
  \bibinfo {pages} {596} (\bibinfo {year} {2025})}\BibitemShut {NoStop}%
\bibitem [{\citenamefont {Grenier}\ \emph {et~al.}(2013)\citenamefont
  {Grenier}, \citenamefont {Dubois}, \citenamefont {Jullien}, \citenamefont
  {Roulleau}, \citenamefont {Glattli},\ and\ \citenamefont
  {Degiovanni}}]{Grenier-2013-1}%
  \BibitemOpen
  \bibfield  {author} {\bibinfo {author} {\bibfnamefont {C.}~\bibnamefont
  {Grenier}}, \bibinfo {author} {\bibfnamefont {J.}~\bibnamefont {Dubois}},
  \bibinfo {author} {\bibfnamefont {T.}~\bibnamefont {Jullien}}, \bibinfo
  {author} {\bibfnamefont {P.}~\bibnamefont {Roulleau}}, \bibinfo {author}
  {\bibfnamefont {D.~C.}\ \bibnamefont {Glattli}},\ and\ \bibinfo {author}
  {\bibfnamefont {P.}~\bibnamefont {Degiovanni}},\ }\bibfield  {title}
  {\bibinfo {title} {Fractionalization of minimal excitations in integer
  quantum hall edge channels},\ }\href
  {https://doi.org/10.1103/PhysRevB.88.085302} {\bibfield  {journal} {\bibinfo
  {journal} {Phys. Rev. B}\ }\textbf {\bibinfo {volume} {88}},\ \bibinfo
  {pages} {085302} (\bibinfo {year} {2013})}\BibitemShut {NoStop}%
\bibitem [{\citenamefont {Cabart}\ \emph {et~al.}(2018)\citenamefont {Cabart},
  \citenamefont {Roussel}, \citenamefont {F{\`e}ve},\ and\ \citenamefont
  {Degiovanni}}]{Cabart-2018-1}%
  \BibitemOpen
  \bibfield  {author} {\bibinfo {author} {\bibfnamefont {C.}~\bibnamefont
  {Cabart}}, \bibinfo {author} {\bibfnamefont {B.}~\bibnamefont {Roussel}},
  \bibinfo {author} {\bibfnamefont {G.}~\bibnamefont {F{\`e}ve}},\ and\
  \bibinfo {author} {\bibfnamefont {P.}~\bibnamefont {Degiovanni}},\ }\bibfield
   {title} {\bibinfo {title} {Taming electronic decoherence in 1d chiral
  ballistic quantum conductors},\ }\href
  {https://doi.org/10.1103/PhysRevB.98.155302} {\bibfield  {journal} {\bibinfo
  {journal} {Phys. Rev. B}\ }\textbf {\bibinfo {volume} {98}},\ \bibinfo
  {pages} {155302} (\bibinfo {year} {2018})}\BibitemShut {NoStop}%
\bibitem [{\citenamefont {Frigerio}\ \emph {et~al.}(2024)\citenamefont
  {Frigerio}, \citenamefont {Rebora}, \citenamefont {Ruelle}, \citenamefont
  {Souquet-Basi{\`e}ge}, \citenamefont {Jin}, \citenamefont {Gennser},
  \citenamefont {Cavanna}, \citenamefont {Pla{\c c}ais}, \citenamefont
  {Baudin}, \citenamefont {Berroir}, \citenamefont {Safi}, \citenamefont
  {Degiovanni}, \citenamefont {F{\`e}ve},\ and\ \citenamefont
  {M{\'e}nard}}]{Frigerio-2024-1}%
  \BibitemOpen
  \bibfield  {author} {\bibinfo {author} {\bibfnamefont {E.}~\bibnamefont
  {Frigerio}}, \bibinfo {author} {\bibfnamefont {G.}~\bibnamefont {Rebora}},
  \bibinfo {author} {\bibfnamefont {M.}~\bibnamefont {Ruelle}}, \bibinfo
  {author} {\bibfnamefont {H.}~\bibnamefont {Souquet-Basi{\`e}ge}}, \bibinfo
  {author} {\bibfnamefont {Y.}~\bibnamefont {Jin}}, \bibinfo {author}
  {\bibfnamefont {U.}~\bibnamefont {Gennser}}, \bibinfo {author} {\bibfnamefont
  {A.}~\bibnamefont {Cavanna}}, \bibinfo {author} {\bibfnamefont
  {B.}~\bibnamefont {Pla{\c c}ais}}, \bibinfo {author} {\bibfnamefont
  {E.}~\bibnamefont {Baudin}}, \bibinfo {author} {\bibfnamefont {J.~M.}\
  \bibnamefont {Berroir}}, \bibinfo {author} {\bibfnamefont {I.}~\bibnamefont
  {Safi}}, \bibinfo {author} {\bibfnamefont {P.}~\bibnamefont {Degiovanni}},
  \bibinfo {author} {\bibfnamefont {G.}~\bibnamefont {F{\`e}ve}},\ and\
  \bibinfo {author} {\bibfnamefont {G.}~\bibnamefont {M{\'e}nard}},\ }\bibfield
   {title} {\bibinfo {title} {Gate tunable edge magnetoplasmon resonators},\
  }\href {https://doi.org/10.1038/s42005-024-01803-6} {\bibfield  {journal}
  {\bibinfo  {journal} {Communications in Physics}\ }\textbf {\bibinfo {volume}
  {7}},\ \bibinfo {pages} {314} (\bibinfo {year} {2024})}\BibitemShut {NoStop}%
\bibitem [{\citenamefont {Delgard}\ \emph {et~al.}(2021)\citenamefont
  {Delgard}, \citenamefont {Chenaud}, \citenamefont {Gennser}, \citenamefont
  {Cavanna}, \citenamefont {Mailly}, \citenamefont {Degiovanni},\ and\
  \citenamefont {Chaubet}}]{Delgard-2021-1}%
  \BibitemOpen
  \bibfield  {author} {\bibinfo {author} {\bibfnamefont {A.}~\bibnamefont
  {Delgard}}, \bibinfo {author} {\bibfnamefont {B.}~\bibnamefont {Chenaud}},
  \bibinfo {author} {\bibfnamefont {U.}~\bibnamefont {Gennser}}, \bibinfo
  {author} {\bibfnamefont {A.}~\bibnamefont {Cavanna}}, \bibinfo {author}
  {\bibfnamefont {D.}~\bibnamefont {Mailly}}, \bibinfo {author} {\bibfnamefont
  {P.}~\bibnamefont {Degiovanni}},\ and\ \bibinfo {author} {\bibfnamefont
  {C.}~\bibnamefont {Chaubet}},\ }\bibfield  {title} {\bibinfo {title} {Coulomb
  interactions and effective quantum inertia of charge carriers in a
  macroscopic conductor},\ }\href
  {https://doi.org/10.1103/PhysRevB.104.L121301} {\bibfield  {journal}
  {\bibinfo  {journal} {Phys. Rev. B}\ }\textbf {\bibinfo {volume} {104}},\
  \bibinfo {pages} {L121301} (\bibinfo {year} {2021})}\BibitemShut {NoStop}%
\bibitem [{Note1()}]{Note1}%
  \BibitemOpen
  \bibinfo {note} {This code has been checked against the closed-form
  analytical expressions that can be obtained for the case of Landau
  excitations.}\BibitemShut {Stop}%
\bibitem [{\citenamefont {Wigner}(1932)}]{Wigner-1932-1}%
  \BibitemOpen
  \bibfield  {author} {\bibinfo {author} {\bibfnamefont {E.}~\bibnamefont
  {Wigner}},\ }\bibfield  {title} {\bibinfo {title} {On the quantum correction
  for thermodynamic equilibrium},\ }\href
  {https://doi.org/10.1103/PhysRev.40.749} {\bibfield  {journal} {\bibinfo
  {journal} {Phys. Rev.}\ }\textbf {\bibinfo {volume} {40}},\ \bibinfo {pages}
  {749} (\bibinfo {year} {1932})}\BibitemShut {NoStop}%
\bibitem [{\citenamefont {Bell}(1987)}]{Bell-EPR-Wigner}%
  \BibitemOpen
  \bibfield  {author} {\bibinfo {author} {\bibfnamefont {J.~S.}\ \bibnamefont
  {Bell}},\ }\bibinfo {title} {Speakable and unspeakable in quantum mechanics}\
  (\bibinfo  {publisher} {Cambridge University Press},\ \bibinfo {year}
  {1987})\ Chap.\ \bibinfo {chapter} {Chapter 21: {EPR} correlations and {EPW}
  distributions}, pp.\ \bibinfo {pages} {196--200}\BibitemShut {NoStop}%
\bibitem [{\citenamefont {Moskalets}\ \emph {et~al.}(2020)\citenamefont
  {Moskalets}, \citenamefont {Kotilahti}, \citenamefont {Burset},\ and\
  \citenamefont {Flindt}}]{Moskalets-2020-1}%
  \BibitemOpen
  \bibfield  {author} {\bibinfo {author} {\bibfnamefont {M.}~\bibnamefont
  {Moskalets}}, \bibinfo {author} {\bibfnamefont {J.}~\bibnamefont
  {Kotilahti}}, \bibinfo {author} {\bibfnamefont {P.}~\bibnamefont {Burset}},\
  and\ \bibinfo {author} {\bibfnamefont {C.}~\bibnamefont {Flindt}},\
  }\bibfield  {title} {\bibinfo {title} {Composite two-particle sources},\
  }\href {https://doi.org/10.1140/epjst/e2019-900121-x} {\bibfield  {journal}
  {\bibinfo  {journal} {Eur. Phys. J. Special Topics}\ }\textbf {\bibinfo
  {volume} {229}},\ \bibinfo {pages} {647} (\bibinfo {year}
  {2020})}\BibitemShut {NoStop}%
\bibitem [{\citenamefont {Shapourian}\ \emph {et~al.}(2017)\citenamefont
  {Shapourian}, \citenamefont {Shiozaki},\ and\ \citenamefont
  {Ryu}}]{Shapourian-2017-1}%
  \BibitemOpen
  \bibfield  {author} {\bibinfo {author} {\bibfnamefont {H.}~\bibnamefont
  {Shapourian}}, \bibinfo {author} {\bibfnamefont {K.}~\bibnamefont
  {Shiozaki}},\ and\ \bibinfo {author} {\bibfnamefont {S.}~\bibnamefont
  {Ryu}},\ }\bibfield  {title} {\bibinfo {title} {Partial time-reversal
  transformation and entanglement negativity in fermionic systems},\ }\href
  {https://doi.org/10.1103/PhysRevB.95.165101} {\bibfield  {journal} {\bibinfo
  {journal} {Phys. Rev. B}\ }\textbf {\bibinfo {volume} {95}},\ \bibinfo
  {pages} {165101} (\bibinfo {year} {2017})}\BibitemShut {NoStop}%
\bibitem [{\citenamefont {Shapourian}\ and\ \citenamefont
  {Ryu}(2019)}]{Shapourian-2019-1}%
  \BibitemOpen
  \bibfield  {author} {\bibinfo {author} {\bibfnamefont {H.}~\bibnamefont
  {Shapourian}}\ and\ \bibinfo {author} {\bibfnamefont {S.}~\bibnamefont
  {Ryu}},\ }\bibfield  {title} {\bibinfo {title} {Entanglement negativity of
  fermions: Monotonicity, separability criterion, and classification of
  few-mode states},\ }\href {https://doi.org/10.1103/PhysRevA.99.022310}
  {\bibfield  {journal} {\bibinfo  {journal} {Phys. Rev. A}\ }\textbf {\bibinfo
  {volume} {99}},\ \bibinfo {pages} {022310} (\bibinfo {year}
  {2019})}\BibitemShut {NoStop}%
\bibitem [{\citenamefont {Schulenborg}\ \emph {et~al.}(2024)\citenamefont
  {Schulenborg}, \citenamefont {Fletcher}, \citenamefont {Kataoka},\ and\
  \citenamefont {Splettstoesser}}]{Schulenborg-2024-1}%
  \BibitemOpen
  \bibfield  {author} {\bibinfo {author} {\bibfnamefont {J.}~\bibnamefont
  {Schulenborg}}, \bibinfo {author} {\bibfnamefont {J.~D.}\ \bibnamefont
  {Fletcher}}, \bibinfo {author} {\bibfnamefont {M.}~\bibnamefont {Kataoka}},\
  and\ \bibinfo {author} {\bibfnamefont {J.}~\bibnamefont {Splettstoesser}},\
  }\bibfield  {title} {\bibinfo {title} {Spectroscopy of hot electron pair
  emission from a driven quantum dot},\ }\href
  {https://doi.org/10.1103/PhysRevB.109.115433} {\bibfield  {journal} {\bibinfo
   {journal} {Phys. Rev. B}\ }\textbf {\bibinfo {volume} {109}},\ \bibinfo
  {pages} {115433} (\bibinfo {year} {2024})}\BibitemShut {NoStop}%
\bibitem [{\citenamefont {Locane}\ \emph {et~al.}(2019)\citenamefont {Locane},
  \citenamefont {Brouwer},\ and\ \citenamefont {Kashcheyevs}}]{Locane-2019-1}%
  \BibitemOpen
  \bibfield  {author} {\bibinfo {author} {\bibfnamefont {E.}~\bibnamefont
  {Locane}}, \bibinfo {author} {\bibfnamefont {P.~W.}\ \bibnamefont
  {Brouwer}},\ and\ \bibinfo {author} {\bibfnamefont {V.}~\bibnamefont
  {Kashcheyevs}},\ }\bibfield  {title} {\bibinfo {title} {Time-energy filtering
  of single electrons in ballistic waveguides},\ }\href
  {https://doi.org/10.1088/1367-2630/ab3fbb} {\bibfield  {journal} {\bibinfo
  {journal} {New Journal of Physics}\ }\textbf {\bibinfo {volume} {21}},\
  \bibinfo {pages} {093042} (\bibinfo {year} {2019})}\BibitemShut {NoStop}%
\bibitem [{\citenamefont {Freise}\ \emph {et~al.}(2020)\citenamefont {Freise},
  \citenamefont {Gerster}, \citenamefont {Reifert}, \citenamefont {Weimann},
  \citenamefont {Pierz}, \citenamefont {Hohls},\ and\ \citenamefont
  {Ubbelohde}}]{Freise-2020-1}%
  \BibitemOpen
  \bibfield  {author} {\bibinfo {author} {\bibfnamefont {L.}~\bibnamefont
  {Freise}}, \bibinfo {author} {\bibfnamefont {T.}~\bibnamefont {Gerster}},
  \bibinfo {author} {\bibfnamefont {D.}~\bibnamefont {Reifert}}, \bibinfo
  {author} {\bibfnamefont {T.}~\bibnamefont {Weimann}}, \bibinfo {author}
  {\bibfnamefont {K.}~\bibnamefont {Pierz}}, \bibinfo {author} {\bibfnamefont
  {F.}~\bibnamefont {Hohls}},\ and\ \bibinfo {author} {\bibfnamefont
  {N.}~\bibnamefont {Ubbelohde}},\ }\bibfield  {title} {\bibinfo {title}
  {Trapping and counting ballistic nonequilibrium electrons},\ }\href
  {https://doi.org/10.1103/PhysRevLett.124.127701} {\bibfield  {journal}
  {\bibinfo  {journal} {Phys. Rev. Lett.}\ }\textbf {\bibinfo {volume} {124}},\
  \bibinfo {pages} {127701} (\bibinfo {year} {2020})}\BibitemShut {NoStop}%
\bibitem [{\citenamefont {Thalineau}\ \emph {et~al.}()\citenamefont
  {Thalineau}, \citenamefont {Weick}, \citenamefont {Bauerle},\ and\
  \citenamefont {Meunier}}]{Thalineau-2014-1}%
  \BibitemOpen
  \bibfield  {author} {\bibinfo {author} {\bibfnamefont {R.}~\bibnamefont
  {Thalineau}}, \bibinfo {author} {\bibfnamefont {A.}~\bibnamefont {Weick}},
  \bibinfo {author} {\bibfnamefont {C.}~\bibnamefont {Bauerle}},\ and\ \bibinfo
  {author} {\bibfnamefont {T.}~\bibnamefont {Meunier}},\ }\bibfield  {title}
  {\bibinfo {title} {Using a two electron spin qubit to detect electrons flying
  above the {Fermi} sea},\ }\bibinfo {note} {arxiv:1403.7770}\BibitemShut
  {NoStop}%
\bibitem [{\citenamefont {Glattli}\ \emph {et~al.}(2020)\citenamefont
  {Glattli}, \citenamefont {Nath}, \citenamefont {Taktak}, \citenamefont
  {Roulleau}, \citenamefont {Bauerle},\ and\ \citenamefont
  {Waintal}}]{Glattli-2020-1}%
  \BibitemOpen
  \bibfield  {author} {\bibinfo {author} {\bibfnamefont {D.}~\bibnamefont
  {Glattli}}, \bibinfo {author} {\bibfnamefont {J.}~\bibnamefont {Nath}},
  \bibinfo {author} {\bibfnamefont {I.}~\bibnamefont {Taktak}}, \bibinfo
  {author} {\bibfnamefont {P.}~\bibnamefont {Roulleau}}, \bibinfo {author}
  {\bibfnamefont {C.}~\bibnamefont {Bauerle}},\ and\ \bibinfo {author}
  {\bibfnamefont {X.}~\bibnamefont {Waintal}},\ }\bibfield  {title} {\bibinfo
  {title} {Design of a single-shot electron detector with sub-electron
  sensitivity for electron flying qubit operation}} (\bibinfo {year} {2020}),\
  \bibinfo {note} {arXiv:2002.03947}\BibitemShut {NoStop}%
\bibitem [{\citenamefont {Shaju}\ \emph {et~al.}(2025)\citenamefont {Shaju},
  \citenamefont {Pavlovska}, \citenamefont {Suba}, \citenamefont {Wang},
  \citenamefont {Ouacel}, \citenamefont {Vasselon}, \citenamefont {Aluffi},
  \citenamefont {Mazzella}, \citenamefont {Geffroy}, \citenamefont {Ludwig},
  \citenamefont {Wieck}, \citenamefont {Urdampiletta}, \citenamefont
  {B{\"a}uerle}, \citenamefont {Kashcheyevs},\ and\ \citenamefont
  {Sellier}}]{Shaju-2025-1}%
  \BibitemOpen
  \bibfield  {author} {\bibinfo {author} {\bibfnamefont {J.}~\bibnamefont
  {Shaju}}, \bibinfo {author} {\bibfnamefont {E.}~\bibnamefont {Pavlovska}},
  \bibinfo {author} {\bibfnamefont {R.}~\bibnamefont {Suba}}, \bibinfo {author}
  {\bibfnamefont {J.}~\bibnamefont {Wang}}, \bibinfo {author} {\bibfnamefont
  {S.}~\bibnamefont {Ouacel}}, \bibinfo {author} {\bibfnamefont
  {T.}~\bibnamefont {Vasselon}}, \bibinfo {author} {\bibfnamefont
  {M.}~\bibnamefont {Aluffi}}, \bibinfo {author} {\bibfnamefont
  {L.}~\bibnamefont {Mazzella}}, \bibinfo {author} {\bibfnamefont
  {C.}~\bibnamefont {Geffroy}}, \bibinfo {author} {\bibfnamefont
  {A.}~\bibnamefont {Ludwig}}, \bibinfo {author} {\bibfnamefont {A.~D.}\
  \bibnamefont {Wieck}}, \bibinfo {author} {\bibfnamefont {M.}~\bibnamefont
  {Urdampiletta}}, \bibinfo {author} {\bibfnamefont {C.}~\bibnamefont
  {B{\"a}uerle}}, \bibinfo {author} {\bibfnamefont {V.}~\bibnamefont
  {Kashcheyevs}},\ and\ \bibinfo {author} {\bibfnamefont {H.}~\bibnamefont
  {Sellier}},\ }\bibfield  {title} {\bibinfo {title} {Evidence of {Coulomb}
  liquid phase in few-electron droplets},\ }\href
  {https://doi.org/10.1038/s41586-025-09139-z} {\bibfield  {journal} {\bibinfo
  {journal} {Nature}\ }\textbf {\bibinfo {volume} {642}},\ \bibinfo {pages}
  {928} (\bibinfo {year} {2025})}\BibitemShut {NoStop}%
\bibitem [{\citenamefont {Chtchelkatchev}\ \emph {et~al.}(2002)\citenamefont
  {Chtchelkatchev}, \citenamefont {Blatter}, \citenamefont {Lesovik},\ and\
  \citenamefont {Martin}}]{Chtchelkatchev-2002-1}%
  \BibitemOpen
  \bibfield  {author} {\bibinfo {author} {\bibfnamefont {N.~M.}\ \bibnamefont
  {Chtchelkatchev}}, \bibinfo {author} {\bibfnamefont {G.}~\bibnamefont
  {Blatter}}, \bibinfo {author} {\bibfnamefont {G.~B.}\ \bibnamefont
  {Lesovik}},\ and\ \bibinfo {author} {\bibfnamefont {T.}~\bibnamefont
  {Martin}},\ }\bibfield  {title} {\bibinfo {title} {Bell inequalities and
  entanglement in solid-state devices},\ }\href
  {https://doi.org/10.1103/PhysRevB.66.161320} {\bibfield  {journal} {\bibinfo
  {journal} {Phys. Rev. B}\ }\textbf {\bibinfo {volume} {66}},\ \bibinfo
  {pages} {161320} (\bibinfo {year} {2002})}\BibitemShut {NoStop}%
\bibitem [{\citenamefont {Lesovik}\ \emph {et~al.}(2001)\citenamefont
  {Lesovik}, \citenamefont {Martin},\ and\ \citenamefont
  {Blatter}}]{Lesovik-2001-1}%
  \BibitemOpen
  \bibfield  {author} {\bibinfo {author} {\bibfnamefont {G.}~\bibnamefont
  {Lesovik}}, \bibinfo {author} {\bibfnamefont {T.}~\bibnamefont {Martin}},\
  and\ \bibinfo {author} {\bibfnamefont {G.}~\bibnamefont {Blatter}},\
  }\bibfield  {title} {\bibinfo {title} {Electronic entanglement in the
  vicinity of a superconductor},\ }\href@noop {} {\bibfield  {journal}
  {\bibinfo  {journal} {Eur. Phys. J. {\bf B}}\ }\textbf {\bibinfo {volume}
  {24}},\ \bibinfo {pages} {287} (\bibinfo {year} {2001})}\BibitemShut
  {NoStop}%
\bibitem [{\citenamefont {Recher}\ \emph {et~al.}(2001)\citenamefont {Recher},
  \citenamefont {Sukhorukov},\ and\ \citenamefont {Loss}}]{Recher-2001-1}%
  \BibitemOpen
  \bibfield  {author} {\bibinfo {author} {\bibfnamefont {P.}~\bibnamefont
  {Recher}}, \bibinfo {author} {\bibfnamefont {E.~V.}\ \bibnamefont
  {Sukhorukov}},\ and\ \bibinfo {author} {\bibfnamefont {D.}~\bibnamefont
  {Loss}},\ }\bibfield  {title} {\bibinfo {title} {Andreev tunneling, coulomb
  blockade, and resonant transport of nonlocal spin-entangled electrons},\
  }\href {https://doi.org/10.1103/PhysRevB.63.165314} {\bibfield  {journal}
  {\bibinfo  {journal} {Phys. Rev. B}\ }\textbf {\bibinfo {volume} {63}},\
  \bibinfo {pages} {165314} (\bibinfo {year} {2001})}\BibitemShut {NoStop}%
\bibitem [{\citenamefont {Samuelsson}\ \emph {et~al.}(2004)\citenamefont
  {Samuelsson}, \citenamefont {Sukhorukov},\ and\ \citenamefont
  {B\"uttiker}}]{Samuelsson-2004-2}%
  \BibitemOpen
  \bibfield  {author} {\bibinfo {author} {\bibfnamefont {P.}~\bibnamefont
  {Samuelsson}}, \bibinfo {author} {\bibfnamefont {E.~V.}\ \bibnamefont
  {Sukhorukov}},\ and\ \bibinfo {author} {\bibfnamefont {M.}~\bibnamefont
  {B\"uttiker}},\ }\bibfield  {title} {\bibinfo {title} {Electrical current
  noise of a beamsplitter as a test of spin entanglement},\ }\href
  {https://doi.org/10.1103/PhysRevB.70.115330} {\bibfield  {journal} {\bibinfo
  {journal} {Phys. Rev. B}\ }\textbf {\bibinfo {volume} {70}},\ \bibinfo
  {pages} {115330} (\bibinfo {year} {2004})}\BibitemShut {NoStop}%
\bibitem [{\citenamefont {Samuelsson}\ \emph {et~al.}(2003)\citenamefont
  {Samuelsson}, \citenamefont {Sukhorukov},\ and\ \citenamefont
  {B\"uttiker}}]{Samuelsson-2003-1}%
  \BibitemOpen
  \bibfield  {author} {\bibinfo {author} {\bibfnamefont {P.}~\bibnamefont
  {Samuelsson}}, \bibinfo {author} {\bibfnamefont {E.~V.}\ \bibnamefont
  {Sukhorukov}},\ and\ \bibinfo {author} {\bibfnamefont {M.}~\bibnamefont
  {B\"uttiker}},\ }\bibfield  {title} {\bibinfo {title} {Orbital entanglement
  and violation of bell inequalities in mesoscopic conductors},\ }\href
  {https://doi.org/10.1103/PhysRevLett.91.157002} {\bibfield  {journal}
  {\bibinfo  {journal} {Phys. Rev. Lett.}\ }\textbf {\bibinfo {volume} {91}},\
  \bibinfo {pages} {157002} (\bibinfo {year} {2003})}\BibitemShut {NoStop}%
\bibitem [{\citenamefont {Beenakker}(2006)}]{Beenakker-2006-1}%
  \BibitemOpen
  \bibfield  {author} {\bibinfo {author} {\bibfnamefont {C.}~\bibnamefont
  {Beenakker}},\ }\bibfield  {title} {\bibinfo {title} {Electron-hole
  entanglement in the {Fermi} sea},\ }in\ \href@noop {} {\emph {\bibinfo
  {booktitle} {Quantum computers, Algorithms and Chaos}}},\ \bibinfo {series}
  {International School of Physics Enrico Fermi}, Vol.\ \bibinfo {volume}
  {162},\ \bibinfo {editor} {edited by\ \bibinfo {editor} {\bibfnamefont
  {G.}~\bibnamefont {Casati}}, \bibinfo {editor} {\bibfnamefont
  {D.}~\bibnamefont {Shepelyansky}}, \bibinfo {editor} {\bibfnamefont
  {P.}~\bibnamefont {Zoller}},\ and\ \bibinfo {editor} {\bibfnamefont
  {G.}~\bibnamefont {Benenti}}}\ (\bibinfo  {publisher} {IOS Press},\ \bibinfo
  {address} {Amsterdam},\ \bibinfo {year} {2006})\ pp.\ \bibinfo {pages}
  {307--347}\BibitemShut {NoStop}%
\bibitem [{\citenamefont {Hofer}\ \emph {et~al.}(2017)\citenamefont {Hofer},
  \citenamefont {Dasenbrook},\ and\ \citenamefont {Flindt}}]{Hofer-2016-2}%
  \BibitemOpen
  \bibfield  {author} {\bibinfo {author} {\bibfnamefont {P.}~\bibnamefont
  {Hofer}}, \bibinfo {author} {\bibfnamefont {D.}~\bibnamefont {Dasenbrook}},\
  and\ \bibinfo {author} {\bibfnamefont {C.}~\bibnamefont {Flindt}},\
  }\bibfield  {title} {\bibinfo {title} {On-demand entanglement generation
  using dynamic single electron sources},\ }\href
  {https://doi.org/10.1002/pssb.201600582} {\bibfield  {journal} {\bibinfo
  {journal} {Physica Status Solidi (b)}\ }\textbf {\bibinfo {volume} {254}},\
  \bibinfo {pages} {1600582} (\bibinfo {year} {2017})}\BibitemShut {NoStop}%
\bibitem [{\citenamefont {Dasenbrook}\ \emph {et~al.}(2016)\citenamefont
  {Dasenbrook}, \citenamefont {Bowles}, \citenamefont {Brask}, \citenamefont
  {Hofer}, \citenamefont {Flindt},\ and\ \citenamefont
  {Brunner}}]{Dasenbrook-2016-1}%
  \BibitemOpen
  \bibfield  {author} {\bibinfo {author} {\bibfnamefont {D.}~\bibnamefont
  {Dasenbrook}}, \bibinfo {author} {\bibfnamefont {J.}~\bibnamefont {Bowles}},
  \bibinfo {author} {\bibfnamefont {J.~B.}\ \bibnamefont {Brask}}, \bibinfo
  {author} {\bibfnamefont {P.~P.}\ \bibnamefont {Hofer}}, \bibinfo {author}
  {\bibfnamefont {C.}~\bibnamefont {Flindt}},\ and\ \bibinfo {author}
  {\bibfnamefont {N.}~\bibnamefont {Brunner}},\ }\bibfield  {title} {\bibinfo
  {title} {Single-electron entanglement and nonlocality},\ }\href
  {http://stacks.iop.org/1367-2630/18/i=4/a=043036} {\bibfield  {journal}
  {\bibinfo  {journal} {New Journal of Physics}\ }\textbf {\bibinfo {volume}
  {18}},\ \bibinfo {pages} {043036} (\bibinfo {year} {2016})}\BibitemShut
  {NoStop}%
\bibitem [{\citenamefont {Dasenbrook}\ and\ \citenamefont
  {Flindt}(2015)}]{Dasenbrook-2015-1}%
  \BibitemOpen
  \bibfield  {author} {\bibinfo {author} {\bibfnamefont {D.}~\bibnamefont
  {Dasenbrook}}\ and\ \bibinfo {author} {\bibfnamefont {C.}~\bibnamefont
  {Flindt}},\ }\bibfield  {title} {\bibinfo {title} {Dynamical generation and
  detection of entanglement in neutral {L}eviton pairs},\ }\href
  {https://doi.org/10.1103/PhysRevB.92.161412} {\bibfield  {journal} {\bibinfo
  {journal} {Phys. Rev. B}\ }\textbf {\bibinfo {volume} {92}},\ \bibinfo
  {pages} {161412} (\bibinfo {year} {2015})}\BibitemShut {NoStop}%
\bibitem [{\citenamefont {Bilitewski}\ and\ \citenamefont
  {Cooper}(2015)}]{Bilitewski-2015-1}%
  \BibitemOpen
  \bibfield  {author} {\bibinfo {author} {\bibfnamefont {T.}~\bibnamefont
  {Bilitewski}}\ and\ \bibinfo {author} {\bibfnamefont {N.~R.}\ \bibnamefont
  {Cooper}},\ }\bibfield  {title} {\bibinfo {title} {Scattering theory for
  {F}loquet-{B}loch states},\ }\href
  {https://doi.org/10.1103/PhysRevA.91.033601} {\bibfield  {journal} {\bibinfo
  {journal} {Phys. Rev. A}\ }\textbf {\bibinfo {volume} {91}},\ \bibinfo
  {pages} {033601} (\bibinfo {year} {2015})}\BibitemShut {NoStop}%
\bibitem [{\citenamefont {Giovannetti}\ \emph {et~al.}(2007)\citenamefont
  {Giovannetti}, \citenamefont {Frustaglia}, \citenamefont {Taddei},\ and\
  \citenamefont {Fazio}}]{Giovannetti-2007-1}%
  \BibitemOpen
  \bibfield  {author} {\bibinfo {author} {\bibfnamefont {V.}~\bibnamefont
  {Giovannetti}}, \bibinfo {author} {\bibfnamefont {D.}~\bibnamefont
  {Frustaglia}}, \bibinfo {author} {\bibfnamefont {F.}~\bibnamefont {Taddei}},\
  and\ \bibinfo {author} {\bibfnamefont {R.}~\bibnamefont {Fazio}},\ }\bibfield
   {title} {\bibinfo {title} {Characterizing electron entanglement in
  multiterminal mesoscopic conductors},\ }\href
  {https://doi.org/10.1103/PhysRevB.75.241305} {\bibfield  {journal} {\bibinfo
  {journal} {Phys. Rev. B}\ }\textbf {\bibinfo {volume} {75}},\ \bibinfo
  {pages} {241305} (\bibinfo {year} {2007})}\BibitemShut {NoStop}%
\bibitem [{\citenamefont {Coleman}(1963)}]{Coleman-1963-1}%
  \BibitemOpen
  \bibfield  {author} {\bibinfo {author} {\bibfnamefont {A.~J.}\ \bibnamefont
  {Coleman}},\ }\bibfield  {title} {\bibinfo {title} {Structure of fermion
  density matrices},\ }\href {https://doi.org/10.1103/RevModPhys.35.668}
  {\bibfield  {journal} {\bibinfo  {journal} {Rev. Mod. Phys.}\ }\textbf
  {\bibinfo {volume} {35}},\ \bibinfo {pages} {668} (\bibinfo {year}
  {1963})}\BibitemShut {NoStop}%
\bibitem [{\citenamefont {Liu}\ \emph {et~al.}(2007)\citenamefont {Liu},
  \citenamefont {Christandl},\ and\ \citenamefont {Verstraete}}]{Liu-2007-1}%
  \BibitemOpen
  \bibfield  {author} {\bibinfo {author} {\bibfnamefont {Y.-K.}\ \bibnamefont
  {Liu}}, \bibinfo {author} {\bibfnamefont {M.}~\bibnamefont {Christandl}},\
  and\ \bibinfo {author} {\bibfnamefont {F.}~\bibnamefont {Verstraete}},\
  }\bibfield  {title} {\bibinfo {title} {Quantum computational complexity of
  the $n$-representability problem: Qma complete},\ }\href
  {https://doi.org/10.1103/PhysRevLett.98.110503} {\bibfield  {journal}
  {\bibinfo  {journal} {Phys. Rev. Lett.}\ }\textbf {\bibinfo {volume} {98}},\
  \bibinfo {pages} {110503} (\bibinfo {year} {2007})}\BibitemShut {NoStop}%
\bibitem [{\citenamefont {Haack}\ \emph {et~al.}(2013)\citenamefont {Haack},
  \citenamefont {Moskalets},\ and\ \citenamefont
  {B{\"u}ttiker}}]{Haack-2012-2}%
  \BibitemOpen
  \bibfield  {author} {\bibinfo {author} {\bibfnamefont {G.}~\bibnamefont
  {Haack}}, \bibinfo {author} {\bibfnamefont {M.}~\bibnamefont {Moskalets}},\
  and\ \bibinfo {author} {\bibfnamefont {M.}~\bibnamefont {B{\"u}ttiker}},\
  }\bibfield  {title} {\bibinfo {title} {Glauber coherence of single electron
  sources},\ }\href {https://doi.org/10.1103/PhysRevB.87.201302} {\bibfield
  {journal} {\bibinfo  {journal} {Phys. Rev. B}\ }\textbf {\bibinfo {volume}
  {87}},\ \bibinfo {pages} {201302(R)} (\bibinfo {year} {2013})}\BibitemShut
  {NoStop}%
\bibitem [{\citenamefont {Souquet-Basi{\`e}ge}\ \emph
  {et~al.}(2026)\citenamefont {Souquet-Basi{\`e}ge}, \citenamefont {Roussel},
  \citenamefont {Rebora}, \citenamefont {M{\'e}nard}, \citenamefont {Safi},
  \citenamefont {F{\`e}ve},\ and\ \citenamefont
  {Degiovanni}}]{Souquet-Basiege-2026-1}%
  \BibitemOpen
  \bibfield  {author} {\bibinfo {author} {\bibfnamefont {H.}~\bibnamefont
  {Souquet-Basi{\`e}ge}}, \bibinfo {author} {\bibfnamefont {B.}~\bibnamefont
  {Roussel}}, \bibinfo {author} {\bibfnamefont {G.}~\bibnamefont {Rebora}},
  \bibinfo {author} {\bibfnamefont {G.}~\bibnamefont {M{\'e}nard}}, \bibinfo
  {author} {\bibfnamefont {I.}~\bibnamefont {Safi}}, \bibinfo {author}
  {\bibfnamefont {G.}~\bibnamefont {F{\`e}ve}},\ and\ \bibinfo {author}
  {\bibfnamefont {P.}~\bibnamefont {Degiovanni}},\ }\bibfield  {title}
  {\bibinfo {title} {The electronic ambiguity function in single electron
  interferometry}} (\bibinfo {year} {2026}),\ \bibinfo {note} {en
  preparation}\BibitemShut {NoStop}%
\bibitem [{\citenamefont {Hoeffding}(1940)}]{Hoeffding-1940-1}%
  \BibitemOpen
  \bibfield  {author} {\bibinfo {author} {\bibfnamefont {W.}~\bibnamefont
  {Hoeffding}},\ }\bibfield  {title} {\bibinfo {title} {Masstabinvariante
  {K}orrelationstheorie},\ }\href@noop {} {\bibfield  {journal} {\bibinfo
  {journal} {Schr. Math. Inst. Univ. Berlin}\ }\textbf {\bibinfo {volume}
  {5}},\ \bibinfo {pages} {181} (\bibinfo {year} {1940})}\BibitemShut {NoStop}%
\bibitem [{\citenamefont {Fr{\'e}chet}(1951)}]{Frechet-1951-1}%
  \BibitemOpen
  \bibfield  {author} {\bibinfo {author} {\bibfnamefont {M.}~\bibnamefont
  {Fr{\'e}chet}},\ }\bibfield  {title} {\bibinfo {title} {Sur les tableaux de
  correlation dont les marges sont donn{{\'e}}es},\ }\href@noop {} {\bibfield
  {journal} {\bibinfo  {journal} {Ann. Univ. Lyon. (Sect. A)}\ }\textbf
  {\bibinfo {volume} {14}},\ \bibinfo {pages} {53} (\bibinfo {year}
  {1951})}\BibitemShut {NoStop}%
\bibitem [{\citenamefont {Martin}\ and\ \citenamefont
  {Landauer}(1992)}]{Martin-1992-1}%
  \BibitemOpen
  \bibfield  {author} {\bibinfo {author} {\bibfnamefont {T.}~\bibnamefont
  {Martin}}\ and\ \bibinfo {author} {\bibfnamefont {R.}~\bibnamefont
  {Landauer}},\ }\bibfield  {title} {\bibinfo {title} {Wave packet approach to
  noise in multichannel mesoscopic systems},\ }\href@noop {} {\bibfield
  {journal} {\bibinfo  {journal} {Phys. Rev. B}\ }\textbf {\bibinfo {volume}
  {45}},\ \bibinfo {pages} {1742 } (\bibinfo {year} {1992})}\BibitemShut
  {NoStop}%
\bibitem [{\citenamefont {Grenier}\ \emph
  {et~al.}(2011{\natexlab{b}})\citenamefont {Grenier}, \citenamefont
  {Herv{\'e}}, \citenamefont {Bocquillon}, \citenamefont {Parmentier},
  \citenamefont {Pla{\c c}ais}, \citenamefont {Berroir}, \citenamefont
  {F{\`e}ve},\ and\ \citenamefont {Degiovanni}}]{Degio-2010-4}%
  \BibitemOpen
  \bibfield  {author} {\bibinfo {author} {\bibfnamefont {C.}~\bibnamefont
  {Grenier}}, \bibinfo {author} {\bibfnamefont {R.}~\bibnamefont {Herv{\'e}}},
  \bibinfo {author} {\bibfnamefont {E.}~\bibnamefont {Bocquillon}}, \bibinfo
  {author} {\bibfnamefont {F.}~\bibnamefont {Parmentier}}, \bibinfo {author}
  {\bibfnamefont {B.}~\bibnamefont {Pla{\c c}ais}}, \bibinfo {author}
  {\bibfnamefont {J.-M.}\ \bibnamefont {Berroir}}, \bibinfo {author}
  {\bibfnamefont {G.}~\bibnamefont {F{\`e}ve}},\ and\ \bibinfo {author}
  {\bibfnamefont {P.}~\bibnamefont {Degiovanni}},\ }\bibfield  {title}
  {\bibinfo {title} {Single electron quantum tomography in quantum {H}all edge
  channels},\ }\href {https://doi.org/10.1088/1367-2630/13/9/093007} {\bibfield
   {journal} {\bibinfo  {journal} {New Journal of Physics}\ }\textbf {\bibinfo
  {volume} {13}},\ \bibinfo {pages} {093007} (\bibinfo {year}
  {2011}{\natexlab{b}})}\BibitemShut {NoStop}%
\bibitem [{\citenamefont {Silvestrov}\ \emph {et~al.}(2025)\citenamefont
  {Silvestrov}, \citenamefont {Kashcheyevs},\ and\ \citenamefont
  {Recher}}]{Silvestrov-2025-1}%
  \BibitemOpen
  \bibfield  {author} {\bibinfo {author} {\bibfnamefont {P.~G.}\ \bibnamefont
  {Silvestrov}}, \bibinfo {author} {\bibfnamefont {V.}~\bibnamefont
  {Kashcheyevs}},\ and\ \bibinfo {author} {\bibfnamefont {P.}~\bibnamefont
  {Recher}},\ }\href {https://arxiv.org/abs/2503.20325} {\bibinfo {title}
  {Theory of two-electrons optics experiments with smooth potentials: Flying
  electron molecules}} (\bibinfo {year} {2025}),\ \Eprint
  {https://arxiv.org/abs/2503.20325} {arXiv:2503.20325 [cond-mat.mes-hall]}
  \BibitemShut {NoStop}%
\end{thebibliography}
\end{document}